\documentclass[3p]{elsarticle}

\usepackage{lineno}
\modulolinenumbers[5]

\usepackage{color}
\usepackage[table]{xcolor}
\usepackage{caption}
\usepackage{subcaption}
\usepackage{lscape}
\usepackage{amsmath, amsfonts, amssymb}
\usepackage{bbold}
\usepackage{amsthm}
\usepackage{longtable}
\usepackage{siunitx}
\usepackage{pifont} 
\usepackage{marvosym}
\usepackage{array}
\usepackage{multirow}
\usepackage{arydshln}
\usepackage{makecell}
\usepackage{textcomp}
\usepackage{subcaption}
\usepackage{url}
\usepackage[colorlinks=true]{hyperref}

\journal{Physics Reports}

\definecolor{orange}{rgb}{1,0.55,0.0}
\definecolor{violet}{rgb}{0.5,0.0,1.0}
\definecolor{darkgreen}{rgb}{0.1,0.4,0.1}

\newcommand{\lra}[1]{\langle #1 \rangle }
\newcommand{\mb}[1]{\mathbf{#1}}

\newcommand{\mc}[1]{\mathcal{#1}}
\newcommand{\dd}{\text{d}}
\newcommand{\Rep}{\mathcal{R}e_{\rm p}}

\newcommand{\yes}{\textcolor{darkgreen}{\ding{52}}}
\newcommand{\no}{\textcolor{red}{\ding{56}}}
\newcommand{\maybe}{\textcolor{orange}{\Flatsteel}}
\newcommand{\near}{\textcolor{violet}{\large{$\mathbf{\approx}$}}}

\begin{document}

\begin{frontmatter}

\title{Particle resuspension: challenges and perspectives for future models}

\author[inria-sop]{Christophe Henry \corref{corres_author}}
\ead{christophe.henry@inria.fr} 
\author[edf-mfee]{Jean-Pierre Minier}
\author[lanl]{Sara Brambilla}

\address[inria-sop]{Universit\'e C\^ote d'Azur, INRIA, CaliSto laboratory, Sophia-Antipolis, France}

\address[edf-mfee]{EDF R$\&$D, M\'{e}canique des Fluides, Energie et Environnement, 6 quai Watier, 
78400 Chatou, France}

\address[lanl]{Los Alamos National Laboratory, Bikini Atoll Rd, Los Alamos NM, 87545, United States}

\cortext[corres_author]{Corresponding author}

\vspace*{2em}

\begin{abstract}

 Using what has become a celebrated catchphrase, Philip W. Anderson once wrote that ``more is different'' (Science, Vol. 177, Issue 4047, pp. 393-396, 1972). First formulated in the context of condensed matter, this statement carries far beyond the sole limits of solid-state physics. It emphasizes that collective behavior can be more than the mere sum of what happens for elementary constituents or the mere collation of the evolution of each degree of freedom. Said otherwise, complex phenomena can arise out of the interplay between multiple sub-phenomena each of which can be relatively simple. The process of particle resuspension, in which discrete particles adhering on a surface are pulled off and carried away by a fluid flow, is another example involving a web of phenomena pertaining to fluid mechanics, particle dynamics and interface chemistry whose cross-effects create an intricate topic.

The purpose of this review is to analyze the physics at play in particle resuspension in order to bring insights into the rich complexity of this common but challenging concern. Following the more-is-different vision, this is performed by starting from a range of practical observations and experimental data. We then work our way through the investigation of the key mechanisms which play a role in the overall process. In turn, these mechanisms reveal an array of fundamental interactions, such as particle-fluid, particle-particle and particle-surface, whose combined effects create the tapestry of current applications. At the core of this analysis are descriptions of these physical phenomena and the different ways through which they are intertwined to build up various models used to provide quantitative assessment of particle resuspension. The physics of particle resuspension implies to hold together processes occurring at extremely different space and time scales and models are key in providing a single vehicle to lead us through such multiscale journeys. This raises questions on what makes up a model and one objective of the present work is to clarify the essence of a modeling approach. In spite of its ubiquitous nature, particle resuspension is still at the early stages of developments. Many extensions need to be worked out and revisiting the art of modeling is not a moot point. The need to consider more complex objects than small and spherical particles and, moreover, to come up with unified descriptions of mono- and multilayer resuspension put the emphasis on solid model foundations if we are to go beyond current limits. This is very much modeling in the making and new ideas are proposed to stimulate interest into this everyday but challenging issue in physics.
\end{abstract}

\begin{keyword}
Resuspension \sep Particle \sep Surfaces \sep Roughness
\end{keyword}

\end{frontmatter}


\clearpage
\tableofcontents

\section{Introduction}
 \label{sec:intro}

Let us imagine the following situation: a set of particles are lying on a surface. Then, a fluid flow blows them off the surface. In a nutshell, this simple image captures the notion of particle resuspension.  

In a more technical formulation, resuspension refers to the physical process by which discrete elements resting on a surface are, first, detached and, then, entrained away through the action of a fluid flow. It follows that this process occurs in wall-bounded two-phase flows, i.e., flows composed of a fluid phase and a discrete phase both contained in a domain bounded by at least one wall. The fluid phase involved in resuspension can be either a liquid (e.g., water, oil, milk) or a gas (like air or any other gas mixture). The discrete elements can be air bubbles, water droplets, organic elements or inorganic solids, and can exhibit a wide range of forms and properties. In this review, we have chosen to focus essentially on inorganic or organic solid particles. This means that some typical features are not addressed, such as the formation of liquid films when droplets merge to form a thin continuous coating along a wall. Nevertheless, the focus on solid particles does not imply an overly simplified problem since a wealth of intricate phenomena does take place. Indeed, as these particles interact with a solid boundary, they can form deposits of various shapes and sizes, and these deposits, or even individual particles, can be complex mechanical objects (e.g., deformable). Deposits with a significant size can alter the fluid flow, and fluid motion over such large-scale deposits can lead to the detachment and entrainment of single particles, or even groups of particles, implying a rich and complex range of physical issues. This corresponds to the all-important topic of particle resuspension. 

Particle resuspension is a process typically encountered in our daily activities. One spectacular example involves sand, dust or leaves, deposited on the ground: these particles can be blown away by a local airflow (e.g., due to the wind, a person walking nearby or even a door being opened/closed) \cite{kok2012physics}. Another notable example is related to the quality of drinking water \cite{liu2016understanding}: it can be compromised by the release of tiny elements that accumulate on the interior of pipes in water distribution systems (e.g., bacteria, nitrates and/or trace metals such as iron or copper). Particle resuspension is a remarkable phenomenon in environmental situations, for instance in hydrology (with the initiation of sediment transport by rivers) \cite{pahtz2020physics} or in oceanography (e.g., re-entrainment of sediments generated by waves) \cite{boegman2019sediment}. Many industrial processes are also concerned by particle resuspension. To name a few examples, resuspension is an undesirable phenomenon when dealing with potentially harmful particles in medical applications (e.g., resuspension of contaminated particles from hospital floors due to human activities or from ventilation systems \cite{boor2013monolayer}), in the energy industry (e.g., radioactive particles blown away by strong wind following nuclear incidents \cite{kissane2012dust}) or even in the automotive industry (e.g., road resuspension of soot particles from exhaust emissions as well as abrasion-derived particles stemming from tires or brakes \cite{amato2016traffic}).

These selected examples bring out that resuspension is a highly multiscale process with challenging and complex issues. More specifically, it involves a variety of flow conditions (e.g., in the atmosphere, rivers, oceans or confined systems) as well as particle properties. For instance, particles can have very different geometrical (e.g., shape, size), physical (e.g., composition, density) as well as mechanical properties (e.g., rigid, extensible, deformable, or flexible). Moreover, resuspension is a process taking place over a wide range of temporal and spatial scales, typically from biofilm resuspension over a few seconds in millimeter-size pipes to sand resuspension over a few minutes/hours in desert dunes. Finally, it is worth pointing out that resuspension has a two-side nature since it can either be required (e.g., to limit the growth of limescale deposits within pipes) or has to be avoided (e.g., when particles are contaminants). 

This brief overview of applications indicates that resuspension is a relevant process in several scientific domains. A first consequence of this ubiquity is that it has led to a rich terminology developed separately and used to refer to what is basically the same physical issue: for example, depending on the application domain, it is called re-entrainment, removal, detachment, reaerosolization, remobilization or even winnowing. A second consequence is that our familiarity with the subject can create the misleading belief that particle resuspension is either a straightforward physical question or that it has been thoroughly investigated, leaving no areas on which to shed new light. However, particle resuspension encompasses different physical mechanisms such as: adhesion between bodies, rupture of static equilibrium, particle motion along and near surfaces, response to near-wall turbulent motions, inter-particle collisions and collective motion. In turn, these mechanisms are a combination of three fundamental interactions, namely particle-fluid, particle-particle, and particle-surface interactions. This means that, despite the complexity of this process and the richness of the applications involved, a more general picture can emerge to describe particle resuspension. Yet, this general picture can only arise from a multidisciplinary and multiphysics approach, since the process is at the crossroads between a range of physics-based fields, from fluid dynamics to physico-chemical interface forces and tribology. New insight is indeed needed to clarify the interplay between these fundamental interactions. Furthermore, it is important to realize that such a description is expressed essentially through the formulation of models. Indeed, we do not have a self-contained and closed set of equations, operating at the same level of physical description, to address particle resuspension. There are subtle quantum effects, as in the inter-molecular origin of van der Waals forces between bodies, while fluid flows are described by continuous mechanics, such as the Navier-Stokes equations and even coarser versions of them for high Reynolds-number turbulent flows. In other words, modeling is an intrinsic and central part of any physics-based description of resuspension.

It follows that modeling particle resuspension represents a multiscale task with, on the one hand, macroscopic effects (such as in fluid mechanics) and, on the other hand, microscopic ones (e.g., physico-chemical forces at the nanoscale between adhering bodies). In fact, modeling particle resuspension remains an open challenge in Statistical Mechanics in relation with dispersed two-phase flow studies~\cite{minier2001pdf, minier2016statistical}. More than in fields where a well-established physics-based modeling framework is already available providing consistent safeguards for the assessment of specific models, it is here important to revisit carefully what a model represents, what is captured, and on which underlying assumptions a given formulation relies on.

Over the years, a range of models have been proposed \cite{henry2014progress, ziskind2006particle}, often with the sole purpose of directly capturing the overall particle resuspension rate. This resuspension rate is a measure of the average number of particles removed from a surface per unit time. As such, it is a very common output of models but which results from the combination of physical phenomena that need to be distinguished and addressed separately before being combined together to provide a comprehensive description of a specific situation. Trying to integrate or bridge over these intricate physical mechanisms without careful consideration can generate confusion or uncertainty with respect to what is truly modeled and, more specifically, to what is actually predicted or assumed in a model. In that sense, a first challenge is to come up with an unambiguous description of what a model of particle resuspension stands for and what we are to expect from well-posed formulations. In addition, uncertainty quantification is currently missing but paramount as the consequences of a particle being resuspended may be non-linear. For instance, infectious agents spread has usually an exponential growth meaning that a certain error in the resuspension prediction can have a much larger error in the forecast of sick population.

\smallskip

In this review, we take up recent experience-based physical analyses of particle resuspension~\cite{henry2014progress, pahtz2020physics} and concentrate on the modeling step, revisiting past and present efforts from a mechanistic standpoint to provide insights into the state of the art. This is helpful to pave the way for extended formulations able to address new problems involving complex-shaped particles while remaining tractable for simulations in realistic situations where resuspension needs to be quantified along with its counterpart phenomenon of particle deposition. More precisely, the objectives of this work are to:
\begin{enumerate}[(1)]
\item Clarify the physical mechanisms and fundamental interactions involved;
\item Provide a general framework to define different classes of models;
\item Distinguish what is assumed and what is predicted by each class of models;
\item Reveal and analyze the origin as well as the physical nature of model inputs;
\item Point out directions of progress where developments are needed along with new well-targeted experiments.
\end{enumerate}
The first three objectives will be used as guidelines to come up with a new classification of existing resuspension models, which will guide interested readers into choosing adequate models depending on the system of interest and the expected results.

\smallskip

To reach these objectives, the paper is organized as follows. The physics of particle resuspension is first recalled in Section~\ref{sec:physics}: an overview of the spectrum of situations and physical contexts where particle resuspension plays an important role is given before introducing a more precise terminology and discussing the physical mechanisms as well as fundamental interactions at play. In Section~\ref{sec:techniques}, existing experimental methods are briefly reviewed together with the corresponding quantities that can be extracted from such measurements. In Section~\ref{sec:models}, the existing models used to measure or predict these quantities of interest are introduced. Then, following basic guidelines for model development recalled in Section~\ref{sec:art_model}, we propose and analyze existing models with respect to a new set of criteria in Section~\ref{sec:analysis}, which provide insights into the contents of particle resuspension models (e.g., inputs, outputs, assumptions). 
Apart from assessing existing propositions, one of the main interests of the approach developed in the first six sections of this review is to pave the way for future extensions. This is done for models in section~\ref{sec:next_model} when more complex particle shapes and particle collective effects are addressed. In this section, the needs to design new experiments aiming at validating key elementary aspects of models and of the physical mechanisms involved are also discussed together with the development of experimental techniques that have the potential to provide new insights into the physics of particle resuspension. Finally, conclusions and perspectives are proposed in section~\ref{sec:concl}.

 \section{The physics of particle resuspension}
  \label{sec:physics}
  
The purpose of this section is to set the stage for the discussions to come by introducing the key notions and terminologies related to the physical aspects of particle resuspension. Since several reviews on particle resuspension are already available (see for instance \cite{pahtz2020physics, henry2014progress, ziskind2006particle, sansone1977redispersion, nicholson1988review, gradon2009resuspension, rienda2021road}), only the most relevant aspects are detailed here.

This section is organized as follows: a range of practical applications is first described in Section~\ref{sec:physics:application}, including a presentation of the terminology used throughout the rest of the paper; second, the phenomenology of particle resuspension is introduced in Section~\ref{sec:physics:phenomenology}, providing an overview of the key interactions at play as well as of the fundamental mechanisms.

 \subsection{Context and applications}
  \label{sec:physics:application}

Etymologically, re-suspension is formed by the combination of the prefix ``re-'' (meaning again/anew) and ``suspension'' (meaning to disperse throughout the bulk of a fluid). Hence, resuspension corresponds to the process that induces a renewed suspension of particles. To put it differently, it means that solid discrete elements, previously at rest on a surface, are pulled off the surface and brought back into the bulk of the fluid in a suspended state. In that sense, this process differs from abrasion or erosion, which may appear very similar but imply the release of tiny elements that were previously parts of larger objects (like freshly formed dust resulting from the erosion of rocks). 

According to this definition, it is evident that particle resuspension is one of the processes at play in bounded dispersed two-phase flows. The reverse process is particle deposition, i.e., the accumulation of particles initially suspended on a surface. The deposition process will only be briefly discussed in Section~\ref{sec:physics:application:deposition} and is not considered in great details. Concentrating on  resuspension, the above definition raises the following issues: 
\begin{enumerate}[(Q1)]
 \item Where does resuspension take place?
 \item What are the conditions required for resuspension to occur? More specifically, what are the nature of the fluid, of the particles, and of the surfaces involved?
 \item What are the driving mechanisms allowing particles to resuspend from surfaces?
 \item What happens to the resuspended particles after being pulled off the surface? 
\end{enumerate}

In the following paragraphs, the key features of particle resuspension and the challenges associated with them are briefly introduced starting from a selection of practical situations taken from natural and industrial applications (see Section~\ref{sec:physics:application:examples}). These selected examples illustrate the issues (Q1), (Q2) and (Q4) (see Section~\ref{sec:physics:phenomenology} for details about the underlying mechanisms related to the issue (Q3)). The reverse process of particle deposition is briefly covered in Section~\ref{sec:physics:application:deposition} while the terminology used in the rest of the manuscript is detailed in Section~\ref{sec:physics:application:terminology}.

   \subsubsection{Illustrations of typical resuspension phenomena}
   \label{sec:physics:application:examples}
  
\begin{figure}[ht]
 \centering
 \captionsetup[subfigure]{justification=centering}
 \begin{subfigure}{0.32 \linewidth}
  \centering
  \includegraphics[width=\textwidth]{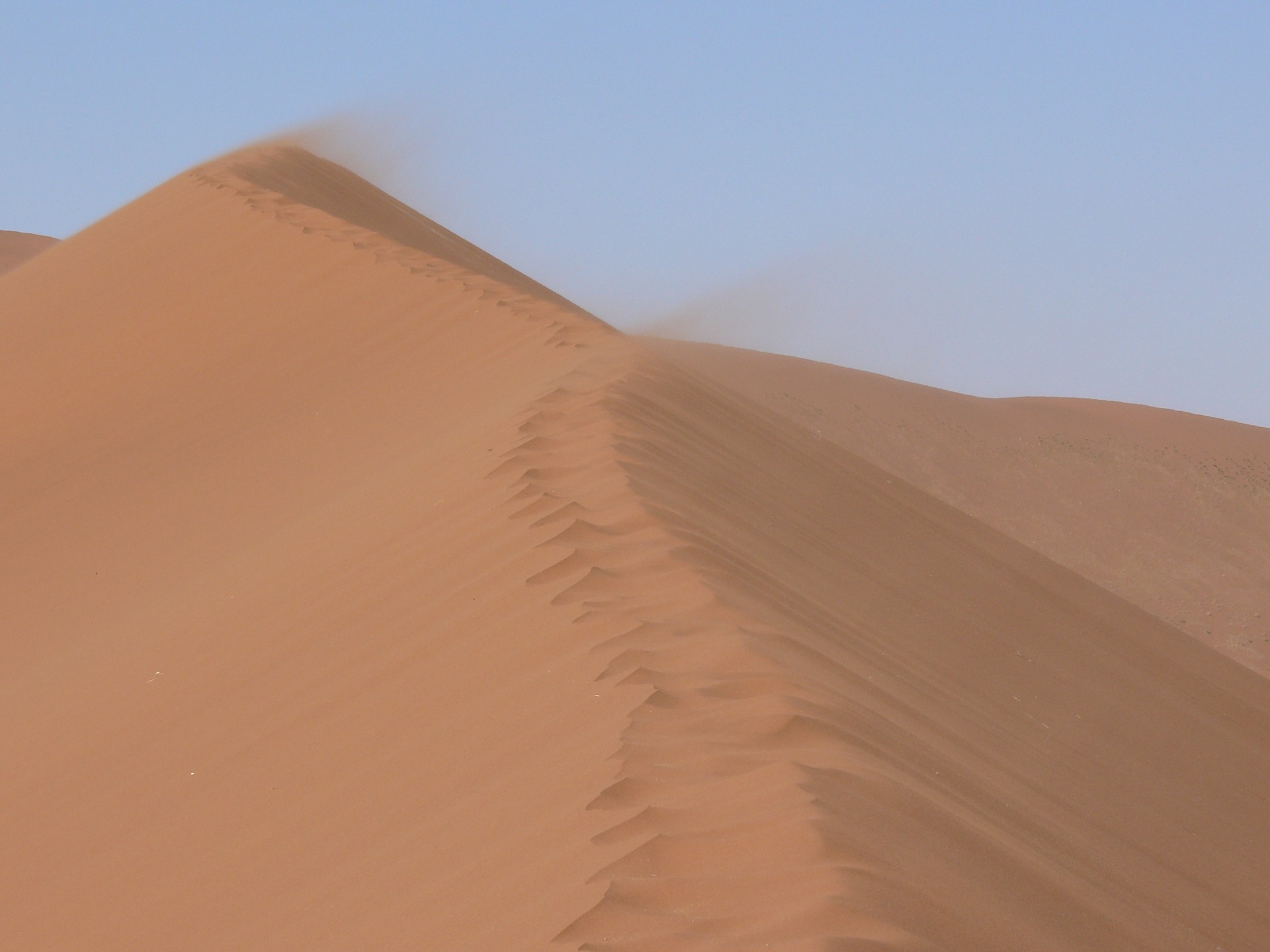}
  \caption{Wind-induced resuspension of sand on a dune. Source: \href{https://pixabay.com/photos/namibia-sossusvlei-sand-dune-sand-1130132/}{Pixabay}.}
  \label{fig:pict_sand_dunes}
 \end{subfigure}
 \hspace{2pt}
 \begin{subfigure}{0.32 \linewidth}
  \centering
  \includegraphics[width=\textwidth]{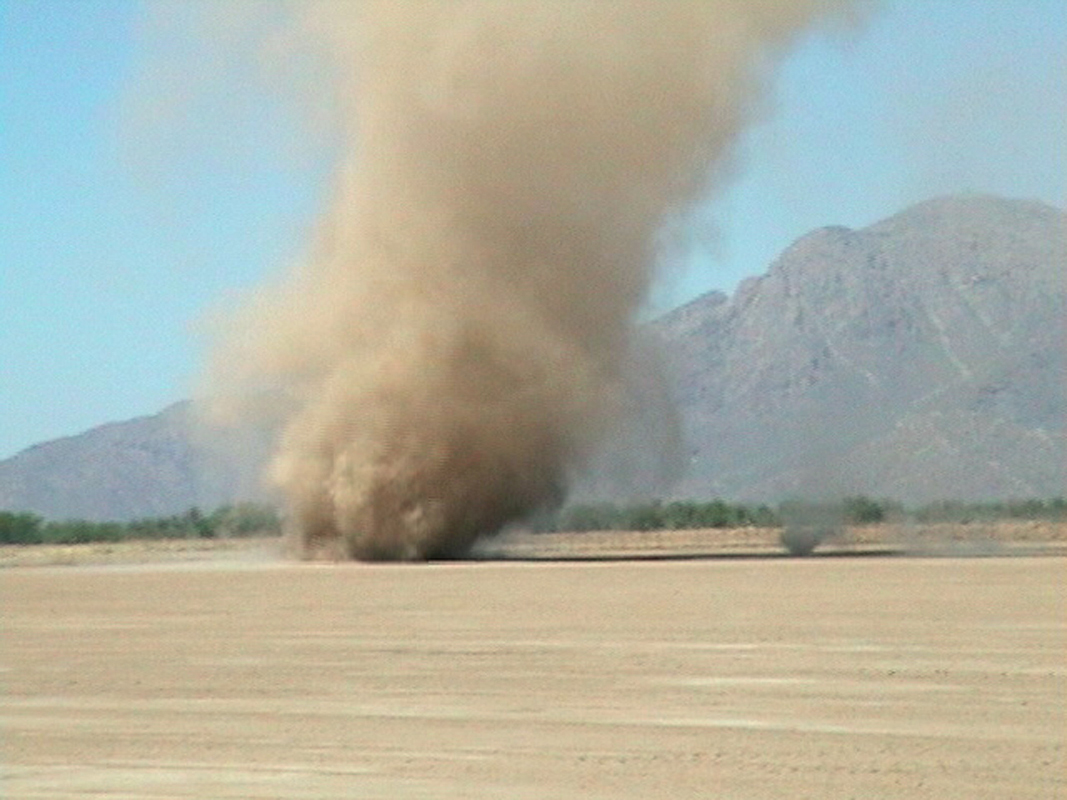}
  \caption{Dust devil in a field. \\ Source: \href{https://www.nasa.gov/vision/universe/solarsystem/2005_dust_devil.html}{NASA}}
  \label{fig:pict_dust_devil}
 \end{subfigure}
 \hspace{2pt}
 \begin{subfigure}{0.32 \linewidth}
  \centering
  \includegraphics[width=\textwidth]{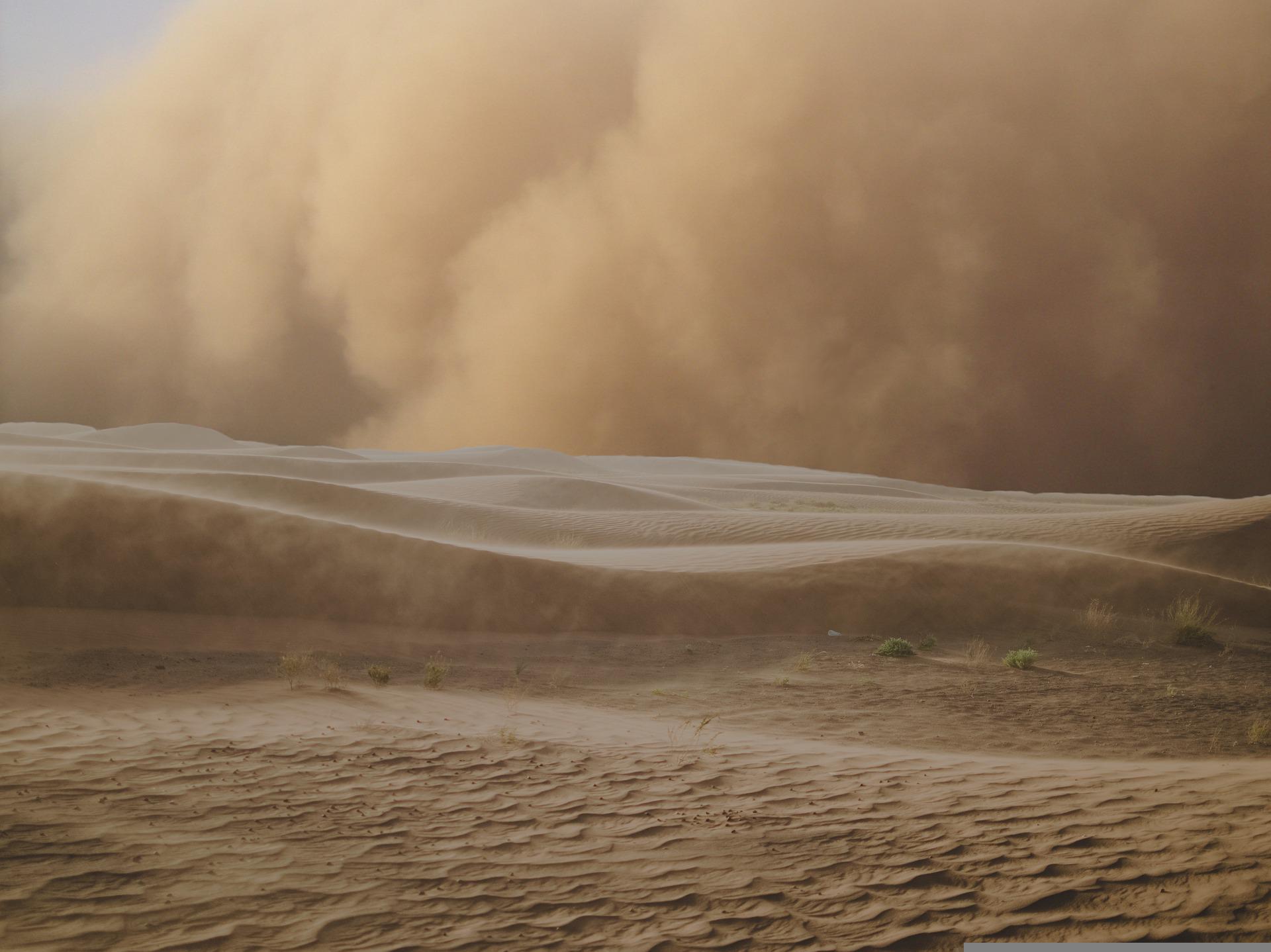}
  \caption{Sand-storm over a desert. \\ Source: \href{https://pixabay.com/photos/desert-sandstorm-sand-dry-dunes-5821732/}{Pixabay}.}
  \label{fig:pict_sand_storm}
 \end{subfigure}
 \caption{Illustrations of wind-driven resuspension of sand by various types of flows: normal wind conditions (left), whirlwinds (middle) and storms (right).}
\end{figure}

\subparagraph{Resuspension in the air:}
When thinking about particle resuspension, one of the first examples that comes to mind is the resuspension of sand/dust through the action of the wind \cite{kok2012physics, bagnold1937transport}. This process is typically at play in the winnowing of a sand dune \cite{bagnold1937size}, in which sand particles located in the upper layer of a dune become suspended when exposed to a strong-enough wind (see Fig.~\ref{fig:pict_sand_dunes}). In the case of sand dunes, this process is generally an intermittent one, meaning that sand particles are brought back in suspension only by intense events due to strong but short-lived fluctuations in the wind velocity and direction. Yet, in the case of extreme wind speeds (as in storms), severe resuspension episodes persist for a significant amount of time, leading to high concentrations of particles in the atmosphere. Such persistent processes are responsible for either very localized phenomena (e.g., dust devils due to local vortices, see Fig.~\ref{fig:pict_dust_devil}) or broader spatial phenomena (e.g., dust storms as shown in Fig.~\ref{fig:pict_sand_storm}). 

\begin{figure}[ht]
 \centering
 \includegraphics[width=0.7\textwidth]{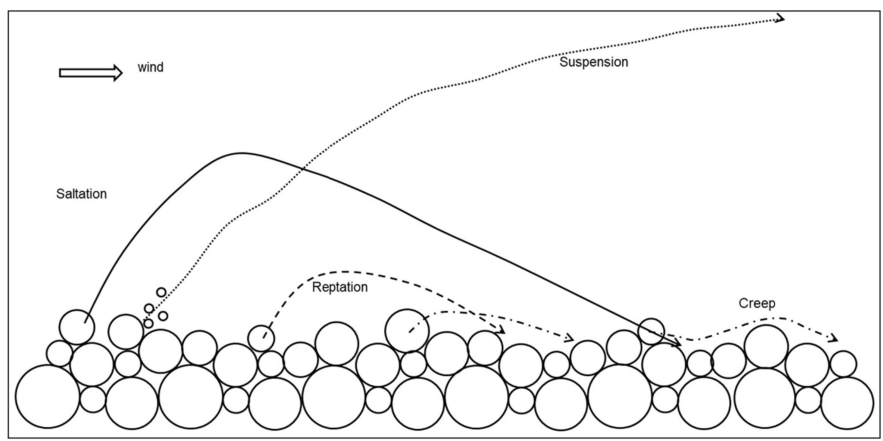}
 \caption{Representation of the different modes of wind-blown sand transport, namely: suspension (i.e., particles remaining suspended in the flow), saltation (long hops above the surface), reptation (short hops above the surface) and creep (migration on the surface). Reprinted with permission from \cite{zhang2021aeolian}. Copyright 2021, Elsevier.}
 \label{fig:zhang_2020_mode_transport}
\end{figure}

Another interesting feature of sand/dust resuspension is that the fate of the resuspended particles depends on both particle properties (size, density, etc.) and fluid flow characteristics (velocity, intermittency, etc.). For instance, as sketched in Figure~\ref{fig:zhang_2020_mode_transport}, large and heavy sand particles can perform short hops above the surface (called reptation) or, alternatively, roll on the surface without actually getting detached from it (creeping). These intermittent, yet persistent, processes are responsible for the constant evolution of sand dunes or the formation/erasure of ripples under the wind. Meanwhile, smaller and lighter particles can remain suspended for much longer times before impacting again the surface: this specific behavior is referred to as saltation. Upon colliding with the surface, saltating particles can eject other particles due to a transfer of kinetic energy. When small dust/sand particles reach high-enough altitudes in the atmosphere, these dust/sand grains can even move across continents (e.g., Sahara sand reaching south-European countries or even South-America). These different modes of transport for wind-blown sand/dust particles are the subject of dedicated studies in the field of geomorphology (interested readers are referred to existing books \cite{bagnold2005physics} or reviews \cite{kok2012physics, pahtz2020physics}). Similar features are also reported on Mars, where dust resuspension occurs under locally strong whirlwind or turbulence \cite{chatain2021seasonal}. Another serious concern related to the exploration of extra-terrestrial planets is the potential contamination of the planet environment by biological particles from Earth, which can resuspend from the rover surface depending on the local atmospheric conditions \cite{mikellides2020modelling, mikellides2020experiments}.

Studies on particle resuspension are not limited to particles of natural origin. For instance, many works concern the resuspension of radioactive particles that can accumulate on the ground following nuclear accidents (see for instance \cite{stempniewicz2008model, reeks2001kinetic, masson2013size}). In this case, particle resuspension is shown to expand the contaminated zone under strong wind conditions. Besides, since even tiny amounts of radioactive particles can jeopardize human health, resuspension can prolong the threat of exposure to radioactive particles even beyond the passage of the initial particle-laden plume. In fact, even though the plume containing radioactive particles is emitted over a limited amount of time following such accidents, these primary particles can deposit on soils and be resuspended later over much longer times. However, quantifying such phenomena is not easy since it requires measuring rare (extreme) events. This is due to the fact that the amount of particles present on a given surface decreases as time goes by, since the initial particle deposit is not replenished with a fresh source of particles. It is indeed observed that loosely-adhering particles are quickly blown away by the wind at short exposure time. Hence, at longer exposure times, only particles adhering strongly to the surface are left on this surface. Removing such particles requires higher wind conditions, which occur rarely. This explains the typical trend observed in experiments, where the amount of resuspended particles is shown to decrease with the inverse of time at long exposure times (e.g., \cite{reeks2001kinetic}). 

When both natural and artificial particles are present, more complex phenomena can take place when these particles interact with each other to form new compounds or aggregates (called secondary particles). In such cases, resuspension can be significantly altered compared to the case with only primary particles since it is not driven by the primary particle properties but rather by the aggregate characteristics (e.g., size, shape, composition). Such phenomena are observed when dealing with artificial hazardous particles that accumulate in soils. For instance, water-soluble $^{137}$Cs ions like those released in the Fukushima accident in 2011 reacted with airborne aerosols and were found attached to mineral particles. After deposition, they were reintroduced into the atmosphere as radioactive dust \cite{kinase2018seasonal}. Identifying the host particles is not always straightforward as many chemical reaction pathways can coexist depending on the soil composition and conditions (e.g., plutonium compounds \cite{xu2014plutonium}). Particles can also be absorbed by vegetation and be resuspended within pollen \cite{kinase2018seasonal} or during wildfires from the combustion of vegetation \cite{kashparovforest}. Similar issues arise in studies of lead poisoning for children in urban environments \cite{laidlaw2008resuspension}, where lead has accumulated in urban soils due to the extensive use of Pb additives in both gasoline and house-paints in the past.

\begin{figure}[ht]
 \centering
 \captionsetup[subfigure]{justification=centering}
 \begin{subfigure}{0.42 \linewidth}
  \centering
  \includegraphics[width=\textwidth, trim=2cm 7cm 8cm 7cm, clip]{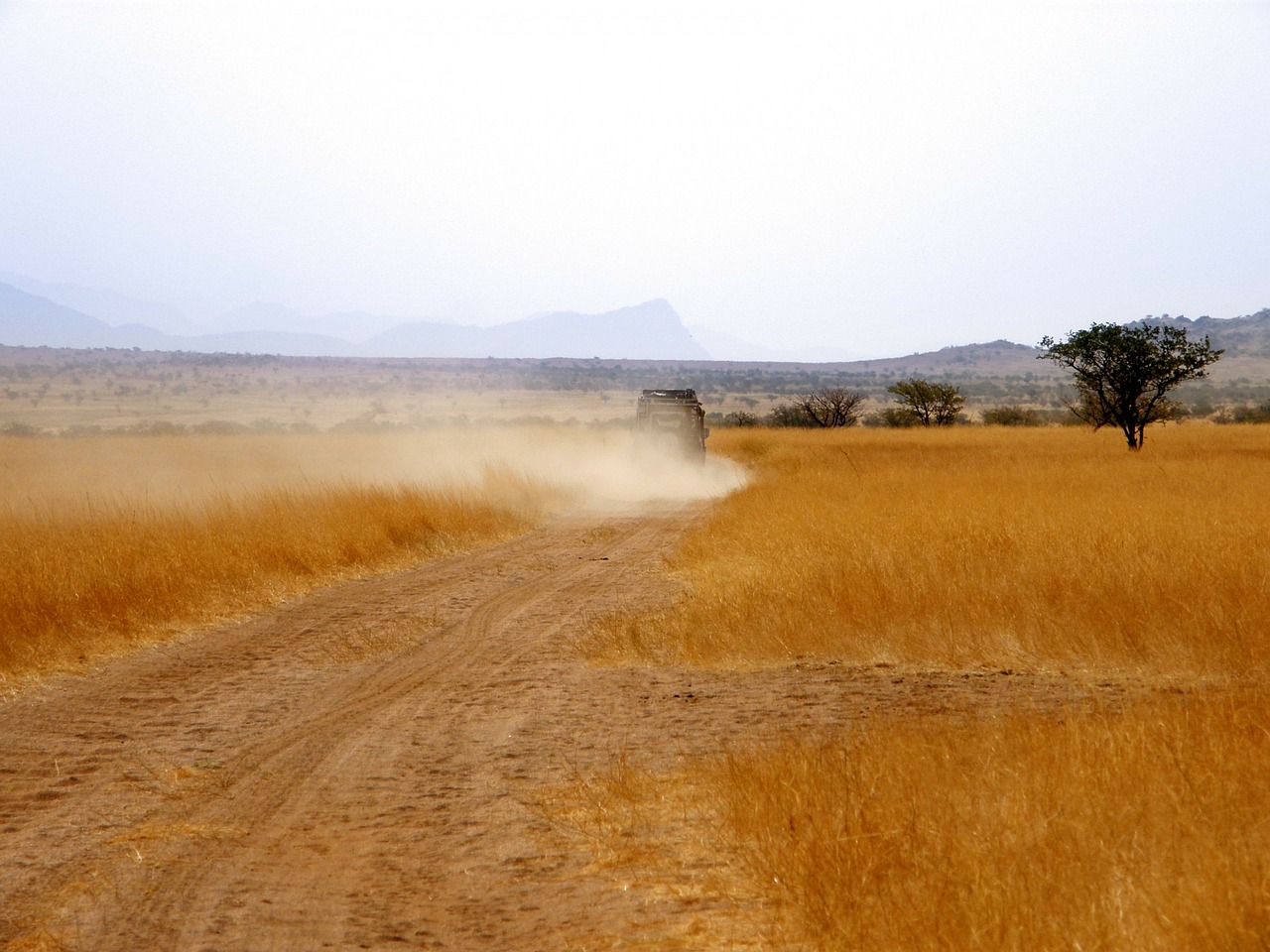}
  \caption{Resuspension of road dust induced by the passing of a single car. Source: \href{https://pixabay.com/fr/photos/range-rover-vehicule-164644/}{Pixabay}.}
  \label{fig:pict_road_dust}
 \end{subfigure}
 \hspace{10pt}
 \begin{subfigure}{0.47 \linewidth}
  \centering
  \includegraphics[width=0.8\textwidth]{fig_kubota_2013_walking_induced}
  \caption{Visualization of walking-induced resuspension during stomping. Reprinted with permission from \cite{kubota2013aerodynamic}. Copyright 2013, Taylor \& Francis.}
  \label{fig:pict_walking_induced}
 \end{subfigure}
 \caption{Illustrations of dust resuspension in the air induced by human activities (traffic on the left, walking on the right).}
\end{figure}

Still referring to atmospheric flows, another example of interest is the resuspension of particulate matter (PM) accumulated on various surfaces. Air pollution from PM has been widely studied in the aerosol science community \cite{friedlander2000smoke} due to their potential impact on human health when inhaled \cite{oberdorster2005nanotoxicology, amato2014urban}. This includes a large variety of particles, such as spores \cite{weis2002secondary}, road dust \cite{rienda2021road}, exhaust emissions of combustion engines \cite{thorpe2007estimation} or abrasion-derived particles from cars (e.g., tires, brake, road-wear) \cite{amato2016traffic, thorpe2008sources}. Conversely to primary radioactive particles, these pollutants are continuously released in the atmosphere and deposited, basically providing an ``infinite'' source of particles to be resuspended. In-situ observations show that resuspension is affected by environmental factors (e.g., local/seasonal variations in wind speed, precipitation, humidity) but also by human activities. For instance, as displayed in Fig.~\ref{fig:pict_road_dust}, the re-emission of road dust is significantly higher in the wake of a vehicle (see for instance \cite{amato2016traffic, sehmel1973particle, sehmel1980particle}). This so-called traffic-induced resuspension of road dust is estimated to contribute to 9-59\% of the PM10 mass concentration from non-exhaust emissions of vehicles (see the recent review \cite{rienda2021road} and references therein). This makes it one of the predominant sources of particulate matter emission, possibly reaching similar levels as the traffic exhaust emissions \cite{thorpe2007estimation}.

This last example on traffic-induced resuspension brings out the question of the pathways followed by particle resuspension in the air. In fact, resuspension events are usually classified in two main categories in the literature \cite{boor2013monolayer, you2015risk}: resuspension induced by an airflow and resuspension induced by human activities. These various pathways have been extensively studied in indoor environments, such as home or office rooms \cite{boor2013monolayer}, due to the growing concerns over air quality over the last few decades \cite{jones1999indoor, world2010guidelines}. This includes both studies on particulate matter (e.g., PM2.5, PM1.0, PM0.1 or lower) \cite{boor2013monolayer} as well as hazardous biological/chemical particles (e.g., anthrax \cite{weis2002secondary} or even flour dust \cite{stobnicka2015exposure, moeller2017adhesion}). These studies have allowed to identify a number of sub-categories briefly listed below (see also Fig.~\ref{fig:draw_indoor_resusp} and previous reviews \cite{boor2013monolayer, sansone1977redispersion}):

\begin{itemize}
 \item Resuspension induced by an airflow:
 \begin{itemize}
  \item Wind-induced resuspension in indoor environments is very similar to the outdoor one. One of the differences is that indoor surfaces can be very diverse (to name a few: wooden floor, concrete floor, metallic surfaces, vinyl floor or carpets) \cite{boor2013monolayer}. As a result, the amount of particles available in each surface can vary significantly while their resuspension depends highly on the local configuration (e.g., lower exposure to the airflow in fibrous environments like carpets, shielding mechanisms). In addition, recent studies have confirmed that the amount of resuspension depends on particles properties, including their size, their shape and their composition and origin (see for instance recent comparisons between dust, cat or dog fur, or bacterial spores \cite{salimifard2017resuspension}).
  \item Ventilation-induced resuspension involves two different aspects related to the impact of ventilation systems. First, some particles can deposit within the ventilation ducts and, as air is flowing through the ducts, these particles can be resuspended and emitted into a room (provided they are not captured by filters) \cite{wang2012experimental, zhou2011particle}. Second, the airflow induced by a ventilation system can affect the amount of particles resuspended from nearby surfaces. This second effect depends naturally on the orientation and flow rate of the ventilation system. For instance, a recent study has shown that the resuspension in impinging jet ventilation (i.e., high-speed airflow perpendicular to the ground with a vent placed at a given height) can be two to three times larger than displacement ventilation (i.e., low-speed airflow parallel to the floor introduced near the ground through a vent), and more than 10 times larger than mixed ventilation (i.e., low-speed airflow through a vent close to the ceiling) \cite{zuo2015experimental}.
 \end{itemize}
 
 \begin{figure}
  \centering
  \includegraphics[width=0.6\textwidth]{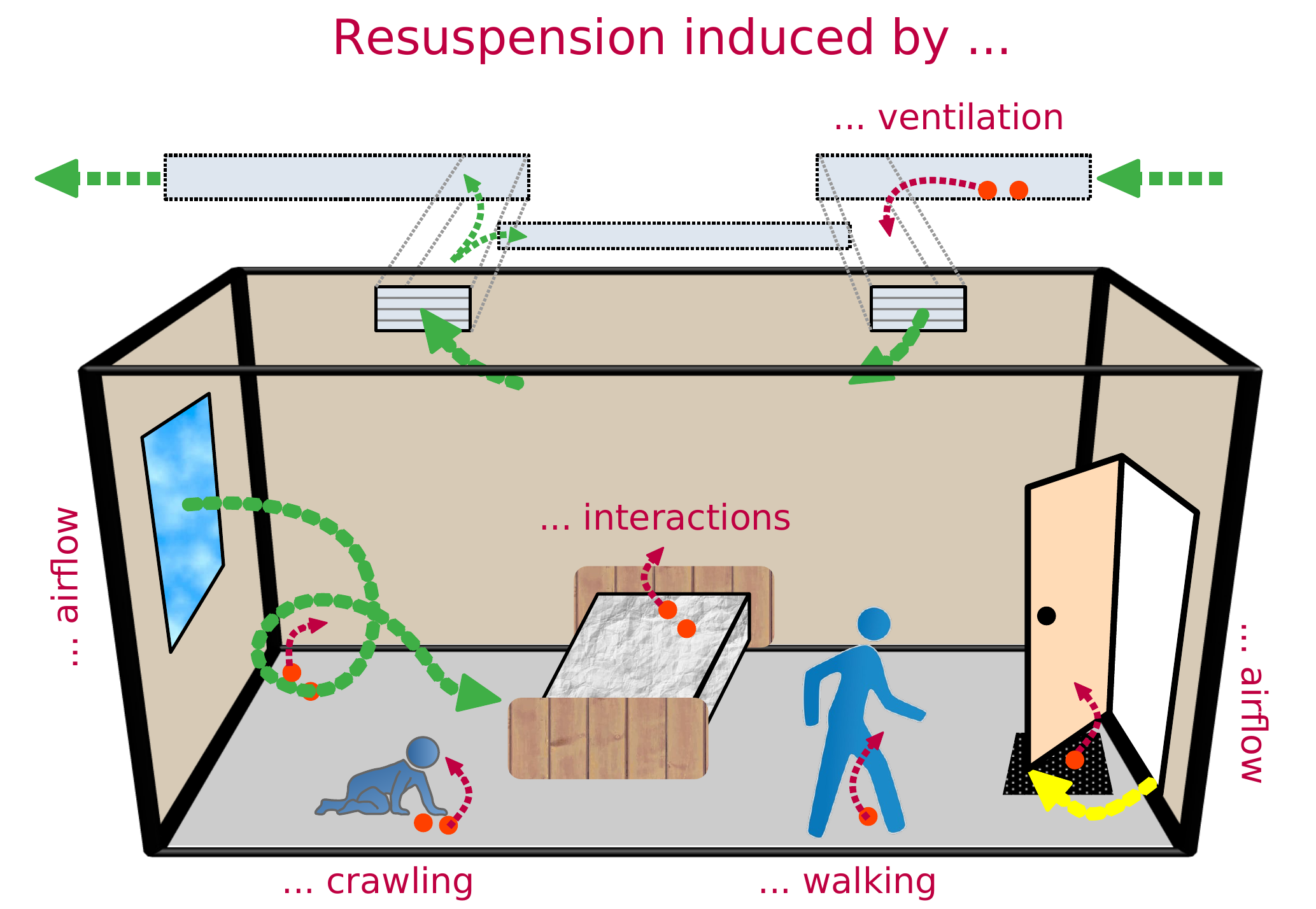}
  \caption{Representation of the different pathways for indoor airborne resuspension, showing resuspension induced by ventilation, crawling, walking, wind/airflows as well as human interactions with surfaces (e.g., scratching).}
  \label{fig:draw_indoor_resusp}
 \end{figure}

 \item Resuspension induced by human activities differs from resuspension induced by airflows since it usually involves sudden changes in the flow conditions (e.g., velocity, turbulence) as well as extra vibrations, mechanical contact and possibly shocks that can enhance particle resuspension. In the following, various indoor human activities are listed following the usual classifications used in the literature (see for instance \cite{you2015risk}):
 
 \begin{itemize}
  \item Walking-induced resuspension is triggered by a combination of several factors (see the recent review \cite{qian2014walking}). As displayed in Fig.~\ref{fig:pict_walking_induced}, the upward and downward foot movement during walking can generate complex air movements in the immediate surroundings and cause particles to resuspend \cite{sehmel1980particle, kubota2013aerodynamic}. Walking motion also induces surface vibrations and electrostatic forces acting between the particles and the shoe bottom surface, which can further affect particle resuspension. In addition, both the pace and style of walking were observed to impact resuspension \cite{qian2008resuspension}. More recently, a study of infant exposure to respiratory pathogens has provided evidence that the type of socks worn by children while walking has profound consequences on resuspension \cite{zhang2022walking}.
  
  \item Crawling-induced resuspension occurs due to the crawling motion of children on the floor, especially on carpets or other fabrics. It has been increasingly studied in the past few years, since children are exposed to higher concentrations of PMs due to their proximity to the surface. In fact, chamber experiments show that infants are exposed to contamination levels that can be one order of magnitude higher than adults \cite{hyytiainen2018crawling, wu2018infant}. These levels are also shown to be lower during walking rather than crawling. 
      
  \item More generally, physical activities are reported to induce higher particle resuspension in indoor environments. It is observed that resuspension intensifies with an increase in the intensity of physical activities (see for instance sports in school gyms \cite{buonanno2012particle}, walking intensities \cite{mcdonagh2014study}, dancing \cite{mcdonagh2014influence} or even motion while sleeping \cite{boor2015characterizing}). This can be related to a combination of effects, including aerodynamic resuspension, surface vibrations, mechanical abrasion, and contact transfer. Besides, particles can resuspend from a variety of surfaces including clothes worn by humans during their activities \cite{luoma2001characterization} and surfaces with which they interact (e.g., mattresses while sleeping \cite{boor2015characterizing}). Recent measurements on clothing have also highlighted that resuspension depends highly on the clothing texture (i.e., roughness, weave pattern) and on the type of motion imposed on the clothes (e.g., stretching with various intensity) while it seems to be slightly affected by the fabric material \cite{mcdonagh2014influence}. 
  
  \item Cleaning activities can also lead to a sudden and drastic change in the amount of particles resuspended. A counter-intuitive example is the case of vacuum cleaners. Vacuum cleaners are designed to collect dust from the floor using a strong airflow that favors their resuspension followed by their aspiration in the device. However, when switched on, vacuum cleaners can also induce PM10 concentrations that can be up to 3 or 4 times higher than under normal conditions \cite{corsi2008particle}. This is explained by the substantial increase in the number of resuspended particles from nearby carpet fibers due to the high-velocity airflow induced around the vacuum cleaner and also to mechanical motion induced by the contact between the wheels and the carpet. In addition, some of the particles captured by the collectors within the vacuum cleaner can be re-entrained during or after the use of the device \cite{trakumas2001comparison}. Such re-entrainment actually depends on the type of collectors used (such as filter bags, wet dust collectors or cyclonic collectors). An alternative cleaning technique is based on mechanical resuspension, as for sweeper trucks with rotating brushes that are used to remove leaves from gutters before aspirating them.
 
  \item Another way to remove particles from a surface is contact transfer. This happens when particles on the host surface are picked up by a different surface during contact between two surfaces. Contact transfer can occur between objects (e.g., a stack of paper placed on a table) but also between the host surface and a person's skin and clothing. This mechanism is only seldom studied in the context of resuspension because particles are not immediately re-emitted into the air. Nevertheless, this process is specifically identified here since it allows particles to be picked up from a contaminated surface in a certain environment and, thanks to their transfer to another surface, to be resuspended at a later time in a different environment (e.g., from one room to the next). This whole process is actually known as translocation. For instance, higher concentration compared to ambient air pollution were observed when people came back from smoke breaks \cite{luoma2001characterization}. Alternatively, contact transfer can also be used to detect the presence of trace explosive particles from fibrous surfaces (e.g., cotton, polyester) \cite{kottapalli2019experimental, kottapalli2021aerodynamic}.
  
 \end{itemize}
\end{itemize}

\subparagraph{Resuspension in marine systems:}
Particle resuspension has also been extensively studied when the carrier phase is a liquid. From a historical perspective, research on particle resuspension has emerged more or less at the same time to address sand transport in the air \cite{bagnold1937transport} and sediment transport in rivers \cite{shields1936anwendung}. Sediments at the bottom of a river are formed by the accumulation of various particles, including natural ones (e.g., clay, sand, gravel) and contaminants (e.g., heavy metals, plastics). As the river discharge fluctuates, various amounts of sediments are resuspended from the riverbed and transported further downstream following similar modes as for airborne particles (i.e., reptation or creep as shown in Fig.~\ref{fig:zhang_2020_mode_transport}). In fact, the combined deposition and resuspension of sediments in rivers plays a key role in the morphodynamics of rivers (i.e., changes in the river path over geological times) and in river deltas. As discussed in recent papers \cite{furbish2021rarefied, ancey2020bedload1, ancey2020bedload2}), decades of research on bedload river transport have revealed that the threshold velocity above which particles start to move depends on a number of parameters, including the particle size and the fluid-to-particle density ratio, thus confirming the early findings of Shields in 1936 \cite{shields1936anwendung}. 

\begin{figure}[htbp]
 \centering
 \captionsetup[subfigure]{justification=centering}
 \begin{subfigure}{0.9 \linewidth}
  \centering
  \includegraphics[width=0.5\textwidth, trim=2cm 0cm 14cm 0cm, clip]{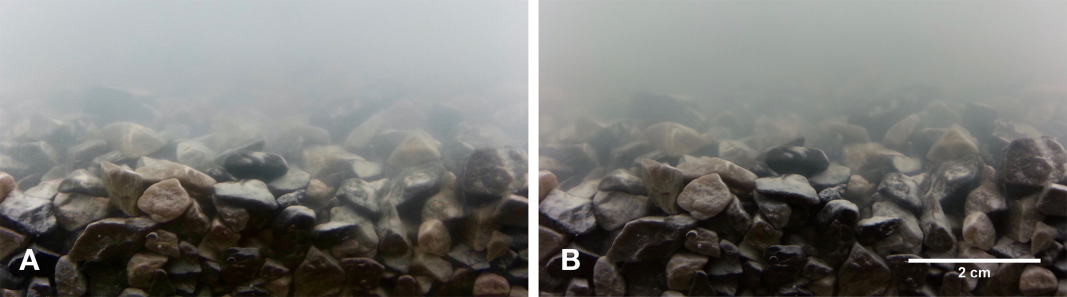}
  \caption{Picture showing the complex deposit formed by gravels at the bottom  of rivers or lakes (which can be later on resuspended by strong flows). Reprinted with permission from \cite{mooneyham2018deposition}. Copyright 2018, John Wiley and Sons.}
  \label{fig:pict_gravel_river}
 \end{subfigure}
 \begin{subfigure}{0.9 \linewidth}
  \centering
  \includegraphics[width=0.49\textwidth, trim=11cm 6cm 3cm 3cm, clip]{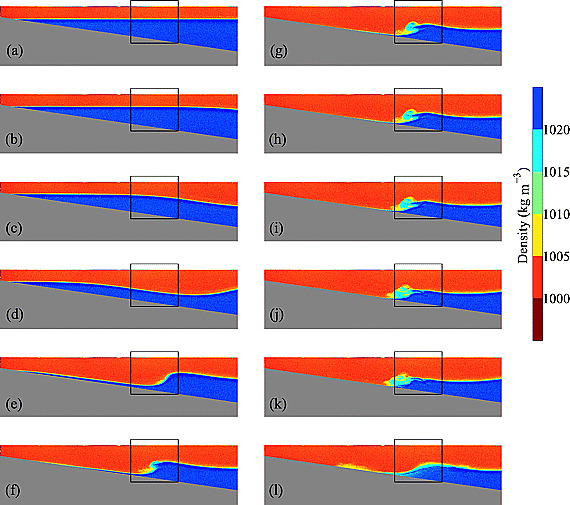}
  \includegraphics[width=0.4\textwidth, trim=10.5cm 9.5cm 5.0cm 10cm, clip]{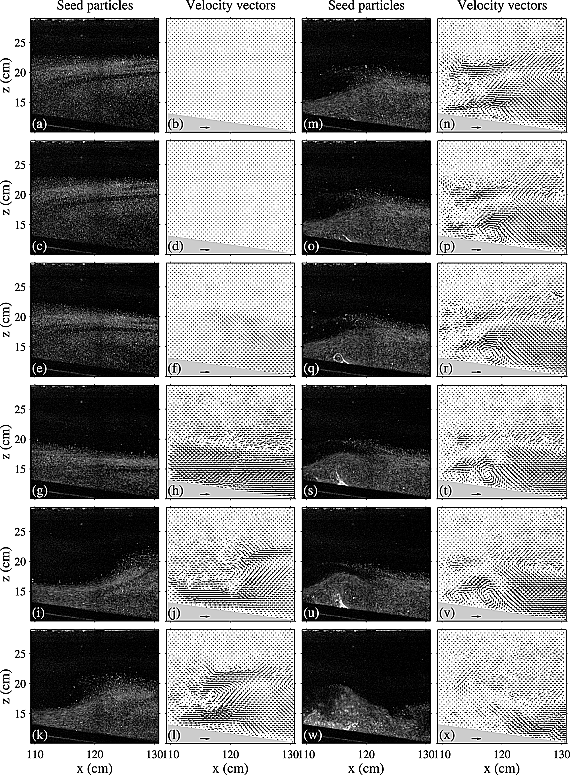}
  \caption{Visualization of internal wave propagation in stratified flows: left images showing the density fields and right images displaying the seed particles being detached from the surface due to internal waves. Reprinted with permission from \cite{boegman2009flow}. Copyright 2009, John Wiley and Sons.}
  \label{fig:pict_internal_wave}
 \end{subfigure}
 \caption{Illustrations showing the sediment beds formed in rivers (a) their resuspension due to waves within lakes/water (b).}
\end{figure}

Resuspension of sediments in marine systems can originate from either natural or artificial processes. For instance, natural resuspension processes in marine systems encompass river currents \cite{bagnold2005physics}, tidal currents or storm surge in oceans \cite{roberts2012causes, diercks2018scales}, the action of surface waves \cite{clarke1982sediment, green2014review} or even internal solitary waves which occur in stratified fluids (see Fig.~\ref{fig:pict_internal_wave}, and the recent review \cite{boegman2019sediment}). Among the various anthropic activities that impact particle resuspension, dredging, shipping and trawling \cite{mengual2016influence} have been confirmed to modify significantly the amount of resuspended sediments. Similarly, other animals like fishes can modify sediment resuspension due to the waves induced by their motion \cite{scheffer2003fish}. These selected examples imply that resuspension can be intermittent in nature, since it can take place over different temporal scales (from less than one hour for dredging operations to weeks for storms (\cite{roberts2012causes}). Besides, as for resuspension of airborne particles, various types of particles can be involved (including natural sediments \cite{pahtz2020physics} or plastic contaminants in riverbeds \cite{liro2020macroplastic}).
 
Studies on sediment transport in rivers have also highlighted the role of the deposit morphology in the resuspension process. In fact, sediments naturally form a multilayer deposit, where particles pile up to form large clusters on the underlying surface (see Fig.~\ref{fig:pict_gravel_river}). In that case, the resuspension process differs from measurements made with sparse particles on the surface. In sparse monolayer deposits, isolated particles tend to resuspend independently of each other, while in multilayer deposits particles collide with each other, possibly leading to cascade phenomena as in avalanches. Hence, resuspension by collision is one of the main modes of resuspension in multilayer systems. This has recently led to a distinction between fluid-driven motion and entrainment-driven motion which are both occurring but with various relative importance depending on the situation considered \cite{pahtz2020physics, pahtz2018cessation}. Collision-induced resuspension has been observed for river sediments \cite{pahtz2020physics} and also for sand particles (where saltation is predominant) \cite{bagnold1937transport}. Since the number and frequency of collisions between particles are driven by the deposit structure, many studies have assessed the role of sediment size, size distribution \cite{sun2002grain}, inter-particle cohesive forces (see for instance \cite{pahtz2020physics, henry2014progress} and references therein). In addition, since very large millimeter-size particles are harder to resuspend than micrometer particles, armoring effects can occur: this corresponds to the presence of less mobile particles preventing resuspension of smaller particles located underneath or in their wake, thereby acting as a sort of shield for nearby mobile particles. Such phenomena have been observed in multilayer deposits formed by particles with a large size distribution \cite{mooneyham2018deposition}. In such cases, the amount of resuspension can display sudden changes with time: it can be initially important until the small particles on the top of the deposit are removed and only large ones are left exposed to the fluid, at which point these large particles act as shelter for the layers of deposit below, thereby severely reducing the resuspension rate. Alternatively, consolidation effects can also be responsible for a decrease in the resuspension at longer times. These processes are actually related to a restructuring of the deposit and sometimes to modifications of the contact area between individual particles, which leads to the formation of more compact and cohesive deposits that are harder to resuspend \cite{tran2019floc}.

\begin{figure}[htbp]
 \centering
 \includegraphics[width=0.9\textwidth]{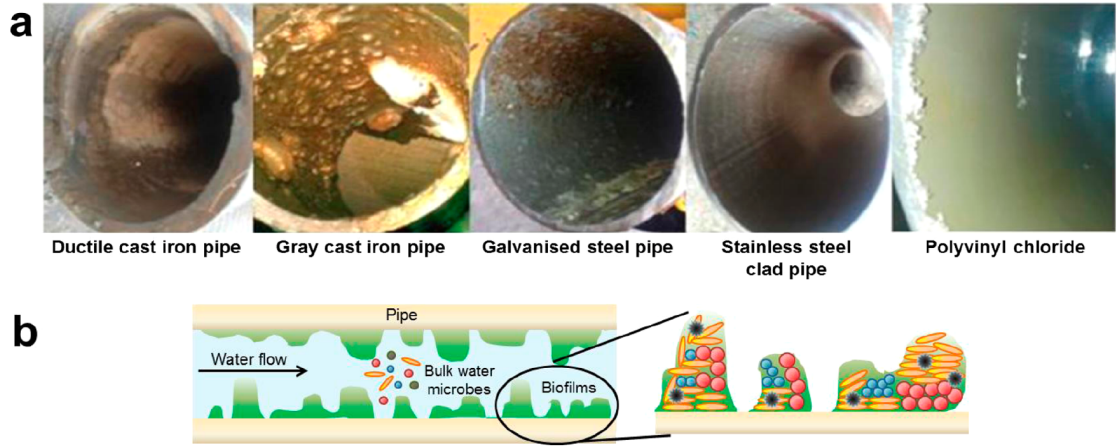}
 \caption{(a) Images showing biofilm growth on different pipe materials. Reprinted with permission from \cite{ren2015pyrosequencing}. Copyright 2015, Springer Nature. (b) Representation of a biofilm life cycle, including their possible resuspension. Reprinted with permission from \cite{liu2016understanding}. Copyright 2016, American Chemical Society.}
 \label{fig:fig_liu_2016_biofilm}
\end{figure}

\subparagraph{Resuspension in confined systems:}
Last but not least, resuspension occurs in many constrained environments, such as pipes or porous media. As discussed in a recent review \cite{liu2016understanding}, this is a matter of concern in drinking water systems, where a biofilm grows on the interior surface of the pipes (see also Fig.~\ref{fig:fig_liu_2016_biofilm}). When the flow through the pipe is strong enough, deposited particles can detach from the surface and be transported in the pipe network. This can lead to the presence of unwanted contaminants in the water (e.g., chemical or biological particles), thus giving rise to poor water quality and increased health hazard for customers \cite{mussared2019origin}. In such cases, the objective is often to limit the accumulation of a deposit on the surfaces (e.g., by imposing high velocities in pipe networks \cite{vreeburg2009velocity}) and, when it cannot be prevented, to limit its resuspension. The difficulty is that the particles involved often span a wide range of sizes \cite{ryan2008particles}. Similar issues arise when studying groundwater quality, which can be affected by the mobilization of contaminants present in the ground (see for instance \cite{ryan1996colloid, johnson2007deposition}). Resuspension in pipes has been also explored in industrial applications. For instance, corrosion products can accumulate on the interior surface of pipes within heat exchangers (leading to fouling), thereby affecting the normal operating conditions \cite{henry2014stochastic}. Resuspension then appears as a desirable process: it has led to the development of cleaning procedures using ultrasonic methods, where cavity bubbles are used to break the deposits formed and then a strong flow through the pipe brings the detached particles out of the system \cite{kieser2011application} (similarly to resuspension in other three-phase flows \cite{lin2019resuspension}). Resuspension is also a matter of concern in nuclear safety \cite{malizia2016review}. For example, in nuclear fusion reactors, radioactive particles (like dust or graphite) can be resuspended and released in the environment during a LOss of Coolant Accident (LOCA, i.e., when the pipe containing the coolant is damaged and fluid is leaking \cite{peng2013graphite, humrickhouse2011htgr, sun2020graphite, andris2020investigations}) or during a Loss Of Vacuum Accident (LOVA, i.e., when the vacuum vessel is damaged \cite{gelain2020cfd, peillon2020dust, rossi2021numerical}). 

\subparagraph{Summary:} Based on these selected examples, several conclusions can be drawn. First, resuspension is observed in a large array of environmental and industrial situations. Second, resuspension involves a variety of fluids (such as liquids, gas or even a mixture of both), surfaces (from smooth glass/metal substrates to fibrous clothes or carpet fibers), and particles (aerosols like spores or radionuclides and sediments). The involved particles have a range of origins (e.g., natural or artificial), compositions (e.g., chemical, biological), shapes (e.g., spheres, spheroids, fibers), and sizes (from nanometric aerosols to millimeter-size sediments). In addition, deposited particles can be arranged either in a monolayer fashion, with isolated individual particles, or in a multilayer configuration with complex particle arrangements. This leads to a distinction between independent resuspension events and correlated resuspension events (possibly with collective motion). Third, resuspension is a highly multiscale process, spanning a wide range of temporal and spatial scales, from the resuspension of biofilms over a few milliseconds in millimeter-size pipes to radioactive resuspension across kilometer-size fields over years. Fourth, resuspension is a multifaceted problem, since the process can occur through a variety of pathways (including flow-induced or collision-induced modes as well as be induced by human activities). This also implies that resuspension is a highly multidisciplinary topic. In fact, the mechanisms and forces at play in resuspension involve physical, chemical and/or biological effects. Hence, studying particle resuspension requires a knowledge at the crossroad between several fields, such as fluid mechanics, interface chemistry, surface sciences, and possibly biology. 

A consequence of this multifaceted, multidisciplinary, and multiscale process is that the important mechanisms and forces at play in a given situation are not necessarily the dominant ones for a different application. This explains why many studies have addressed resuspension within a specific sub-domain and avoided tackling the overall phenomenon. As will be seen later, this represents a challenge when trying to come up with generic models (see Section~\ref{sec:models:approach}). Nevertheless, despite these difficulties, a common and universal description of the resuspension process is slowly but surely emerging (see also the recent review \cite{pahtz2020physics}).

  \subsubsection{The reverse process of particle deposition}
  \label{sec:physics:application:deposition}
  
\begin{figure}[h]
 \centering
 \includegraphics[width=0.9\textwidth]{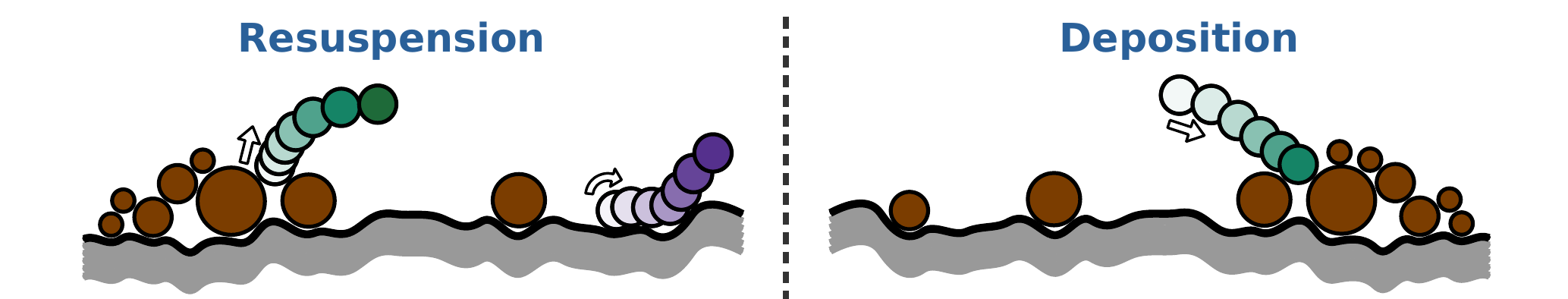}
 \caption{Illustration of the resuspension process and of the reverse deposition process}
 \label{fig:sketch_depo_resusp}
\end{figure}

Previous examples have revealed that resuspension occurs often concomitantly with the reverse process of particle deposition. As illustrated in Fig.~\ref{fig:sketch_depo_resusp}, the deposition process corresponds to the migration of suspended particles toward a surface on which they can adhere and eventually come to rest. From a physical point of view, it is best to describe particle deposition as a two-step process (see for instance \cite{yiantsios1998effect}): first, a transport step where suspended particles are carried by the flow towards the immediate vicinity of a surface and, second, an attachment step in which these particles can stick to the surface or rebound from it. As such, this process is the opposite of the resuspension process, where particles adhering to a substrate are brought back in suspension and transported by the action of a fluid flow. 

In fact, deposition and resuspension are deeply intertwined. For example, once particles are detached from a surface and resuspended, their subsequent transport by the flow can either lead them to remain suspended in the fluid or to redeposit later on the surface (as in reptation and saltation). This continuous switch from one process to another is responsible for the slow and intermittent migration of sand dunes \cite{kok2012physics, bagnold2005physics}. Another example of the intricate coupling between these two processes is the formation of large complex-shaped deposits on a surface. This happens due to the continuous accumulation of deposited particles on a surface, which leads to the formation of multilayer deposits (e.g., sand dunes in desert, sediments in riverbeds or fouling of heat exchangers by corrosion products). In turn, these deposits can modify the velocity of the flow near the surface, especially when their size is large enough (typically with a size comparable to the smallest scale in the fluid, i.e., the Kolmogorov length scale $\eta_K$ in a turbulent flow). This modified fluid velocity can then lead to spatial variations in the amount of resuspended particles, giving rise sometimes to regular shapes (e.g., ripples in sand dunes). 

Since particle deposition usually occurs before resuspension start to kick in, it has been extensively studied in the multiphase flow community and interested readers are referred to existing reviews on particle deposition for further information (see for instance \cite{wang2021deposition, henry2012towards}). Given their correspondence, it is not surprising that the key factors affecting deposition are similar to the ones identified for resuspension, namely:
\begin{itemize}
 \item Particle characteristics, including: particle size distribution, density, shape, surface area, hygroscopicity and hydrophobicity. For instance, atmospheric aerosols can be highly hygroscopic and absorb water vapor at high relative humidity thus changing dimension, density, and optical properties (\cite{donateo2014correlation}).
 \item Surface characteristics, including: surface roughness, adhesiveness with particles, electric charge or magnetic properties. For instance, roughness features much smaller than the particle size can significantly modify the adhesive forces, while roughness features with a size comparable or bigger than the particle size can lead to sheltering effects \cite{henry2012towards}. When the size of roughness features becomes comparable or larger than the smallest fluid scale, it also modifies near-wall fluid velocities (and, consequently, particle transport in the vicinity of the surface).
 \item Flow characteristics, including: fluid properties (like density, viscosity), turbulence intensity (e.g., intermittency, coherent structures near boundaries). For deposition in the atmosphere, atmospheric stability is also important and in general, the ratio of the deposition velocity and the friction velocity appears to be smaller in stable atmospheric conditions (\cite{donateo2014correlation}, \cite{gallagher2002measurements}).
\end{itemize}

  \subsubsection{Terminology suggested and used here}
  \label{sec:physics:application:terminology}

To depict accurately any field of physics and develop sound physical analyses as well as proper theories, it is important to start by naming precisely objects and, especially, processes. In the field of particle resuspension, this is not a step to overlook. Indeed, from the various examples introduced in Section~\ref{sec:physics:application:examples}, it appears that the terminology currently in use varies depending on the specific application which is addressed. To provide a few instances, this includes terms like reaerosolization of airborne particles (such as viruses, bacteria, spores) \cite{krauter2007reaerosolization}, remobilization of plastics in rivers \cite{liro2020macroplastic} or during floods \cite{roebroek2021plastic}, winnowing of sand in ripples and mega-ripples \cite{tholen2022megaripple}, etc. Particle resuspension is also frequently named particle removal \cite{kottapalli2019experimental}, particle detachment \cite{ibrahim2003microparticle, benito2016validation} or particle re-entrainment \cite{matsusaka1996particle}. Alternatively, in the context of sediment in rivers, it is common to find studies focused on the onset of motion or the incipient motion of particles \cite{pahtz2020physics}. Obviously, this terminology does not provide a clear-cut indication as to which physical mechanism is being addressed specifically. In the literature, the only distinction currently available has been suggested in both aeolian transport and sediment transport: it consists in trying to identify specifically the subsequent transport of particles once they start moving. More precisely, this has led to the definition of creep motion (i.e., continuous or near-continuous contact with the surface through rolling/sliding motion), reptation (particles undergoing short hops above the surface), saltation (i.e., particles undergoing long hops with splashing during subsequent collision events) and suspension (i.e., particles which do not deposit on the surface, at least during the observation time). Clearly, this is not a satisfactory state of affairs: such vague formulations can quickly lead not only to confusion but blurred physical descriptions.

\begin{figure}
 \centering
 \includegraphics[width=0.85\textwidth]{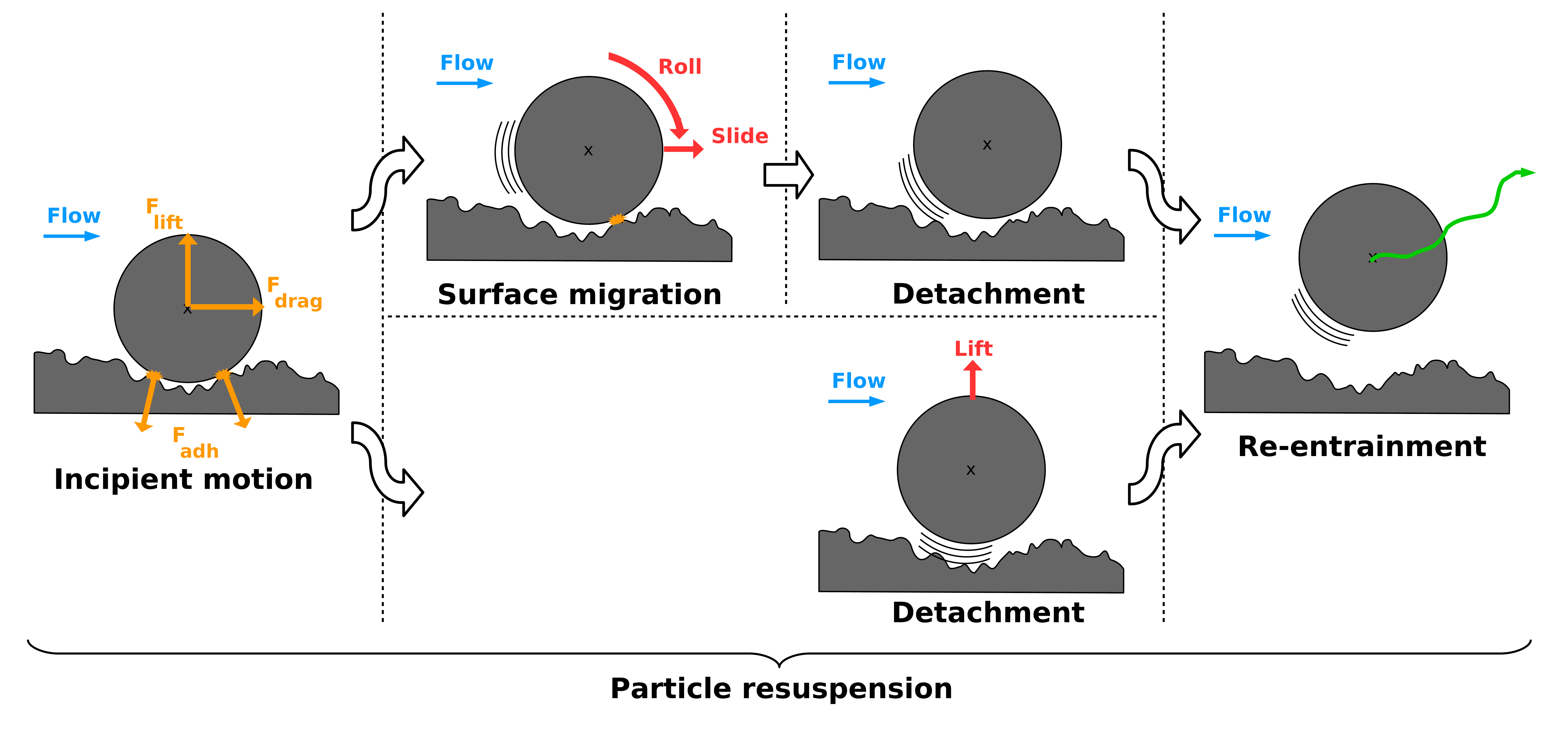}
 \caption{Sketch summarizing the terminology used in the present paper for particle resuspension: the incipient motion corresponds to the instant when a particle starts moving, leading to its detachment from the surface (possibly with migration on the surface beforehand) and its subsequent re-entrainment by the flow in the near-wall region. This makes up the whole resuspension process.}
 \label{fig:sketch_def}
\end{figure}

To avoid these pitfalls, we introduce here specific definitions to distinguish between the different phenomena at play. These notions are then used throughout the present paper regardless of the application domain. First, we start by making a distinction for particles that are in contact with a surface or a deposit:
\begin{enumerate}[(i)]
 \item Sticking/deposited/adhering particles correspond to particles that are in contact with a substrate (i.e., either a surface or other deposited particles) at a given location (no motion).
 \item Attached particles are particles in contact with the substrate but that can be moving.
 \item Detached particles refers to previously attached particles whose contact with the surface has been broken.
 \item Adhesion forces refer to the contact forces between a particle and a surface.
 \item Cohesion forces designate the contact forces between particles.
\end{enumerate}
In addition to these fundamental definitions, we introduce another distinction regarding the resuspension process (see also Fig.~\ref{fig:sketch_def}):
\begin{enumerate}
 \item[(def-1)] \textbf{Particle incipient motion} refers to the moment at which a sticking particle starts moving. It is related to the rupture of the balance between forces preventing its motion and forces inducing its motion. The dislodgement of sticking particles triggers their motion either on the surface (attached particles) or in the flow (detached particles).
 \item[(def-2)] \textbf{Particle migration} is related to the motion of attached particles until they are actually \textbf{detached} from the surface.
 \item[(def-3)] \textbf{Particle re-entrainment} refers to the motion of particles after detachment, i.e., in the near-wall region.
 \item[(def-4)] \textbf{Particle resuspension} is a generic term that encompasses all phenomena (i.e., incipient motion, detachment, migration and re-entrainment).
\end{enumerate}
Hence, particle resuspension can be defined here as the phenomenon that leads to the detachment and sub-sequent re-entrainment of a sticking particle from a surface.

These definitions allow to distinguish between attached and detached particles while making a clearer separation between the three motion encountered in particle resuspension (i.e., incipient motion, migration and re-entrainment). These definitions remain broad enough to encompass phenomena observed in other fields. For instance, in the context of geological flows, abrasion refers to the erosion of a surface by exposure to scraping or other mechanical constraints on the surface \cite{shao2008physics, knight2008environmental}: it can thus be interpreted here as a specific case of particle detachment due to hydro-mechanical forces. Yet, it does not necessarily imply resuspension (unless particles were suspended previously). More generally, an interesting characteristic of this terminology is that it allows to bring out the difference between particle resuspension and particle release (also called seeding for pollens). In fact, the two processes might sometimes appear to be similar (e.g., release of pollen by plants due to the wind). However, according to the present definitions, particle release does not involve particles that were previously suspended, while particle resuspension does. In addition, these definitions encompass the previous distinctions introduced earlier between creep and saltation/reptation. More precisely, creep corresponds now to a specific case of particle migration (in which particles remain in contact with the surface and perform rolling/sliding motion) while reptation and saltation correspond now to particle resuspension. It is worth noting that these definitions do not allow to distinguish between reptation and saltation. The main reason for this choice is that they both involve particles that have resuspended and, after a certain time in suspension, collide with the surface again. While reptation does not trigger the resuspension of other particles during this collision, saltation leads to multiple subsequent resuspension events through splashing (see Section~\ref{sec:physics:phenomenology} for more details). This is sometimes referred to as particle entrainment in the aeolian and fluvial communities \cite{pahtz2020physics} (in the sense of motion induced by collisions as in avalanche processes).  However, we do not wish to introduce such a distinction here since similar phenomena can occur for particles that were simply in suspension and collide with the surface (i.e., not necessarily resuspended). Hence, we believe that such a distinction should rather be handled by proper definitions introduced in the description of the deposition phenomenon.

 \subsection{The phenomenology of particle resuspension}
  \label{sec:physics:phenomenology}

The overview of applications in Section~\ref{sec:physics:application} was meant to illustrate the relevance of particle resuspension in an array of practical situations as well as to point to the intricate physical phenomena which are involved. Drawing on these observations, the questions that now come to mind are: 
\begin{enumerate}[(a)]
\item How are deposited particles resuspended from a surface? 
\item What are the fundamental interactions acting in this process? 
\item What are the underlying mechanisms? 
\end{enumerate}

The purpose of this section is to provide answers to these questions. A simple picture meant to clarify point (a) is first given in Section~\ref{sec:physics:phenomenology:how}, after which an outline of the fundamental physical interactions with respect to point (b) is provided in Section~\ref{sec:physics:phenomenology:interactions} (before being developed later in Section~\ref{sec:models:forces}). Then, the underlying mechanisms mentioned in point (c) are analyzed in Section~\ref{sec:physics:phenomenology:mechanisms}.

\subsubsection{A competition between motion-inducing versus motion-preventing factors}
   \label{sec:physics:phenomenology:how}

As it transpires from the terminology introduced in Section~\ref{sec:physics:application:terminology}, particle resuspension results from a subtle balance between forces that keep particles attached to a deposit and forces that induce motion through detachment/migration. Therefore, in spite of its inherent complexity, it is worth starting with a simple but leading image. In this crudely cut image, it can be said that particle resuspension boils down to a competition between, on the one hand, elements preventing particle motion and, on the other hand, factors acting to dislodge particles from the deposit (either the wall surface or layers of deposited particles). 

\subsubsection{Fundamental physical interactions}
   \label{sec:physics:phenomenology:interactions}

In turn, these two main categories of forces, acting either to counteract or facilitate particle motion, result from three fundamental physical interactions, namely \textbf{particle-fluid}, \textbf{particle-particle} and \textbf{particle-wall} interactions. As will be seen later in Section~\ref{sec:models:forces}, these interactions are mostly treated using a continuum assumption, i.e. considering that materials/fluids are continuous media (by contrast with discrete assumptions made in molecular descriptions). We are therefore dealing with a process at the crossroads between different subjects of physics: this includes turbulence, interface chemistry and material properties, whose combined effects create the tapestry of observed phenomena. Actually, by gathering particle-particle and particle-wall forces into a general \textbf{particle-surface} interaction, it is possible to consider two main types of interactions. This does not mean, however, that there is a simple one-to-one correspondence with the two categories of motion-inducing and motion-preventing factors. Indeed, while particle-fluid interactions are essentially acting to induce particle motion, gravity forces (when present) can either be motion-inducing or motion-preventing depending on the geometry considered. This means that we need to describe specifically the key resuspension mechanisms at play in particle resuspension to better grasp this complexity.

\subsubsection{Key resuspension mechanisms}
   \label{sec:physics:phenomenology:mechanisms}
  
To describe the key physical mechanisms at play in the resuspension process, we can rely on a range of experimental observations which have provided ample evidence over the years. Since comprehensive reviews have already described these mechanisms (see for instance \cite{pahtz2020physics, henry2014progress}), we limit ourselves to recalling key points, with the difference that these mechanisms are summarized here following an original presentation that paves the way for a unified description of particle resuspension. In this new presentation, we distinguish between individual resuspension events, in which particle-particle interactions are negligible, and collective resuspension events, in which particle-particle collisions and forces play a role. Note that this is different from considering resuspension from mono- and multilayer deposits, which explains why collective events are further divided into two effects.

\begin{figure}[ht]
 \centering
 \begin{subfigure}[c]{0.9\textwidth}
  \centering
  \includegraphics[width=0.85\textwidth, trim= 0cm 0cm 0cm 0cm, clip]{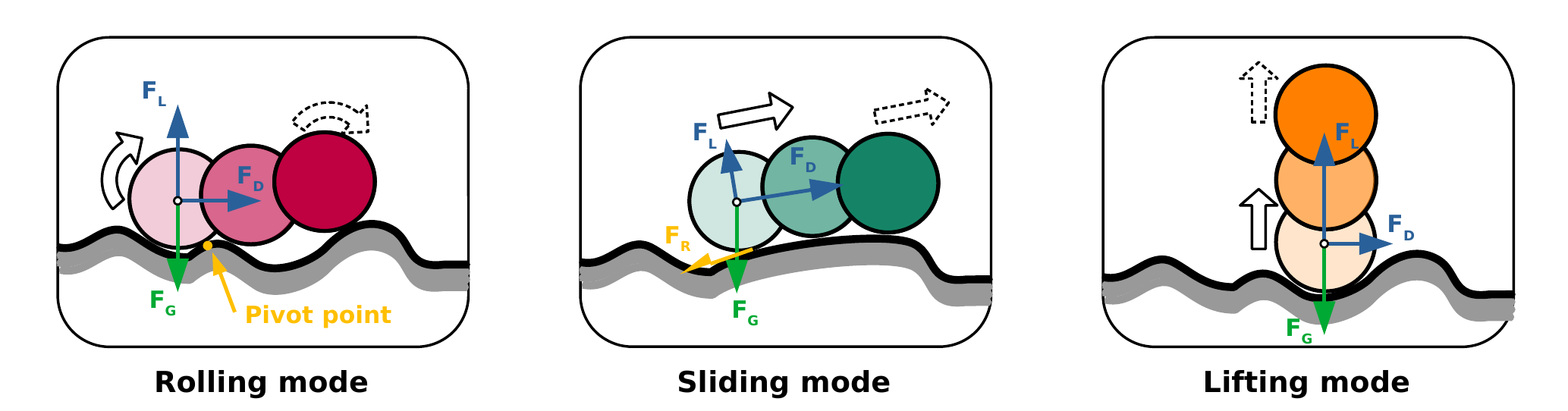}
  \caption{Illustration of single particle entrainment from a surface: (a) rolling mode, (b) sliding mode and (c) lifting mode.}
  \label{fig:sketch_3mechanisms}
 \end{subfigure}
 \vspace{10pt}
 \begin{subfigure}[c]{0.9\textwidth}
  \centering
  \includegraphics[width=0.6\textwidth, trim=3.0cm 0cm 0cm 0cm, clip]{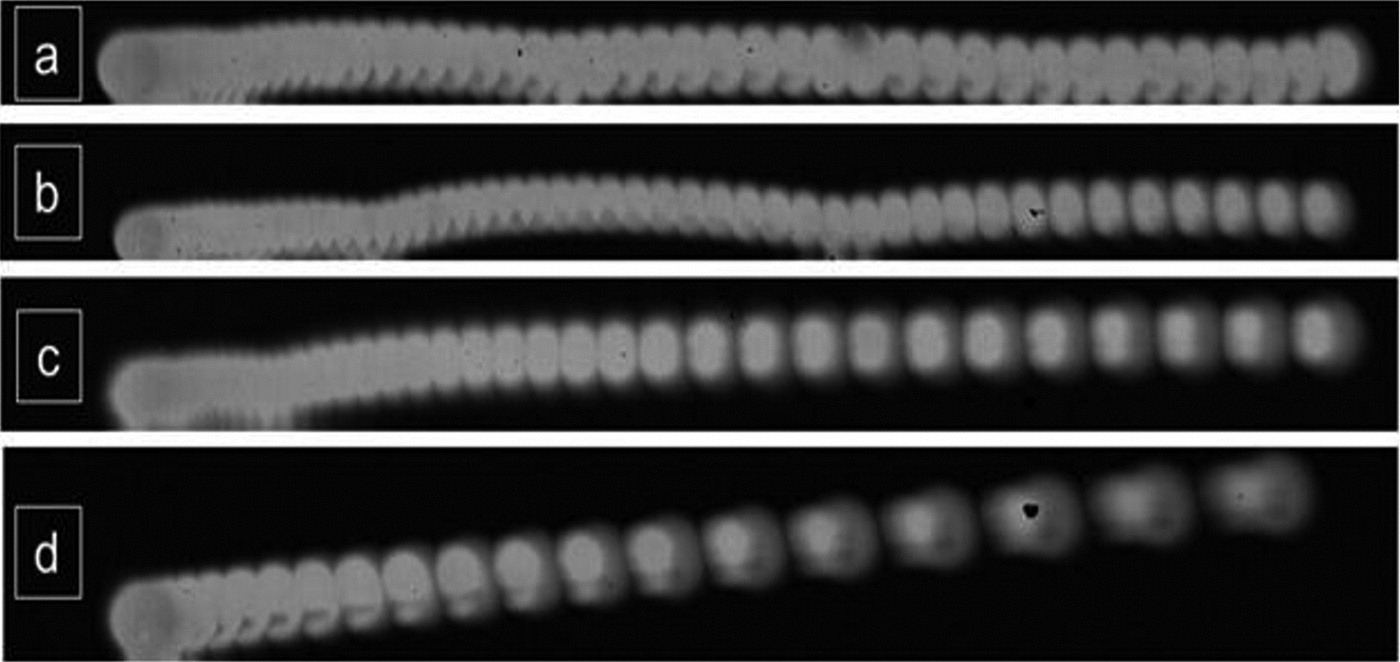}
  \caption{Snapshots showing the different types of induced motion for micrometer-size glass particles on a ceramic substrate: rolling/bouncing (top), rolling/bouncing and potential liftoff (2nd), rolling/bouncing and late liftoff (3rd) and immediate liftoff without rolling/bouncing (bottom). Reprinted with permission from \cite{kassab2013high}. Copyright 2013, Taylor \& Francis.}
  \label{fig:fig_kassab_2013_mechanisms}
 \end{subfigure}
 \caption{Sketch of the fundamental mechanisms for the resuspension of individual particles from surfaces and the corresponding images obtained from experimental observations.}
\end{figure}

\subparagraph{Individual resuspension events} First, studies of individual particles deposited on a surface have revealed that the incipient particle motion is triggered by a rupture of balance between forces/torques acting on the particle. This rupture of balance can lead to three different types of incipient motion sketched in Fig.~\ref{fig:sketch_3mechanisms}: \textbf{sliding} (a type of surface migration in which the particle has a purely translational motion along the surface), \textbf{rolling} (another type of surface migration where the particle has a purely rotational motion along the surface) and \textbf{lift-off} (when the particle is immediately detached from the surface due to pulling forces). These three modes induce a variety of complex particle motion near the surface that have been observed experimentally (see  Fig.~\ref{fig:fig_kassab_2013_mechanisms}): a particle can start rolling/sliding on the surface before being actually detached from it, while another one can be immediately detached from the surface due to direct lift-off \cite{cleaver1973mechanism}. Once resuspended, particles can collide with the surface again, where they can be recaptured (either deposited or migrating on the surface) or just bounce on it.

Following extensive experimental observations, there is now an emerging consensus in the multiphase flow community on how such particle dynamics depends on particle and fluid properties. This is formulated in a recent review \cite{henry2014progress} which has shown that particles small enough to be fully embedded within the viscous sublayer are more prone to display rolling/sliding motion, while larger particles are more inclined to be directly lifted-off the surface as they interact with the near-wall turbulent structures which exist only at some distances away and not in the immediate vicinity of the wall (where viscosity effects smooth things out). Such individual resuspension events are thus induced by the interaction between the fluid flow and deposited particles (through the drag and lift forces discussed in more details in Section~\ref{sec:models:forces}), and involve only particle-fluid and particle-wall forces but not particle-particle interactions.

\begin{figure}[ht]
 \centering
 \begin{subfigure}[c]{0.35\textwidth}
  \centering
  \includegraphics[width=\textwidth, trim=0cm 0cm 0cm 1.1cm, clip]{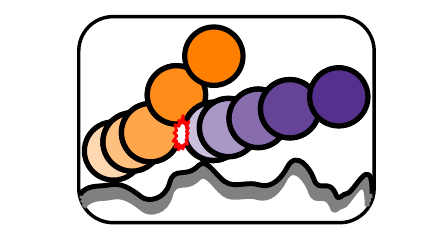}
  \caption{Sketch of collision-induced resuspension.}
  \label{fig:sketch_part_coll}
 \end{subfigure}
 \hspace{15pt}
 \begin{subfigure}[c]{0.6\textwidth}
  \centering
  \includegraphics[width=\textwidth, trim=0cm 2.9cm 0cm 0cm, clip]{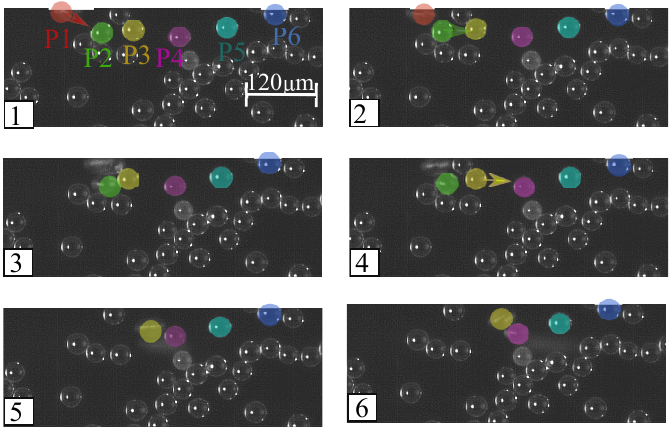}
  \caption{Consecutive snapshots of \SI{40}{\mu m} glass beads showing how collisions trigger a series of resuspension events in dense monolayer deposits. Reprinted with permission from \cite{banari2021evidence}. Copyright 2021, American Physical Society.}
  \label{fig:fig_banari_2021_coll}
 \end{subfigure}
 \caption{Illustration of the role of inter-particle collisions in triggering successive resuspension events.}
 \label{fig:fig_part_coll}
\end{figure}

\subparagraph{Collective resuspension events} Second, more recent studies on dense monolayer deposits have highlighted the role of inter-particle collisions in resuspension events \cite{banari2021evidence, rondeau2021evidence}. As shown in Fig.~\ref{fig:fig_part_coll}, once a particle starts migrating on the surface or moving near it, it can collide with other deposited particles, thereby setting these other particles into motion. 
As a result, when the distance between deposited particles is small enough, the incipient motion of a single particle is sufficient to trigger the motion of several other particles due to a series of inter-particle collisions (as collision propagation in avalanches). In such cases, this means that the resuspension of individual particles are not independent of each other but appear rather as correlated events. Such studies have bridged the gap between studies of individual particle resuspension from sparse monolayer deposits and studies of the collective resuspension events in multilayer deposits.

\begin{figure}[ht]
 \centering
 \begin{subfigure}[c]{0.9\textwidth}
  \centering
  \includegraphics[width=0.42\textwidth, trim=0cm 0.0cm 0cm 0.0cm, clip]{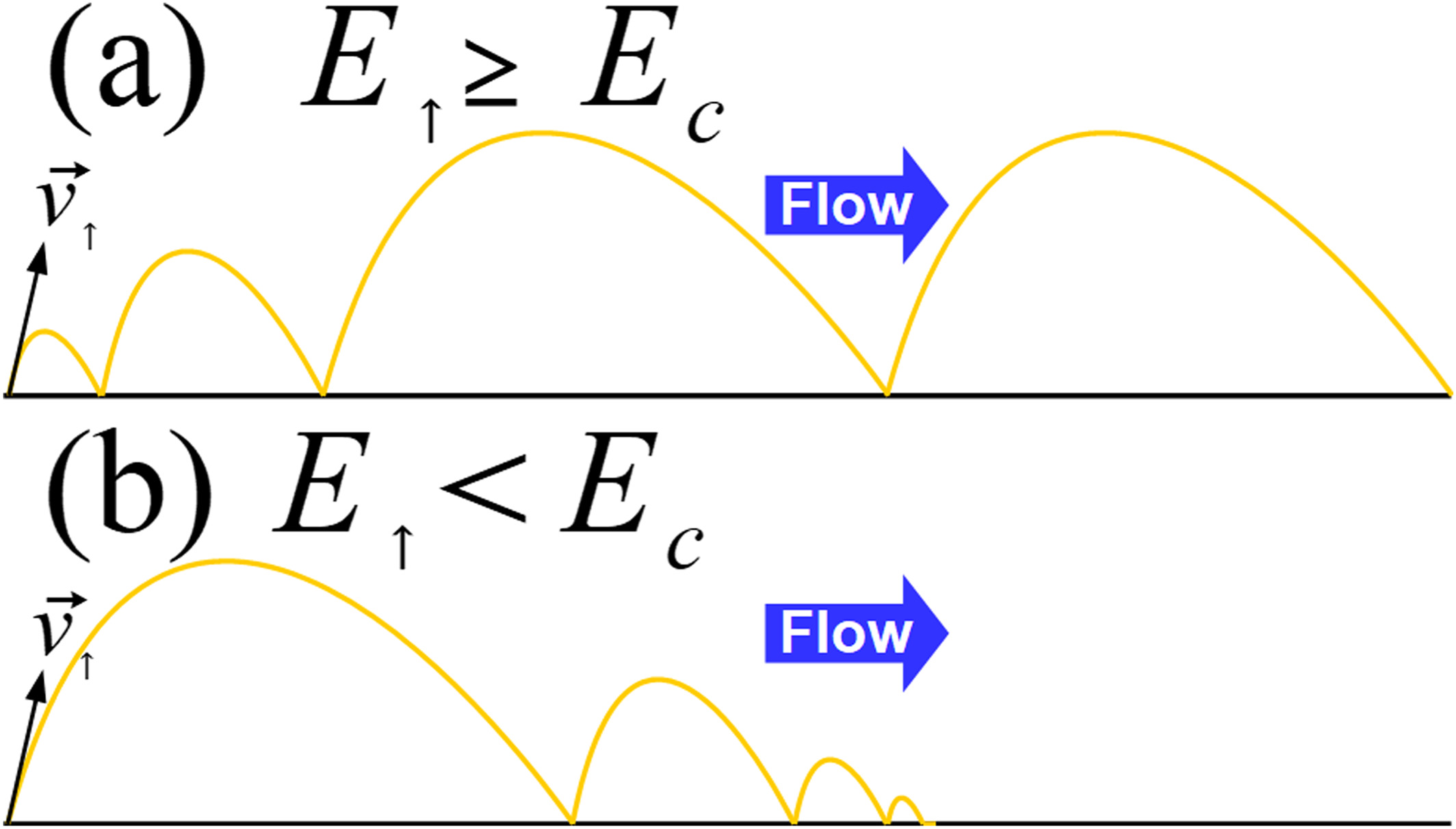}
  \caption{Sketch of the continuous rebound mechanism: the motion of a particle (yellow lines) can either lead to periodic hopping motion (when the kinetic energy gained after each impact $E_{\uparrow}$ compensates the energy lost during the collision $E_c$) or to the particle stopping on the surface after a few collisions (when the energy loss during each rebound $E_c$ overcomes the energy gained during each hop $E_{\uparrow}$). Reprinted with permission from \cite{pahtz2020physics}. Copyright 2020, John Wiley and Sons.}
  \label{fig:fig_pahtz_2020_cont_rebound}
 \end{subfigure}
 
 \vspace{10pt}
 \begin{subfigure}[c]{0.99\textwidth}
  \centering
  \includegraphics[width=0.9\textwidth, trim=0cm 0cm 0cm 0cm, clip]{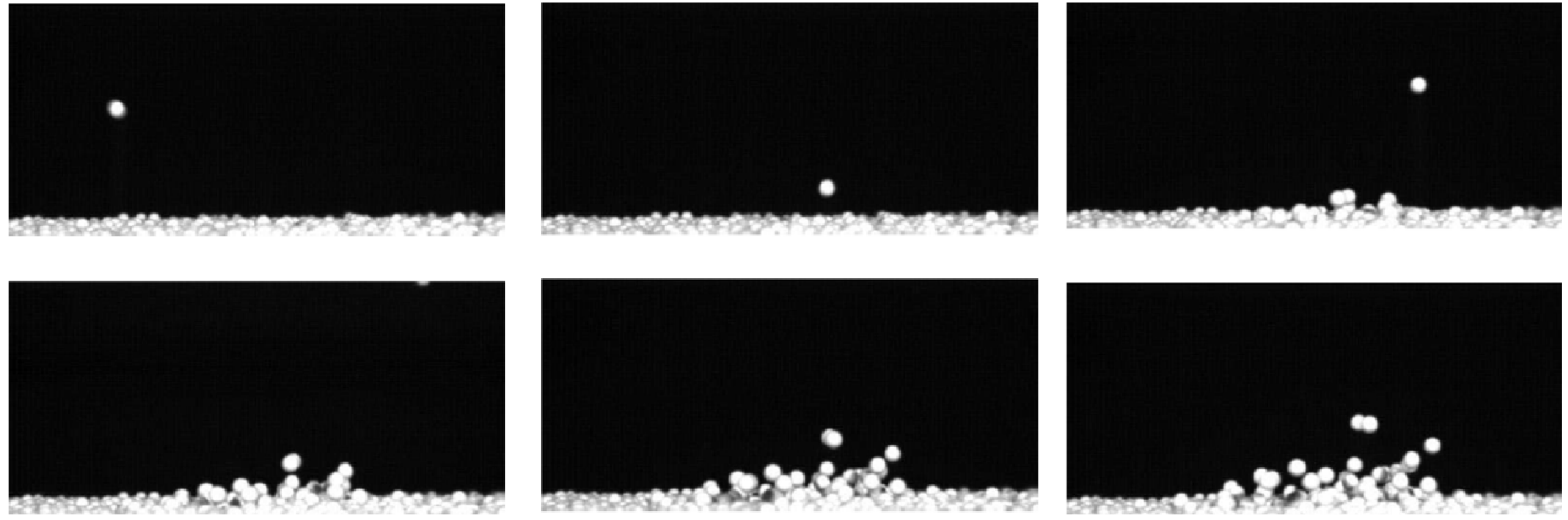}
  \caption{Consecutive snapshots of \SI{6}{mm} PVC beads showing how impacting particles can induce splashing events (i.e., multiple deposited particles are resuspended). Reprinted with permission from \cite{beladjine2007collision}. Copyright 2007, American Physical Society.}
  \label{fig:fig_beladjine_2007_splash}
 \end{subfigure}
 \caption{Illustrations of the fundamental mechanisms at play in particle resuspension from multilayer deposits.}
\end{figure}

Third, decades of experimental observations on resuspension from complex multilayer systems have been performed within the fluvial and aeolian transport communities. Recently compiled in a review paper \cite{pahtz2020physics}, these data provide a unified description of the resuspension process in both aeolian and fluvial research. In particular, they emphasize that, when the flow intensity is small enough, only a few particles located on the top of the deposit start moving over relatively short distances. This gives rise to the so-called creep motion (where moving particles remain in contact with the rest of the deposit). According to the current terminology given in Section~\ref{sec:physics:application:terminology}, this type of motion is thus a sub-category of surface migration. Such resuspension events are actually intermittent (due to near-wall turbulence) and uncorrelated to each other. In that sense, this is similar to the hydrodynamic-induced resuspension over sparse monolayers, the only difference being the type of surface upon which the particles are moving. At higher fluid velocities, particles start displaying long-lasting rolling/sliding/hoping motion over the surface since near-wall fluid velocities are large enough to counteract the potential loss of energy that can happen upon colliding with the surface (for reptating particles), leading to the so-called continuous rebound mechanisms sketched in Fig.~\ref{fig:fig_pahtz_2020_cont_rebound}. During such sustained motion, the dynamics of particles becomes more and more correlated as the particles are moving closer to each other. Finally, when the flow intensity is even stronger, the kinetic energy of particles impacting the surface becomes high enough to trigger the resuspension of other deposited particles. In fact, resuspension becomes sustained by the impaction of moving particles on the surface, which can lead to splashing events (where several particles are resuspended following the collision of a single one, as displayed in Fig.~\ref{fig:fig_beladjine_2007_splash}). This has been referred to as a threshold for impact entrainment, in the sense that new resuspension events are continuously triggered by particles impacting the surface due to their inertia. In that case, the resuspension events are largely correlated, resulting in a continuous transport of the upper layers of the bed. 

\begin{figure}[ht]
  \centering
  \includegraphics[width=0.85\textwidth]{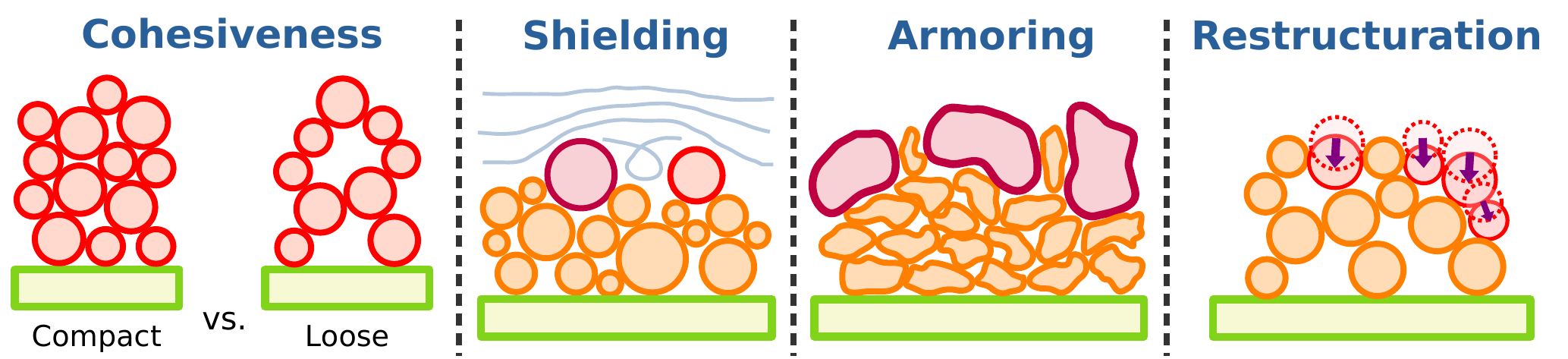}
  \caption{Illustration of the possible roles of the deposit structure on particle resuspension: the strength of cohesion within clusters is related to the deposit morphology (left); upstream particles can act as shelters for other particles in their wake ($2^{nd}$ image); large and heavy particles can prevent the resuspension of smaller particles located beneath ($3^{rd}$ image); the deposit structure can evolve with time ($4^{th}$ image).}
  \label{fig:sketch_morphology}
 \end{figure}

The phenomena related to impact entrainment imply that the deposit morphology and structure should be carefully characterized. In fact, as depicted in Fig.~\ref{fig:sketch_morphology}, the deposit structure has a direct impact on: 
 
 \begin{enumerate}[a - ]
  \item \textit{Cohesion}: The geometrical arrangement of particles is responsible for the existence of cohesion forces within the deposit whose intensity vary depending on the loose or compact nature of the deposit. Indeed, at the particle level, the number of contact of a single particle with its neighbors (which is referred to as the coordination number) varies depending on the local arrangement of nearby particles. This typically makes loose deposit easier to remove than compact deposits. 
  \item \textit{Shielding effects}: This happens when a given particle is in the wake of another one placed upstream, thereby reducing its exposure to the fluid flow (see Fig.~\ref{fig:sketch_morphology}). Such shielding effects can become significant when the upstream particle/cluster has a size larger than, or comparable to, the Kolmogorov scale, meaning that the fluid motion is modified by its presence. 
  \item \textit{Armoring effects:} This occurs when a layer of particles harder to resuspend is located on top of the deposit, thereby protecting layers located underneath (even if they contain particles easier to resuspend). Such effects have been observed when dealing with particles having a range of sizes (i.e., polydisperse) \cite{mooneyham2018deposition}. In such cases, since large particles are harder to remove than small ones, the deposit structure evolves with time as follows: the small particles are quickly resuspended, leaving mostly large particles in the upper layer of the deposit after a finite time. At that point, the number of resuspension events drops since large particles act as a shelter for all the particles located underneath.
  \item \textit{Restructuration:} This can occur when a deposit is exposed to a strong enough flow, leading to the rearrangement of particles within the deposit. Restructuring can help form more compact deposits over time, meaning that particles become harder to remove due to the increased cohesion within the deposit. Such restructuring has been mostly characterized in the context of nanoparticle aggregates due to its impact on the fragmentation process (see for instance \cite{weber1997situ, eggersdorfer2010fragmentation}). 
 \end{enumerate}
 
\begin{figure}[ht]
 \centering
 \includegraphics[width=0.90\textwidth,trim=0cm 0cm 0cm 0cm,clip]{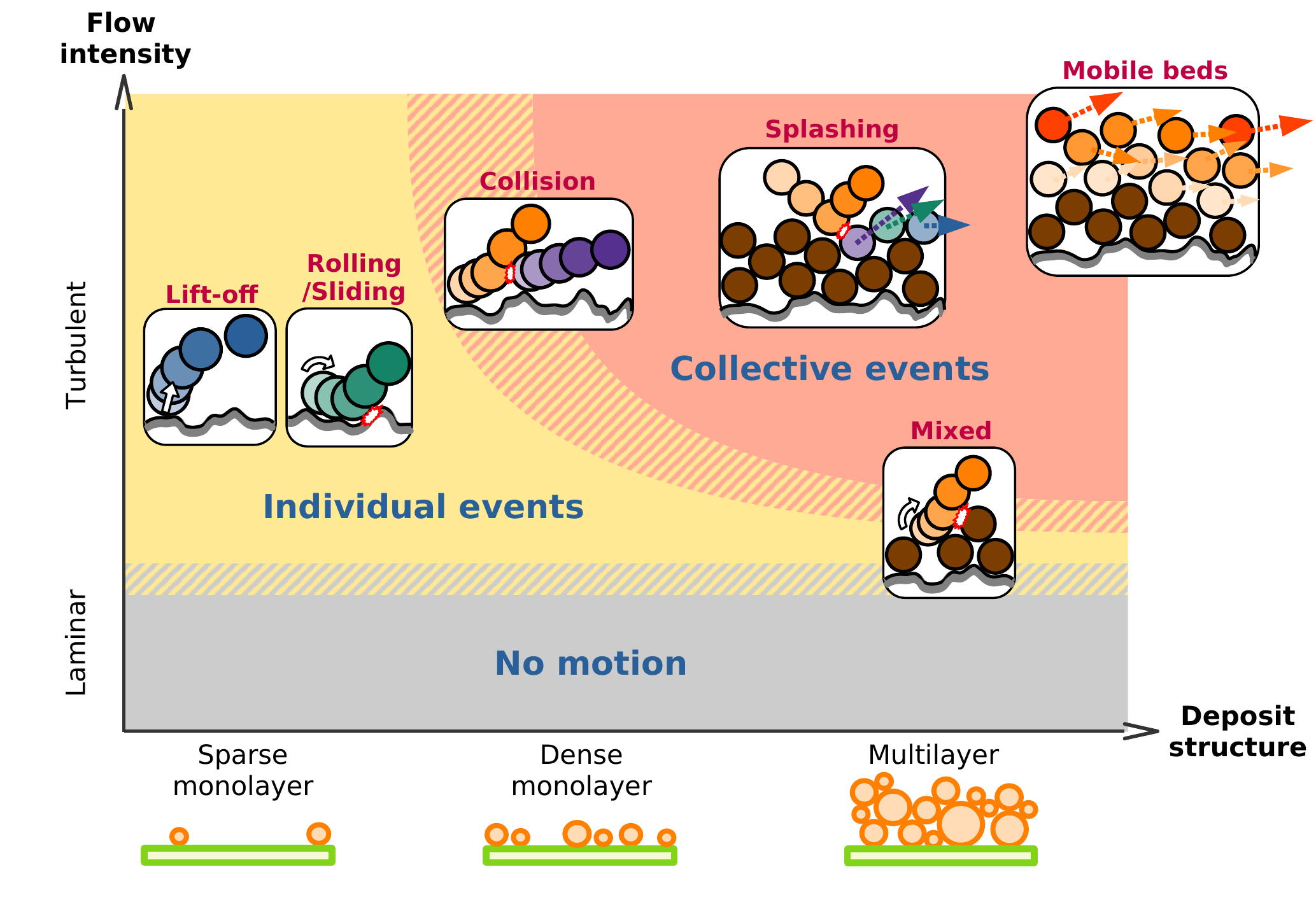}
 \caption{Sketch showing how the various mechanisms at play in particle resuspension evolve as a function of the deposit structure and flow intensity. For the sake of simplicity, this illustration considers a fixed particle size, shape and density (which could be added to form N-dimensional graphs instead of this 2D representation).}
 \label{fig:sketch_mechanisms}
\end{figure}

\subparagraph{Overall picture of resuspension mechanisms:} Drawing on the similarities between the phenomenology of resuspension from monolayer and multilayer deposits, a new overall picture is now proposed. This is sketched in Fig.~\ref{fig:sketch_mechanisms}, which represents the mechanisms at play in particle resuspension as a function of both the deposit structure (monolayer versus multilayer deposits) and of the flow intensity. In this attempt towards a unified description of particle resuspension, the main types of incipient motion for particles (namely rolling, sliding or lifting) appear as predominant in the case of sparse monolayers, where particles are so distant from each other that each event is independent from other ones (i.e., we are dealing with uncorrelated events). As the density of deposited particles is increased, these types of incipient motion are still relevant but the subsequent motion of particles on/near the surface can trigger additional resuspension events due to inter-particle collisions. In the case of multilayer systems, the impaction of particles with a high enough inertia can result in more complex splashing events (where several particles are resuspended). When the resuspension becomes highly correlated (due to a combination of hydrodynamic-induced events and sustained collision-induced effects), the upper layers of the deposit can display complex collective motion. 

At this stage, it is worth mentioning that Figure~\ref{fig:sketch_mechanisms} only pictures the evolution of the type of mechanisms as a function of the deposit structure and of the flow intensity. In reality, it also depends on the particle properties (size, shape, density). For the sake of clarity, only a 2D sketch has been displayed here but similar drawings can be made to show the evolution with respect to other parameters (like the particle size, shape, fluid to particle density ratio).


 \section{Estimating resuspension: experimental measurements and quantities}
 \label{sec:techniques}

Having introduced the basic physical concepts behind the resuspension process in the previous Section, the purpose of the present section is to describe how particle resuspension can be estimated. To that effect, the various quantities used to quantify particle resuspension are first introduced in Section~\ref{sec:technique:quantities}. Then, the experimental techniques developed to measure such quantities are reviewed in Section~\ref{sec:technique:measurement}. 

 \subsection{Indicators for particle resuspension}
  \label{sec:technique:quantities}

When trying to quantify particle resuspension, two main questions appear:
\begin{itemize}
 \item What are the physical quantities that characterize the resuspension process?
 \item What is the information contained in these quantities (especially the level of information)? How are they related to each other and how can they be interpreted?
\end{itemize}
The aim of the next paragraphs is to set forth a list of the physical quantities used to measure resuspension (see Section~\ref{sec:technique:quantities:variable}), then to analyze the meaning / level of information contained within these quantities (see Section~\ref{sec:technique:quantities:interp}).

  \subsubsection{Physical quantities of interest}
  \label{sec:technique:quantities:variable}

A basic, yet efficient, way to characterize resuspension can be inferred from its definition. Recalling that resuspension corresponds to the process by which particles adhering to a surface are detached and then entrained away through the action of a flow, a crude quantification would consist in estimating the amount of particles that are detached from a surface. In practice, this can be achieved by monitoring the evolution of various quantities with time, such as: the number of particles deposited on a surface, the motion of particles near the surface, or even the amount of detached particles that are collected downstream of a deposit. In the literature, this translates into a number of quantities listed in the following (see also Fig.~\ref{fig:sketch_indicators}):

\begin{figure}[ht]
 \centering
 \includegraphics[width=0.95\textwidth]{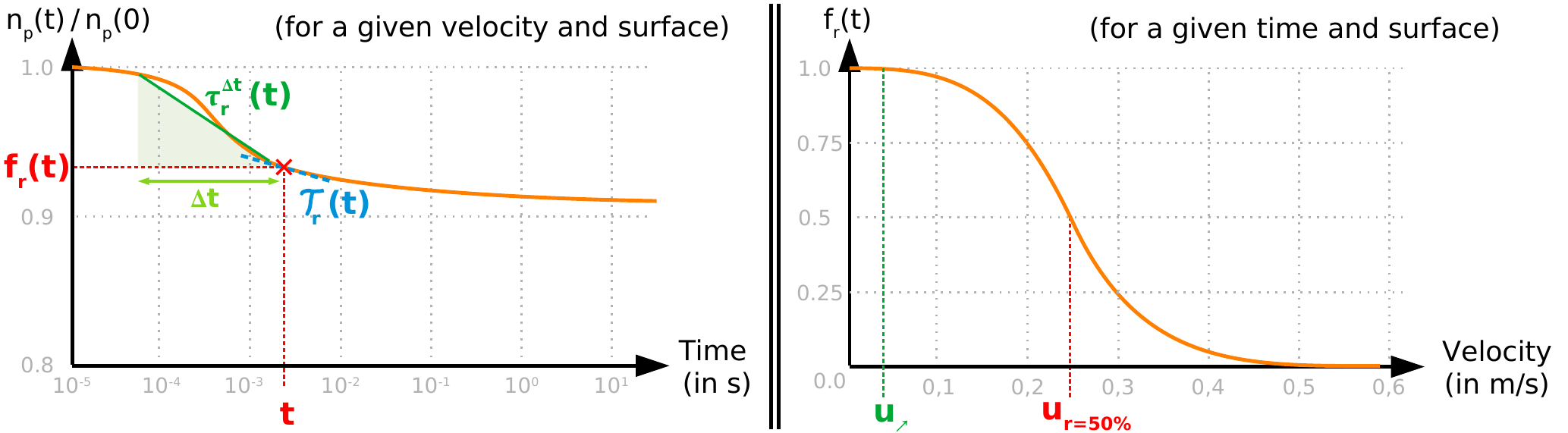}
 \caption{Sketch summarizing some of the physical quantities of interest in particle resuspension.}
 \label{fig:sketch_indicators}
\end{figure}

\begin{itemize}
 \item Quantities based on surface fluxes/rates:
 
 To quantify resuspension, the first idea that comes to mind is to measure the flux of particles leaving the surface. For that purpose, one can count the number of deposited particles $N_{\rm p}$ on a given surface $\mc{S}$ before and after exposure to a given fluid flow `intensity' $\mc{F}$ for a certain time $t$. The notion of the fluid flow intensity actually encompasses various properties (such as the friction velocity, the shear rate, the flow rate or the turbulent kinetic energy). By comparing this number to the initial amount of deposited particles $N_{\rm p}(t=0)$, one obtains the number fraction $n_{\rm p}(t) = N_{\rm p}(t)/N_{\rm p}(t=0)$ from which various quantities can be extracted, including:

 \begin{itemize}
  \item The instantaneous resuspension rate $\mc{T}_r(t)$ 
  
  This is a measure of the rate of change of the number of particles deposited on a surface per unit time (in \SI{}{s^{-1}}). It is defined by the derivative of the particle number fraction at a specified time $t$:
  \begin{equation}
   \mc{T}_r(t) = \left. \frac{\dd n_{\rm p}}{\dd t} (t)\right|_{\mc{S},\mc{F}=\text{cte}}
  \end{equation}

  Alternatively, it can be expressed as a mass rate (i.e., resuspended mass per unit time, in \SI{}{kg.s^{-1}}), or even as a flux (i.e., the resuspended mass per unit time and per unit surface area, in \SI{}{kg.m^{-2}.s^{-1}}). 
  
  \item The time-averaged resuspension rate $\tau_r^{\Delta t}(t)$
  
  This corresponds to the rate of change of the particle number fraction due to resuspension over a certain time interval $\Delta t$ (in \SI{}{s^{-1}}):
  \begin{equation}
   \tau_r^{\Delta t}(t) =  \left. \frac{n_{\rm p}(t+\Delta t) - n_{\rm p}(t)}{\Delta t} \right|_{\mc{S},\mc{F}=\text{cte}}
  \end{equation}
  It is usually called the resuspension rate (without referring to the averaging process over a time interval $\Delta t$), and is related to the instantaneous resuspension rate by simple integration:
  \begin{equation}
   \tau_r^{\Delta t}(t) =  \frac{1}{\Delta t} \ \int_{t}^{t+\Delta t} \mc{T}_r(s) ds
  \end{equation}
  
  Alternatively, as for the instantaneous resuspension rate, it can be expressed as a mass rate (in \SI{}{kg.s^{-1}}) or as a flux (in \SI{}{kg.m^{-2}.s^{-1}}).
  
  \item The resuspended fraction $f_r$:
  
  This refers to the fraction of particles that are resuspended after a given exposure time $t$ \cite{ibrahim2004microparticle}:
  \begin{equation}
   f_r(t) =  \left. n_{\rm p}(t) \right|_{\mc{S},\mc{F}=\text{cte}}
  \end{equation}
  It can also be referred to as the release fraction \cite{rossi2021numerical}, the resuspension efficiency \cite{keedy2012removal}, or sometimes the resuspension factor \cite{ren2022experimental}.
  
  \item The remaining fraction $f_l$:
  
  This indicates the fraction of particles left on the surface after a given exposure time $t$. It is the opposite of the resuspended fraction:
  \begin{equation}
   f_l(t) = 1-f_r(t)
  \end{equation}
  
  \item Median critical velocity $u_{r=50\%}$:
  
  A median critical velocity is sometimes defined as the fluid velocity at which \SI{50}{\%} of the particles are removed from the initial deposit \cite{mikellides2020modelling, mikellides2020experiments}. This means that:
  \begin{equation}
   f_r(u_{r=50\%}) = 50 \%
  \end{equation}
  More generally, using the notion of quantiles, it is possible to define a critical velocity $u_{r=k\%}$ at which $k$ percent of the particles have been removed from the initial deposit after a fixed exposure time $t$ (i.e., such that $f_r(u_{r=k\%})=k$). 
  
  In most experiments, this fluid velocity corresponds to the friction velocity (also called the shear velocity) $u_{\star}$, which is related to the shear stress near the surface $\tau_{\rm p}$ through $u_{\star} = \sqrt{\tau_{\rm f}/\rho_{\rm f}}$ ($\rho_{\rm f}$ is the fluid density). Alternatively, critical shear stresses or critical Reynolds numbers can also be defined instead of relying on a criteria based on the fluid velocity.
  
 \end{itemize}
  
 \item Quantities based on particle motion:
  
 Another way to detect resuspension is to record the motion of particles initially deposited on a surface. This can be achieved using high-speed cameras (see the measurement techniques described in Section~\ref{sec:technique:measurement:protocol}). By analyzing the images acquired, the following quantities can be extracted:
 
 \begin{itemize}
  \item Threshold fluid velocity for onset of motion $u_{\nearrow}$
  
  This corresponds to the fluid velocity at which a particle starts moving. To put it differently, it is the extremum (minimum) value of the fluid velocity which sets a particle in motion \cite{charru2007motion}. Similarly to the median critical velocity $u_{r=50\%}$, most experimental studies provide information on the shear velocity $u_{\star}$. Hence, this threshold criteria $u_{\star,\nearrow}$ is largely based on the shear velocity at which a particle starts moving and is expressed as follows: 
  \begin{equation}
   \|\mathbf{U}_{\rm p}(u_{\star,\nearrow})\| > 0
  \end{equation}
  where $\mathbf{U}_{\rm p}$ is the particle velocity. This quantity is also called the threshold velocity for incipient particle motion. 
  
  \item Median threshold fluid velocity for onset of motion $u_{\nearrow=50\%}$
  
  When studying a large number of particles, the velocity at which each particle starts moving can vary from one particle to another. This can be due to variations in the instantaneous hydrodynamic force, to local differences in adhesive forces or in the particle shape and size. As a result, similarly to the median critical velocity $u_{r=50\%}$, one can define a median threshold velocity at which \SI{50}{\%} of the particles start moving $u_{\nearrow=50\%}$.
  
  More generally, using the notion of quantiles, it is possible to define a threshold velocity $u_{\nearrow=k\%}$ at which $k$ percent of the particles start to move. 
  
  \item Particle velocity on/near the surface $\mathbf{U}_{\rm p}$
  
  Recent advances in experimental techniques allow to extract directly information on the particle velocity and acceleration throughout time. By doing so, it becomes possible to distinguish between surface migration (such as rolling/sliding motion that can occur during creeping) and particles undergoing hops/jumps above the surface (such as saltating particles in aeolian research \cite{sehmel1980particle}). In the latter case, one can even extract statistical information on saltation events (like the average saltation time, distance or kinetic energy upon re-impact on the surface).
  
 \end{itemize}
 
 \item Quantities based on volume concentrations/fluxes:
 
 Another method for estimating resuspension is to measure the concentration of particles downstream of a deposit (in close proximity to it). By doing so, the following quantities can be extracted:
 
 \begin{itemize}
  \item Resuspension factor $\mc{R}_f$:
  
  This is a measure of the volumetric to surface (or soil) concentration of particles ( in \SI{}{m^{-1}}). To put it differently, it is defined as the ratio between the airborne mass concentration of particles sampled in a region near the surface $\rho_{\mc{V}}$ and the surface (mass) concentration of particles initially on the surface $\rho_{\mc{S}}$ \cite{stewart1964resuspension, nicholson1988review}. Therefore, it provides an estimate of the concentration of resuspended particles in the vicinity of a deposit conditioned on the amount of particles initially present on the surface.
  
  \item Flux of moving particles $\Phi_{\rm p}$:
  
  By sampling the particle concentration at various distances from the initial deposit, it is possible to reconstruct the flux of particles around the deposit \cite{harris2009monte} (as in airborne concentration measurements). Hence, such measurements provide information on the average particle motion along the streamwise, wall-normal, and tangential directions. 
  
  These fluxes can be expressed either as mass fluxes (based on the mass concentration in \SI{}{kg.m^{-3}}), as volumetric fluxes (based on the volume concentration) or as numeral fluxes (based on number concentration in \SI{}{m^{-3}}). When dealing with resuspension induced by human activities (e.g., dancing, walking, driving), these fluxes can also be related to the intensity of the activity \cite{licina2018clothing, ren2022experimental}.
  
 \end{itemize}
 
\end{itemize}

In the rest of the paper, unless explicitly specified, the terminology ``resuspension rate'' and ``resuspended fraction'' is used to refer to measurements based on numeral fluxes. It is worth noting that these numeral rates and fractions are the same as those obtained based on mass flux or area flux when dealing with monodispersed particle sizes. However, significant differences would appear when dealing with polydispersed particle sizes. This immediately brings out the following question: which quantity is best suited to study particle resuspension? The next paragraphs provide some guidelines to help selecting a relevant quantity based on the context and objectives of each study.

 \subsubsection{Usage and interpretation of these quantities}
 \label{sec:technique:quantities:interp}
 
When choosing the relevant quantity to characterize particle resuspension in a given study, one has to keep in mind the constraints coming from the objectives of that study. In particular, this choice is driven by a combination of factors in relation with the level of information required as well as existing technical restrictions (e.g., in-situ versus laboratory measurements). In the following, we illustrate some of the choices made by presenting the evolution of the measured quantities from a historical point of view.

\begin{figure}[ht]
 \centering
 \includegraphics[width=0.48\textwidth]{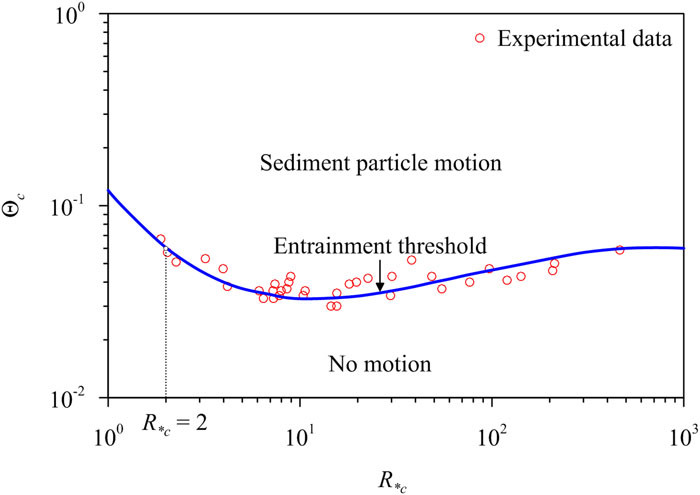}
 \caption{Shields diagram showing the evolution of the critical shear stress $\Theta_c$ (or Shields number) as a function of the particle-based Reynolds number (here $R_{\star_c}$). Reprinted with permission from \cite{dey2018advances}. Copyright 2018, AIP Publishing.}
 \label{fig:fig_dey_2018_shields}
\end{figure}

\subparagraph{Threshold velocities:} The pioneering studies carried out in the early 1940's on river transport \cite{shields1936anwendung} and aeolian transport \cite{bagnold1937transport} were conducted by measuring the threshold fluid velocity for incipient motion $u_{\star,\nearrow}$. At that time, this threshold velocity provided the first insights into the necessary condition for particle motion and this helped to bring out the following picture: a particle starts moving when the forces inducing motion exceed the resisting ones \cite{pahtz2020physics}. Ultimately, this led to the introduction of the well-known Shields number $\Theta_c$ \cite{shields1936anwendung}, defined as:
\begin{equation}
 \Theta_c = \frac{\tau_{\nearrow}}{(\rho_{\rm p}-\rho_{\rm f})\ g\ d_{\rm p}}
\end{equation}
where $\tau_{\nearrow}$ is the shear stress for the incipient motion of a particle, $\rho_{\rm f}$ (resp. $\rho_{\rm p}$) the fluid density (resp. the particle density), $g$ the gravity constant, and $d_{\rm p}$ the particle diameter. The Shields number measures the ratio between the fluid shear stress needed to move a particle and the particle's immersed weight. By plotting the evolution of the Shields number with respect to the particle-based Reynolds number $Re_{\rm p} = u_{\tau} d_{\rm p} / \nu_{\rm f}$, one obtains a master curve, the so-called Shields diagram (see Fig.~\ref{fig:fig_dey_2018_shields}).

The main difficulty is that the threshold for incipient motion is not exactly the same for two identical particles exposed to a given fluid flow intensity. This is due a combination of two factors: first, motion-inducing hydrodynamic forces fluctuates in time and space (especially due to the intermittency of turbulent flows \cite{lavelle1987critical}); second, motion-preventing forces depend on the local configuration (roughness features on the surface at scales much smaller than the particle size, as well as the deposit structure in multilayer deposits). As a result, the threshold value for the incipient motion of identical particles is not unique but exhibits instead a range of values, even when fluid flow characteristics and particle properties are fixed. Therefore, this questions the significance of a threshold value for a group of identical particles exposed to a similar flow (see also \cite{pahtz2020physics}). At this point, since no unique value exists, one can consider statistical quantities, such as the median threshold value $u_{\nearrow=50\%}$ (or more generally any quantile threshold value $u_{\nearrow=k\%}$). However, regardless of the threshold value chosen, one has to remember that such quantile threshold values do not provide information on the whole distribution of values where particles start moving. Hence, by resorting to such threshold values, some of the information about resuspension is definitely lost. 

Another issue corresponds to the time-dependence of these threshold values. In fact, the number of resuspended particles changes with time in turbulent flows. This is due to the intermittency in turbulence and especially to near-wall turbulent structures, which can lead to particles being exposed to intense structures for a short period of time. This means that particles with the lowest motion-preventing forces tend to remain deposited on the surface only for a short time. Meawhile, particles with the highest motion-preventing forces can remain deposited for much longer times. To avoid the dependency of these threshold values on the observation time, one can measure the number of particles set in motion after an observation time long-enough to reach a stationary value. This is all the more relevant since several studies have shown a quick and intense resuspension at short-term (where all loosely deposited particles are resuspended) followed by a much slower resuspension at longer times (with a resuspension rate decaying with time) \cite{reeks1988resuspension, brambilla2018glass}, as displayed in Fig.~\ref{fig:sketch_indicators} (left panel).

As more phenomena related to particle resuspension were unveiled by these pioneering studies, new threshold criteria emerged. For instance, a distinction was introduced between the incipient motion and the cessation of motion (i.e., when particles stop moving as the fluid velocity decreases). In fact, it is observed that, due to their inertia, resuspended particles continue their motion even when the fluid velocity is decreased below the threshold for incipient motion. As a result, this leads to a sort of hysteresis effect. More recently, specific thresholds have been suggested to distinguish between rolling, sliding, and lifting motion \cite{dey2018advances}. In the case of multilayer deposits, another threshold has been introduced: the threshold for impact entrainment, which corresponds to the velocity at which saltating particles induce extra resuspension upon colliding with the deposit \cite{martin2018distinct}. This threshold allows to quantify when collisions between particles result in enhanced resuspension rates, similar to collision propagation effects observed in avalanche phenomena. Recently, it has even been argued that most of the so-called Shields diagrams do not plot the evolution of the threshold for incipient motion but rather the threshold for impact entrainment \cite{pahtz2020physics}. This is attributed to the measurement protocol used but the actual phenomena involved depend also on the definition of the threshold value. To be more specific, let us consider a multilayer bed composed of a few hundreds of sediments resting on a bed and exposed to a fixed flow rate. For the \SI{10}{\%} quantile threshold $u_{\nearrow=10\%}$, a few particles set in motion are enough to reach that quantile. In that case, particle motion is essentially triggered through individual, and independent, events. This means that the threshold measured correspond to the incipient motion of individual particles. However, if one considers the median threshold, many more particles must bet set in motion to reach this threshold: this is probably achieved by a combination of individual resuspension events and collision-induced effects. Therefore, the choice of a given threshold value leads to different information being accessible. In other words, depending on the selection of a given threshold value, we may not measure the same resuspension mechanism.

In spite of these issues, researchers in fluvial and aeolian transport are still relying on threshold values which are easy to use and remain convenient in many practical applications. This explains why, since the pioneering work of Shields, many studies have evaluated how such threshold values depend on fluid properties (such as its density or velocity) and on particle properties (e.g., diameter and density). 

\begin{figure}[ht]
 \centering
 \includegraphics[width=0.55\textwidth]{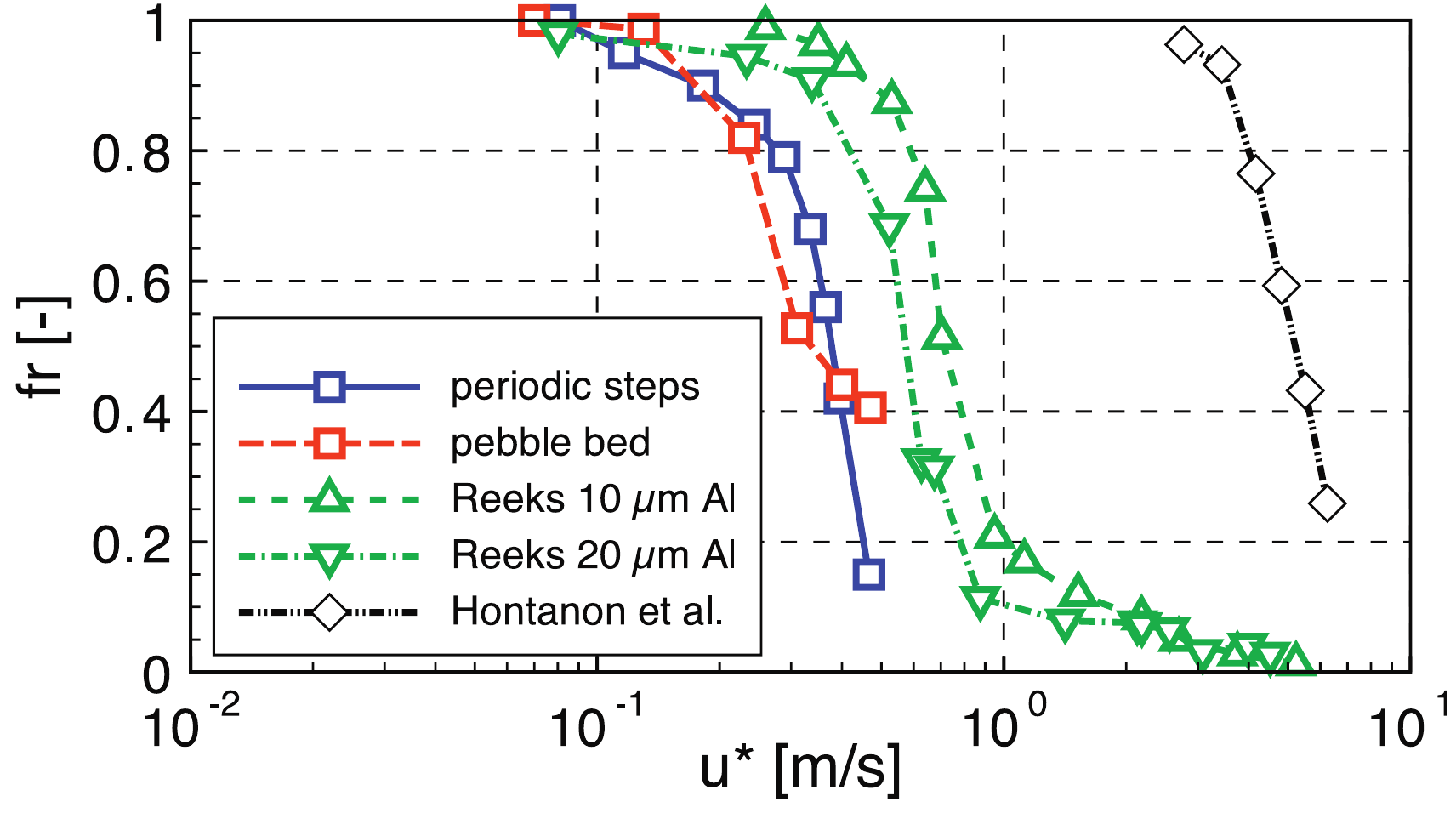}
 \caption{Comparison of the fraction of particle remaining on the surface $fr$ against the fluid friction velocity $u^*$ for various experiments. It shows the typical decay of $fr$ as the velocity increases (with a sharp decrease near the median velocity $u_{\rm r=50\%}^*$). Reprinted with permission from \cite{barth2015particle}. Copyright 2015, Elsevier.}
 \label{fig:fig_barth_2015_fremain}
\end{figure}

\subparagraph{Surface fluxes/rates:} By counting/weighting particles on a given surface (see Section~\ref{sec:technique:measurement:protocol} for measurement techniques), one can determine the number/mass of particles remaining on the surface after being exposed to a given flow for a certain amount of time (see the left panel of Fig.~\ref{fig:sketch_indicators}). Then, by fixing the observation time $\Delta t$, one can reconstruct the evolution of the remaining fraction $f_r(t)$ as a function of the fluid velocity (see also the right panel of Fig.~\ref{fig:sketch_indicators}). As displayed in Fig.~\ref{fig:fig_barth_2015_fremain}, surface fluxes are widely used in the multiphase flow community (see for instance \cite{henry2014progress, ziskind2006particle} and references therein). 

The main interest of these measures is to provide information on the whole distribution of velocities required to resuspend particles from a substrate. This information is complementary to the one contained in threshold values for incipient motion. In the case where all particles that start moving get resuspended, the threshold value for incipient motion can actually be extracted from the evolution of the fraction remaining with the fluid velocity (as shown in the right panel of Fig.~\ref{fig:sketch_indicators}). The information can be further simplified to give a critical velocity (like $u_{r=50\%}$), which is commonly used in studies dealing with potentially harmful particles (like pollen, allergens, radioactive particles) \cite{qian2014walking}. 

Yet, surface fluxes/rates should be handled carefully due to a number of issues:
\begin{itemize}
 \item These quantities depend both on the observation time $\Delta t$ and on the time $t$. More precisely, this means that the time-averaged resuspension rate $\tau_r^{\Delta t}(t)$ changes with time when the observation time $\Delta t$ is fixed, and that it also varies with the observation time $\Delta t$ for a fixed time $t$. In fact, laboratory experiments have shown that, when a fresh deposit is exposed to a constant flow, a large quantity of particles is resuspended from the surface over a very short period of time (typically less than a second, as sketched in the left panel of Fig.~\ref{fig:sketch_indicators}). This intense short-term resuspension rate is attributed to the resuspension of all particles that are loosely adhering to the surface and, hence, are easily resuspended by the flow. At longer exposure times, the number of particles being resuspended is much lower since most of the particles likely to be resuspended have already left the surface (in reality, long-term resuspension is related to the interaction with high velocity turbulent structures). Experimental measurements have indicated that this long-term resuspension rate is usually inversely proportional to time \cite{reeks1988resuspension}. 
 
 This distinction between short-term and long-term resuspension is widely used in practice. In fact, after a few seconds of exposure to a given flow, the long-term resuspension rate is reached, meaning that the number of resuspended particles evolves very slowly. This gives the impression that a nearly steady state is reached once all the particles that can resuspend at this fluid velocity are gone (see right panel of Fig.~\ref{fig:sketch_indicators}). Drawing on such observations, measurements in the multiphase flow community are usually performed by evaluating this nearly steady-state value \cite{barth2014single} (i.e., typically after a few seconds of exposure to a given flow).
 
 \item Surface fluxes/rates do not necessarily provide information on the type of motion of each individual particle. Hence, the only information that can be extracted is the total amount of particle resuspended, regardless of the mode involved (e.g., rolling, sliding, or lifting). More recently, new experimental techniques have allowed to classify particles with respect to their mode of motion \cite{agudo2017detection}: this allows to precisely know how much particles are resuspended through rolling, sliding or lifting motion.
 
 \item Another issue is that the value obtained can overestimate the real resuspension rate. In fact, unless specific techniques are used, it is impossible to distinguish between particles that have been resuspended from the surface (i.e., detached from the surface) and migrating particles that have left the measurement area (e.g., through rolling or sliding motion). Therefore, when counting techniques are used, all particles leaving the observation area are counted as resuspended particles (including migrating ones).
 
\end{itemize}

\begin{figure}[ht]
 \centering
 \includegraphics[width=0.95\textwidth, trim=0cm 3.5cm 0cm 19.1cm,clip]{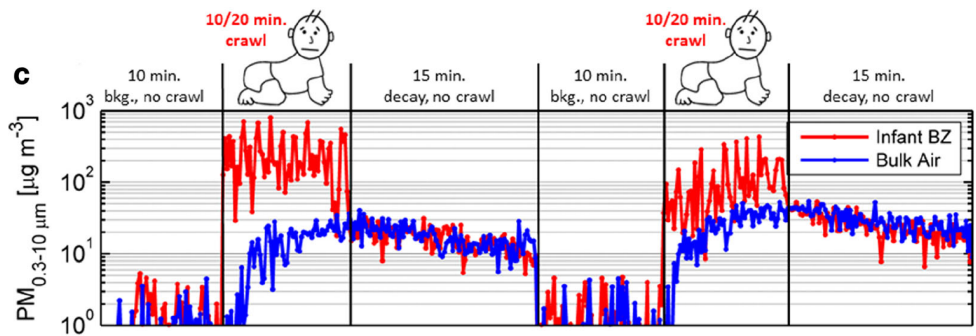}
 \caption{Evolution of the mass concentrations of particulate matter ((PM$_{0.3-10\ \mu m}$, i.e., with sizes between 0.3 and \SI{10}{\mu m}) monitored in the infant breathing zone (BZ) and in the bulk air (at \SI{1.5}{m} above ground). It shows the increase in the concentration when an infant crawls on a carpet and that the concentration can be up to one order of magnitude higher in the infant breathing zone than in the bulk air region. Reprinted with permission from \cite{hyytiainen2018crawling}. Copyright 2018, Springer Nature.}
 \label{fig:fig_hyytiainen_2018_crawling}
\end{figure}

\subparagraph{Volume fluxes:} The main interest of these quantities is that they deliver information on where the resuspended particles can be transported once they are detached from the surface. For instance, by placing sensors at a target height representative of an adult person, one obtains direct access to the exposure of an individual to potentially harmful particles (e.g., pollution from roads \cite{rienda2021road}, pollen from plants, spores from soil \cite{kim2010source}, dust from clothing \cite{ren2022experimental}). A typical result is displayed in Fig.~\ref{fig:fig_hyytiainen_2018_crawling}: it shows the evolution of the concentration of particles in the bulk air (i.e., far above the floor) and in the breathing zone of a crawling infant as a function of time and human activities. It appears that the particle concentration increases by two orders of magnitude in the infant breathing region when infants are crawling on the carpet. Meanwhile, it only increases by one order of magnitude when sampled at \SI{1.5}{m} from the floor (labeled here bulk air but it can also be seen as the adult breathing zone).

This simple example actually brings out the difficulties encountered when using such quantities:
\begin{itemize}
 \item Since particles are often naturally present in environmental flows, the actual information of interest is not exactly to measure a volume concentration but rather variations of this volume concentration as a function of time or environmental conditions (e.g., wind speed, human activities, etc.). These variations are usually obtained by comparing the values at two different times and/or two different conditions.
 
 \item Another drawback is that volume concentrations depend heavily on the measurement protocol (see Section~\ref{sec:technique:measurement}). For instance, it is evident that the concentration of particles resuspended from the floor varies with respect to the distance from the floor. Hence, sensor placement is key when designing such experiments and it is hard to compare two experiments relying on different protocols. In addition, an implicit assumption is that the measurement at a certain height represents the concentration of the plume of airborne particles from the deposit, in a top-hat fashion. The example in Fig.~\ref{fig:fig_hyytiainen_2018_crawling} challenges this assumption, which may lead to an over- or under-estimation of the amount resuspended.
 
 \item Finally, volume fluxes actually contain information on the environmental conditions that act both on particle resuspension and on their subsequent transport (e.g., terrain topology, target height, intensity of wind and/or human activities) \cite{nicholson1988review}. For that reason, it is very common to express such volume fluxes relative to the intensity of the activity \cite{licina2018clothing, ren2022experimental}. For example, in road dust resuspension \cite{ren2022experimental}, the emission factor is defined as the mass of particles resuspended due to the passage of a single vehicle per unit distance traveled by the vehicle (expressed in \SI{}{g.km^{-1}.veh^{-1}}).
  
\end{itemize}

\subparagraph{Individual particle tracking} More recently, the development of high-frequency recording techniques has paved the way for precise measurements of particle motion near the surface. While such measurements are still somewhat limited to large millimeter-size particles (see for instance \cite{agudo2017detection}), they provide very detailed information on the statistics of particle properties (such as the translational or rotational velocities, the acceleration). The main interest of such measurements is that they contain a large amount of information: it is possible to reconstruct all the previous threshold quantities and surface fluxes/rates (volume quantities can be accessed depending on the location of the last tracked particle position with respect to the initial deposit). Yet, such quantities require high capacity storage facilities due to the large amounts of data generated upon short observation times. 

 \subsubsection{Summary on physical quantities}

This overview of the physical quantities measured for particle resuspension has shown that the level of information differs widely. This ranges from the most basic information on threshold velocities for incipient motion to detailed statistical information on particle velocities. Yet, the choice of one quantity or another depends on the specific objective which is pursued: studies focused on human exposure to contaminants tend to rely on volume fluxes/rates while investigations on surface contamination prefer to use surface fluxes/rates. Some studies even consider several quantities simultaneously. For instance, in river research about sediment transport, it is frequent to measure both the threshold velocity and a transport rate \cite{furbish2021rarefied, ancey2020bedload1, ancey2020bedload2}. Similarly, studies related to airborne particle often characterize the evolution of the resuspended fraction with the friction velocity together with the median critical velocity \cite{mikellides2020experiments, qian2014walking}. In addition, the choice of a given quantity is also driven by the method of measurement, which can be suitable for either in-situ and/or laboratory studies. These techniques together with the corresponding procedure are described in the following Section~\ref{sec:technique:measurement}.

 \subsection{Measuring particle resuspension}
  \label{sec:technique:measurement}

Having introduced the physical quantities used to characterize particle resuspension, the next issue is: how can these physical quantities be measured experimentally? The aim of the following paragraphs is to describe the corresponding techniques and protocols designed to access these quantities (see Section~\ref{sec:technique:measurement:protocol}) before discussing the limitations of these experimental methodologies (see Section~\ref{sec:technique:measurement:limitations}).

 \subsubsection{Measurement techniques and protocols}
 \label{sec:technique:measurement:protocol}
 
\begin{figure}[ht]
 \centering
 \includegraphics[width=0.87\textwidth, trim=0cm 0cm 0cm 1.5cm, clip]{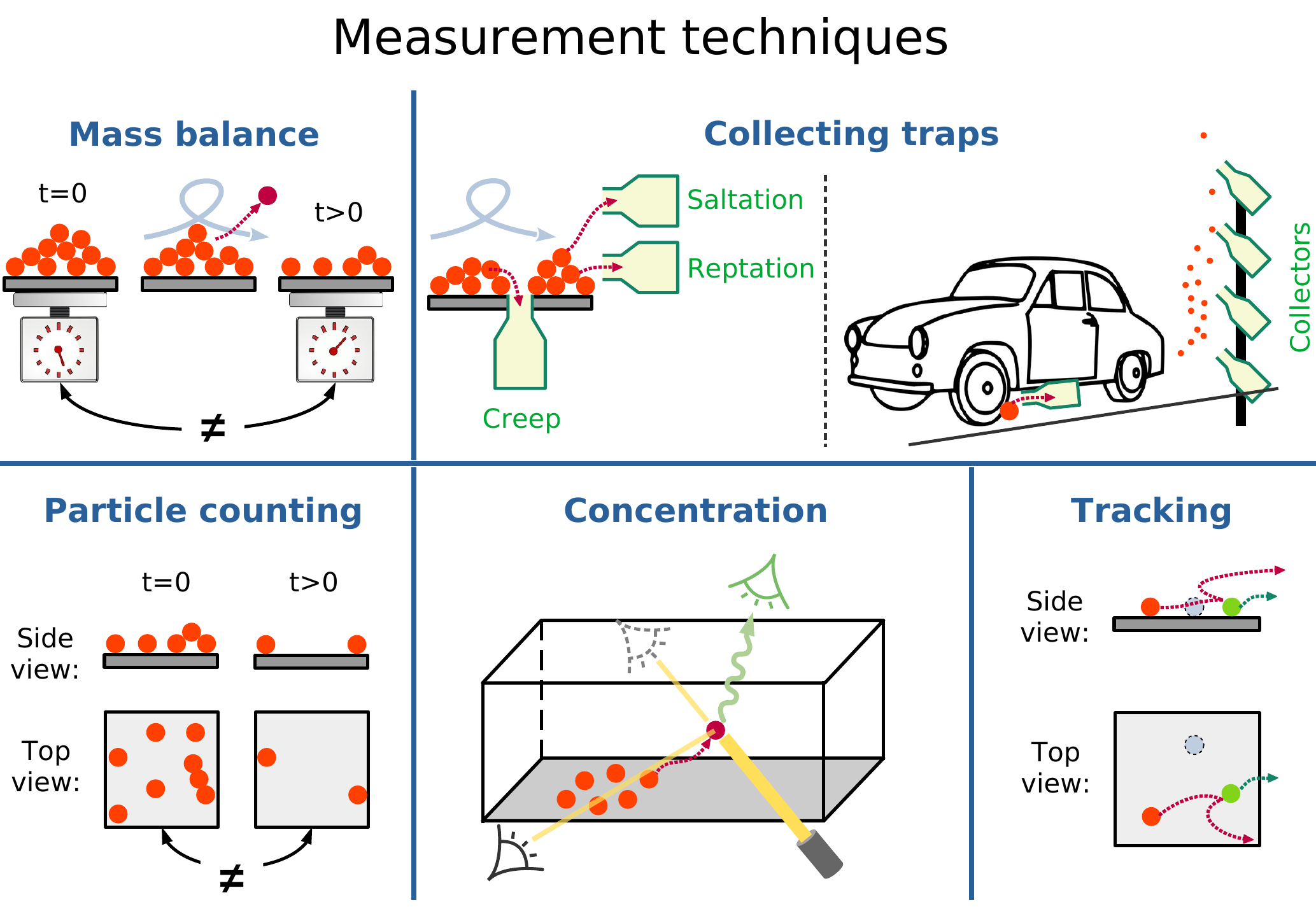}
 \caption{Sketch of the main available measurement techniques to quantify particle resuspension. It shows how mass balance, collecting traps, particle counting, volume concentration measurements and tracking techniques can be used to monitor resuspension.}
 \label{fig:sketch_measurement_techniques}
\end{figure}

Various methods to set up experimental protocols to measure resuspension factors of interest have been suggested by combining specific procedures and measurement techniques. These methods are briefly recalled in the following (interested readers are referred to detailed reviews dedicated to measurements, e.g., \cite{bloesch1994review, xu2015light}). These techniques are also illustrated in Figure~\ref{fig:sketch_measurement_techniques} and summarized in Table~\ref{tab:measurement_techniques}.

\begin{itemize}
 \item Mass-balance measurement: 
 
 \textit{Principle:} Historically, one of the first techniques consisted in measuring variations in the deposit mass before and after exposure to a given flow \cite{sehmel1973particle}. This simple procedure provides macroscopic information on the portion of materials that has been resuspended by applying a given mean flow over the deposit. Yet, it does not provide microscopic information on individual particles nor insights into the resuspension mechanisms. 
 
 More recently, the procedure has been extended to provide information not only on the amount of particles but also on their chemical composition. This is achieved by combining weighting techniques with other intrusive methods where the deposits are further analyzed. For instance, when dealing with organic compounds, chemical analysis techniques can be employed such as chromatography (which allows to separate constituents in a complex gas or liquid mixture) and mass spectroscopy. Such combinations have been used to quantify the amount of trace explosive residues resuspended from fibrous surfaces \cite{kottapalli2021aerodynamic}. 
 
 \textit{Usage:} Due to its simplicity and to its non-intrusiveness, it has been widely employed to quantify resuspension by airflows \cite{sehmel1980particle}, especially in laboratory measurement. It is still used in combination with other optical counting techniques, especially when dealing with resuspension in complex and confined environments \cite{lin2019resuspension}. 
 
 \item Collecting traps or chambers:
 
 \textit{Principle:} The idea here is to collect particles at specific locations by resorting to traps \cite{bloesch1994review}. Compared to simple mass-balance measurements, these techniques can provide some information on the motion of the particles. For instance, particles that are slowly moving along the surface (e.g., creeping motion) can be collected by traps buried within the deposit with the orifice oriented vertically (see Figure~\ref{fig:sketch_measurement_techniques}). Alternatively, traps placed at various heights above the deposit with the orifice oriented horizontally allow to collect particles that are moving on the surface (e.g., through creeping motion) as well as particles moving above the surface (like reptating or saltating particles).
 
 These techniques have been later refined for atmospheric measurements using collection chambers with filters on which particles pile up. These collected particles can then be analyzed with various methods including image analysis (for particle size and shape), gravimetric determination (for mass and density), and chemical speciation (for composition). 
 
 \textit{Usage:} Historically, such techniques were used by Bagnold \cite{bagnold1937size} to quantify creeping motion. Such techniques are still widely employed in aeolian transport to measure the amount of creep, reptation, and saltation of sand particles (see the recent review \cite{zhang2021aeolian}). These techniques are also used to measure the motion of sediments in rivers or in lakes \cite{bloesch1994review, bloesch1995mechanisms}. The popularity of this technique comes from the fact that the procedure can be easily set up for in-situ field measurements since it only requires to install a few traps and to collect the accumulated particles from time to time. 
 
 Collection chambers with filters are also widely used to monitor road resuspension of dust and particulate matter due to human activities and traffic (see the recent review on this topic \cite{rienda2021road}). For instance, such passive sensors are placed at various heights from the ground near roads to quantify traffic-induced resuspension \cite{amato2016traffic}. Another approach consists in placing a sensor behind the tires of a vehicle and to compare the amount of particles collected to the ambient concentration (measured with sensors placed in the front or at the top of the vehicle) \cite{rienda2021road}.
 
 \item Particle counting:
 
 \textit{Principle:} Another common technique relies on taking a sequence of images to record the amount of particles deposited on a sampling area. The analysis of these pictures consists in counting the number of particles that were initially deposited on a sample surface and then comparing it to the number of particles remaining on the same sample surface after exposure to a given flow for a certain duration. Hence, this provides direct access to the amount of particles resuspended. The images can be taken using various techniques, including:
 
 \begin{itemize}
  \item Optical high-resolution cameras:
  
  Cameras placed under/over the sampling area allow taking pictures of the particles deposited on the surface (provided the substrate is transparent to allow the light to go through it). For instance, it is commonly used to investigate how particle resuspension evolves with time by taking pictures at regular time intervals. Alternatively, this technique is frequently used to measure the evolution of resuspension with respect to the flow rate by taking pictures after exposing a deposit to a given flow rate for a specified time \cite{barth2014single}. 
  
  \item Scanning Electron Microscopy (SEM):
  
  SEM produces images of a sampling area by scanning the surface with a focused beam of electrons. Compared to traditional optical microscopy, the resolution of an electron microscope is much higher since it allows to visualize objects below the micrometer size. Yet, this higher resolution often implies that the area of the scanned surface is smaller than the one using an optical microscope. Hence, only a limited number of particles is accessible through SEM images, making the technique more prone to statistical errors on the concentration. However, SEM images provide essential information on the particle shape and size \cite{mcdonagh2014influence, kottapalli2019experimental, kottapalli2021aerodynamic}. For that reason, this procedure is less frequently used than other optical methods.
  
 \end{itemize}

 \textit{Usage:} Particle counting techniques from optical images are still widely used to determine the amount of particles remaining on a surface after exposure to a given flow for a certain duration (as in \cite{barth2014single}). However, experimental measurements are increasingly relying on SEM images to provide information on the geometry and surface properties of both particles and substrates (see for instance \cite{brambilla2018glass}).
 
 \item Measurements of concentrations in the fluid: 
 
 \textit{Principle:} Techniques based on the measurement of particle concentration within the fluid (and not on the surface) have been quickly adapted to quantify particle resuspension. Such methods are very similar to collecting traps since they consist in sampling the particle concentration at a given location in the domain. For instance, one can measure the concentration at various heights above the surface or in the wake of a deposit. The key difference with collecting traps is that such sensors are non-intrusive in the sense that they do not necessarily modify the flow, thus allowing for in-line measurements. As a result, it becomes possible to monitor how particle resuspension evolves with time as well as to capture how the local instantaneous flow velocity affects the resuspension rate. In fact, the acquisition frequency for particle concentration depends on the modus operandi. Various procedures exist, including:
 
 \begin{itemize}
  \item Radioactive materials and activity measurements:
  
  A first technique consists in marking particles using either radionuclides \cite{bloesch1995mechanisms} or by exposing particles to a source of neutron (Neutron Activation Analysis) \cite{mcdonagh2014study}. By recording the radioactivity (e.g., with X-ray or gamma ray detectors), it is then possible to estimate the local concentration of particles in certain areas of interest. The main advantage of radioactive particles is that such techniques can be applied even in domains with opaque boundaries, whereas other optical techniques (e.g., turbidity) require transparent boundaries near the measurement regions.
   
  \item Fluorescent particles and stereoscopic measurements:
  
  More recently, researchers have started using fluorescent particles to measure the local concentration of particles in certain areas of interest thanks to fluorescence stereoscopes \cite{ren2022experimental}. This technique is very similar to the one relying on radioactive particles. In fact, it removes the need to handle potentially harmful radioactive particles but requires to work with transparent boundaries in order to measure particle concentrations.
  
  \item Optical turbidity measurements:
  
  The presence of suspended particles in a fluid affects the reflection, absorption, and scattering properties of a light beam passing through the fluid. The intensity of the scattering effect is actually directly influenced by the particle concentration, the particle size, the particle shape as well as by the wavelength of the light \cite{xu2015light}. This effect is caused by the fact that particles have a different refractive index than the surrounding medium. As a result, the intensity of the light transmitted through the suspension is reduced: this is called turbidity \cite{elimelech2013particle}. Hence, turbidity measurements provide information on the concentration of particles in a fluid, provided that some information is already known on their size and shape. In practice, this is achieved by measuring light scattering at different angles, light transmission through the fluid and/or light absorption. This has led to the development of static light scattering and dynamic light scattering methods (interested readers are referred to the recent review \cite{xu2015light}).
  
  \item Acoustic backscatter measurements:
  
  Similarly to the optical turbidity, particle concentrations can be measured using acoustic methods. This means that a sound is propagated within the solution and its scattering due to the presence of particles is recorded \cite{boegman2019sediment}. 
  
 \end{itemize} 
 
 \textit{Usage:} Optical sensors have been widely used to measure particle resuspension induced by currents and waves in marine systems (e.g., lakes or shallow water, see \cite{kularatne2008turbulent, chung2009sediment, valipour2017sediment} and references therein) as well as resuspension in atmospheric environments (see for instance \cite{benabed2020human, boor2015characterizing}). Due to their ease-of-use for in-situ observations, acoustic backscatter methods are also widely used to measure resuspension in marine systems (see the short review \cite{bloesch1994review}), especially in lakes \cite{valipour2017sediment} and in stratified fluids \cite{boegman2019sediment} (sometimes in combination with optical methods). These instruments are often coupled to other tools to provide information on the local fluid velocity. This is achieved either by resorting to acoustic Doppler velocimeter (i.e., based on acoustic sensors, see \cite{kularatne2008turbulent, valipour2017sediment} and references therein) or on Laser Doppler Anemometry (i.e., measuring the shift between the incident and scattered light frequencies \cite{sechet1999bursting}). Furthermore, these sensors can be coupled to Aerodynamic Particle Sizer (APS) spectrometers \cite{buonanno2012particle, mcdonagh2014influence, mcdonagh2014study}, which allow to measure the size distribution of particles suspended in airflows (by accelerating particles through a constricted nozzle and recording their velocity as they pass through two laser beams).

 \item Individual particle tracking:
 
 \textit{Principle:} Recent measurements are increasingly relying on instruments to track the motion of individual particles. Such methods provide very detailed information on the time-evolution of the particle position, velocity, and acceleration. Various techniques can be used, including:
 
 \begin{itemize}
 \item Laser Distance Sensor (LDS):
  
  LDS corresponds to the tracking of an individual particle placed on a surface using a laser beam focused on the surface of the particle of interest. By measuring the reflected and/or scattered light, it is possible to accurately record the position of an individual particle and to reconstruct the streamwise and/or vertical components of the instantaneous velocity vector. Such methods have been used to measure the initial motion of millimeter-size particles \cite{diplas2010nonintrusive, xiao2021rock}. 
  
  \item Particle Tracking Velocimetry (PTV):
  
  PTV techniques consist in illuminating particles as they cross laser-light planes and then reconstructing the trajectory of each individual particle by correlating their positions between sequential frames (see the review of PTV techniques in atmospheric sciences \cite{obrien2016ptv}). Hence, PTV naturally allows to obtain detailed information on the local position, velocity \cite{salevan2017determining}, and acceleration \cite{traugott2017experimental} of each tracked particle. In addition, depending on where the cameras are placed, PTV can provide information not only on the streamwise velocities \cite{charru2007motion} but also on the wall-normal velocities \cite{kassab2013high}. Yet, as for Particle Image Velocimetry (PIV) which tracks the motion of fluid tracers, PTV methods perform poorly in dense particle suspensions. This is due to the fact that, when the number of particles in suspension increases, it becomes harder to determine the trajectory of each individual particle.

  Recent advances in PTV techniques gave rise to the use of transparent millimeter-size particles with dots drawn at specific locations on the surface \cite{agudo2017detection}. By tracking the combined motion of these dots on each particle, it is possible to extract both the translational and rotational motion. Hence, these new procedures pave the way for precise measurements of rolling and sliding motion of particles on surfaces. This method is still restricted to large-enough particles (typically a few millimeters), but with the progress in optical techniques, similar measurements for micrometer-size particles might emerge in the near future.
  
  \item Tomographic Particle Tracking Velocimetry (TOMO-PTV) or 3D-PTV:
  
  TOMO-PTV (also called 3D-PTV) is an extension of the PTV technique that allows to illuminate, record, and reconstruct the motion of any particles within a 3D volume. This is achieved using several (at least four) high-speed cameras to record simultaneous views of the whole volume investigated \cite{schobesberger2021role}. The key difference with PTV lies in the ability to reconstruct the particle trajectory across a whole volume and not only on preestablished 2D light planes.

  \item X-ray 3D tomography:
  
  Another recent extension of classical PTV techniques relies on the use of X-ray tomography instead of optical tomography \cite{stannarius2019high}. The main interest of the X-ray tomography is that it allows seeing right through opaque surfaces and through particles. Hence, when dealing with complex multilayer deposits, it provides key information on the deposit morphology that cannot be accessed using optical techniques. It has been successfully used to monitor the motion of a 3D aggregate composed of millimeter-size particles (with close to $1000$ particles in the cluster) \cite{hodge2020xray}.
  
 \end{itemize}

 \textit{Usage:} PTV techniques are increasingly used to measure the motion of individual particles on the surface (see the review on creep by sand particles \cite{zhang2021aeolian}). In addition, PTV is a very practical tool to track motion of particles once they are detached from the surface, especially to investigate the role of near-wall turbulent structures on their subsequent entrainment \cite{shnapp2015comparative, traugott2017experimental}. More recently, TOMO-PTV has been successfully used to characterize particle entrainment velocities within the near-wall turbulent structures \cite{schobesberger2021role} and also in near fixed obstacles \cite{tominaga2018wind, bhamitipadi2021incipient}. 
 
 These techniques are often combined with measurement of the local fluid velocities. In fact, most studies rely on a coupling between similar approaches (e.g., PIV+PTV \cite{sechet1999bursting} or TOMO-PIV+TOMO-PTV \cite{schobesberger2021role}) to track the motion of both tracers and solid particles. Other studies have been carried out using a combination of particle tracking and Laser Doppler Anemometry \cite{sechet1999bursting}.
 
\end{itemize}

\subparagraph{Summary}
This overview of experimental protocols illustrates that various quantities can be measured to characterize particle resuspension and that the associated experimental techniques differ depending on the specific objectives and context (see also Table~\ref{tab:measurement_techniques}). In fact, the level of information that can be extracted from these measurements spans a wide range of scales, from the most basic information on a threshold velocity for incipient motion to more detailed statistical information on particle velocities. In addition, it has been shown that the choice of an experimental procedure is based on a number of criteria including: the quantity to be measured, the properties of the fluid, the particles, and the surfaces involved. One additional piece of information that has not been discussed previously is related to the type of motion used to resuspend particles. In the case of field measurements in aeolian or fluvial research, particles are naturally exposed to the wind or the water flow (see for instance \cite{pahtz2020physics, boegman2019sediment}). The flow configuration can either be relatively simple (e.g., sand dunes \cite{bagnold1937size}, wave-induced resuspension \cite{farzi2016critical, boegman2019sediment}) or more complex (e.g., flow in vacuum cleaners \cite{trakumas2001comparison}). Yet, in the case of laboratory measurements, the vast majority of studies rely on particles exposed to a flow within a channel (see for instance \cite{barth2014single}), near obstacles in wind tunnels \cite{tominaga2018wind, bhamitipadi2021incipient}, within an impinging-jet, within a nozzle-jet (see for instance \cite{kottapalli2019experimental}), or within a rotating shear flow (see for instance \cite{agudo2017detection}). More recently, an increasing interest in the resuspension of particles due to vibrations has emerged due to their application in granular beds \cite{aracena2018frequency}, in dust resuspension due to human foot tapping \cite{benabed2020human}, in resuspension from mattresses \cite{boor2015characterizing, chatoutsidou2021resuspension} and from clothing \cite{licina2018clothing}.

\begin{landscape}
\begin{table}
 \begin{footnotesize}
 \centering
 \begin{tabular}{|>{\centering}m{0.5cm}>{\centering}m{0.7cm}|>{\centering}m{0.25cm}|>{\centering}m{0.25cm}|>{\centering}m{0.25cm}|>{\centering}m{0.25cm}|>{\centering}m{0.8cm}|>{\centering}m{0.35cm}|>{\centering}m{0.35cm}|>{\centering}m{0.35cm}|>{\centering}m{0.95cm}|>{\centering}m{0.35cm}|>{\centering}m{0.35cm}|>{\centering}m{0.7cm}|>{\centering}m{0.5cm}|>{\centering}m{1.5cm}|>{\centering}m{0.8cm}|>{\centering}m{0.65cm}|>{\centering}m{0.65cm}|>{\centering}m{1.0cm}|>{\centering\arraybackslash}m{1.5cm}|}
  
  \hline
  
  \multicolumn{2}{|c|}{\multirow{4}{1.5cm}{\centering Measurement techniques}} & \multicolumn{19}{c|}{Characteristics} \\
  
  \cline{3-21}
  
  & & \multicolumn{10}{c|}{Variable accessible} & \multicolumn{2}{c|}{Fluid}  & \multicolumn{2}{c|}{Particle} & Surface & \multicolumn{2}{c|}{Deposit} & \multicolumn{2}{c|}{Specificities} \\
  
  \cdashline{3-21}
  
  & & \multicolumn{5}{c|}{Surface} & \multicolumn{2}{c|}{Volume} & \multicolumn{3}{c|}{Motion} & \multirow{2}{0.5cm}{\centering Gas} & \multirow{2}{0.7cm}{\centering Liquid} & \multirow{2}{0.6cm}{\centering Size} & \multirow{2}{1.8cm}{Composition} & \multirow{2}{1.0cm}{Nature}  & \multirow{2}{0.7cm}{\centering Monolayer} & \multirow{2}{0.7cm}{\centering Multilayer} & \multirow{2}{1.0cm}{\centering Non-intrusive} & \multirow{2}{1.4cm}{\centering Acquisition frequency} \\
  
  \cdashline{3-12}
  
  & & $\mc{T}_r$ & $\tau_r$ & $f_r$ & $f_l$ & $u_{r=50\%}$ & $\mc{R}_f$ & $\Phi_{\rm p}$ & $u_{\nearrow}$ & $u_{\nearrow=50\%}$ & $\mathbf{U}_{\rm p}$ & & & & & & & & & \\
  
  \hline
  
  \multicolumn{2}{|c|}{\rotatebox[origin=c]{-90}{\parbox[c]{1.2cm}{\centering Mass-balance}}} & \no & \no & \yes & \yes & \yes & \no & \no & \no & \yes & \no & \yes & \yes & $>$\SI{}{\mu m} & Any & Any & \yes & \yes & \yes & Low ($\sim$ 1 per min, h or day) \\
  
  \hline
  
  \multicolumn{2}{|c|}{\rotatebox[origin=c]{-90}{\parbox[c]{1.5cm}{\centering Collecting traps}}} & \no & \no & \no & \no & \no & \yes & \yes & \no & \yes & \no & \yes & \yes & $>$\SI{}{\mu m} & Any & Any & \yes & \yes & \no & Low ($\sim$ 1 per min, h or day) \\
  
  \hline
  
  \multirow{2}{2.4cm}{\rotatebox[origin=c]{-90}{\parbox[c]{1.5cm}{\centering Particle counting}}} & \rotatebox[origin=c]{-90}{\parbox[c]{1.2cm}{\centering Optical images}} & \no & \yes & \yes & \yes & \yes & \no & \no & \no & \near & \no & \yes & \yes & $>$\SI{}{\mu m} & Any & Any & \yes & \no & \maybe & Low to high (0.01--1\,000~\SI{}{Hz}) \\
  
  \cdashline{2-21}
  
  & \rotatebox[origin=c]{-90}{\parbox[c]{1.0cm}{\centering SEM}} & \no & \yes & \yes & \yes & \yes & \no & \no & \no & \near & \no & \yes & \yes & $>$\SI{}{nm} & Any & Any & \yes & \no & \no & Low ($\sim$ 1 per min, h or day) \\
  
  \hline 
  
  \multirow{3}{2.4cm}{\rotatebox[origin=c]{-90}{\parbox[c]{2.5cm}{\centering Concentration measurements}}} & \rotatebox[origin=c]{-90}{\parbox[c]{0.8cm}{\centering Radio-active}} & \no & \no & \no & \no & \no & \yes & \yes & \maybe & \yes & \maybe & \yes & \yes & $>$\SI{}{nm} & Any (tagged) & Any & \yes & \yes & \yes & Medium to high (1--1000~\SI{}{Hz}) \\
  
  \cdashline{2-21}
  
  & \rotatebox[origin=c]{-90}{\parbox[c]{0.9cm}{\centering Fluore-scence}} & \no & \no & \no & \no & \no & \yes & \yes & \maybe & \yes & \maybe & \yes & \yes & $>$\SI{}{\mu m} & Any (tagged) & Trans-parent & \yes & \yes & \yes & Medium to high (1--1\,000~\SI{}{Hz}) \\
  
  \cdashline{2-21}
  
  & \rotatebox[origin=c]{-90}{\parbox[c]{1.3cm}{\centering Turbidity or Acoustic}} & \no & \no & \no & \no & \no & \yes & \yes & \no & \yes & \no & \yes & \yes & $>$\SI{}{\mu m} & Any & Any & \yes & \yes & \no & Medium (1--10~\SI{}{Hz}) \\
  
  \hline 
    
  \multirow{4}{2.9cm}{\rotatebox[origin=c]{-90}{\parbox[c]{3.2cm}{\centering Individual tracking}}} & \rotatebox[origin=c]{-90}{\parbox[c]{0.6cm}{\centering LDS}} & \no & \no & \no & \no & \no & \no & \no & \yes & \yes & \yes & \yes & \yes & $>$\SI{}{mm} & Any & Any & \yes & \maybe $\,$ (top) & \yes & High (100--10\,000~\SI{}{Hz}) \\
  
  \cdashline{2-21}
  
  & \rotatebox[origin=c]{-90}{\parbox[c]{0.6cm}{\centering PTV}} & \maybe  $\,$ 2D & \maybe  $\,$ 2D & \maybe  $\,$ 2D & \maybe  $\,$ 2D & \yes & \maybe  $\,$ 2D & \maybe  $\,$ 2D & \yes & \yes & \maybe  $\,$ 2D & \yes & \yes & $>$\SI{}{\mu m} & Any & Trans-parent & \yes & \maybe  $\,$ 2D & \yes & High (100--10\,000~\SI{}{Hz}) \\
  
  \cdashline{2-21}
  
  & \rotatebox[origin=c]{-90}{\parbox[c]{0.8cm}{\centering 3D-PTV}} & \yes & \yes & \yes & \yes & \yes & \yes & \yes & \yes & \yes & \yes & \yes & \yes & $>$\SI{}{\mu m} & Any & Trans-parent  & \yes & \maybe $\,$ (top) & \yes & High (100--10\,000~\SI{}{Hz}) \\
  
  \cdashline{2-21}
  
  & \rotatebox[origin=c]{-90}{\parbox[c]{1.1cm}{\centering X-ray TOMO}} & \yes & \yes & \yes & \yes & \yes & \yes & \yes & \yes & \yes & \yes & \yes & \yes & $>$\SI{}{mm} & Any (tagged) & Any & \yes & \yes & \yes & High (100--1\,000~\SI{}{Hz}) \\
  
  \hline
  
 \end{tabular}
 \caption{Summary of the main measurement techniques to quantify particle resuspension and of their characteristics (sorted in terms of variable accessible, range of applicability regarding fluid/particle/surface/deposit properties and specificities).}
\label{tab:measurement_techniques}
\end{footnotesize}
\end{table}
\end{landscape}

 \subsubsection{Limitations of experimental methodologies}
 \label{sec:technique:measurement:limitations}

Despite considerable progress in measurement techniques, there are still a number of limitations and uncertainties that require specific attention to avoid meaningless measurements. The main sources of uncertainties are related to:

\begin{itemize}
 \item The reverse deposition process:
 
 One of the most frequent issues encountered in measurements of particle resuspension is related to the reverse-process of particle deposition which often occur simultaneously \cite{francia2015role}. In fact, this issue was already identified in the early 1980s \cite{cleaver1976effect} and it has remained a matter of concern ever since, especially for in-situ observations (where particles are both deposited and in suspension at the start of the measurement). Laboratory measurements of particle resuspension can also be hampered by initially suspended particles. For instance, when simply counting the number of particles before and after exposure to a given flow, no information is available to determine if some (and how many) of the particles present on the surface at the end of the procedure actually come from suspended particles that have deposited on the area of observation.
 
 Various attempts have been made to mitigate this issue. In laboratory studies, a wide range of measurements have been performed using fluids without any suspended particles. This means that the only particles within the domain considered are the particles that are initially deposited on the surface, hence effectively removing any contamination due to the deposition of particles initially in suspension. With the advent of new measurement techniques such as 3D-PTV, laboratory studies or in-situ observations can now explicitly track the motion of each individual particle near the boundary. Hence, these new methods allow to identify particles depositing on the surface and particles resuspending from it. Alternatively, some studies actually rely on the combined effects of deposition and resuspension processes in environments initially devoid of suspended particles. For instance, investigations of resuspension from clothes often characterize the ratio between the amount of particles resuspended from clothing substrates and the portion of these resuspended particles that reach a target surface on the ground due to subsequent deposition (this is referred to as the clothing release fraction) \cite{licina2018clothing, ren2022experimental}. In such cases, this measure is relevant since it corresponds to the portion of resuspended particles that reach the ground (where they are less prone to be inhaled by an adult person and where they can be collected later in vacuum cleaners).

 \item The procedure to form the initial deposit:
 
 Even when particle resuspension is studied in environments free of any other sources of particles (meaning that only deposited particles are present in the domain studied), the procedure used for the deposition can have significant consequences on resuspension. This can be due to a variety of reasons including the dry-vs.-wet deposition (e.g., spores from dry deposition are easier to remove than from wet deposition  \cite{epa2014determination, kesavan2017deposition}), the flexibility of the particles (which affects the cohesive forces between particles), and even how the particles were deposited. In many wind-tunnel studies, the deposit is usually formed by sedimenting particles within a quiescent flow (i.e., without any airflow) \cite{barth2014single}. Once fully formed, the deposit is then exposed to a fluid flow in order to study the resuspension process. In reality, deposits are usually formed by a subtle balance between particle deposition and resuspension since both processes happen simultaneously. This means that the particles loosely adhering to the surface would be removed shortly after deposition. Hence, if the flow velocity is suddenly increased, these particles would not be available for resuspension and the total amount resuspended would be less than the one computed considering that all particles were present initially on the surface (i.e., considering the full adhesion force distribution). This is a matter of concern when trying to distinguish between short-term and long-term resuspension. As mentioned previously, short-term resuspension corresponds to the large amount of particles that leaves the surface over a very short time as soon as the deposit is exposed to a flow \cite{reeks1988resuspension}. Yet, it appears that this short-term resuspension rate may not be relevant to all practical applications and can even be an artifact from the protocols used to create deposits on a surface. 
 
 \item Spatial/temporal resolution:
 
 Another limitation is related to the spatial/temporal resolution. In fact, regardless of the measurement technique, particle resuspension is always characterized over a finite surface and finite time: the size of this surface can be the size of the weighting plate, the surface observed in particle counting methods, the surface (resp. volume) considered in PTV (resp. 3D-PTV), or the area behind which collectors are placed. Due to this finite-size, it is not possible to ascertain if some of the particles initially deposited on the surface have actually moved on the surface (resp. detached from the surface) before coming to a halt on the surface outside of the observation area (resp. redeposited outside of the area, like saltating particles). As a result, measurements of particle resuspension always overestimate the number of resuspended particles that would be obtained considering an infinite surface. In fact, this can be even more complicated when relying on counting techniques. In that case, the number of particles still on the surface after being exposed to a flow for a certain duration is simply compared to the initial number of particles. However, some of the particles might have been rolling/sliding on the surface before coming to a stop later on: if they remain within the observation area, these particles are considered as not resuspended (regardless of their motion) while, if they leave the observation area, they are counted as resuspended materials (strictly speaking, they are migrating on the surface). 
 
 The temporal resolution used when quantifying particle resuspension can have similar effects as the spatial resolution. In fact, in the case where particles undergo small hops over the surface (reptation), the result obtained with simple counting techniques varies with the observation time. In addition, as was mentioned previously, the resuspension rates extracted from such measurements depends both on the observation time $\Delta t$ and on the time $t$ at which the measure is made. Hence, it is paramount to keep this information in mind to avoid drawing inappropriate conclusions by comparing two experimental data (especially if they were made using different observation times). 
 
 \item Complex surfaces:
 
 Another experimental artifact is related to studies concerning resuspension from vegetation. Experiments in wind tunnels may use an unrealistic unidirectional wind flow compared to the conditions in the actual atmosphere. Resuspension may therefore be underestimated because vegetation like grass can bend in the direction of the flow, thereby protecting the deposited particles. For instance, resuspension from 25-\SI{30}{cm} grass doubled when the grass trays were rotated during the experiment \cite{giess1997factors}. 
  
 \item Incomplete measurements of surface topology and adhesive properties:
 
 Despite recent advances in measurement techniques which now provide information on near-wall particle velocities, a complete picture of a single particle resuspension event is still missing. This is mostly related to the difficulty in characterizing precisely what is the exact adhesive force between a particle and a surface at a given location. This is due to the chaotic nature of the surface profiles, which usually display roughness at the nanoscale. Such uncertainties severely impact the resuspension of small colloidal particles (which are highly sensitive to adhesive forces). Nowadays, statistical information on roughness profiles and the corresponding adhesive force are available thanks to AFM (atomic force microscopy, see existing reviews \cite{cappella1999force, butt2005force}), but there are still no procedure allowing to measure both the adhesion force of a single particle located at a certain position and its resuspension as it is exposed to a given flow. 
 
 \item Lack of uncertainty quantification:
 
 Despite the number of experimental studies performed, there has been no attempt to quantify the sources of uncertainty, at least to the authors' knowledge. Uncertainty quantification in experiments consists in quantifying the sources of uncertainty and errors using probabilistic and statistical tools \cite{smith2013uncertainty, moffat1988describing}. To put it differently, ``uncertainty quantification offers a rational basis to interpret the scatter on repeated observations, thus providing the means to quantify the measurement inaccuracies and imprecision'' \cite{sciacchitano2019uncertainty}. In practice, there are two main sources of uncertainty and errors in experiments: limited/incomplete data and limited accuracy or resolution of sensors. In the present case, a number of parameters have been shown to result in variations of the measured value for particle resuspension (see the previous paragraphs). The interest of uncertainty quantification is that it allows to sort these various sources of errors with respect to their impact on the measured quantity. As a result, it can provide useful information as to which parameters should be carefully set up (typically those with a high impact on the measured quantity). However, uncertainty quantification requires a thorough examination of available experimental data with a strict procedure (where each parameter is successfully varied over the range of possible values while all the other ones are fixed). For that reason, it has only been seldom carried out (see for instance a recent paper on PIV measurements \cite{sciacchitano2019uncertainty}).
 
\end{itemize}

Drawing on these limitations, several suggestions can be made to go beyond the current state-of-the-art measurements. These propositions are detailed in Section~\ref{sec:next_model}.

\section{Modeling particle resuspension}
 \label{sec:models}

In parallel to experimental measurements, a range of theories have been developed to predict the physical quantities of interest (see the list in Section~\ref{sec:technique:quantities}). Once terminology issues are clarified, it appears that some of these models are equivalent. There are, however, often structural differences between them due to: (a) the different physical quantities being sought, (b) the various levels of description at which they operate and (c) the range of time/spatial scales considered in each of these theories. It is therefore useful to provide a general framework in which different classes of models can be set. 

To that effect, we review here the various types of models developed in the literature, starting with the different types of forces that can either induce or prevent the motion of deposited particles in Section~\ref{sec:models:forces}, along with the existing expressions for these forces. The different levels of description and strategies that can be followed to generate the information required to compute these forces, especially the modeling approaches for the turbulent flow field and for surface roughness, are recalled in Section~\ref{sec:models:turb_rough}. A number of representative resuspension models are then detailed in Section~\ref{sec:models:approach}. This overall description of models paves the way for the more in-depth analysis of resuspension models to be presented in Section~\ref{sec:analysis}.

\subsection{Key forces at play in resuspension}
   \label{sec:models:forces}

As introduced in Section~\ref{sec:physics:phenomenology:interactions}, particle resuspension results from the rupture of balance between forces inducing motion and forces preventing motion. These forces are often classified in terms of the particle-fluid, particle-particle and particle-surface forces (see previous reviews \cite{henry2014progress, ziskind2006particle}). Here, we adopt another standpoint related to the `system' considered and to the nature of the forces involved. More precisely, we consider that the system under consideration is composed of three elements: (a) a fluid, (b) a physical domain with solid boundaries (or walls), and (c) particles immersed in the fluid. From this simple definition, one can then distinguish between external forces (related to interactions with physical objects, or force fields, external to the system) and internal forces (related to interactions between elements within the system). The former include gravity forces or electro-magnetic interactions with external fields while the latter include particle-fluid, particle-wall and particle-particle forces. In the following, we first review external forces (see Section~\ref{sec:models:forces:ext}) before listing internal forces within the system, which are classified as: particle-fluid forces (see Section~\ref{sec:models:forces:hydro}), non-contact forces (see Section~\ref{sec:models:forces:non-contact}) and contact forces (see Section~\ref{sec:models:forces:contact}). The main reason for choosing to depict separately non-contact and contact forces is to better identify which forces occur when particles are deposited (i.e., in contact with a substrate) and which forces act on detached particles.

  \subsubsection{Interactions with external fields}
   \label{sec:models:forces:ext}

\begin{figure}[ht]
 \centering
 \includegraphics[width=0.9\textwidth, trim=0cm 0.0cm 0cm 1.0cm, clip]{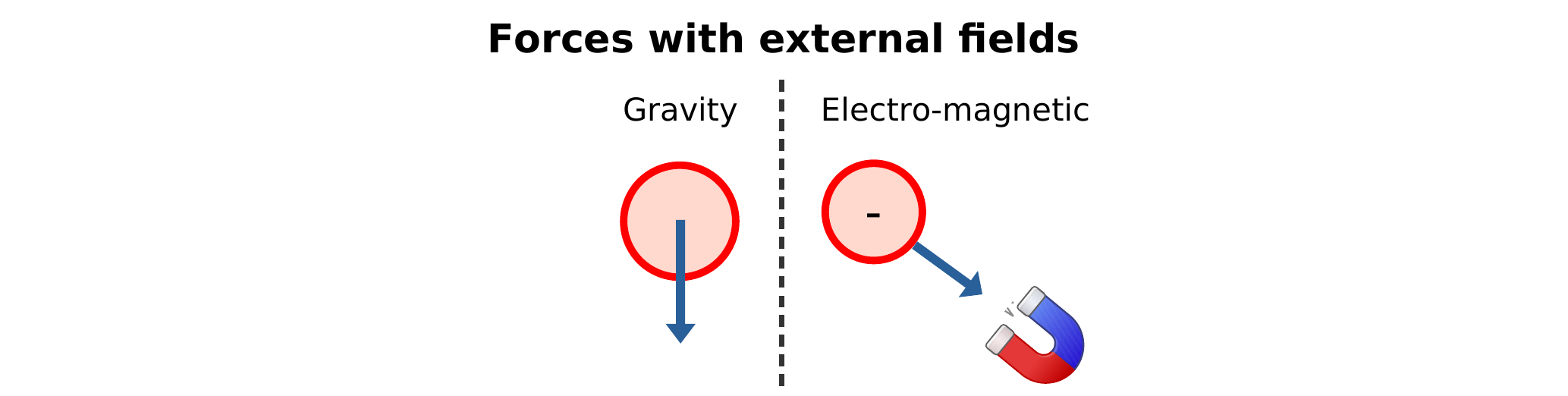}
 \caption{Illustration of the most important forces with external fields at play in particle resuspension.}
 \label{fig:sketch_forces_ext}
\end{figure}

\subparagraph{Physical origin:} The system considered here (i.e., composed of a fluid, walls and particles) can interact with external fields. In particular, the dynamics of the fluid as well as the dynamics of particles are affected by their interaction with external fields, such as (see also Fig.~\ref{fig:sketch_forces_ext}): gravity forces or electro-magnetic forces. 

\subparagraph{Expressions:} Focusing on the dynamics of particles, the interaction with external fields is given by well-known formulas for each contribution.
 \begin{enumerate}[A.]
  \item Gravitational forces occur in most applications since particles are deposited on substrates that are often at the surface of a planet. Gravity forces arise due to the weight of the particle and are given by:
  \begin{equation}
   \mathbf{F}_{\rm grav} = \mathcal{V}_{\rm p} \ \rho_{\rm p} \ \mathbf{g}~,
   \label{eq:Fgrav}
  \end{equation}
  where $\rho_{\rm p}$ is the particle density, $\mathcal{V}_{\rm p}$ its volume and $\mathbf{g}$ the gravity acceleration. In many cases, gravity tends to prevent the motion of large particles since it acts to keep particles in place (like rocks or gravels at the bottom of a river).
  
  \item Electro-magnetic forces occur when the two bodies display electro-magnetic properties (such as ferromagnetic particles) \cite{viota2005stabilization}. In that case, the force on each particle is actually given by Lorentz force:
  \begin{equation}
   \mathbf{F}_{\rm magn} (h) = q\, \left( \mb{E} + \mb{U}_{\rm p} \times \mb{B}\right)~,
   \label{eq:FLorentz}
  \end{equation}
  where $q$ is the particle charge, $\mb{E}$ the ambient electric field, $\mb{B}$ the ambient magnetic field and $\mb{U}_{\rm p}$ the particle velocity. Although these effects are rarely present in the applications mentioned until now, their contribution to particle resuspension can yield rich and complex collective phenomena due to the induced magnetic dipoles and their effects on the local magnetic fields.
   
 \end{enumerate}

\subparagraph{Role in resuspension:} External forces can impact particle resuspension in different ways. First, gravity forces tend to prevent the motion of particles that have sedimented on a surface (like gravels on river beds). On some rare occasions where particles have accumulated on the upper regions within pipes, gravity can favor resuspension. Second, most electromagnetic forces induce a motion of both the fluid and the particles. Hence, they rather tend to favor the onset of particle motion. 

  \subsubsection{Fluid forces on particles}
   \label{sec:models:forces:hydro}

\subparagraph{Physical origin:} Any body placed in a fluid flow undergoes aero- or hydro-dynamic forces related to the interactions between the fluid and the particle. Using a continuum description of fluids (which will resurface in Section~\ref{sec:models:turb_rough:turb}), these forces are either functions of the fluid pressure or of the relative velocity between the fluid and the object.

\begin{figure}[ht]
 \centering
 \includegraphics[width=0.8\textwidth]{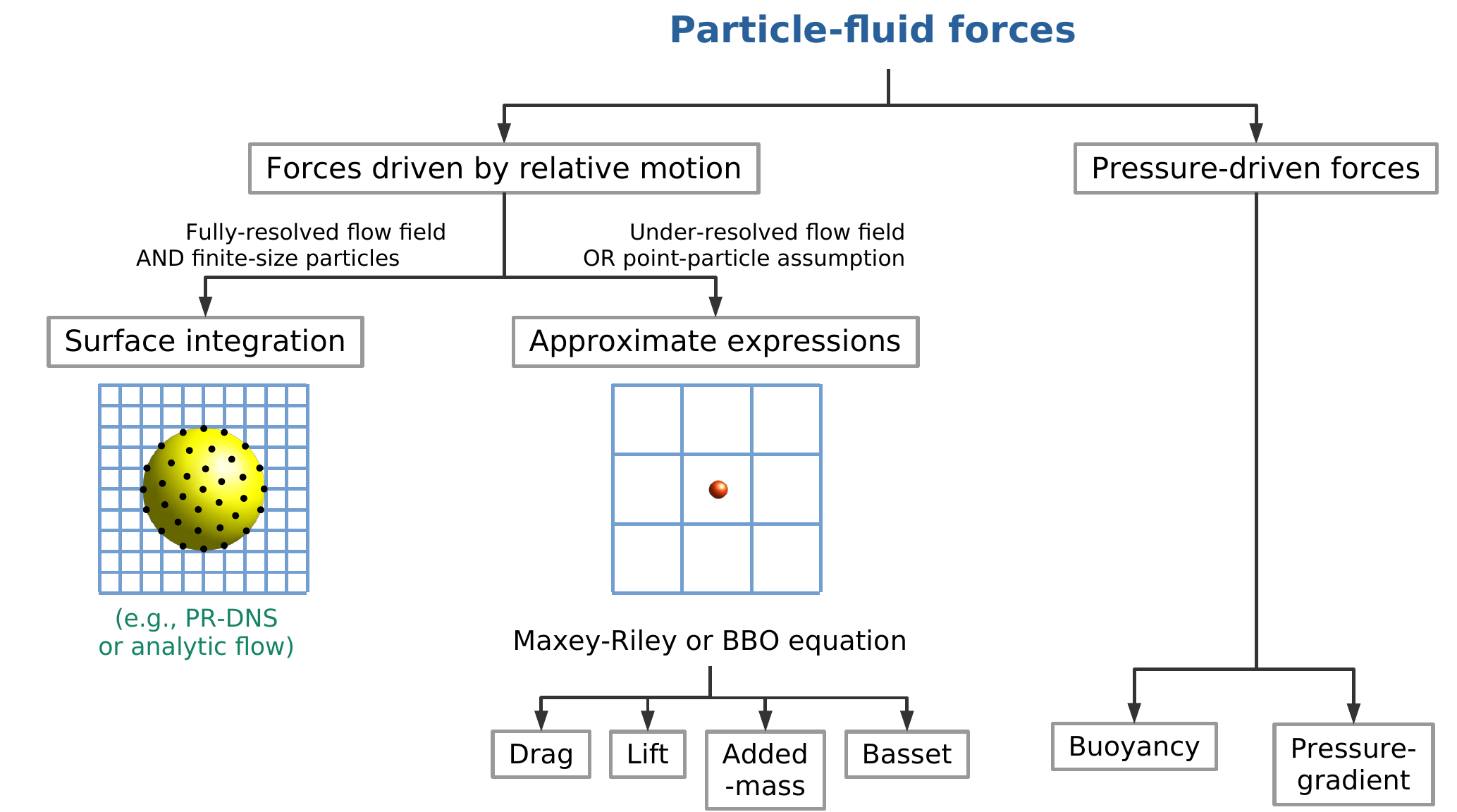}
 \caption{Illustration of the approaches to evaluate the hydrodynamic force exerted on object immersed in a fluid.}
 \label{fig:sketch_forces_hydro}
\end{figure}

\subparagraph{Expressions:} As depicted in Fig.~\ref{fig:sketch_forces_hydro}, particle-fluid interactions have two origins:
\begin{equation}
 \mb{F}_{\rm f\rightarrow p}  = \, \underbrace{\mb{F}_{\rm PG} \, + \, \mb{F}_{\rm buoy}}_{\displaystyle \mb{F}_{\rm pressure-driven}}  + \underbrace{\mb{F}_{\rm drag} + \mb{F}_{\rm lift} + \mb{F}_{\rm AM} + \mb{F}_{\rm hist}}_{\displaystyle \mb{F}_{\rm relative-motion}}~.
 \label{eq:Ff-p}
\end{equation} 

\begin{enumerate}[A.]
 \item Pressure-driven forces are actually decomposed in two contributions:
 \begin{enumerate}[i -]
  \item Buoyancy is the force exerted by the fluid that opposes the weight of an object immersed in a fluid. Strictly speaking, it comes from the pressure difference in a fluid: since pressure increases with depth (due to the weight of the fluid in the upper layers), the pressure at the bottom of a particle is slightly greater than at its top. This pressure difference results in a net upward force on the object. Following Archimedes' principle, this force is equivalent to the weight of the displaced fluid (i.e., the fluid that would otherwise occupy the submerged volume of the object). It is simply given by:
  \begin{equation}
   \mathbf{F}_{\rm buoy} = - \mathcal{V}_{\rm p} \, \rho_{\rm f} \, \mathbf{g}~,
   \label{eq:Fbuoy}
  \end{equation}
where $\rho_{\rm f}$ is the fluid density. If we combine buoyancy with the gravity force, Eq.~\eqref{eq:Fgrav}, we retrieve the well-known expression of the apparent weight of an immersed body: 
	\begin{equation}
	\mathbf{F}_{\rm grav} + \mathbf{F}_{\rm buoy} = \mathcal{V}_{\rm p} \, \left( \rho_{\rm p} - \rho_{\rm f} \right) \, \mathbf{g}~.
	\end{equation}
  
  \item The pressure-gradient force $\mb{F}_{\rm PG}$ corresponds to the force exerted by the velocity field undisturbed by the presence of the particle \cite{kuerten2016point}. It is usually expressed as:
  \begin{equation}
   \mathbf{F}_{\rm PG} = \mathcal{V}_{\rm p} \, \rho_{\rm f} \, \frac{D\mb{U}_{\rm f}}{Dt}~,
  \end{equation}
  where $D/Dt = \partial/\partial t \, + \mathbf{U}\cdotp \nabla$ is the material derivative.

 \end{enumerate}

 \item Hydrodynamic forces related to velocity differences between particles and fluid can be evaluated using two approaches:
 \begin{enumerate}[i -]
  \item Integration on a fully-resolved flow field:
 
  When the instantaneous flow field around the object is fully resolved, one can directly estimate the strength of these hydrodynamic forces by integrating the extra fluid stress tensor $\delta \mathcal{T}_{\rm f}$ due to the perturbation induced by the particle along the entire surface of the object $\mathcal{S}_{\rm p}$. This gives:
  \begin{equation}
   \mb{F}_{\rm relative-motion} = \int_{\mathcal{S}_{\rm p}} \delta \mathcal{T}_{\rm f} \cdot \mathbf{n}_{\mathcal{S}} \, dA
  \label{eq:fhydro1}
  \end{equation}
  where $\mathbf{n}_{\mathcal{S}}$ is the outgoing normal to the local surface. As one can expect, such integrations require very detailed information on the flow field around each individual particle. This can be obtained either with prescribed flow fields in simple configurations or with fine numerical simulations of turbulent flows (see Section~\ref{sec:models:turb_rough} for more details).
 
 \item Approximate formulas:
 
 When direct integration of the hydrodynamic force is not possible, one can resort to approximate expressions to estimate the hydrodynamic forces acting on a particle. This usually occurs when particles are much smaller than the grid size used in the computation of the flow field, leading to the so-called point-particle approximation. The interest of this point-wise particle approximation is that the hydrodynamic forces are expressed using information only on the fluid velocity at the particle position, which defines the velocity of the fluid seen $\mb{U}_{\rm s}(t)=\mathbf{U}_{\rm f}(t,\mathbf{X}_{\rm p}(t))$. 

Despite decades of research on the subject (see the work of Stokes in the mid-19th century and those of Basset-Boussinesq-Oseen between 1888 and 1927), the exact expression of these hydrodynamic forces is still a subject of research today. In the case of point particles, the hydrodynamic forces are generally given by the Maxey-Riley-Gatignol equation \cite{maxey1983equation, gatignol1983faxen} (also referred to as the BBO equation). This equation takes into account several contributions \cite{minier2001pdf, kuerten2016point, brandt2022particle}:
 \begin{equation}
  \mb{F}_{\rm relative-motion} = \mb{F}_{\rm drag} + \mb{F}_{\rm lift} + \mb{F}_{\rm AM} + \mb{F}_{\rm hist}~.
  \label{eq:Fhydro2}
 \end{equation} 
 \begin{enumerate}[a $\cdot$]
  \item Drag forces $\mb{F}_{\rm drag}$ arise from the friction due to the relative velocity between the fluid and particle velocities. Drag forces act along the direction of the relative velocity and tend to move particles in the same direction as the flow (such that the relative velocity becomes zero). Hence, these forces are directly expressed in terms of the relative velocity $\mb{U}_{\rm r}=\mb{U}_{\rm p} - \mb{U}_{\rm s}$ \cite{clift2005bubbles, marshall2014adhesive}:
  \begin{equation}
   \mb{F}_{\rm drag} = \frac{1}{2}\rho_{\rm f} C_D A_{\rm p} \, \vert \mb{U}_{\rm r} \vert \, \mb{U}_{\rm r}~,
   \label{Eq_Drag}
  \end{equation}
  with $A_{\rm p}$ the particle cross-sectional area exposed to the flow and $C_D$ the drag coefficient.  The particle cross-sectional area $A_{\rm p}$ is straightforward to compute for simple mathematical shapes (like spheres, spheroids) but can be more cumbersome when dealing with irregular particles. The difficulty comes from the expression of the drag coefficient, which is a non-linear function that depends on the particle shape and on the particle Reynolds number $\Rep = \|\mb{U}_{\rm r}\|d_{\rm p}/\nu_{\rm f}$ ($\nu_{\rm f}$ being the fluid kinematic viscosity and $d_{\rm p}$ the particle size). Nevertheless, semi-empirical expressions for $C_D$ can be found in simple cases. For instance, for spherical particles, the Stokes drag coefficient is rigorously given by $C_D = 24/\Rep$ (derived in the Stokes regime where $\Rep \ll 1$), leading to the exact formula for the Stokes drag force \cite{clift2005bubbles, marshall2014adhesive}:
  \begin{equation}
   \mb{F}_{\rm drag} = 3 \, \pi \, d_{\rm p} \rho_{\rm f} \nu_{\rm f} \left( \mb{U}_{\rm s} - \mb{U}_{\rm p} \right)~.
   \label{eq:Fdrag}
  \end{equation}
  Extended formulations of the Stokes drag coefficient have been proposed by adding a correction factor $C_D=24/\Rep\,f(\Rep)$ which extends the range of validity of the formula. For example, a frequently used correction factor is $f(\Rep)=1+0.15\,\Rep^{0.687}$ (valid up to $\Rep\le1000$, at which point the drag coefficient is usually fixed to a constant equal to $0.44$) \cite{henry2014progress, clift2005bubbles}. More recently, general approximations of the drag coefficient have been pursued using machine-learning techniques (as in \cite{hwang2022machine} for particles with arbitrary shapes or in \cite{balachandar2020toward} for dense suspensions).
  
  In the context of particle resuspension, the Stokes drag formula should be also corrected to account for the presence of the wall, giving:
  \begin{equation}
   \mb{F}_{\rm drag} = 3 \, \pi \, d_{\rm p} \rho_{\rm f} \nu_{\rm f} \left( \mb{U}_{\rm s} - \mb{U}_{\rm p} \right) f_{\rm wall}~,
   \label{eq:Fdrag_mod}
  \end{equation}
  with $f_{\rm wall}$ a correction factor (equal to $1.7009$ according to \cite{o1968sphere}). Another consequence of the fluid motion anisotropy in the near-wall region is that it produces a non-zero torque (since the velocity is smaller near the surface). In the Stokes regime, the torque is usually approximated with the O'Neill formula \cite{o1968sphere}:
  \begin{equation}
   \mb{M}_{\rm drag} = 1.4 \, r_{\rm p} \, \mb{F}_{\rm drag}~,
   \label{eq:Mdrag}
  \end{equation}
	with $r_{\rm p}$ the particle radius.

  \item Lift forces $\mb{F}_{\rm lift}$ correspond to the force exerted by the fluid flow in the direction perpendicular to the oncoming flow direction (contrary to the drag forces that act along the direction of the relative velocity). The lift force depends on the relative velocity between the particle and the fluid (termed Saffman lift) and on the particle rotation rate (labelled Magnus lift) \cite{marshall2014adhesive}.
  
  As for drag forces, several analytic or empirical formulas have been presented in the literature. Yet, there is no consensus on a valid formulation (or taken as such) regardless of the particle Reynolds number $\Rep$. For instance, empirical formulas have been written for spherical particles in a fully developed turbulent boundary layer \cite{hall1988measurements, mollinger1996measurement}, which reads:
  \begin{equation}
   \mb{F}_{\rm lift} = \alpha \nu_{\rm f}^2\rho_{\rm f} \left(\Rep\right)^{\beta}~,
   \label{eq:Flift}
  \end{equation}
  with $\alpha$ and $\beta$ two constants. Hall proposed to use $\alpha=20\pm1.57$ and $\beta=2.31\pm0.02$ for large particles in the range $3.6\le\Rep\le140$ \cite{hall1988measurements}, while Mollinger et al. suggested $\alpha=56.9\pm1.1$ and $\beta=1.87\pm0.04$ for small particles such that $0.6\le\Rep\le4$ \cite{mollinger1996measurement}
  
  Yet, these formulas should be carefully chosen and assessed, especially since discrepancies have been found between theoretical and experimental works leading to estimates that differ by orders of magnitude \cite{brambilla2017adhesion}. It should be highlighted that the approach to determine the drag and lift forces has to be consistent and based on the same assumptions. For instance, if the O'Neill formulation for drag is chosen, the lift force is zero and cannot be replaced with just any other equation. In addition, as for the drag force, the length scale in the lift force computation may depend on the particle orientation for non-spherical particles.
  
  \item Added-mass forces $\mb{F}_{\rm AM}$ are associated with the additional force that is needed to accelerate/decelerate a body immersed in a dense fluid due to the fact that the body also moves some of the nearby surrounding fluid as it moves through it. Hence, the added-mass force is directly proportional to the difference between the particle acceleration and the fluid one at the particle location. It is given by:  
  \begin{equation}
   \mathbf{F}_{\rm AM} = \frac{1}{2} \, \rho_{\rm f} \, \mathcal{V}_{\rm p} \, \left[ \left(\frac{D\mb{U}_{\rm f}}{Dt}\right)(t,\mathbf{X}_{\rm p}(t)) - \frac{d\mb{U}_{\rm p}}{dt} \right]
   \label{eq:Fadded_mass}~.
  \end{equation}
  While added-mass forces can severely impact the dynamics of small particles with a density close to the fluid density, it is often negligible compared to drag forces for particles much heavier than the fluid, i.e. $\rho_{\rm f} \ll \rho_{\rm p}$ (like sand in air).
  
  \item Basset history forces are related to the response of the fluid boundary layer to an acceleration of the relative velocity. Due to viscous effects, there is indeed a temporal delay in the development of the boundary layer around a particle when the relative velocity changes, resulting in a `history' effect \cite{basset1888treatise}. Hence, the Basset force is usually expressed by an integration of the relative velocity over the trajectory of the particle:
  \begin{equation}
   \mathbf{F}_{\rm hist} = \frac{3}{2} \, d_{\rm p}^2 \, \rho_{\rm f} \, \sqrt{\pi\nu_{\rm f}} \int_{-\infty}^{t} \frac{d}{ds}\left(\mb{U}_{\rm s}(s)-\mb{U}_{\rm p}(s)\right) \frac{ds}{\sqrt{t-s}}~.
  \end{equation}
  The Basset force is negligible in stationary flows \cite{basset1888treatise} but not in unsteady flows. Yet, in practice, it is often neglected due to its cumbersome evaluation (with a singularity in the term $1/\sqrt{t-s}$). 
  \end{enumerate}
 \end{enumerate}

  \smallskip
  At this stage, it is worth noting that the Maxey-Riley equation is, strictly speaking, only valid for particles such that $\Rep\ll 1$. However, they are also used when $\Rep\gg1$ (but resorting to generalized formulas for the drag and lift forces). The particle equation of motion can be further extended to include effects related to Brownian motion. This corresponds to the random motion of particles suspended in a fluid due to their collision with molecules of the fluid. While Brownian motion has been shown to be relevant for colloidal particles (i.e., with a size smaller than a few micrometers), it quickly becomes negligible when dealing with larger particles (which are more sensitive to gravity forces) \cite{henry2014progress}. A usual and straightforward way to include Brownian motion in a particle equation of motion is to adopt a stochastic description of the particle dynamics and to add a white noise term (to account for the random Brownian motion). Neglecting other contributions, this gives the following Langevin equation (more details in \cite{minier2016statistical}):
  \begin{subequations}
   \begin{align}
    d\mb{X}_{\rm p} &= \mb{U}_{\rm p} \, dt \\
    d\mb{U}_{\rm p} &= (\text{other accelerations}) + K_{\rm Br} \, d\mb{W}
   \end{align}
   \label{eq:Langevin}
  \end{subequations}
  where $K_{\rm Br}$ is the diffusion coefficient for Brownian motion and $\mb{W}$ is a vector of independent Wiener process (independent noise terms). 
\end{enumerate}

\subparagraph{Role in resuspension:} Hydrodynamic drag and lift forces are key ingredients in particle resuspension. They are directly involved in particle detachment, since their action tends to set in motion particles initially at rest on a surface. In addition, they play a role in the dynamics of particles once they are detached from the surface (i.e., the re-entrainment phase). Due to turbulence effects (see Section~\ref{sec:models:turb_rough:turb}), these forces exhibit fluctuations and are therefore characterized by a distribution of values. This distribution is responsible for the range of values at which resuspension occurs for a given type of particle and flow conditions.

  \subsubsection{Non-contact forces}
   \label{sec:models:forces:non-contact}

\subparagraph{Physical origin:} Having described particle-fluid forces, we now turn our attention to particle-wall and particle-particle interactions. These forces are usually referred to as inter-surface forces (or simply surface forces), since they act between two macroscopic bodies (more details in \cite{israelachvili2011intermolecular}). From a physical point of view, they arise due to the intermolecular (weak) forces acting between every atoms/molecules composing each body. Hence, surface forces are short-ranged and important when dealing with small particles (typically colloids). We focus specifically on forces when there is no contact between surfaces, which means that these forces affect the dynamics of particles detached from the surface (contact forces are described in Section~\ref{sec:models:forces:contact}).

\begin{figure}[ht]
 \centering
 \includegraphics[width=0.86\textwidth, trim=0cm 0.0cm 0cm 1.0cm, clip]{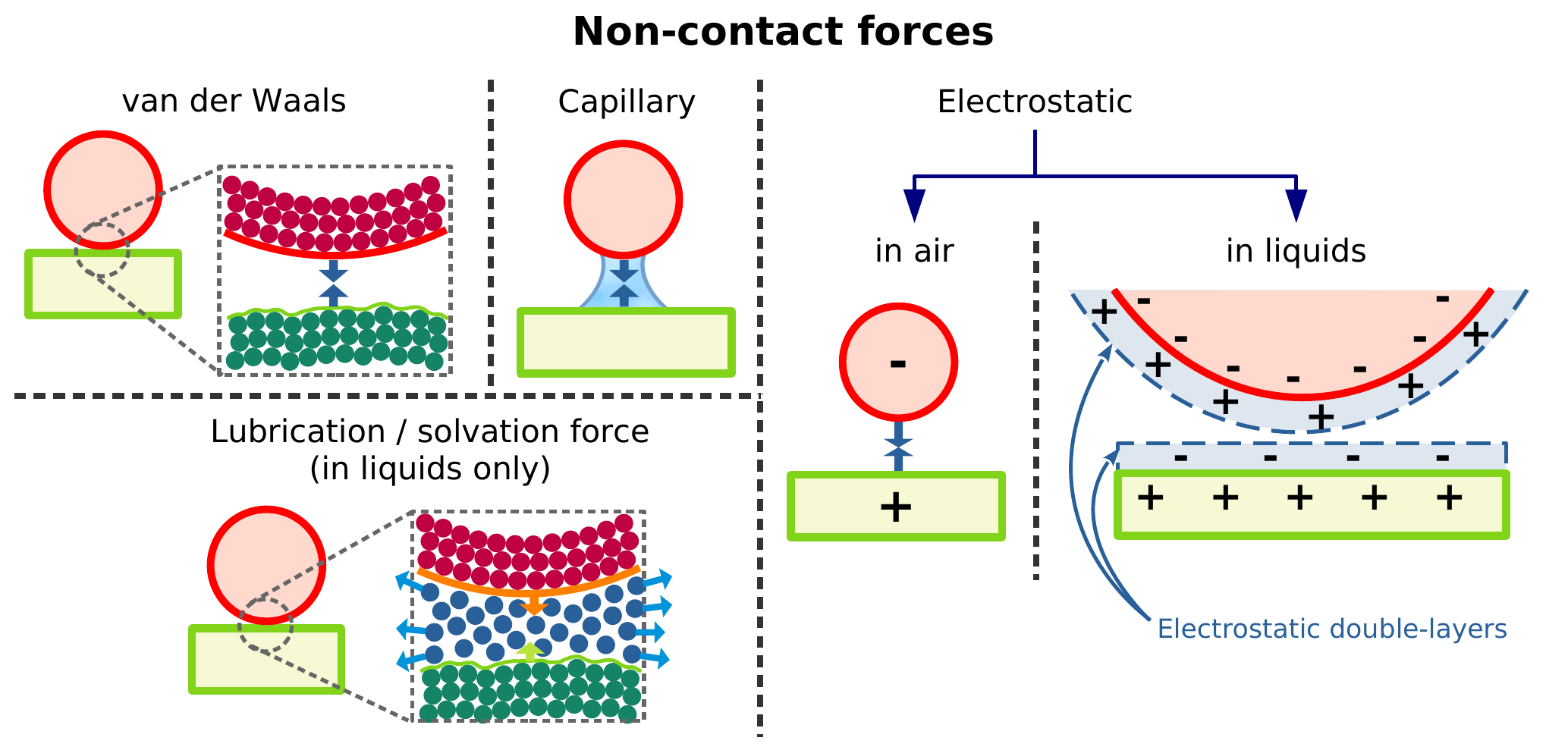}
 \caption{Illustration of some of the non-contact forces at play in particle resuspension.}
 \label{fig:sketch_forces1}
\end{figure}

\subparagraph{Expressions:} Non-contact surface forces encompass various contributions, including (see also Fig.~\ref{fig:sketch_forces1}):
\begin{enumerate}[A.]
 \item van der Waals (VDW) interactions result from the correlations in the fluctuating polarization of nearby dipoles \cite{israelachvili2011intermolecular}. This includes contributions from two permanent dipoles (Keesom force), a permanent dipole and an induced one (Debye force), as well as two induced instantaneous dipoles (London dispersion forces). In most cases, VDW forces are attractive and responsible for adhesion between macroscopic bodies. Yet, they quickly vanish when distances exceed a few nanometers/micrometers. As a result, VDW interactions are playing a significant role in the adhesion of small particles, with a size ranging from a few nanometers up to a few hundreds of micrometers. Above this threshold, VDW forces are still present but their magnitude becomes negligible compared to other macroscopic forces (such as gravity). 

 Methodologies to evaluate VDW forces are based on three different levels of description:
 \begin{itemize}
  \item First, direct numerical integration are based on Surface Element Integration methods (SEI). As depicted in Fig.~\ref{fig:sketch_forces1}, these approaches describe each body as an ensemble of molecules (i.e., using a molecular level of description). Then, the potential energy $E_{\rm VDW}$ is simply obtained by integrating the elementary molecule-molecule interactions $E_{\rm vdw}$ over the exact topography of the two bodies involved \cite{bhattacharjee1997surface}. This gives:
  \begin{equation}
   E_{\rm VDW} = \int_{\mathcal{V}_1} \, \int_{\mathcal{V}_2} \, E_{\rm vdw}(\rm{mol\,1-mol\,2}) \,d\mathcal{V}_1 \,d\mathcal{V}_2
   \label{eq:Fadh_int}
  \end{equation}
  Note that we use capital letters for interactions between macroscopic bodies and lowercase letters for molecular interactions. The resulting VDW force is then simply obtained by deriving the potential energy with respect to the separation distance $h$: 
	\begin{equation}
	\mb{F}_{\rm VDW}=-\frac{dE_{\rm VDW}}{dh}~.
	\end{equation}
   
  The main interest of this approach is that it can be used regardless of the particle shape and topography. It only requires to define the inter-molecular potential energy: this is often taken as a Lennard-Jones potential $E_{\rm L-J} = C_1/h^{12}-C_2/h^6$ (with $C_1$ and $C_2$ two constants), which combines a van der Waals attractive term ($1/h^6$ law) and a repulsive term due to Pauli exclusion principle \cite{israelachvili2011intermolecular}. Yet, the computational costs are very high since VDW forces are obtained from a molecular integration (similar to Molecular Dynamics in terms of the level of description). 
   
  \item Second, approximate expressions can also be obtained using the Hamaker approach. This approach still adopts a microscopic point of view, where each macroscopic body is described in terms of its molecules. However, instead of performing direct numerical integration, simplified formulas are obtained with the following assumptions: (a) dispersion forces between molecules inside the same body are neglected; (b) only simple geometries are considered (where mathematical integration is possible). For instance, assuming that the basic molecule-molecule interactions is given by the unretarded energy $E_{\rm vdw} = -C/h^6$ (with $C$ a constant), one obtains the unretarded VDW interactions between two spheres \cite{hamaker1937london}:
  \begin{align}
   E_{\rm VDW, Sph1-Sph2} =  -\frac{A_{\rm Ham}}{6} & \left[ \frac{2\,r_{\rm p,1}\,r_{\rm p,2}}{\left(2\,r_{\rm p,1}+2\,r_{\rm p,2}+h\right)\,h}+\frac{2\,r_{\rm p,1}\,r_{\rm p,2}}{\left(2\,r_{\rm p,1}+h\right)\,\left(2\,r_{\rm p,2}+h\right)} \right. \nonumber \\
  & \, \left.+\, {\rm ln}\frac{\left(2\,r_{\rm p,1}+2\,r_{\rm p,2}+h\right)\,h}{\left(2\,r_{\rm p,1}+h\right)\,\left(2\,r_{\rm p,2}+h\right)}\right]~,
  \end{align}
  with $A_{\rm Ham} = \pi^2 \,C\,\rho_1\rho_2$ the Hamaker constant (related to the constant $C$ for molecular vdw interactions and to the number of atoms per unit volume $\rho$).
   
  In general, these approximate formulas are written in the form of a Hamaker constant $A_{\rm Ham}$ multiplied by a geometrical term (accounting for the separation distance $h$ between bodies and for their shape). For instance, using the Derjaguin approximation (which assumes that the separation distance is much smaller than the radius of both bodies), a simplified formula for the unretarded VDW forces between spheres is obtained \cite{elimelech2013particle}:
  \begin{equation}
   E_{\rm VDW, Sph1-Sph2} =  -\frac{A_{\rm Ham}}{6h} \frac{r_{\rm p,1}\,r_{\rm p,2}}{r_{\rm p,1}+r_{\rm p,2}}
   \label{eq:eq_VDW_sphsph}
  \end{equation}
  The formulas can be more complex depending on whether retardation effects are accounted for (more details in \cite{elimelech2013particle, parsegian2005van}). These retardation effects result from the finite propagation velocity of electromagnetic waves between dipoles, which induces vdw energies that decay as $1/h^7$ at larger separation distances between molecules. Such effects are often introduced into the Hamaker expressions by adding a correction term proportional to a characteristic retardation wavelength $\lambda$ (that measures the distance above which retardation effects play a role).
  
  \item Third, a macroscopic point of view can be adopted using the Lifshitz approach \cite{parsegian2005van}: it consists in treating each body as a continuous media instead of considering atomic details. As a result, VDW forces between macroscopic bodies are expressed with the electromagnetic properties of the interacting media (such as dielectric constants and refractive indices). In fact, the formulas are often very similar to those obtained with a Hamaker approach, except that Hamaker constants are here calculated considering the whole dielectric spectra of each material (including retardation effects).
  \end{itemize}

 \item Electrostatic forces occur when particles are electrically charged (e.g., soot particles from combustion engines \cite{maricq2004size}). Expressions for these forces depend on the medium in which the particles are immersed.
 \begin{enumerate}[i -]
  \item When particles are immersed in the air (or a gas), electrostatic forces occur between each charged body. In the case of punctual spherical particles, this force is inversely proportional to the square of the distance between the objects $h$:
  \begin{equation}
   \mathbf{F}_{\rm elect} (h) = \frac{k_0 \, q_{\rm p,1} \, q_{\rm p,2} \, e^2}{h^2} \, \mb{x}_{1-2}
   \label{eq:Felect}
  \end{equation}
  where $q_{\rm p,1}$ and $q_{\rm p,2}$ are the number of elementary charges for each particle, $e=$\SI{1.602176634 e-19}{C} the elementary charge, $k_0 = $\SI{8.987551792e9}{kg.m^3/s^2/C^2} the Coulomb constant and $\mb{x}_{1-2}$ the unit vector connecting the two particles.
  
  \item Electrostatic double-layer forces arise between charged bodies immersed in a liquid. In that case, the surface charge of each particle is balanced by an excess of oppositely-charged ions present in the solution near the surface \cite{israelachvili2011intermolecular}: this results in the formation of a so-called electrostatic double-layer (i.e., a volume of fluid in close proximity of the body where counter-ions are more abundant than co-ions). When two bodies are in close proximity, the electrical double-layer of each body can overlap with each other. In turn, this overlap induces either an attractive force (when the two bodies are oppositely charged) or a repulsive interaction (for bodies with the same sign of charges). This force is called the electrostatic double-layer (EDL) interaction. Like VDW forces, EDL forces are relatively short-ranged (with a range that depends on the concentration of ions in the solution, typically from a few tenths of nanometers up to a few hundreds of micrometers). Hence, EDL forces are not acting at distances as large as electrostatic forces in air. 
  
  Expressions of EDL forces between two charged bodies can only be obtained by solving the Poisson-Boltzmann equation with appropriate boundary conditions at the surfaces. For instance, assuming that the surface potential remains constant when the two bodies come closer and that the potentials are small enough to use a linearized Poisson-Boltzmann equation, one obtains the following equation for EDL potential energy between two spheres \cite{elimelech2013particle}:
  \begin{equation}
   \label{eq:EDL}
    U_{\rm EDL,Sph1-Sph2} = 64\pi \varepsilon_{0}\varepsilon_{R} \left( \frac{k_B\,T_{\rm f}}{z\,e} \right)^{2} \text{tanh}^2\left(\frac{z\,e\,\Psi_0}{4\,k_B\,T_{\rm f}}\right) \, \left(\frac{r_{\rm p,1}\,r_{\rm p,2}}{r_{\rm p,1}+r_{\rm p,2}}\right) \, e^{-\kappa h}~,
   \end{equation}
   with $\Psi_0$ the surface potential, $k_B=$\SI{1.38 e-23}{J.K^{-1}} the Boltzmann constant, $T_{\rm f}$ the fluid temperature, $\varepsilon_{0}=$\SI{8.854e-12}{C.V^{-1}.m^{-1}} the dielectric permittivity of vacuum, $\varepsilon_{R}$ the dielectric constant of the media (e.g., $\varepsilon_{R}=78.5$ for water), and z the valence of ions present in the solution. Equation~\eqref{eq:EDL} involves the inverse Debye length $\kappa$, defined by:
   \begin{equation}
    \kappa = \sqrt{\frac{e^{2}\sum_{i}n_{i}^{0}z_{i}^{2}}{\epsilon_{0}\epsilon_{r}k_{B}T_{\rm f}}} = \sqrt{\frac{2e^{2}I}{\epsilon_{0}\epsilon_{r}k_{B}T_{\rm f}}}~,
   \end{equation}
   $I$ being the solution ionic strength.
 \end{enumerate}
  
 \item For particles immersed in a viscous fluid, an additional force comes from the extra pressure generated by the interstitial fluid that is being squeezed between the two surfaces approaching each other. Due to viscous effects at such small scales, an extra repulsive force is induced, called the lubrication force in the multiphase flow community \cite{uhlmann2005immersed, kempe2012collision}. When the flow velocity field is known even at scales much smaller than the separation distance between particles $h$, this force is fully determined. Yet, in simulations of particle resuspension, the fluid flow is not resolved at such microscale distances (see Section~\ref{sec:models:turb_rough:turb}). Hence, additional models for lubrication forces are usually introduced. For instance, in the case of two spherical particles nearing each other, lubrication forces are often expressed by \cite{cox1967slow}:
 \begin{equation}
  \mb{F}_{\rm lub} = \left\{
  \begin{aligned}
   & -\frac{6\pi\nu_{\rm f}\rho_{\rm f}\,\mb{U}_{\rm r}\cdot\mb{n}_{\perp}}{h} \, \frac{r_{\rm p,1}\,r_{\rm p,2}}{r_{\rm p,1}+r_{\rm p,2}} \, \mb{n}_{\perp}, \quad & \text{if } h_{min} \le h\le h_{max} \\
   & \quad 0, & \text{elsewhere}  
  \end{aligned}
  \right.
 \end{equation}
 where $\mb{U}_r\cdot\mb{n}_{\perp}$ is the projection of the relative velocity between the two particles along the direction connecting their center of gravity $\mb{n}_{\perp}$. The force is set to $0$ when the separation distance is larger than $h_{max}$, which is usually taken as twice the smallest scale solved by the solver for the flow field (see Section~\ref{sec:models:turb_rough:turb}): in fact, at higher distances, such effects would already be accounted for and do not need to be modeled. Similarly, another minimum cutoff distance $h_{min}$ is introduced to avoid the singularity at $h=0$. The main reason for this cutoff distance $h_{min}$ is that real surfaces are rough (see Section~\ref{sec:models:turb_rough:rough}), meaning that, at the macroscopic scale, contact happens at a short (but non-zero) distance. Another justification for this cutoff distance could arise even without accounting for surface roughness by bridging the fields of multiphase flow and interface science. In fact, in the latter field, specific forces account for the repulsion arising from the ordering of solvent molecules when only a few layers are left in the small interstitial space between the two surface (see Fig.~\ref{fig:sketch_forces1}). These forces are referred to as solvation or structural forces (or even hydration forces when the medium is water). When dealing with perfectly smooth surfaces, the solvation potential is usually approximated by an exponentially decaying cos-function of the form \cite{israelachvili2011intermolecular}:
 \begin{equation}
  E_{\rm solv} \approx - 2 \, \Delta\gamma \, \text{cos}\left(\frac{2\pi h}{\sigma}\right) \, e^{-h/\sigma}~,
  \label{eq:Fsolv}
 \end{equation}
 where $\sigma$ is the size of solvent molecules, and $\Delta\gamma$ is the interfacial energy between the two surfaces (which is a a function of the interfacial energy of each surface taken alone and of the medium). This gives rise to a number of potential wells at different separation distances at the molecular scale (one of which could be related to the cutoff distance used in more macroscopic descriptions for lubrication forces).
  
 \item Capillary forces develop when a capillary bridge forms between two bodies \cite{israelachvili2011intermolecular}. This effect induces higher adhesion between the two bodies, since separating them requires breaking the capillary bridge. Such capillary adhesion is especially important when dealing with particles in the air with a high relative humidity (as for bacteria in indoor surfaces \cite{salimifard2017resuspension} or sand dunes \cite{fecan1998parametrization}). Alternatively, it can also occur for particles immersed in a liquid but with a high relative saturation (meaning that air bubbles may spontaneously form). The strength of the capillary bridge also depends on the surface properties of the two bodies: due to their affinity with water, it is easier to form capillary bridges with hydrophilic surfaces (e.g., dust mite) than with hydrophobic particles (such as cat fur, dog fur) \cite{salimifard2017resuspension}. 
  
 In general, the force due to a capillary bridge between a sphere and a smooth plate can be expressed using the Laplace pressure in the liquid, which acts on a small area between the two surfaces \cite{israelachvili2011intermolecular}:
 \begin{equation}
  \mb{F}_{\rm capil} = - 4 \, \pi \, r_{\rm p} \, \gamma_{\rm L} \, {\rm cos}\, \theta \, \left(1-\frac{h}{2\,r_{\rm menisc}\,{\rm cos}\theta}\right) \mb{n}_{\perp}~,
 \end{equation}
 where $\gamma_{\rm L}$ is the liquid interfacial surface tension (i.e. between the air and the liquid), $\theta$ the liquid contact angle and $r_{\rm menisc}$ the radius of curvature of the liquid meniscus formed between the two surfaces.
  
 \item Specific chemical bindings arise due to strong intermolecular bonds between atoms, ions or molecules (such as covalent, ionic or metallic bonds) \cite{israelachvili2011intermolecular}. Other specific surface forces can arise between inorganic bodies due to the presence of polymers or surfactants in the solution. In fact, the presence of polymers attached to each body can result in more complex polymer-mediated interactions between the two bodies. This includes both repulsive steric forces (due to Pauli exclusion principle between polymer-covered surfaces) as well as attractive intersegment and bridging forces (when the polymers form bonds like hook-and-loop fasteners). The expression of such forces is not straightforward since it depends on how polymers are arranged near the surface. In general, the interaction potential is often considered to be an exponentially decaying function of the separation distance $h$ (more details in \cite{israelachvili2011intermolecular}).
  
 \item Singular biological forces can take place when dealing with organic or biological bodies (such as pollen or bacteria). In addition to the forces mentioned previously, selective forces can significantly increase the adhesion between two bodies (e.g., due to ligand-receptor interactions or due to bridging effects when polymers get bounded to both surfaces) \cite{israelachvili2011intermolecular}. Another specificity of the biological forces is related to their non-equilibrium properties, mostly due to the fact that organic surfaces continuously change according to their interactions with the local environment. For instance, organisms like bacteria can grow on the surface of water distribution systems \cite{liu2016understanding} (provided they find sources of proteins, lipids or other compounds on the surface). Hence, while the previously mentioned forces can be viewed as `static', biological interactions are by essence `dynamic'. For the sake of simplicity, biological particles are not specifically addressed in the rest of this review (unless specifically mentioned to insist on the similarities with non-organic materials).

\end{enumerate}

\subparagraph{Role in resuspension:} Non-contact forces play a double role in particle resuspension. On the one hand, many of these forces act on particles deposited on the surface (together with contact surfaces described in the next Section~\ref{sec:models:forces:contact}). On the other hand, these forces affect the dynamics of particles once they are detached from the surface. Hence, like hydrodynamic forces, they play a key role in the re-entrainment phase. At this stage, it is important to realize that short-ranged surface forces are often characterized by a distribution of possible values (due to effects of surface roughness, see Section~\ref{sec:models:turb_rough:rough}), which is responsible for the range of values at which resuspension occurs for a given type of particles and flow condition.

  \subsubsection{Contact forces}
   \label{sec:models:forces:contact}

\begin{figure}[ht]
 \centering
 \includegraphics[width=\textwidth, trim=0cm 0cm 0cm 12.0cm, clip]{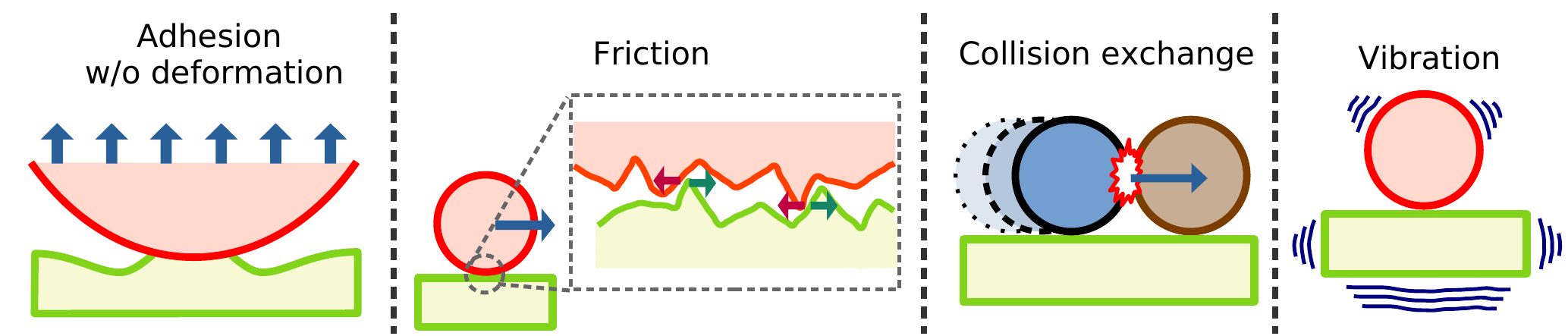}
 \caption{Illustration of contact forces at play in particle resuspension.}
 \label{fig:sketch_forces2}
\end{figure}

\subparagraph{Physical origin:} Various physico-chemical effects play a role in the forces acting between two objects in contact (see also Fig.~\ref{fig:sketch_forces2}). This includes contributions from: (a) adhesion forces, which belong to the category of surface forces and are responsible for the adhesion between two surfaces (with or without surface deformations); (b) friction forces, which resist the lateral motion between two surfaces in contact due to the non-trivial contact between them; (c) momentum/energy exchanges during inter-particle collisions (e.g., due to action-reaction principles and conservation laws); (d) vibrations, which can induce extra mechanical effects that can favor particle resuspension.

\subparagraph{Expressions:} Contact forces can be evaluated using a range of descriptions.
\begin{enumerate}[A.]
 \item Adhesion of solids arise from short-ranged intermolecular interactions between the molecules composing each solid. These forces can be calculated using various models, which can be classified in two main categories sketched in Fig.~\ref{fig:sketch_adhesion_models} (see \cite{israelachvili2011intermolecular, henry2014progress, prokopovich2011adhesion} for more details): those based on Hamaker's approach for van der Waals forces \cite{hamaker1937london} and those based on contact mechanics theories \cite{johnson1987contact}.
 
  \begin{enumerate}[i -]
   \item The Hamaker model describes adhesion as a result of the short-range attractive force between surfaces at very close separation distances. For instance, many models are based on a Lennard-Jones potential for inter-molecular interactions, which includes two contributions: an attractive term that scales as $1/h^6$ (due to the London dispersion force) and a repulsive term that scales as $1/h^{12}$ (due to Pauli repulsion effect that prevents the overlap of electron orbitals). The interest of such descriptions is that it allows to grasp what is meant by the notion of `contact'. In fact, when two macroscopic bodies are in contact, this means that the two surfaces are separated by a distance typically smaller than or within the nanometer range (since the molecules from each body cannot overlap due to Pauli exclusion principle). 
   
   Using the same Hamaker approach introduced previously for VDW forces (see Section~\ref{sec:models:forces:non-contact}), formulas between simple geometrical objects have been obtained by integrating inter-molecular forces over the volume of both objects. In fact, the adhesion force obtained is usually very similar to the VDW force obtained for a separation distance equal to the ``contact distance'' $z_0$ (since repulsive forces decay extremely rapidly with the separation distance). At this stage, it is worth noting that other inter-surface contributions (such as electric double-layer forces, solvation forces, steric forces) can sometimes be non-negligible and are easy to incorporate in such descriptions of adhesion forces based on a Hamaker model. 
   
   \begin{figure}[ht]
    \centering
    \includegraphics[width=0.75\textwidth]{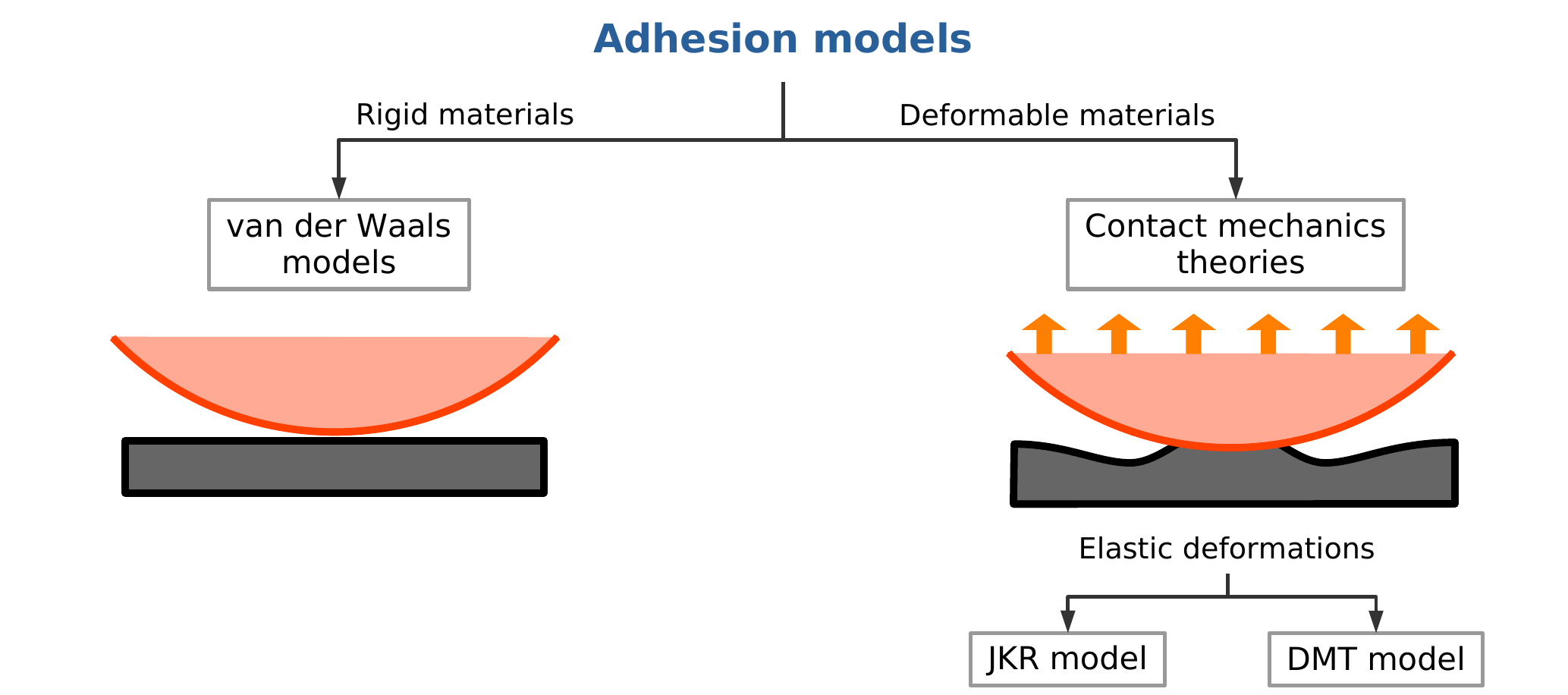}
    \caption{Illustration of the various modeling approaches to evaluate the adhesive force between surfaces. Models can be either based on descriptions using van der Waals forces or contact mechanics theories (with models based on Hertz theory for elastic deformation, such as JKR or DMT models).}
    \label{fig:sketch_adhesion_models}
   \end{figure}
   
   \item Contact-mechanics models are concerned with surface deformation of the two bodies in contact \cite{johnson1987contact} (see also Fig.~\ref{fig:sketch_adhesion_models}). In practice, various models have been suggested in the literature to account for the elastic deformations of surfaces upon contact. This includes the JKR theory (named after the work of Johnson Kendall and Roberts \cite{johnson1971surface}) and the DMT theory (named after the work of Derjaguin, Muller, and Toporov \cite{derjaguin1975effect}). 
   \begin{itemize}
    \item The JKR theory is based on the assumption that surfaces deform elastically upon contact according to the Hertz theory (i.e., for elastic deformations) and that adhesion occurs only within the contact area. As a result, the pull-off force (i.e. the force required to separate two surfaces) predicted by the JKR theory between a sphere and a plate is given by:
    \begin{equation}
     F_{\rm JKR} = 3\pi\, \Delta\gamma \, r_{\rm p}~,
     \label{eq:adh_JKR}
    \end{equation} 
    with $\Delta\gamma$ the surface energy between the two objects. The JKR theory is valid only for large and `soft' particles (i.e., highly deformable). 
    
    \item The DMT theory is based on the assumption that surfaces deform elastically upon contact according to the Hertz theory but that adhesion includes also non-contact forces (van der Waals contribution) which act across the gap between the two surfaces. As a result, the pull-off force predicted by the DMT theory between a sphere and a plate is given by:
    \begin{equation}
     F_{\rm DMT} = 4\pi\, \Delta\gamma \, r_{\rm p}~.
     \label{eq:adh_DMT}
    \end{equation}
    The value obtained is larger than the one predicted with the JKR theory (due to the additional contribution from VDW forces outside the contact area). The DMT theory is valid for small or `hard' particles (i.e., slightly deformable).

   \end{itemize}
  
   More recently, extended theories for elastic deformations have been proposed to encompass both JKR and DMT theories, while naturally recovering both theories in the limit cases of soft or hard particles. Other formulations include the role of plastic deformations in adhesion forces (e.g. the Maugis-Pollock model \cite{maugis1984surface}). In fact, all materials can deform to a certain extent under the action of an external load/force \cite{callister2000fundamentals, bertram2012elasticity}. When the load is small, the surface on which this load is applied undergoes an elastic deformation. During this phase, the induced deformation (also called strain in material sciences) depends linearly on the magnitude of the force applied (called stress in material sciences). This relation is actually given by Hooke's law and the constant of proportionality is referred to as the modulus of elasticity (also called Young's modulus). Such elastic deformations are reversible (meaning that the surface returns to its original shape once the stress is removed) until the yield point. When the force is so high that the resulting stress becomes higher than the yield stress, plastic deformation happens. At that point, some of the deformations induced in the structure become permanent and non-reversible. During plastic deformations, the relationship between the stress and the deformation does not follow a linear trend anymore. At even higher values of the stress, fracture occurs resulting in permanent damage to the surface structure (see Fig.~\ref{fig:sketch_deform}).
   \begin{figure}[ht]
    \centering
    \includegraphics[width=0.85\textwidth]{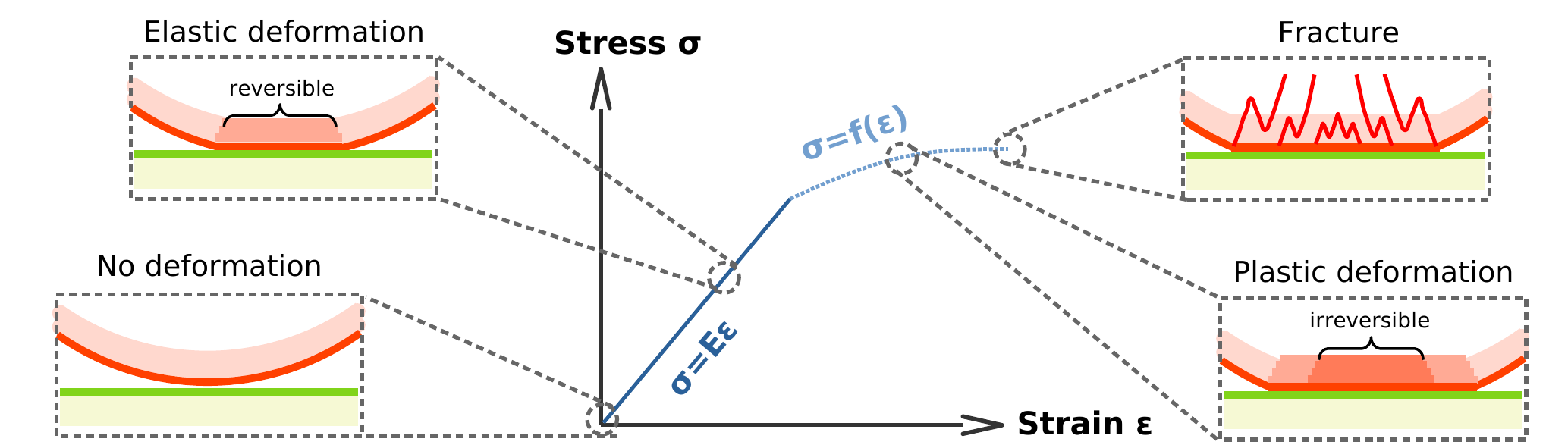}
    \caption{Illustration of the stress-strain curve showing the deformation of a particle in contact with a smooth surface (all the way from elastic deformation to plastic deformation and eventually fracture).}
    \label{fig:sketch_deform}
   \end{figure}

  \end{enumerate}
  
  \item Friction forces correspond to the resistance to lateral motion when two bodies are in contact. Unlike previously mentioned forces, these forces do not follow a force-versus-distance law but rather arise only as a reaction to motion or to another force (interested readers can find more details in the literature related to tribology, see \cite{israelachvili2011intermolecular, persson2013physics, persson2013sliding, popova2015research}). From a microscopic point of view, friction forces emerge from the contact between two irregular surfaces, i.e., surfaces that are not smooth (which is always the case at the molecular level). As sketched in Fig.~\ref{fig:sketch_forces2}, the presence of `protruding' elements on the surface of each body leads to a sort of entrapment which resists lateral motion. To put it differently, the two irregular surfaces can only move freely in the lateral direction once they are separated by a distance larger than the size of the irregularities present on the surfaces. This means that lateral motion is preceded by a short motion along the perpendicular direction. Yet, this perpendicular motion is prevented by adhesive forces (such as VDW contributions) and/or load forces (such as gravity). Hence, lateral motion is only possible provided that the lateral force is high enough to overcome the forces that oppose this short separation along the perpendicular direction. From this simplified picture of friction forces, it becomes evident that the stronger the adhesion between two bodies, the stronger the resistance to motion and hence the higher is the friction force. In practice, friction forces are expressed with the ratio between lateral and perpendicular forces: this is the so-called friction coefficient $\mu_{\rm frict}$. This gives:
  \begin{equation}
   F_{\rm resist,||} = \mu_{\rm frict} \, F_{\perp}~.
  \end{equation}
  A distinction is usually introduced between the static friction coefficient (i.e., when the two surfaces are not moving yet) and the dynamic friction coefficient (i.e., when the two surfaces are already moving). The values of these coefficients are often tabulated \cite{tetb_2004_friccoef}.
  
 \medskip
 \item Exchanges during inter-particle collisions:
 
 Deposited particles can also be set into motion when they are hit by other moving particles (either in suspension or migrating on the surface) \cite{rondeau2021evidence, banari2021evidence}. This brings out two important questions (see Fig.~\ref{fig:sketch_collision}):
 \begin{enumerate}[i -]
  \item How to detect collisions?
  
  Detecting particle collisions implies to take into account that particles have a finite size. This means that specific algorithms are used to detect when and where two particles collide, such as:
  \begin{itemize}
   \item When detailed information is available on the position and velocity of each particle, one can resort to deterministic algorithms. For instance, assuming particles to be non-deformable `hard' spheres, a collision happens if two particles overlap, i.e. if the inter-particle distance $h$ becomes smaller than the sum of the two radii at any time $h(t)\le r_{\rm p,1}+r_{\rm p,2}$ \cite{chen1999direct}. Otherwise, assuming a fixed velocity during each small time interval, one can also use geometric criteria to detect if two particles collide during this time interval \cite{sigurgeirsson2001algorithms}.
   \item Alternatively, when the exact trajectory of each particle is not known exactly during an observation time, one can resort to probabilistic algorithms to detect collisions. These algorithm use a-priori information on the relative motion between each pair of particle to estimate the probability that a pair of particle does collide during a certain time (as in \cite{sommerfeld2019advances, henry2014astochastic, o1981collective}).
  \end{itemize}
  
  \begin{figure}[ht]
   \centering
   \includegraphics[width=0.8\textwidth,trim=1.5cm 0cm 1.3cm 0cm, clip]{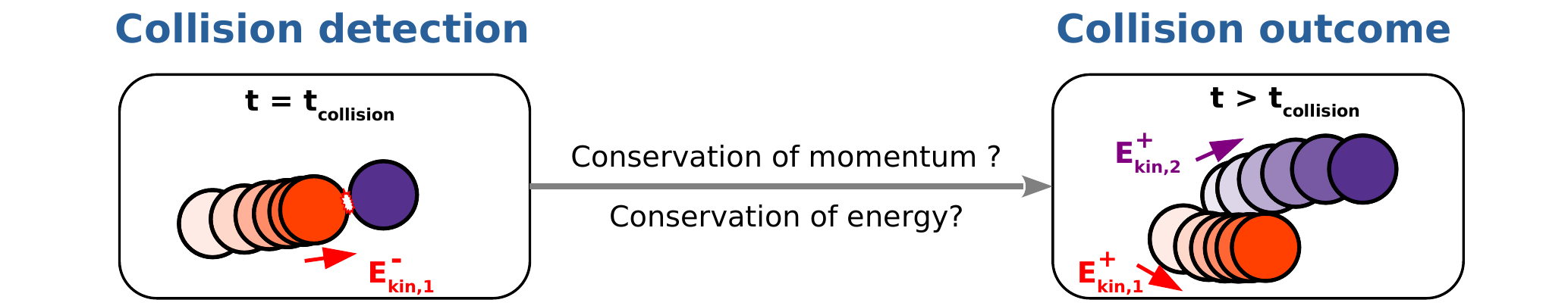}
   \caption{Illustration of the issues related to inter-particle collisions: the detection of such collisions (left) and the treatment of each collision outcome (right).}
   \label{fig:sketch_collision}
  \end{figure}
   
  \item How to treat collisions?
  
  Once a collision is detected, the outcome of each event has to be determined. In practice, the outcome of a collision depends on the conservation laws that apply to the system made up by the two particles and the surface. In a closed system, momentum is conserved but not necessarily kinetic energy. For instance, elastic collisions correspond to collisions where kinetic energy is conserved during the collision while inelastic collisions occur when a fraction of the energy is dissipated (e.g., due to plastic deformations). These phenomena have been extensively studied in the field of impact mechanics, which is concerned with the ``reaction forces that develop during a collision and the dynamic response of structures to these reaction forces'' \cite{stronge2018impact}. In the context of particle collisions away from boundaries, they are traditionally modeled by a restitution coefficient $e_{coll}$, which measures the ratio between the kinetic energy just before impact $E_{\rm kin}^-$ and the kinetic energy right after it $E_{\rm kin}^+$. For particles impacting a surface, the same ideas are used although one makes a difference between the normal and tangential components \cite{beladjine2007collision}:
  \begin{subequations}
   \begin{align}
    E_{\rm kin,\perp}^+ &= e_{\rm coll,\perp} \, E_{\rm kin,\perp}^+ \\
    E_{\rm kin,||}^+    &= e_{\rm coll,||} \, E_{\rm kin,||}^+
   \end{align}
  \end{subequations}
  Hence, such formulations can be used to determine the motion of each particle after the collision, thereby allowing to capture successive collisions and avalanche phenomena.
 \end{enumerate}

 \item Mechanical forces (like those induced by vibration) can be relevant in some applications \cite{aracena2018frequency, chatoutsidou2021resuspension}. For instance, in walking-induced resuspension, vibration is induced when a foot hits the ground. Part of the vibrating energy can be transferred to the particle, thanks to the contact between surfaces. Yet, when surfaces are deformable, vibration can facilitate/hinder the breakage of the contact depending on the vibrating frequency. Vibration can also be important in the context of outdoor resuspension from vegetation, where the wind can shake grass blades and tree branches. 
  
\end{enumerate}

\subparagraph{Role in resuspension:} Contact forces are important only for particle detachment since they vanish when the particle/substrate contact is ruptured. They actually play a key role in the detachment of small colloidal particles, since adhesion forces are the most significant ones preventing particle motion.

  \subsubsection{Summary of forces}
   \label{sec:models:forces:summary}

From this overview of the expressions for the forces/torques acting on particles, it appears that a closed set of equations can be defined to describe particle dynamics. However, this requires to carry out a careful analysis of the input variables (available from experiments) and the output ones (those that should be predicted). This partially explains the profusion of models available in the literature, since each model can introduce slight changes either in the equations of motion or in the expressions of the forces/torques involved.

This brief catalog illustrates not only the diversity of fields involved in particle resuspension but also the complexity of the forces at play. In fact, it appears that the forces listed previously can sometimes act to hinder particle motion or favor it, depending on the conditions. Table~\ref{tab:force_contrib} summarizes how each of these forces contribute to the resuspension of particles.
\begin{table}[ht]
 \centering
 \begin{footnotesize}
 \centering
 \begin{tabular}{|>{\centering}m{4.5cm}>{\centering}m{4.5cm}|>{\centering}m{2.5cm}|>{\centering\arraybackslash}m{2.5cm}|}
  
  \hline
  \multicolumn{2}{|c|}{\multirow{2}{3.5cm}{\centering Type of force}} & \multicolumn{2}{c|}{Role in resuspension} \\
  
  \cdashline{3-4}
  & & Inducing motion & Resisting motion \\
  
  \hline
  
  \multirow{2}{4.5cm}{\centering \makecell{Interactions with \\external fields}} & Gravity & \makecell{\yes \\ \textit{(ceiling)}} & \makecell{\yes \\ \textit{(floor)}}  \\
  
  \cdashline{2-4}  
  & Electro-magnetic & \yes & \maybe \\
  
  \hline
  
  \multirow{8}{4.5cm}{\centering Hydrodynamic forces} & Pressure gradient & \maybe & \maybe  \\
  
  \cdashline{2-4}  
  & Buoyancy & \makecell{\yes \\ \textit{(lighter particles)}} & \makecell{\yes \\ \textit{(heavier particles)}} \\
  
  \cdashline{2-4}  
  & Drag & \yes & \no  \\
  
  \cdashline{2-4}  
  & Lift & \yes & \no  \\
  
  \cdashline{2-4}  
  & Added-mass & \yes & \no  \\
  
  \cdashline{2-4}  
  & Basset history & \yes & \no  \\
  
  \cdashline{2-4}  
  & Brownian motion & \yes & \maybe  \\
  
  \hline
  \multirow{9}{4.5cm}{\centering \makecell{Non-contact forces \\ \\ \textit{(incl. particle-substrate} \& \\ \textit{particle-particle interactions)}}} & van der Waals & \no & \yes  \\
  
  \cdashline{2-4}
  & Electrostatic (in air/liquid) & \makecell{\yes \\ \textit{(Opposite charges)}} & \makecell{\yes \\ \textit{(Similar charges)}} \\

  \cdashline{2-4}
  & Lubrication / solvation & \maybe & \maybe \\

  \cdashline{2-4}  
  & Chemical bindings & \makecell{\yes \\ \textit{(repulsive)}} & \makecell{\yes \\ \textit{(attractive)}}  \\
  
  \cdashline{2-4}  
  & Biological & \makecell{\yes \\ \textit{(repulsive)}} & \makecell{\yes \\ \textit{(attractive)}} \\
  
  \cdashline{2-4}  
  & Capillary & \no & \yes  \\
  
  \hline
  
  \multirow{4}{4.5cm}{\centering  \makecell{Contact forces \\ \\ \textit{(incl. particle-substrate} \& \\ \textit{particle-particle interactions)}}} & Adhesion & \no & \yes \\
  
  \cdashline{2-4}  
  & Friction & \no & \yes  \\
  
  \cdashline{2-4}
  & Collision exchange & \yes & \no  \\
  
  \cdashline{2-4}  
  & Mechanical (vibration) & \yes & \no  \\
  
  \hline
  
 \end{tabular}
 \caption{Summary of the forces and their effect on resuspension (especially whether each force favors or hinders resuspension).}
\label{tab:force_contrib}
\end{footnotesize}
\end{table}

 \subsection{Accounting for turbulence and surface roughness}
  \label{sec:models:turb_rough}

The hydrodynamical forces detailed in Section~\ref{sec:models:forces} appear as deterministic expressions but require to have access to the exact velocity near/at particle locations. The same is true for adhesion forces which require to know the exact topographies of the surfaces involved. In practice, such information is not always available or is provided in a statistical sense. For instance, in turbulent flows, the velocity at a given point is often described only in a probabilistic manner. Similarly, surface profiles can be depicted as `rough' and characterized only by some statistical measures. This implies that hydrodynamical and adhesion forces are to be addressed also in a statistical sense. To understand the impact of turbulence and of surface roughness on the key forces at play in particle resuspension, it is therefore useful to recall how turbulent flows and rough surfaces are modeled in practice. This is done here by considering, first, turbulence models in Section~\ref{sec:models:turb_rough:turb} and, second, how surface roughness is represented in Section~\ref{sec:models:turb_rough:rough}.

  \subsubsection{Capturing the impact of turbulence on resuspension}
   \label{sec:models:turb_rough:turb}
  
\begin{figure}[ht]
 \centering
 \includegraphics[width=0.8\textwidth]{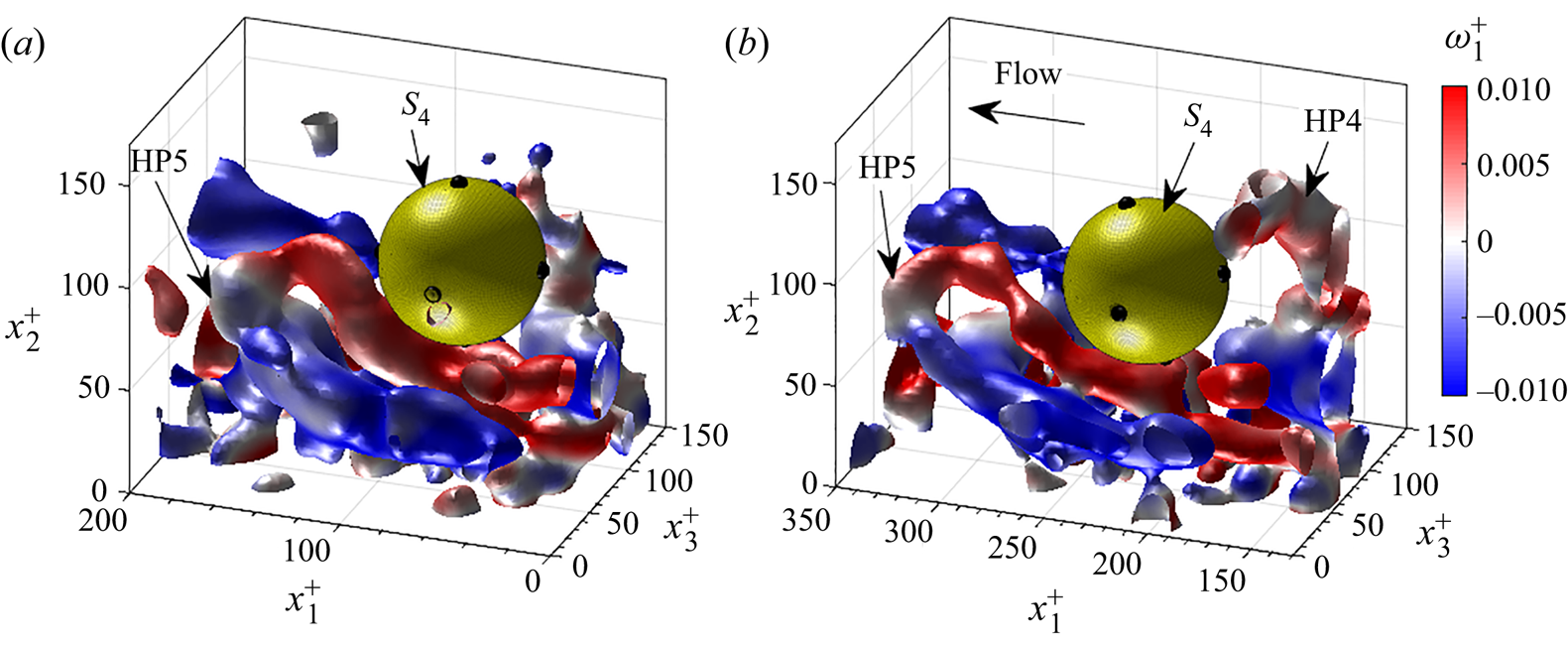}
 \caption{Sequence of two images showing the instantaneous vortical structures in the vicinity of a hydrogel sphere (labelled $S_4$, in yellow with black markers to indicate its change in orientation). The two images are separated by a time $\Delta t^{+}=16.7$. Vortical structures are visualized by iso-surfaces of the second invariant of the velocity gradient tensor (Q-criterion) that were overlaid by the fluid vorticity $\omega _1^+$. Reprinted with permission from \cite{van2022combined}. Copyright 2022, Cambridge University Press. }
 \label{fig:sketch_coherent_struct}
\end{figure}

\subparagraph{Physical origin and characteristics:} Whereas molecular viscosity is a property of a fluid, turbulence is a property of a flow and is manifested by the fact that characteristic fields, such as the velocity, pressure or temperature fields, exhibit variations in space and in time which cover continuously a wide range of scales (interested readers are referred to books dedicated to turbulence \cite{frisch1995turbulence, pope2000turbulent}). This explains that turbulent flows are traditionally treated as random fields (thus the terminology of `fluctuations' to describe apparently chaotic, or rather random, behavior). Classical descriptions of turbulent flows have essentially relied on statistical, or probabilistic, formulations but, in recent decades, experimental and numerical studies have revealed that these random fluctuations can be consistent with the existence of some specific geometrical features, referred to as coherent structures. This is especially the case in the near-wall region of turbulent flows, where the presence of a wall surface induces a highly anisotropic flow, and in which the dynamical role played by long-lasting and intense vortices often named hairpin-vortices (see also Fig.~\ref{fig:sketch_coherent_struct}) has been studied in depth. This is an illustration that the distinction between mean values and fluctuations does not correspond to a clear-cut separation between order and disorder and one of the intricate characteristics of turbulent flows is, precisely, that these fluctuations have non-zero space and time correlations which makes them inherently different from `heat bathes' as in classical thermodynamics. 

\subparagraph{Role in resuspension:} These hairpin vortices induce near-wall velocity fluctuations, which are especially strong in the logarithmic layer but are also found within the viscous sublayer (their intensity being smaller than further away from the wall). Therefore, particles large enough to interact with the regions of strong fluctuations associated to these near-wall coherent structures are more easily set in motion (see for instance \cite{dey2018advances} and references therein). This is typically the case for particles larger than a few tens of micrometers while colloidal particles (i.e., with sizes smaller than a few micrometers) are rather deeply embedded within the viscous sublayer so that their resuspension is much less dependent on near-wall structures. Such phenomena correspond to the so-called episodic bursts in resuspension events \cite{corino1969visual, cleaver1973mechanism, yung1989role, kaftori1995particle1, kaftori1995particle2}. 
 
Another aspect related to turbulence is that it can be induced by thermodynamic (essentially heat) effects, i.e., when surfaces are warmer/cooler than the fluid. For instance, in wildfires, the creation of an updraft at the flame location induces a convergent wind field, resulting in winds close to the surface that differ from the undisturbed ones if the fire was not present. Turbulence is also more intense at the fire front \cite{clements2008first}. The net result is that prescribed fires and wildfires promote resuspension, especially of soil and dust particles, due to the increased wind speeds and turbulence as well as because of the presence of a vertical updraft.

\begin{figure}[ht]
 \centering
 \includegraphics[width=0.8\textwidth]{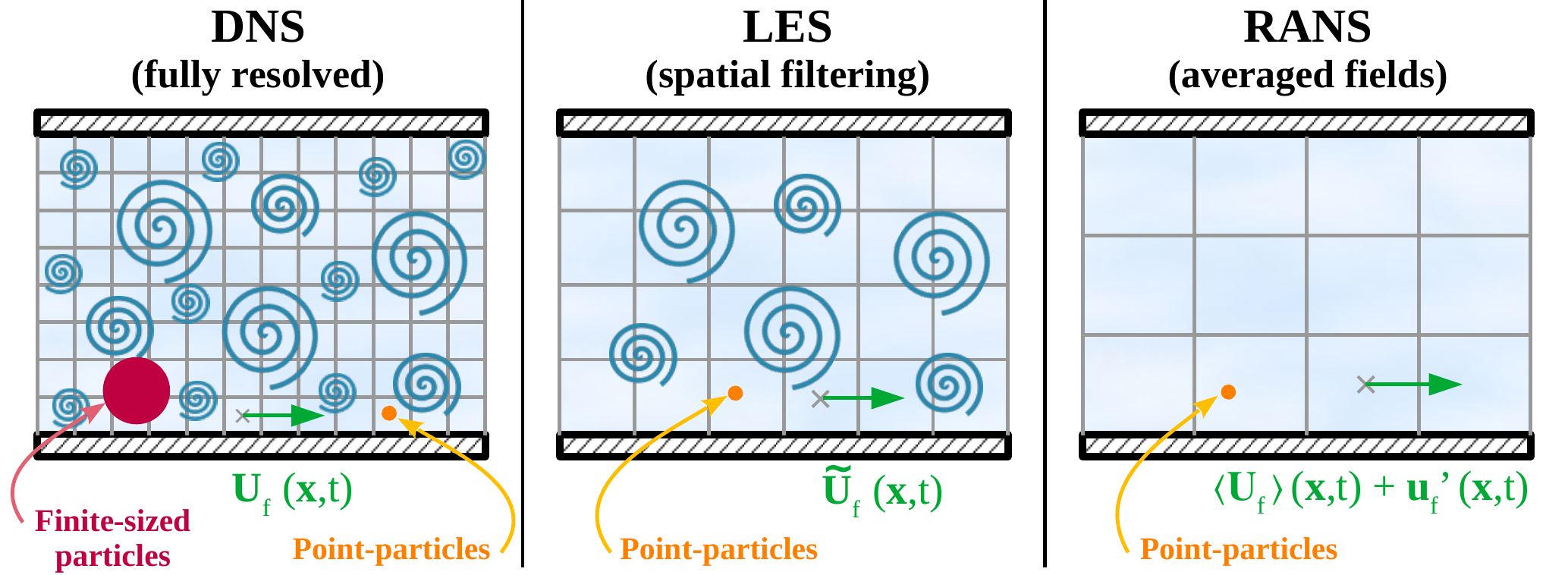}
 \caption{Summary of the main various levels of description for the computation of a turbulent flow.}
 \label{fig:sketch_CFD_fluid}
\end{figure}

\subparagraph{Modeling turbulent flows:} The basic equations governing fluid flows are well-established: these are the Navier-Stokes equations. In turbulence modeling, the issue is therefore to come up with reduced statistical descriptions, formulated in terms of a tractable number of degrees of freedom, still able to capture relevant flow characteristics. The conventional approach consists in applying a statistical operator to the Navier-Stokes equations to derive exact but open equations and, then, to introduce closure proposals. This leads naturally to field-based methods which are solved by using mesh-based numerical methods. In recent years, particle-based models, such as probability density function (PDF) or smooth particle hydrodynamics (SPH) approaches, have also been developed. Their similarities and differences with respect to classical ones have been analyzed in a recent review \cite{minier2016statistical} and are not repeated here. Indeed, to provide the necessary background for the introduction of particle resuspension models in Section~\ref{sec:models:approach}, it is sufficient to limit ourselves to the most usual turbulence models, which are illustrated in Fig.~\ref{fig:sketch_CFD_fluid}. Interested readers are referred to other reviews \cite{minier2016statistical}, as well as existing books on the topic among which \cite{pope2000turbulent, brennen2005fundamentals, frisch1995turbulence, monin2013statistical}. 

The classical turbulence models pictured in Fig.~\ref{fig:sketch_CFD_fluid} can be outlined as follows:
\begin{enumerate}[A.]
 \item In single-phase turbulence, the most detailed level of description consists in solving numerically the Navier-Stokes equations without any turbulence model. This implies that all the degrees of freedom of a turbulent flow must be explicitly captured \cite{moin1998direct} and this approach is referred to as direct numerical simulations (DNS). Given initial and boundary conditions, each DNS computation corresponds to a single realization of the flow and one can extract statistics either by averaging over several realizations or by using time-averages for locally stationary flows or space-averages for homogeneous flows. In the case of an incompressible flow, where the fluid density $\rho_{\rm f}$ is constant, the Navier-Stokes equations read:
 \begin{subequations}
 \label{eq:NavierStokes}
  \begin{align}
   \nabla \cdot \mb{U}_{\rm f} &= 0~, \label{eq:NavierStokes_gen1} \\
   \frac{\partial \mb{U}_{\rm f}}{\partial t} + \mb{U}_{\rm f}\cdotp\nabla\mb{U}_{\rm f} & = -\frac{1}{\rho_{\rm f}}\nabla P_{\rm f} + \nu_{\rm f}\nabla^{2}\mb{U}_{\rm f}+\textbf{F}~, \label{eq:NavierStokes_gen2}
  \end{align}
 \end{subequations}
where $P_{\rm f}$ is the fluid pressure field and $\mb{F}$ possible external forces. The first equation, Eq.~\eqref{eq:NavierStokes_gen1}, corresponds to the conservation of mass written for an incompressible fluid. The second equation, Eq.~\eqref{eq:NavierStokes_gen2}, corresponds to the conservation of momentum and relates the fluid acceleration to all forces acting on the fluid (including pressure, viscous constraints and external forces, like gravity). Additional boundary conditions are applied when dealing with wall-bounded flows (e.g., by imposing a no-slip velocity condition on wall surfaces). 
 
 In dispersed two-phase flow, an additional complexity comes from the treatment of solid particles and two different levels of descriptions can be found:
 \begin{enumerate}[i -]
  \item When the particle size is larger than the Kolmogorov scales, finite-sized particles can be explicitly considered as such in the simulation. This means that the grid size used in the computation of the fluid flow is smaller than the characteristic particle size. Hence, additional boundary conditions are imposed along the grid points where particles are present (to avoid penetration of the fluid inside the particle). As a result, the flow around each particle is resolved (as shown in the left panel of Fig.~\ref{fig:sketch_forces_hydro}): such methods are called PR-DNS (for ``Particle-Resolved Direct Numerical Simulation'').
  
  This approach is consistent with fine descriptions of the particle dynamics, where the hydrodynamic forces due to the relative motion between the fluid and the particle are explicitly calculated by integrating over all the grid points constituting the particle surface (i.e., without resorting to approximate formulas for these hydrodynamic forces) \cite{uhlmann2005immersed, kempe2012collision}. The only force related to the fluid that has to be modeled is the lubrication force, which occurs at scales smaller than the grid size (thus not resolved here, see \cite{uhlmann2005immersed, vowinckel2016entrainment, jain2019collision}). Other non-contact and contact forces can also be added to these descriptions (such as hard-sphere models to avoid the inter-penetration of particles).
  
  \item The point-particle approximation is widely used for particles smaller than the Kolmogorov scales, leading to the so-called PP-DNS \cite{kuerten2016point} (see also the right panel of Fig.~\ref{fig:sketch_forces_hydro}). In that case, the flow around each individual particle is not available. Nevertheless, it is possible to calculate two-phase turbulent flows, including modifications due to the presence of a large number of particles in the flow. This is achieved by including in the term $\mb{F}$ in the Navier-Stokes equations, cf. Eq.~\eqref{eq:NavierStokes_gen2}, the forces exerted by the particles on the fluid which can be obtained by averaging the individual force exerted by a single particle on the fluid (which, according to the action-reaction principle, is simply the opposite of the hydrodynamic forces) over all particles present in each grid cell. In that case, the flow is said to include particle feedback on the fluid or two-way coupling.
  
  Such PP-DNS are consistent with approximate expressions for the hydrodynamic forces (i.e. those due to the relative motion between the particle and the fluid). In fact, a PP-DNS provides the exact instantaneous fluid velocity at any point in space (including at the position of each individual particle) and no additional model is required to evaluate the hydrodynamic forces acting on particles.
 \end{enumerate}

 \item At a less detailed level, Large Eddy-Simulation (LES) consists in solving the Navier-Stokes equations but only down to a certain cut-off length. This cut-off length correspond to a spatial filtering operation (the cut-off length happens to be the size of the grid used). This means that ``large-scales'' motions of the fluid are explicitly computed while residual ``small-scales'' motions are modeled. Hence, each variable is decomposed into a filtered value $\langle\cdot\rangle_{ls}$ and a residual one $\cdot'$, which gives for the velocity field:
 \begin{equation}
  \mb{U}_{\rm f}(t,\mb{x}) = \langle\mb{U}_{\rm f}\rangle_{ls}(t,\mb{x}) + \mb{u}'_{\rm f} (t,\mb{x})~.
 \end{equation}
 Applying this filtering operation, one obtains the spatially filtered Navier-Stokes equation (neglecting particle feedback on the flow):
 \begin{equation}
  \frac{\partial \langle\mb{U}_{\rm f}\rangle_{ls}}{\partial t} +  \nabla\cdotp \left(\langle\mb{U}_{\rm f}\mb{U}_{\rm f}\rangle_{ls}\right) = -\frac{1}{\rho_{\rm f}}\nabla \langle P_{\rm f}\rangle_{ls} + \nu_{\rm f}\nabla^{2} \langle\mb{U}_{\rm f}\rangle_{ls}.
  \label{eq:NS_LES}
 \end{equation}
 This equation differs from the exact Navier-Stokes equation because the filtered product $\langle\mb{U}_{\rm f}\mb{U}_{\rm f}\rangle_{ls}$ is not the same as the product of the filtered velocities $\langle\mb{U}_{\rm f}\rangle_{ls}\langle\mb{U}_{\rm f}\rangle_{ls}$. In fact, the difference is the residual stress tensor $\tau_{\rm f}$ (also called the subgrid-stress tensor). This tensor has to be modeled and a range of approaches exist in the literature (such as the famous Smagorinsky model which is an eddy-viscosity model \cite{pope2000turbulent}).
 
 LES methods are consistent with point-particle approximations but have to be adapted since the exact instantaneous velocity at the position of each particle is not known. In fact, the unresolved part of the velocity field has to be modeled for each particle, using an approach that has to be consistent with the subgrid-scale model used in the fluid simulation.
 
 \item At a more macroscopic level, Reynolds-Averaged Navier-Stokes (RANS) simulations consist in writing mean-flow equations. These methods are based on the Reynolds decomposition of the fluid velocity into its mean and fluctuating parts:
 \begin{equation}
  \mb{U}_{\rm f}(t,\mb{x}) = \langle\mb{U}_{\rm f}\rangle(t,\mb{x}) + \mb{u}'_{\rm f} (t,\mb{x})~.
 \end{equation}
 Applying this true probabilistic averaging operator to the Navier-Stokes equations gives the Reynolds (or mean Navier-Stokes) equation:
 \begin{equation}
 \label{eq:NS_RANS}
  \frac{\partial \langle\mb{U}_{\rm f}\rangle}{\partial t} +  \langle\mb{U}_{\rm f}\rangle\cdotp\nabla\ \langle\mb{U}_{\rm f}\rangle = - \nabla \cdotp \left( \langle \mb{u}'_{\rm f} \otimes \mb{u}'_{\rm f} \rangle \right) -\frac{1}{\rho_{\rm f}}\nabla \langle P\rangle + \nu_{\rm f}\nabla^{2} \langle\mb{U}_{\rm f}\rangle +\langle\textbf{F}\rangle.
 \end{equation}
where the first term on the right hand side is the Reynolds stress tensor which has to be modeled. In practice, two classes of models are traditionally used: (i) local models, where the Reynolds stress tensor is approximated using an eddy-viscosity concept in which this turbulent viscosity is based on known local values of the fluid (such as the turbulent kinetic energy $k$ and the turbulent dissipation rate $\epsilon$ in the famous $k-\epsilon$ models); (ii) non-local models, which solve additional equations for the transport of each component of the Reynolds stress tensor (like Reynolds-stress models, RSM).
 
 RANS approaches are consistent with point-particle approximations but require an additional model to simulate the instantaneous fluid velocity at each particle position. Indeed, RANS simulations provide only one-point statistical information on the flow fields. This means that the instantaneous fluid velocity at any point in space and time is not directly available but can be reconstructed - in a statistical sense - using information on the average velocity and its fluctuating part (more details in \cite{minier2001pdf, minier2016statistical, minier2014guidelines}). 
 
\end{enumerate}

  \subsubsection{Surface roughness}
   \label{sec:models:turb_rough:rough}
  
\begin{figure}[ht]
 \centering
 \begin{subfigure}[c]{0.45\textwidth}
  \centering
  \includegraphics[width=0.78\textwidth, trim=0cm 0cm 1.2cm 7.5cm, clip]{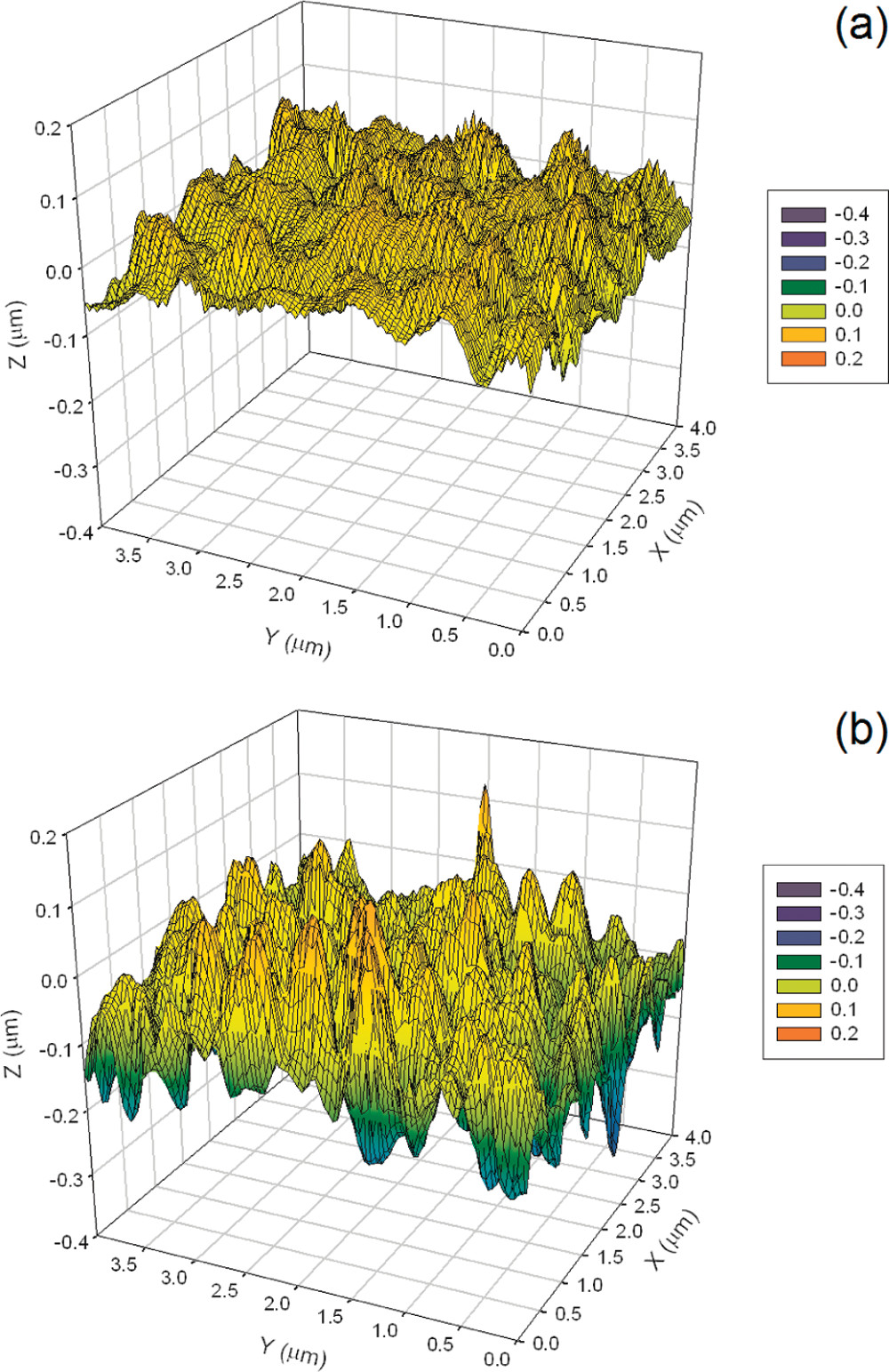}
  \caption{AFM image showing the surface topography of stainless steel. Reprinted with permission from \cite{prokopovich2010multiasperity}. Copyright 2010, American Chemical Society.}
  \label{fig:surf_roughness_steel}
 \end{subfigure}
 \hspace{15pt}
 \begin{subfigure}[c]{0.48\textwidth}
  \centering
  \includegraphics[width=0.85\textwidth, trim=0cm 0cm 7.cm 5.9cm, clip]{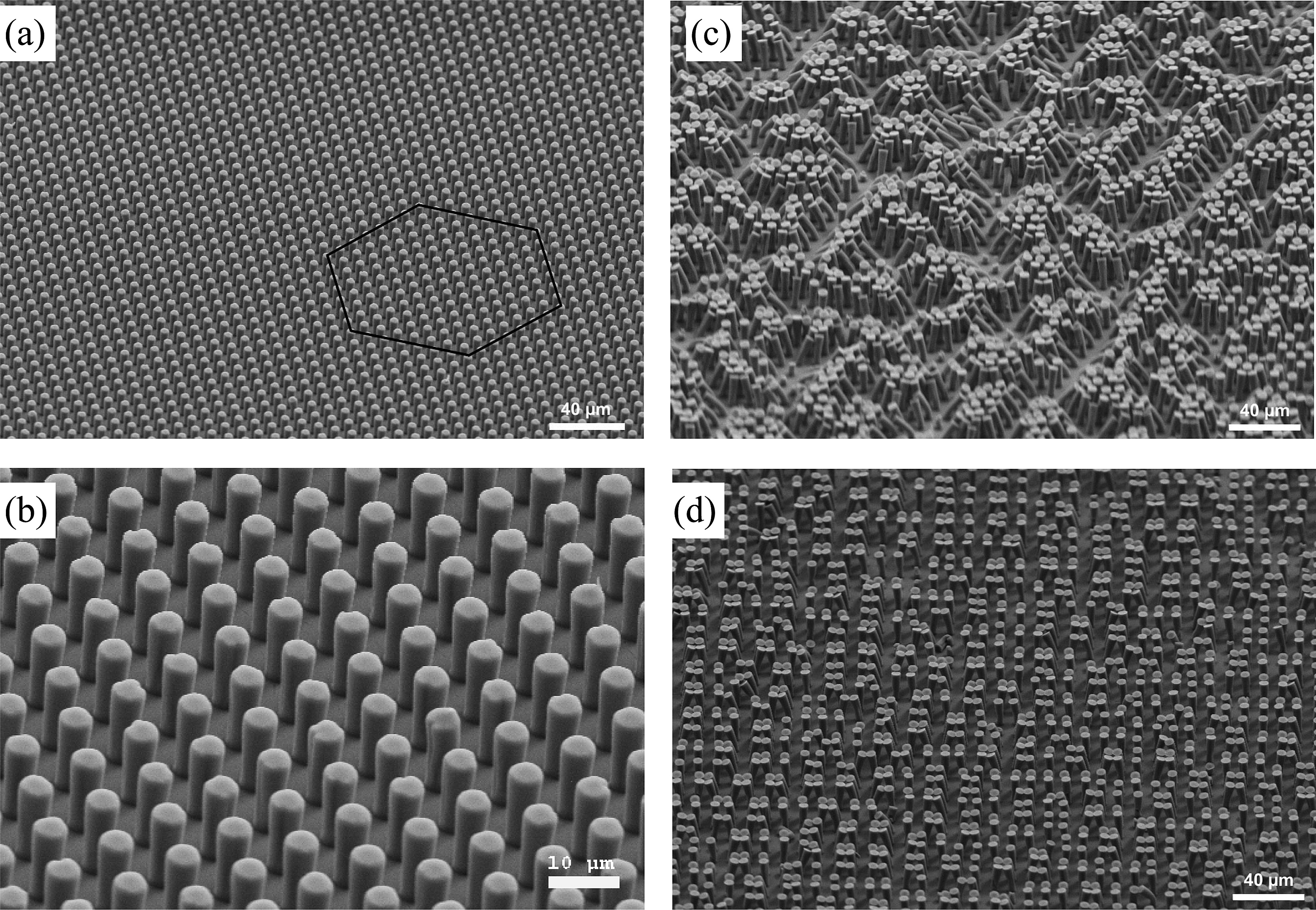}
  \caption{SEM image showing arrays of pillars made using photolithographic techniques (hexagonal packing of pillars with a radius of 2.5 $\mu$m and a height of 20 $\mu$m). Reprinted with permission from \cite{greiner2007adhesion}. Copyright 2007, American Chemical Society.}
  \label{fig:surf_roughness_litho}
 \end{subfigure}
 \begin{subfigure}[c]{0.45\textwidth}
  \centering
  \includegraphics[width=0.65\textwidth, trim=1cm 6.8cm 1.0cm 0.0cm, clip]{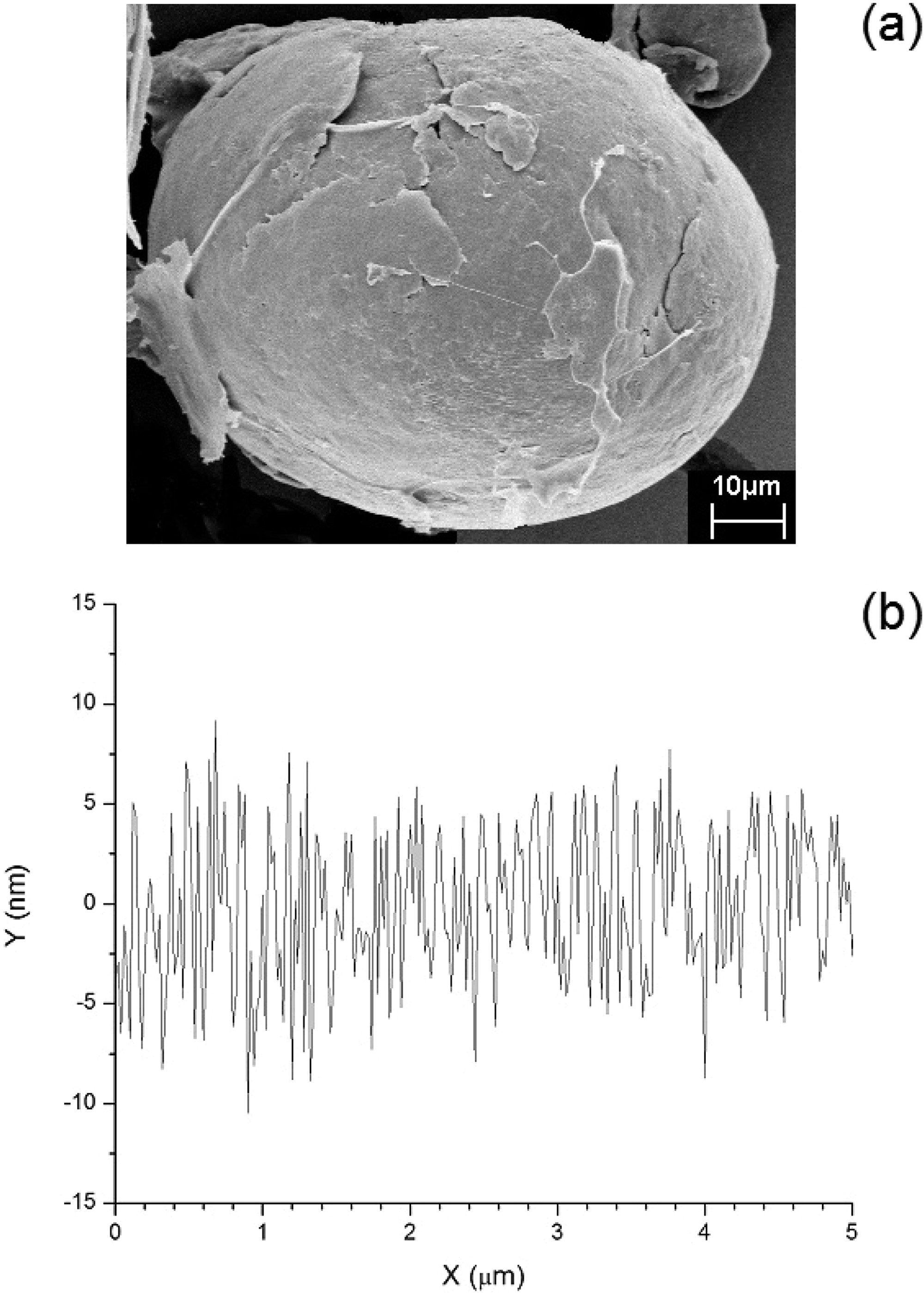}
  \caption{SEM image of a polybutylene terephthalate particle. Reprinted with permission from \cite{prokopovich2010multiasperity}. Copyright 2010, American Chemical Society.}
  \label{fig:part_roughness_pbt}
 \end{subfigure}
 \hspace{15pt}
 \begin{subfigure}[c]{0.48\textwidth}
  \centering
  \includegraphics[width=0.55\textwidth, trim=0cm 9.4cm 5.5cm 0.0cm, clip]{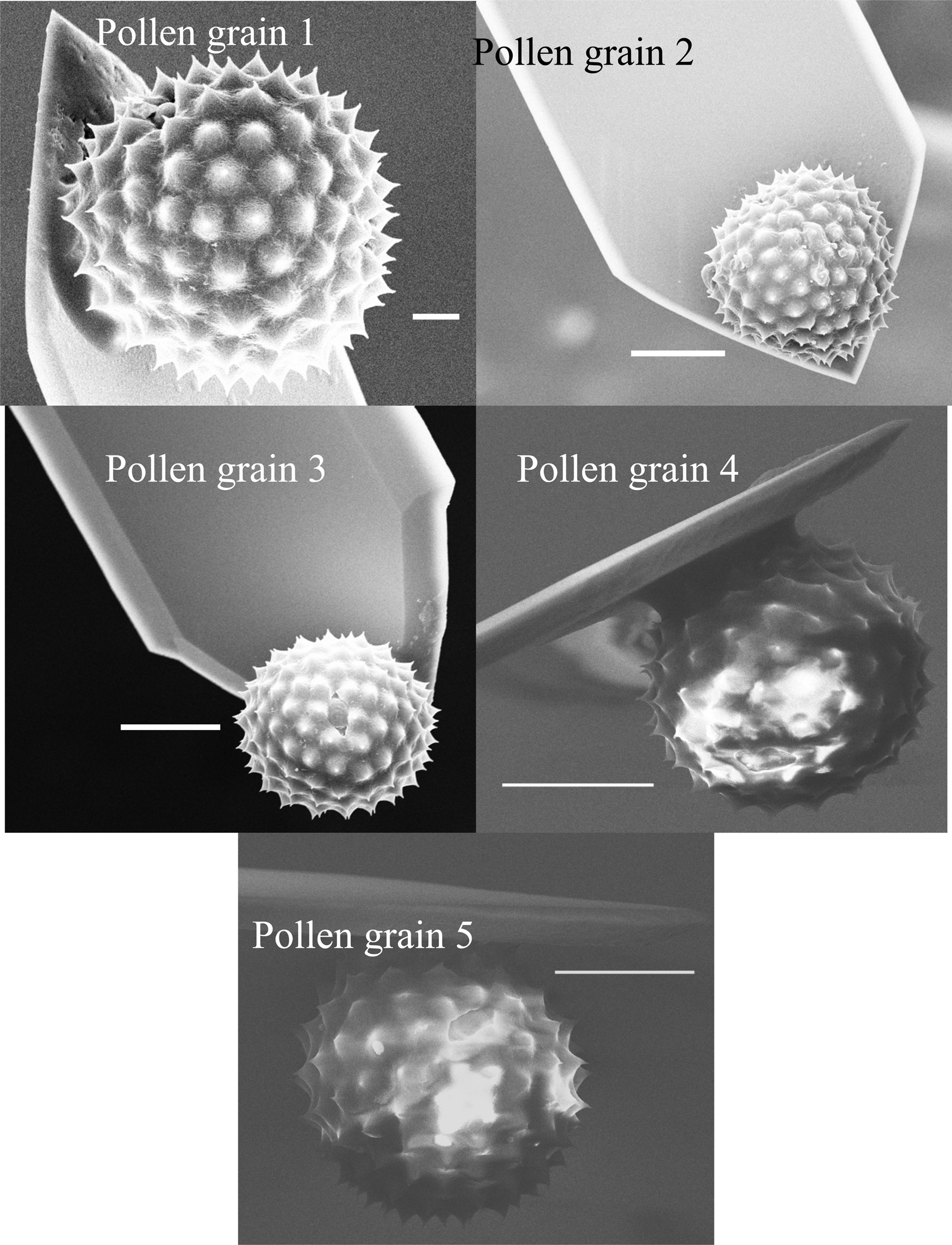}
  \caption{SEM image of a ragweed pollen grain glued to the end of a tipless AFM. Reprinted with permission from \cite{thio2009characterization}. Copyright 2009, American Chemical Society.}
  \label{fig:part_roughness_pollen}
 \end{subfigure}
 \caption{Images showing the roughness of two substrates (top) and two particles (bottom). Surface roughness can either appear as very chaotic (left, with raw substrate or particles) or highly regular (right, with engineered surfaces or pollen grain).}
 \label{fig:surf_roughness}
\end{figure}

\subparagraph{Physical origin:} As displayed in Fig.~\ref{fig:surf_roughness}, the surfaces of particles and substrates are rarely smooth and usually present irregularities. In fact, the surfaces of raw materials (e.g., on a stainless steel substrate or on a micrometer-size plastic particle) appear as quite chaotic with roughness features covering a range of sizes (from a few nanometers up to a few micrometers here), whereas more regular roughness, or `smooth surfaces', can only be obtained by using special surface engineering techniques (e.g., micro-patterned surfaces obtained by photolithography) or when handling biological particles (like pollen grains). This explains that surface topologies are regarded as random and are, at best, characterized by a few averaged dimensions such as \cite{henry2018colloidal}: the mean roughness (i.e., the arithmetic average height above a reference plane), the rms roughness (i.e., root-mean-square roughness) and the peak-to-peak distance. However, more detailed statistical information is needed to properly account for the effect of surface roughness and this point is taken up in Section~\ref{sec:next_model:complete:surf}. 

\subparagraph{Role in resuspension:}  
Surface roughness plays a central role in resuspension through the following phenomena: 
 \begin{itemize}
  \item First, as mentioned previously, surface irregularities lead to the existence of friction forces.
  
  \item Second, roughness features with sizes smaller or comparable to the particle size can have a profound impact on adhesion forces (especially VDW forces). Actually, small asperities on the surface lead often to a significant reduction of the actual adhesion force compared to the one that would exist if we were dealing with perfectly smooth surfaces. This effect of surface roughness on adhesion forces is now well documented (see previous reviews, e.g., \cite{persson2006contact, ziskind2006particle, henry2014progress} and specific papers on the topic \cite{gotzinger2004particle, prokopovich2010multiasperity, prokopovich2011adhesion}, as well as reference therein) and can be summarized as follows (see also \cite{rush2018glass}): protruding roughness features keep particles further away from the rest of the surface, often at separation distances over which van der Waals forces are significantly reduced. Hence, the particle-substrate interaction is mostly governed by roughness features that are in close contact, giving rise to adhesion forces much weaker than the ones obtained when assuming perfectly smooth surfaces (which allow to have large surface areas in close proximity). The overall effect is sketched in Fig.~\ref{fig:sketch_roughness_Fadh}, which illustrates that surface roughness is responsible for a distribution of adhesion forces between surfaces. Given that surface roughness is described as random, this explains that adhesion forces are also naturally treated as random variables, characterized by their probabilistic distribution or PDF. In practice, various types of distributions have been measured for the adhesion forces between rough surfaces, such as Gaussian or Weibull distributions (e.g., \cite{prokopovich2011adhesion}) or even multimodal distributions (as in \cite{audry2009adhesion, rush2018glass, brambilla2018glass}). 
 \begin{figure}[ht]
  \centering
  \includegraphics[width=0.8\textwidth]{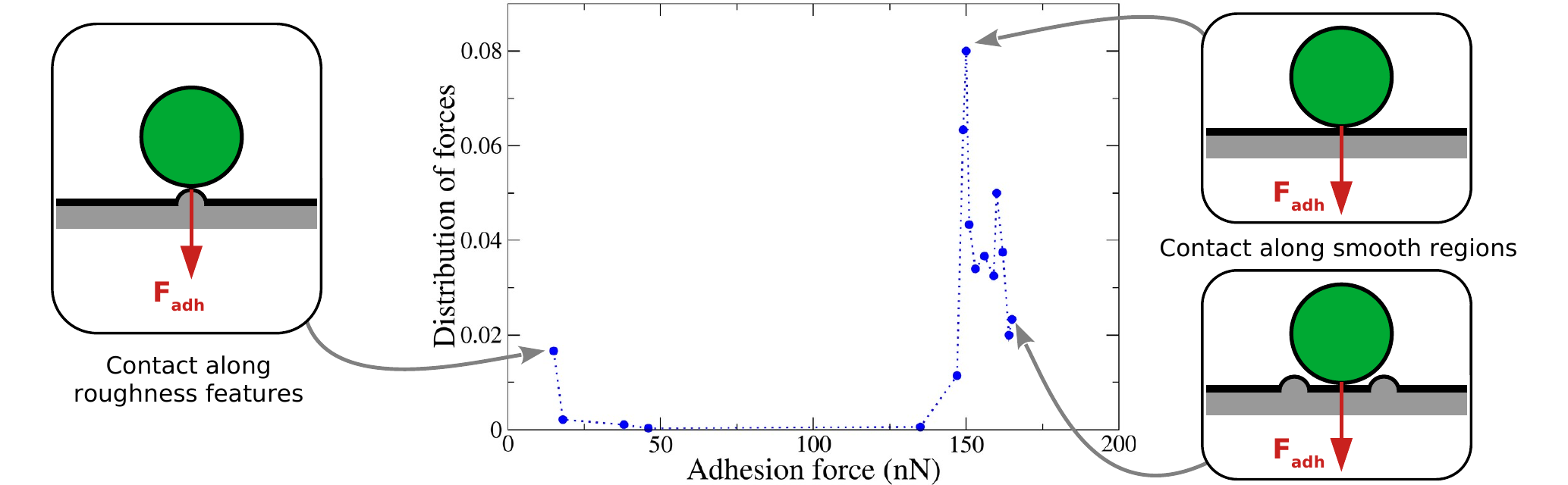}
  \caption{Illustration of the effect of surface roughness on adhesion forces (which can be significantly reduced when roughness features are present between the bulk of each materials). Experimental data taken from \cite{zhou2003influence}.}
  \label{fig:sketch_roughness_Fadh}
 \end{figure}
 
  \item Third, roughness features with sizes comparable to or larger than the particle size can modify the hydrodynamic forces acting on particles. More precisely, two separate effects can take place. First, large-scale roughness features can shelter particles located in their wake from the flow, effectively reducing resuspension. Second, when roughness features display sizes comparable to the smallest fluid scale (typically the Kolmogorov scale in a turbulent flow), the fluid flow around such geometrical features is modified (e.g., the wind flow around sand dunes \cite{kok2012physics}). Depending on the relative size between surface roughness and particle diameter, this can either facilitate resuspension or hinder it \cite{nasr2020model}. 
  
 \end{itemize}

\subparagraph{Modeling surface roughness:} When developing resuspension models, it follows from the previous arguments that the effects of surface roughness on adhesion forces have to be carefully included. This means that information at the nano- and micro- scales have to be accounted for, even if these scales are much smaller than the typical ones of a fluid flow (which range from a few millimeters to a few meters). As a result, surface roughness is often simplified in resuspension models and various descriptions can be found. In the following, we briefly review the existing approaches to represent rough surfaces and how these descriptions are involved in the evaluation of adhesion forces (see Fig.~\ref{fig:sketch_represent_roughness} and \cite{henry2014progress, prokopovich2011adhesion, prokopovich2010multiasperity} for more details): 
\begin{figure}[ht]
 \centering
 \includegraphics[width=0.75\textwidth]{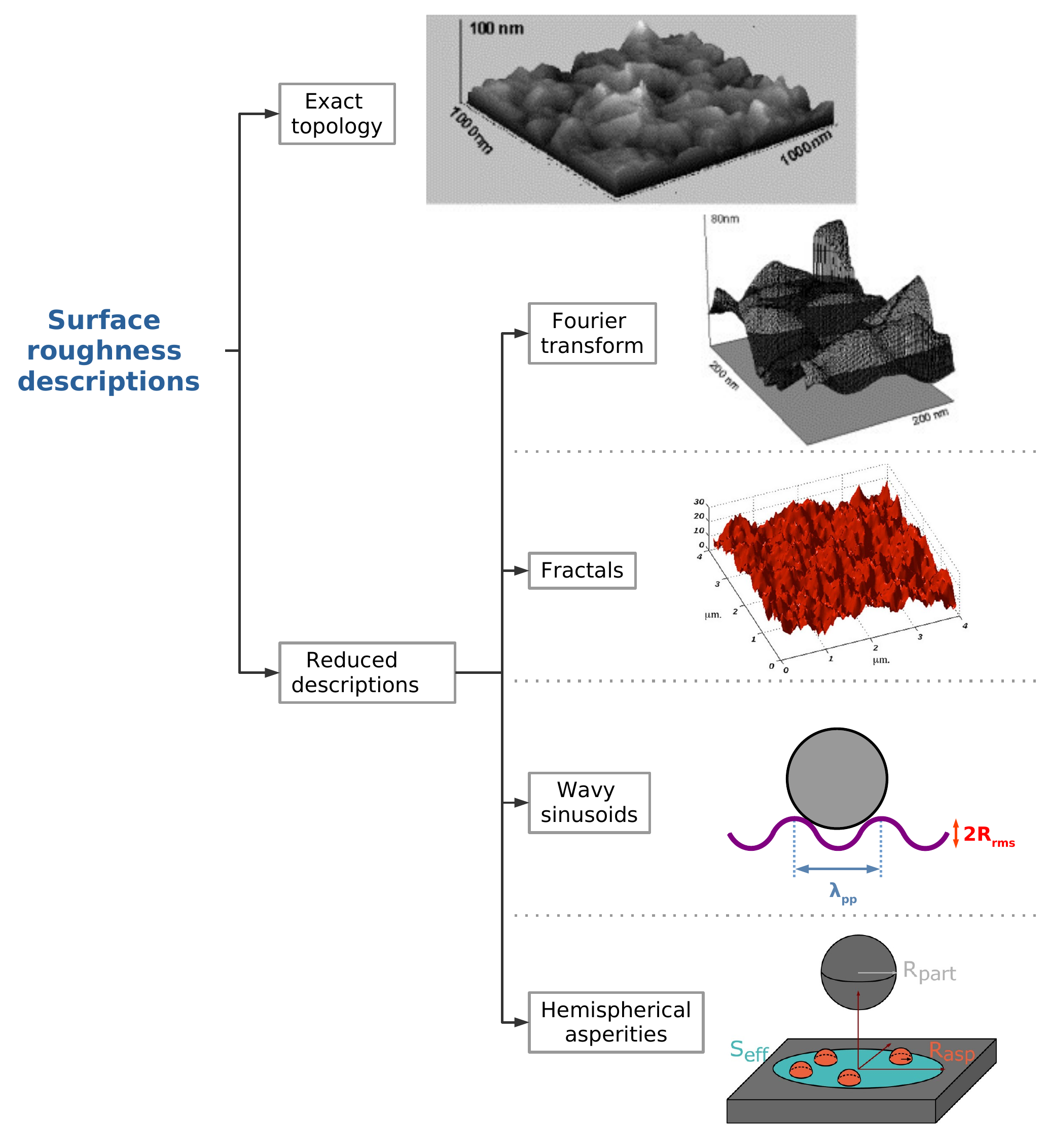}
 \caption{Sketch of existing descriptions of surface roughness, showing both exact topologies and reduced descriptions (sorted according to the level of information contained). Top 2 figures reprinted with permission from \cite{eichenlaub2004roughness}. Copyright 2004, Elsevier. Middle figure reprinted with permission from \cite{rezvanian2007surface}. Copyright 2007, IOP Publishing.}
 \label{fig:sketch_represent_roughness}
\end{figure}
\begin{itemize}
 \item A first nanoscopic approach consists in using the exact information on the surface topography to compute the resulting forces explicitly. This is only possible if the surface topography has been fully characterized using detailed measurement techniques (such as SEM of AFM methods). The exact topography can then be used to compute the adhesion force $\mb{F}_{\rm adh}$ at any contact point between a spherical particle and this rough substrate \cite{zhou2008modeling}. This is usually achieved by discretizing the volume of each body (e.g., using Finite Element Method) and then performing a direct numerical integration of van der Waals forces similar to the Surface Element Integration technique given in Eq.~\eqref{eq:Fadh_int} \cite{bhattacharjee1997surface}. 
 \item Other descriptions of surface roughness rely on more limited information on the surface topology. One way to represent a rough surface, still with a relatively high level of information, is to resort to a Fourier transform of the surface \cite{jaiswal2009modeling}. This allows to generate numerical rough substrates that closely approach the real ones. The adhesion force between such a rough substrate and a particle can be computed using similar numerical integration techniques (such as SEI methods).
 \item A coarser description of rough surfaces can be made by resorting to fractal surfaces (especially with Weierstrass-Mandelbrot's functions \cite{eichenlaub2004roughness, rezvanian2007surface}). Once again, this allows to randomly generate various numerically rough surfaces \cite{rezvanian2007surface} and, using numerical integration techniques, to estimate the distribution of adhesion forces.
 \item In practice, it is very frequent to have access only to average information on surface roughness (such as the rms roughness $r_{\rm a,rms}$, the peak-to-peak distance, the extremum sizes of roughness features). In that case, the issue is how to estimate the adhesion force. Depending on the level of information required, two types of model have been suggested:
 \begin{itemize}
  \item When the objective is to approximate the complete distribution of adhesion forces, various models have proposed to represent rough surfaces by a smooth plate covered by hemispherical asperities randomly distributed on the surface \cite{gotzinger2004particle}. The size of these asperities (resp. their spacing) is determined by the available data on the rms roughness (resp. the peak-to-peak distance). Then, using the Hamaker approach, the total adhesion force is simply given by summing the interaction between the particle and all the elements composing the surface (i.e., all the hemispherical asperities plus the smooth substrate beneath):
  \begin{equation}
   F_{\rm adh} ({\rm sphere-roughS})= F_{\rm adh}({\rm sphere-smoothS}) + \sum_{\rm all \,asperities} F_{\rm adh}({\rm sphere-asperity}). 
  \end{equation}
  By repeating the procedure to generate different realizations of the rough surface, the distribution of adhesion forces can be estimated through Monte-Carlo methods \cite{henry2014stochastic, henry2018colloidal}.
  \item When the aim is to predict only the average value of the adhesion force, one can rely directly on analytic formulas. These analytic formulas are obtained by summing the interaction between a spherical particle and all the elements forming the rough surface (i.e., the hemispherical asperities and the smooth surface underneath). Yet, instead of placing asperities randomly, the analytic formulas are obtained either by placing one asperity at a fixed position (e.g., right beneath the particle as in \cite{rumpf2012particle}) or by regularly placing asperities on the surface (as in \cite{rabinovich2000adhesion1, rabinovich2000adhesion2}). For instance, in the case of multiple asperities present on the surface, the average value of the adhesion force can be estimated with:
  \begin{equation}
   F_{\rm adh} = \frac{A_{\rm Ham}r_{\rm p}}{6 z_{0}^{2}}\left( \frac{1}{1+\frac{58.144r_{\rm p}r_{\rm a,rms}}{\lambda_{\rm a}^2}} + \frac{1}{\left(1+\frac{1.817r_{\rm a,rms}}{z_{0}}\right)^2} \right)
  \label{Eq_Rabinovich}
  \end{equation}
  where $A_{\rm Ham}$ is the Hamaker constant (which reflects the strength of the adhesion between two objects), $z_0$ the contact distance, $r_{\rm p}$ the particle radius, $r_{\rm a,rms}$ the rms roughness, and $\lambda_{\rm a}$ the peak-to-peak distance between consecutive asperities.  
   
  More recently, a similar model has been developed by describing the rough surface as a sinusoidal wave surface \cite{nasr2020model}. The amplitude of the sinusoidal function is given by the average roughness size while the frequency is proportional to the peak-to-peak distance. The adhesion force is then simply obtained using contact mechanics theories (like the JKR model in \cite{nasr2020model}) and considering that the particle is resting on a stable position on the sinusoidal wave surface (i.e., two points of contact in 2D cases or 3 in 3D cases). Similarly, other formulations based on contact mechanics theories have been proposed to describe single or multiple asperity contacts (more details in \cite{prokopovich2011adhesion} and references therein).
 \end{itemize}

\end{itemize}

 A few caveats are in order. First, these adhesion models only approximate the real force. For example, in the case of non-spherical particles having irregular shapes and their own roughness (which are more common in realistic situations, as for sand beads), the particle and the surface can come in contact at multiple points \cite{zhang2015afm}, making the resulting adhesive force dependent on the orientation of the particle \cite{zhao2015four}. Second, contact mechanics theories assume that deformation is much smaller than any characteristic dimension of the body. Yet, this assumption is not always true for small colloidal particles and small roughness elements \cite{brambilla2020impact}. Third, the simplified representations based on hemispherical asperities have been derived from information extracted from 2D profiles of surfaces. Yet, a recent study \cite{brambilla2020impact} has shown that it is not always straightforward to use 3D information, as protuberances on the surface do not necessarily resemble parabolas. Fourth, more general theories that account for plastic deformations are still not available \cite{xiao2006elastoplasticity}.

 \subsection{A wide range of approaches}
  \label{sec:models:approach}

Although various attempts have been made to suggest classifications of these models (see recent reviews \cite{ziskind2006particle, henry2014progress, dey2018advances, ziskind1997adhesion}), no consensus has emerged since different points of views can be adopted to categorize modeling approaches. In the following, we draw on the classification proposed in \cite{henry2014progress} to propose a hierarchy of existing models. These models are sorted according to the level of information contained, starting with the most detailed models up to the more macroscopic ones (interested readers are directed to this previous review for a comprehensive discussion of different model classes). Note however that this classification is extended to include other models that were not accounted for in the previous review. 

  \subsubsection{Dynamic models based on spatio-temporal integration}
   \label{sec:models:approach:dyn_model}
  
\subparagraph{General description} The leading idea of exact dynamic models is to track a large number of individual particles within a domain by solving explicitly the particle equations of motion. 

\subparagraph{Model formulation} These models are designed in a step-by-step manner summarized as follows: 
 \begin{itemize}
  \item The first step consists in defining the particle state vector $\mb{Z}_{\rm p}$ which gathers the variables attached to each particle and which are selected to describe each particle as a mechanical object. For instance, one can choose to retain the position $\mb{X}_{\rm p}$ and the velocity $\mb{U}_{\rm p}$ of each individual particle. This corresponds to having $\mb{Z}_{\rm p}=(\mb{X}_{\rm p}, \mb{U}_{\rm p})$, which is the classical state vector in Statistical Physics when dealing with kinetic approaches. While this choice appears relevant to capture the translational motion of rigid spherical particles (e.g., sliding or lifting motion), it does not allow to track rolling motion of solid spheres on a surface since the rotational velocity $\mb{\Omega}_{\rm p}$ is missing. For that reason, most resuspension models are based on the state vector $\mb{Z}_{\rm p}=(\mb{X}_{\rm p}, \mb{U}_{\rm p}, \mb{\Omega}_{\rm p})$. Additional variables can be introduced in the state vector, such as the particle size $r_{\rm p}$ when dealing with aggregates or the particle orientation $\mb{p}_{\rm p}$ when dealing with non-spherical particles (see Section~\ref{sec:next_model}). 
  \item Once the particle state vector $\mb{Z}_{\rm p}$ is chosen, the second step consists in writing an equation for the evolution of these quantities in time. For the sake of clarity, we consider only here the simplest case of rigid spherical particles where the state vector is $\mb{Z}_{\rm p}=(\mb{X}_{\rm p}, \mb{U}_{\rm p}, \mb{\Omega}_{\rm p})$ (the particle diameter or radius is a constant for each particle and is left out of the state vector). In that case, the equations of translational and rotational motion are given by Newton's second law and write:
 \begin{subequations}
  \label{eq:eq_motion}
  \begin{align}
  \frac{\dd \mb{X}_{\rm p}}{\dd t} = & \mb{U}_{\rm p}~,  \\
   m_{\rm p} \frac{\dd\mb{U}_{\rm p}}{\dd t} = & \mb{F}_{\rm f \to p} + \mb{F}_{\rm s \to p} + \mb{F}_{\rm p \to p} + \mb{F}_{\rm ext}~, \\
   I_{\rm p} \frac{\dd \mb{\Omega}_{\rm p}}{\dd t} = & \mb{M}_{\rm f \to p} + \mb{M}_{\rm s \to p} + \mb{M}_{\rm p \to p} + \mb{M}_{\rm ext}~,
  \end{align}
 \end{subequations}
 with $m_{\rm p}$ the particle mass and $I_{\rm p}$ its moment of inertia. The terms on the right hand side of Eqs.~\eqref{eq:eq_motion} correspond to the forces $\mb{F}$ and torques $\mb{M}$ acting on the particles. This includes terms related to fluid-particle interactions $(.)_{\rm f\to p}$, surface-particle interactions $(.)_{\rm s\to p}$, particle-particle interactions $(.)_{\rm p\to p}$, and external interactions $(.)_{\rm ext}$ (see also Section~\ref{sec:models:forces} for a physical interpretation of the forces at play).

\end{itemize}
  
   In the present hierarchy of models, dynamics models embodied by Eqs.~\eqref{eq:eq_motion} represent the most detailed level of description or, in other words, fine-grained formulations. In order to obtain a closed set of equations, they require however that all the forces and torques appearing on the rhs of Eqs.~\eqref{eq:eq_motion} be explicitly known. Given the expressions of these forces and torques given in Section~\ref{sec:models:forces}, it is useful to emphasize that they rely on detailed information on the instantaneous fluid velocity around each particle for fluid-particle interactions while accurate expressions for the adhesive forces between rough substrate can only be obtained if information on the nanoscale surface topology is available.

Eqs.~\eqref{eq:eq_motion} are expressed for the variables entering the one-particle state vector $\mb{Z}_{\rm p}$, but, since we are tracking N particles simultaneously, the actual state vector is $\mb{Z}=(\mb{Z}_{\rm p}^{(1)}, \mb{Z}_{\rm p}^{(2)}, \dots, \mb{Z}_{\rm p}^{({\rm N})})$, where $\mb{Z}_{\rm p}^{({\rm i})}$ stands for the state vector of the particle labeled $({\rm i})$. The governing equations can then be written in a compact form for the N-particle state vector $\mb{Z}$ as a general ordinary differential equation (ODE) in a ${\rm N} \times {\rm N_p}$ dimensional space (${\rm N_p}$ being the number of variables in the one-particle state vector $\mb{Z}_{\rm p}$)
\begin{equation}
\frac{\dd \mb{Z}}{\dd t} = \mc{A}\left(t,\mb{Z}(t)\right)~, 
\end{equation}
where the function $\mc{A}$ gathers all the terms on the rhs of Eqs.~\eqref{eq:eq_motion}. In a statistical formulation, we are therefore dealing with a N-particle PDF approach, in which the PDF $p(t;\mb{z})$ (where $\mb{z}$ represents the possible values taken by the state vector $\mb{Z}$) is the solution of a Liouville equation in its corresponding sample space:
\begin{equation}
\label{eq:Liouville eq}
\frac{\partial \, p(t;\mb{z})}{\partial t}= - \frac{\partial \left[ \mc{A}_k(t;\mb{z}) \, p(t;\mb{z}) \right]}{\partial z_k}~, \quad k=1, \dots, {\rm N}\times {\rm N_p}~.
\end{equation}
 
\subparagraph{Typical output and characteristics}
As it transpires from the previous discussion on the choice of a state vector, exact dynamic models can provide very detailed information at the particle scale (e.g., position, translational and rotational velocity of each individual particle). They allow naturally to predict the history of individual particles (see Section~\ref{sec:technique:quantities:variable}) and, by calculating ensemble averages or using other adequate statistical analysis, to obtain various statistical quantities or factors of interest in each situation (see the list in Section~\ref{sec:technique:quantities:variable}). In addition, when specific models are used to compute the adhesion force between rough objects, these approaches yield also microscopic predictions, such as the distribution of adhesion forces. For example, the 10~\SI{}{\%} quantile threshold velocity for incipient motion  can be estimated by running numerical simulations with various fluid velocities and detecting when more than 10~\SI{}{\%} of the particles have started moving.

More generally, such dynamic models are naturally consistent with Lagrangian approaches for particle transport by fluid flows, which consist in solving the same equations for the dynamics of particles (but not necessarily with particle-surface interactions which act only close to wall boundaries). Another motivation is that they are very flexible. For example, it is relatively straightforward to extend Eqs.~\eqref{eq:eq_motion} to treat the case of non-spherical particles: this requires to add an extra equation for the particle orientation $\mb{p}_{\rm p}$ and to modify the expressions for the hydrodynamic forces in order to account for the changes in the drag and lift forces due to the particle non-sphericity and its orientation (see developments in Section~\ref{sec:next_model:complete:part}).
 
 The main drawback of dynamic approaches is the high computational cost, as well as the near-impossibility to have access to the exact topography of wall surfaces in most practical cases. In fact, simulations can hardly be performed in realistic environmental applications (such as resuspension from sand dunes), since they would require to follow the trajectories of an extremely large number of particles (typically $10^{12}$ or even larger), while still suffering from uncertainties due to choices made in the description of surface characteristic geometrical features. Besides, the time step used has to be small enough to capture fluctuations in adhesive forces and the models for the adhesion force between rough objects often require extensive computations.

\subparagraph{Current use} 
Given that these models are naturally compatible with N-particle tracking within a flow, most of them are coupled to the Distinct Element Method (DEM), which solves equations of motion similar to Eqs.~\eqref{eq:eq_motion} anywhere in the physical domain (and not only on the surface). Due to the high level of information contained in these formulations, they are usually coupled to a Direct Numerical Simulations (DNS) of the turbulent flow carrying these particles. In practice, the two methods described in Section~\ref{sec:models:turb_rough:turb}, namely the PR-DNS and PP-DNS, have been used:
 \begin{itemize}
  \item Using the PR-DNS formulation, an Immersed Boundary Method has been used in a recent paper \cite{vowinckel2016entrainment} to discretize the surface of each particle and apply a zero-flux condition for the fluid velocity on each particle surface. As displayed in Fig~\ref{fig:fig_vowinckel_2016_struct}, these simulations allow to obtain precise information on the flow around 13\,500 particles forming a single monolayer. Numerical results showed that a particle in a dense monolayer bed is resuspended by the combination of two processes: first, it can be temporarily dislocated from the surface when collided by another particle transported by the flow; second, as it interacts with a sweep event, this particle can be entirely dislodged from its initial position. 
  
  \begin{figure}[ht]
   \centering
   \includegraphics[width = 0.8\textwidth]{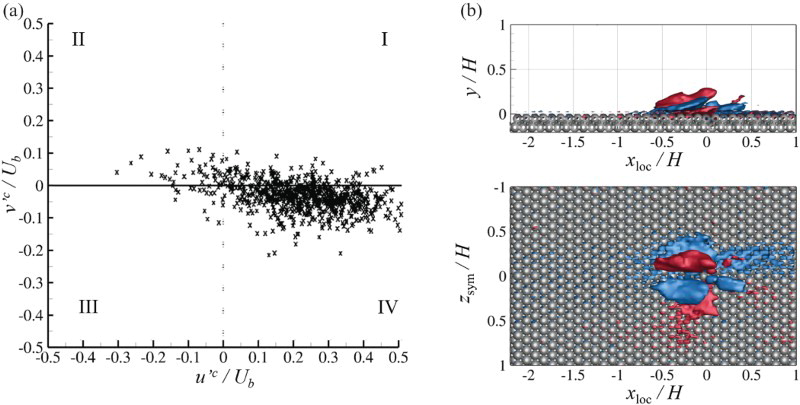}
   \caption{Analysis of the fluid velocity around a dense monolayer deposit obtained using finite-size DNS revealing the role of coherent structures in resuspension: (a) Quadrant analysis showing all events reported for the fluctuating velocities (streamwise $u'^c$ and wall-normal $v'^c$ components); (b) iso-surfaces of the vorticity $\omega_x$ for the side view (top) and the top view (bottom); blue: $\omega_x$ = \SI{-0.4}{}, red: $\omega_x$ = \SI{0.4}{}. Reprinted with permission from \cite{vowinckel2016entrainment}. Copyright 2016, Taylor \& Francis.}
   \label{fig:fig_vowinckel_2016_struct}
  \end{figure} 
  
  Similar results were obtained using a Lattice-Boltzmann approach to compute the fluid flow around finite-size particles \cite{cui2022lattice}. Considering that resuspension is triggered by a simple rupture of balance between the forces acting on each particle, the authors highlighted the role played by surface roughness in the modifications of the near-wall velocity and its potential consequences on particle resuspension.
  
  \item In the frame of the PP-DNS formulation, the hydrodynamic drag and lift forces are evaluated using one of the existing formulas presented in Section~\ref{sec:models:forces}. At this stage, it is worth mentioning that the precision of the value for the velocity of the fluid seen $\mb{U}_{\rm s}(t)$ (with $\mb{U}_{\rm s}(t)=\mb{U}_{\rm f}(t,\mb{X}_{\rm p}(t))$) depends on how it is evaluated numerically from the DNS simulations. In fact, particles are distributed within the computational domain and are not necessarily located at one of the grid points used in the DNS simulation. Hence, an exact value can be estimated by resorting to the Fourier transform (the same as the one used to solve the Navier-Stokes equation in the DNS simulation). However, only an approximate value is obtained when resorting to an interpolation method: in that case, the accuracy is driven by the precision of the interpolation algorithm (e.g., tri-linear interpolation or tri-cubic interpolation \cite{homann2007impact}). 
  
  Recently, similar N-body simulations have been carried out to investigate the role of impact entrainment in complex multilayer deposits. For that purpose, 15\,000 particles were initially deposited to form a deposit with more than 10 layers (see Fig~\ref{fig:fig_pahtz_2017_saltation}). Particle motion is then computed neglecting buoyancy, lift, and shielding effects while inter-particle collisions are taken into account using a simple model based on the friction coefficient and the restitution coefficient (i.e., the amount of energy of the impacting particle remaining after the collision). This has allowed to confirm that reptation and saltation motion mainly differ due to differences in the kinetic energy upon collision: saltating particles display much more pronounced hoping motion, with higher energies upon collision with the surface, that can result in much more dislocations of deposited particles. Statistical analysis tools can then be used to extract more macroscopic information such as the threshold for particle incipient motion or the Shields parameter (see for instance \cite{pahtz2021unified}).
  
  \begin{figure}[ht]
   \centering
   \captionsetup[subfigure]{justification=centering}
   \begin{subfigure}{0.47 \linewidth}
    \includegraphics[width=\textwidth, trim=0cm 12.7cm 0cm 0cm, clip]{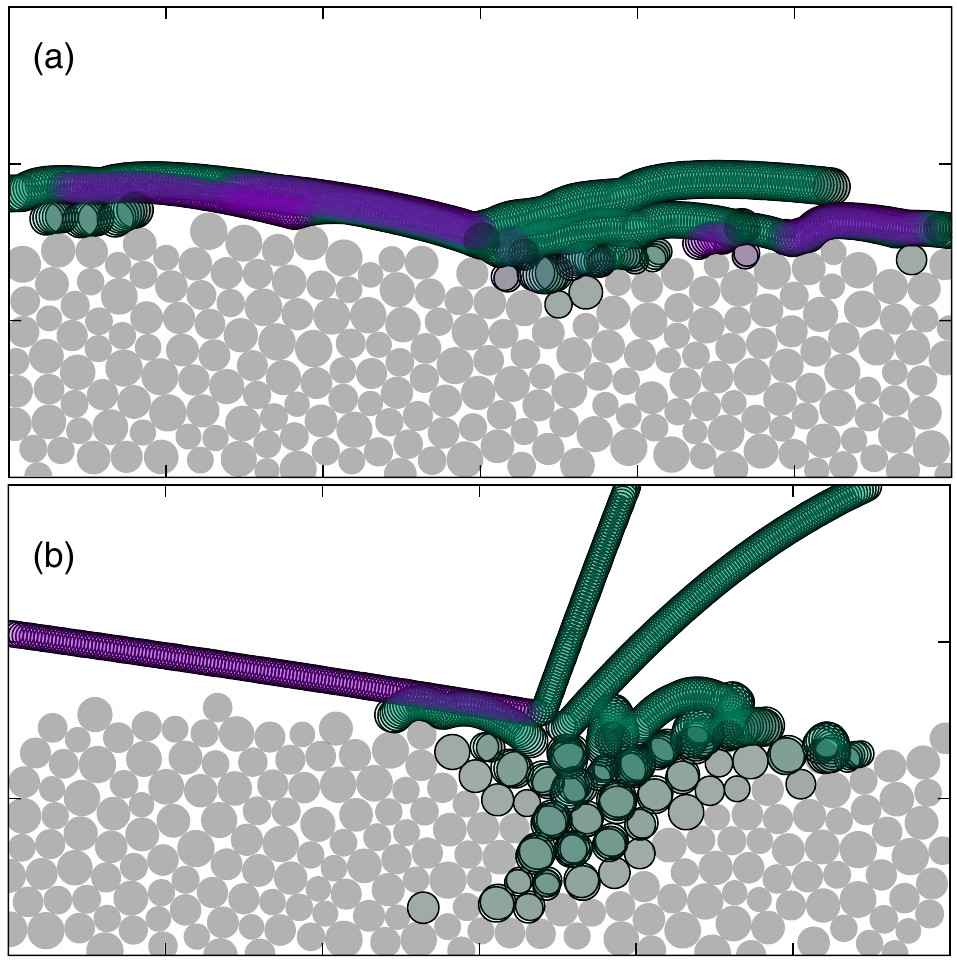}
    \caption{Turbulent bedload transport}
   \end{subfigure}
   \hspace{15pt}
   \begin{subfigure}{0.47 \linewidth}
    \includegraphics[width=\textwidth, trim=0cm 0cm 0cm 12.7cm, clip]{fig_pahtz_2017_saltation}
    \caption{Saltation transport}
   \end{subfigure}
   \caption{Visualizations of impact entrainment events for turbulent bedload (left) and saltation transport (right) from direct sediment transport simulations near the threshold conditions. The purple and green colors indicate the trajectory of particles before and after impact, respectively. Particles move from the left to the right. Reprinted with permission from \cite{pahtz2017fluid}. Copyright 2017, American Physical Society.}
   \label{fig:fig_pahtz_2017_saltation}
  \end{figure}

 \end{itemize}

It is essential to remember that these models require very detailed information on all the forces acting on particles. In terms of level of information, it is thus more consistent to couple these approaches with DNS simulations of the fluid flow, but statistical information can also be derived from them provided that data is available on velocity distributions (this point is taken up in Section~\ref{sec:analysis:cross}). 

  \subsubsection{Dynamic PDF approaches}
   \label{sec:models:approach:dyn_PDF}

\subparagraph{General description} Broadly speaking, dynamic PDF models consist in coarse-grained descriptions of exact dynamic models in which particle variables are modeled as stochastic processes. In line with the general ideas of Synergetics, they correspond to reduced statistical descriptions where the fast degrees of freedom are eliminated to leave white-noise type of stochastic processes as their trace in the evolution equations of the remaining slow degrees of freedom \cite{minier2001pdf,minier2015kinetic}. This amounts to adopting a probabilistic point of view \cite{peirano2006mean, minier2001pdf, minier2016statistical}, where the objective is to approximate relevant particle-related statistics rather than compute the exact trajectories of individual particles. In a weak formulation, this is equivalent to approximating the probability density function (PDF) of the variables selected to describe particle dynamics, hence the name of the approach. These methods retain a Lagrangian formulation but we are now dealing with stochastic particles which represent samples of this underlying PDF and are to be regarded as random versions of real ones. In principle, it is possible to track a large number of particles, or pairs of particles, or triplets of particles, etc., which corresponds to handling a one-particle PDF, or a two-particle PDF, or a three-particle PDF, etc. (respectively). Yet, for practical applications in general non-homogeneous and wall-bounded flows, only one-particle PDF are well-developed at the moment. This means that they produce only one-point statistical outcomes (note that this point will resurface in Section~\ref{sec:analysis:cross}).

\subparagraph{Model formulation} 
Like the exact dynamical model described above, the formulation of PDF approaches is achieved through a two-step construction. However, both steps are less straightforward and involve specific issues that need to be addressed to clarify what these PDF models can and cannot capture. 

\begin{itemize}
\item The first step concerns the selection of the variables entering the particle state vector $\mb{Z}_{\rm p}$ for particles transported by a flow described by a turbulent model, which means that only limited statistical information, such as the one-point first- and second-order velocity moments, is available. To bring out the issues at stake, it is sufficient to consider the simple, but reference, case of small discrete particles heavier than the carrier fluid and described as point-wise particles. Then, the particle equations of motion have the form (retaining only drag and gravity forces)
\begin{subequations}
\label{eq:simplified_eq_motion}
\begin{align}
\frac{\dd \mb{X}_{\rm p}}{\dd t} = & \, \mb{U}_{\rm p}~,  \label{eq:simplified_eq_motion xp}\\
\frac{\dd \mb{U}_{\rm p}}{\dd t} = & \, \frac{ \mb{U}_{\rm s} - \mb{U}_{\rm p}}{\tau_{\rm p}} + \mb{g}~, \label{eq:simplified_eq_motion Up}
\end{align}
\end{subequations} 
where $\tau_{\rm p}$ is the particle relaxation timescale. The drag force on the rhs of Eq.~\eqref{eq:simplified_eq_motion Up} is written as a return-to-equilibrium term towards the fluid velocity seen $\mb{U}_{\rm s}$, which is the instantaneous fluid velocity `sampled' by discrete particles as they move across the fluid flow (\textit{i.e.}, $\mb{U}_{\rm s}(t)=\mb{U}_{\rm f}(t,\mb{X}_{\rm p}(t))$ where $\mb{U}_{\rm f}(t,\mb{x})$ is the instantaneous fluid velocity field). If we retain the classical kinetic particle state vector, $\mb{Z}_{\rm p}=(\mb{X}_{\rm p},\mb{U}_{\rm p})$, then $\mb{U}_{\rm s}$ appears as an external driving force. In the context of DNS and the exact dynamical models considered before, $\mb{U}_{\rm s}$ is a known deterministic signal, leading to the well-posed Liouville formulation. In the frame of modeled turbulent flows (like LES or RANS), $\mb{U}_{\rm s}$ appears, however, as a stochastic process with non-zero correlation timescales (due to the non-zero space and time correlations of turbulent fluctuations). This deviation from `heat-bath properties' leads to severe difficulties and results even in ill-based probabilistic formulations (this key issue is addressed in detail in \cite{minier2016statistical, minier2015kinetic}). In line with the notion of slow/fast variables mentioned above, it is much better to include the fluid velocity seen in the particle state vector, so that $\mb{Z}_{\rm p}=(\mb{X}_{\rm p},\mb{U}_{\rm p},\mb{U}_{\rm s})$. This means that the external driving force is now the fluid acceleration which, according to Kolmogorov turbulence theory, can be regarded as a fast process and has thus better chances to be safely replaced by a stochastic diffusion process, thereby ensuring well-posed probabilistic formulations.

\item With the selection of the particle state vector, the second step consists in devising a stochastic model for the velocity of the fluid seen. This has been discussed in several reviews \cite{minier2001pdf, minier2016statistical, minier2015lagrangian} and present state-of-the-art formulations consist in modeling $\mb{U}_{\rm s}$ as the solution of a generalized Langevin equation. Basically, these models are extensions of the ones developed to simulate the dynamics of fluid-like particles for turbulent reactive flows \cite{pope2000turbulent,pope1994lagrangian}, providing a unified framework \cite{minier2014guidelines} which allows also new methodologies to be proposed \cite{minier2021methodology}. The specific details of such Langevin equations can be found in previous works and, for our present concern, it is sufficient to write the evolution equations for the retained particle state vector as
\begin{subequations}
\label{eq:stochastic_eq_motion}
\begin{align}
\dd \mb{X}_{\rm p} = & \, \mb{U}_{\rm p}\, \dd t~,  \\
\dd \mb{U}_{\rm p} = & \, \mb{D}_{\rm p}(t,\mb{Z}_{\rm p})\, \dd t~, \label{eq:stochastic_eq_motion Up} \\
\dd \mb{U}_{\rm s} = & \, \mb{D}_{\rm s}(t,\mb{Z}_{\rm p}, \mc{F}[ \lra{\mb{Z}_{\rm p}} ], \lra{\Phi})\, \dd t + \mb{B}_{\rm s}(t,\mb{Z}_{\rm p}, \mc{F}[ \lra{\mb{Z}_{\rm p}} ], \lra{\Phi})\, \dd \mb{W}~.  \label{eq:stochastic_eq_motion Us}
\end{align}
\end{subequations} 
In these equations, $\mb{D}_{\rm p}$ typically represents the drag and gravity forces, as in Eq.~\eqref{eq:simplified_eq_motion Up}, but other forces can also be included. The vector $\mb{D}_{\rm s}$ and the matrix $\mb{B}_{\rm s}$ in Eq.~\eqref{eq:stochastic_eq_motion Us} are the drift and diffusion coefficients of the stochastic diffusion process retained for the velocity of the fluid seen, while $\mb{W}$ is a vector of independent Wiener processes. In the drift and diffusion coefficients in Eq.~\eqref{eq:stochastic_eq_motion Us}, a general notation has been used to indicate that these coefficients can depend on the value of the state-vector $\mb{Z}_{\rm p}$ but also on functionals of the mean fields which are calculated from the simulation of that state-vector, written as $\mc{F}[ \lra{\mb{Z}_{\rm p}} ]$, as well as on some external fields represented by $\lra{\Phi}$. A typical example of $\mc{F}[ \lra{\mb{Z}_{\rm p}} ]$ is the particle mean-velocity field while the fluid mean-pressure is another example of what $\lra{\Phi}$ can stand for. From Eqs.~\eqref{eq:stochastic_eq_motion}, it follows that $\mb{Z}_{\rm p}$ is a stochastic diffusion process whose trajectories are nowhere differentiable but still continuous. This is in line with the idea of modeling discrete particle agitated by the actions of random but continuous fluid motions. Note that for variables undergoing abrupt changes and having therefore discontinuous trajectories in sample space, such as particle velocities upon collisions, Poisson stochastic processes are more likely model candidates but this aspect is outside the scope of this presentation (interested readers are referred to \cite{gardiner1985handbook, chibbaro2014stochastic} among other references). Using then a compact notation, the general evolution equations in dynamical PDF models can therefore be formulated as:
 \begin{equation}
 \label{General stochastic diffusion process}
  \dd \mb{Z}_{\rm p} = \mc{A}(t,\mb{Z}_{\rm p}, \mc{F}[ \lra{\mb{Z}_{\rm p}} ], \lra{\Phi})\, \dd t + \mc{B}(t,\mb{Z}_{\rm p}, \mc{F}[ \lra{\mb{Z}_{\rm p}} ], \lra{\Phi})\, \dd\mb{W}_t~,
 \end{equation}
 where $\mc{A}$ is the drift vector and $\mc{B}$ the diffusion matrix modeling the time-rate-of-change of $\mb{Z}_{\rm p}$. 
\end{itemize}

Contrary to the exact dynamical model where we are considering a N-particle PDF evolving through the Liouville equation, cf. Eq.~\eqref{eq:Liouville eq}, we are handling here a one-particle PDF and, in that sense, the actual particle state vector $\mb{Z}$ is identical to the particle one, \textit{i.e.} $\mb{Z}=\mb{Z}_{\rm p}$. In the present PDF formalism, the true quantity being simulated is the (one-particle) Lagrangian mass density function (MDF) which, for the chosen state vector $\mb{Z}_{\rm p}=(\mb{X}_{\rm p},\mb{U}_{\rm p},\mb{U}_{\rm s})$, is defined as:
\begin{equation}
F_{\rm p}^{\rm L}(t; \mb{Y}_{\rm p},\mb{V}_{\rm p},\mb{V}_{\rm s})=\sum_{i=1}^{\rm N} m_{\rm p}^{(i)} \delta(\mb{Y}_{\rm p} - \mb{X}_{\rm p})\, \delta(\mb{V}_{\rm p} - \mb{U}_{\rm p})\, \delta(\mb{V}_{\rm s} - \mb{U}_{\rm s})~,
\end{equation}
where $\mb{Y}_{\rm p}$, $\mb{V}_{\rm p}$ and $\mb{V}_{\rm s}$ stand for the sample space variables corresponding to $\mb{X}_{\rm p}$, $\mb{U}_{\rm p}$ and $\mb{U}_{\rm s}$, respectively. The Eulerian MDF is simply defined as the Lagrangian one at a given location, which writes
\begin{equation}
F_{\rm p}^{\rm E}(t, \mb{x};\mb{V}_{\rm p},\mb{V}_{\rm s})= F_{\rm p}^{\rm L}(t; \mb{Y}_{\rm p}=\mb{x},\mb{V}_{\rm p},\mb{V}_{\rm s})= \int F_{\rm p}^{\rm L}(t; \mb{Y}_{\rm p},\mb{V}_{\rm p},\mb{V}_{\rm s})\, \delta(\mb{Y}_{\rm p} - \mb{x})\, \dd \mb{Y}_{\rm p}~.
\end{equation}
It is important to note that these are one-point MDFs and, therefore, only one-point mean fields can be extracted. Using the same compact notation as in Eq.\eqref{General stochastic diffusion process} but leaving out the functional dependencies for the sake of keeping simple notations, the evolution equation in sample space followed by both $F_{\rm p}^{\rm L}$ and $F_{\rm p}^{\rm E}$ is now a Fokker-Planck equation
\begin{equation}
\frac{\partial \, F_{\rm p}^{\rm E}}{\partial t}= - \frac{\partial \left[ \mc{A}_k \, F_{\rm p}^{\rm E} \right]}{\partial z_k} + \frac{\partial^2 \left[ (\mc{B} \mc{B}^{\rm T})_{kl} \, F_{\rm p}^{\rm E} \right]}{\partial z_k \, \partial z_l}~, \quad k,l=1, \dots, {\rm N_p}~,
\end{equation}
where $\mc{B}^{\rm T}$ is the transpose of the matrix $\mc{B}$.

\begin{figure}[ht]
  \centering
  \includegraphics[scale=0.15]{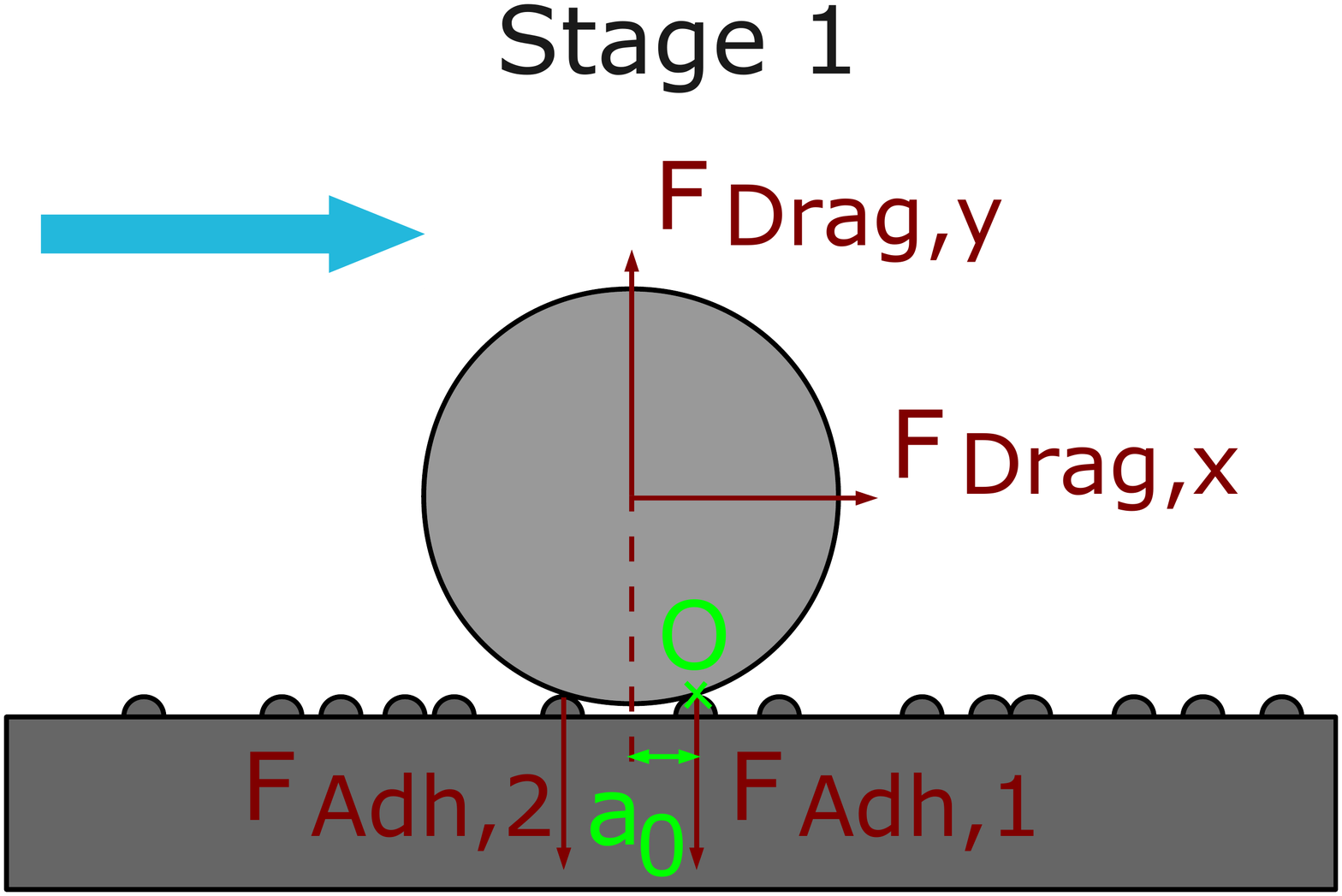}
  \includegraphics[scale=0.15]{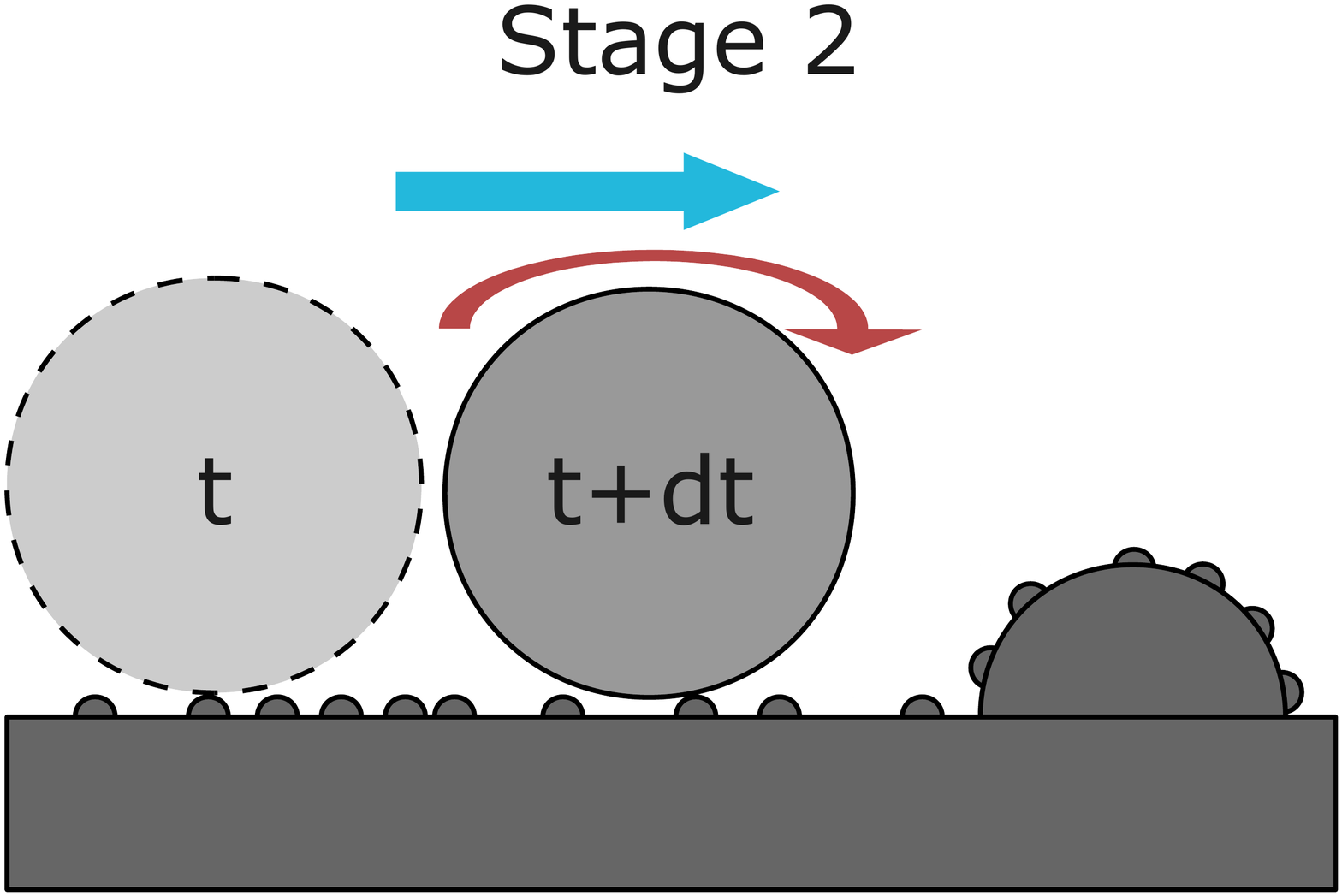}
  \includegraphics[scale=0.15]{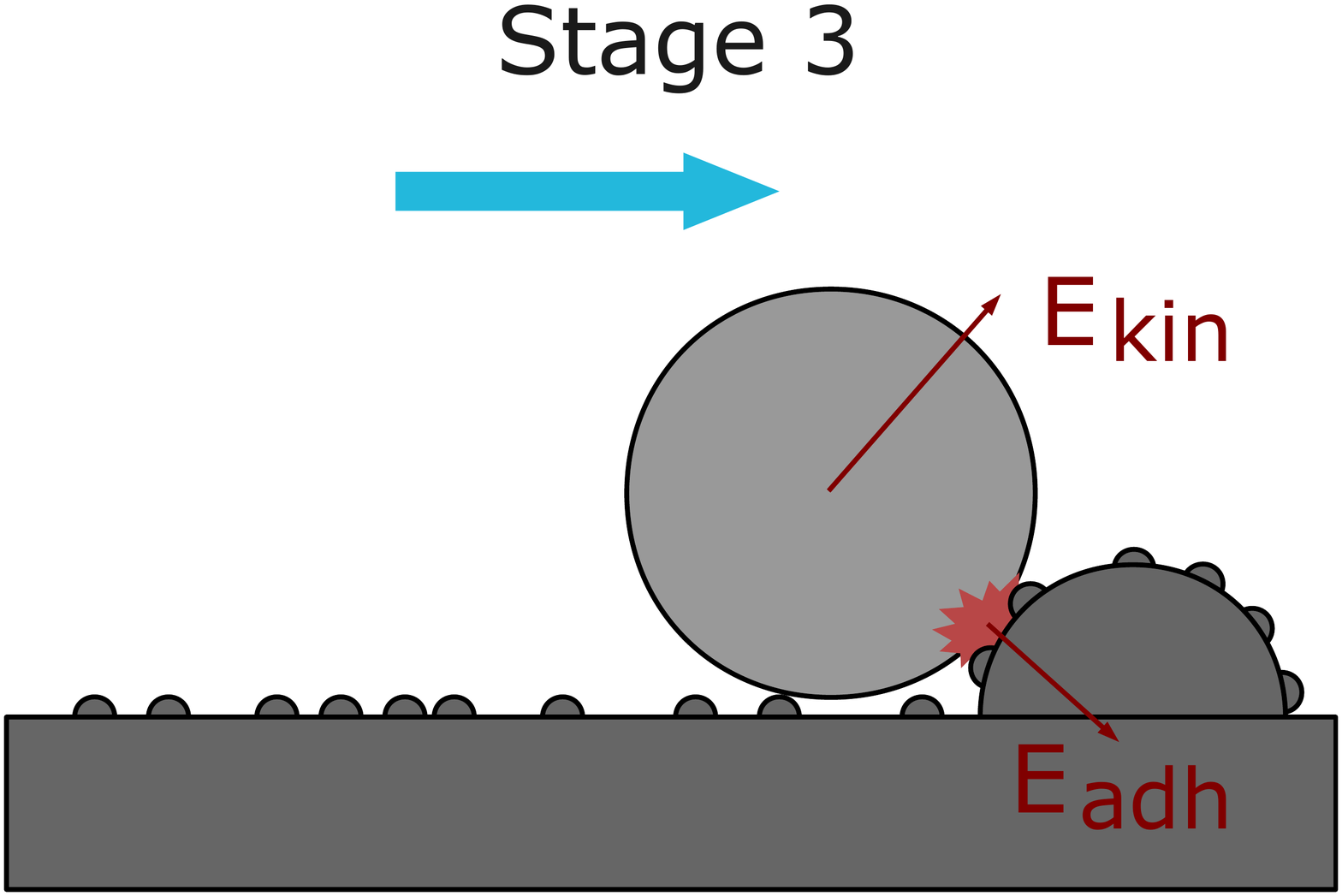}
  \caption{Sketch showing the three-stage scenario used in a recent dynamic PDF approach, where resuspension is assumed to be driven by an incipient motion (when the balance of torques is ruptured) followed by particle migration on the surface and rocking events (where detachment occurs if the kinetic energy is higher than the adhesion energy). Reprinted with permission from \cite{henry2012numerical}. Copyright 2012, American Chemical Society.}
 \label{fig:sketch_model_dynPDF}
\end{figure}

 To exemplify how this general formalism is put in practice and to illustrate PDF dynamic models, it is instructive to consider a recent dynamic PDF approach for particle resuspension whose main aspects are briefly recalled here (more details can be found in \cite{henry2012numerical, henry2014stochastic, henry2018colloidal}). This model has been developed to describe the motion of colloidal particles on a rough substrate in the case of a dispersed monolayer. These particles are small enough to be embedded in the viscous sublayer and, consequently, only rolling motion is considered. This rolling motion is assumed to be triggered by hydrodynamic drag forces only, while adhesive forces are acting to prevent it. The small amount of particles on the surface allows to treat each particle independently from one another (i.e., without accounting for particle-particle interactions). The model is based on a three-step scenario for particle resuspension (see also Fig.~\ref{fig:sketch_model_dynPDF}): first, deposited particles are set in motion when the balance of torques is ruptured; second, particles are migrating on the surface (possibly accelerating/decelerating); third, particles can detach from the surface during rocking events (where particles impact large-scale roughness features that act as a ramp for detachment). 

Assuming a linear relation between the particle angular velocity $\mb{\Omega}_{\rm p}$ and the streamwise translational velocity $U_{\rm p,||}\simeq d_p \Omega/2$, there is no need to extend to explicitly include $\mb{\Omega}_{\rm p}$ in the state vector, which is kept equal to $\mb{Z}_{\rm p}=(\mb{X}_{\rm p}, \mb{U}_{\rm p}, \mb{U}_{\rm s})$. The translational motion of particles is then given by:
 \begin{equation}
  I_{\rm p}\frac{\dd U_{\rm p,||}}{\dd t} \simeq R_{\rm p} M_{\rm drag,||} - \frac{d_{\rm p}}{2} M_{\rm adh}~.
  \label{eq:eq_Roll}
 \end{equation}
 Solving Eq.~\eqref{eq:eq_Roll} requires to know the torques due to hydrodynamic drag and to adhesion forces but, contrary to the exact dynamic models, only statistical information are needed in dynamic PDF formulations. In this model, $\mb{M}_{\rm drag}$ is evaluated by resorting to a stochastic Langevin model to simulate the instantaneous fluid velocity encountered by particles near the boundary \cite{guingo2008new, henry2012numerical} while the adhesive forces are approximated with Hamaker approach coupled to a simple model for substrate roughness (represented by a smooth surface covered by hemispherical asperities) \cite{henry2014stochastic, henry2018colloidal}. Hence, only statistical information on the fluid velocity (average velocity) and on the surface roughness (mean roughness size and surface coverage) are necessary to obtain a closed formulation and run numerical simulations. The third step in the scenario corresponds to the detachment criteria: when the particle kinetic energy is higher than the adhesive one during a rocking event, a particle is considered as being detached from the surface and, hence, resuspended. This third step requires, therefore, to determine the probability of occurrence of these rocking events. This probability is simply calculated using the surface 'seen' by the particle as it migrates on the surface together with experimental information on the percentage of large-scale roughness features covering the substrate. 
 
\subparagraph{Typical output and characteristics} Although the exact velocity of each particle cannot be evaluated in dynamic PDF methods, the fact that particle velocity is explicitly retained in the particle state vector means that statistical information on particle velocities is readily available. As a result, these formulations provide access to a wide range of factors, like velocity statistics, any quantile threshold velocity for incipient motion, any quantile critical velocity, and, of course, resuspension rates, or remaining/left fractions.
 
 Similarly to the dynamic models, a further interest in resorting to dynamic PDF approaches is that they are flexible enough to include additional information. For instance, following recent observations of collision propagation events in dense monolayer deposits, dynamic PDF approaches are now extended to account for the role of inter-particle collision in triggering the motion of individual particles \cite{banari2021evidence}. For that purpose, the state vector is left unchanged $\mb{Z}_{\rm p}=(\mb{X}_{\rm p}, \mb{U}_{\rm p}, \mb{U}_{\rm s})$ but an additional treatment is introduced to evaluate the probability that a migrating particle collides with another deposited particle and to handle the outcome of the collision. Interestingly, this leads us to evolve beyond the classical Boltzmann-like treatment of collisions since the velocities of such colliding particles are naturally correlated by the same underlying fluid driving force (this point is discussed in more details in \cite[Section 10.6]{minier2016statistical}). In addition, it is believed that these approaches can be extended to address multilayer deposits (where inter-particle collisions and collective motion are much more frequent), though such extensions are still at the early stages of development.

 \begin{figure}[ht]
  \centering
  \captionsetup[subfigure]{justification=centering}
  \begin{subfigure}{0.47 \linewidth}
   \centering
   \includegraphics[width=1.0\textwidth, trim=0cm 0.6cm 7.1cm 0cm, clip]{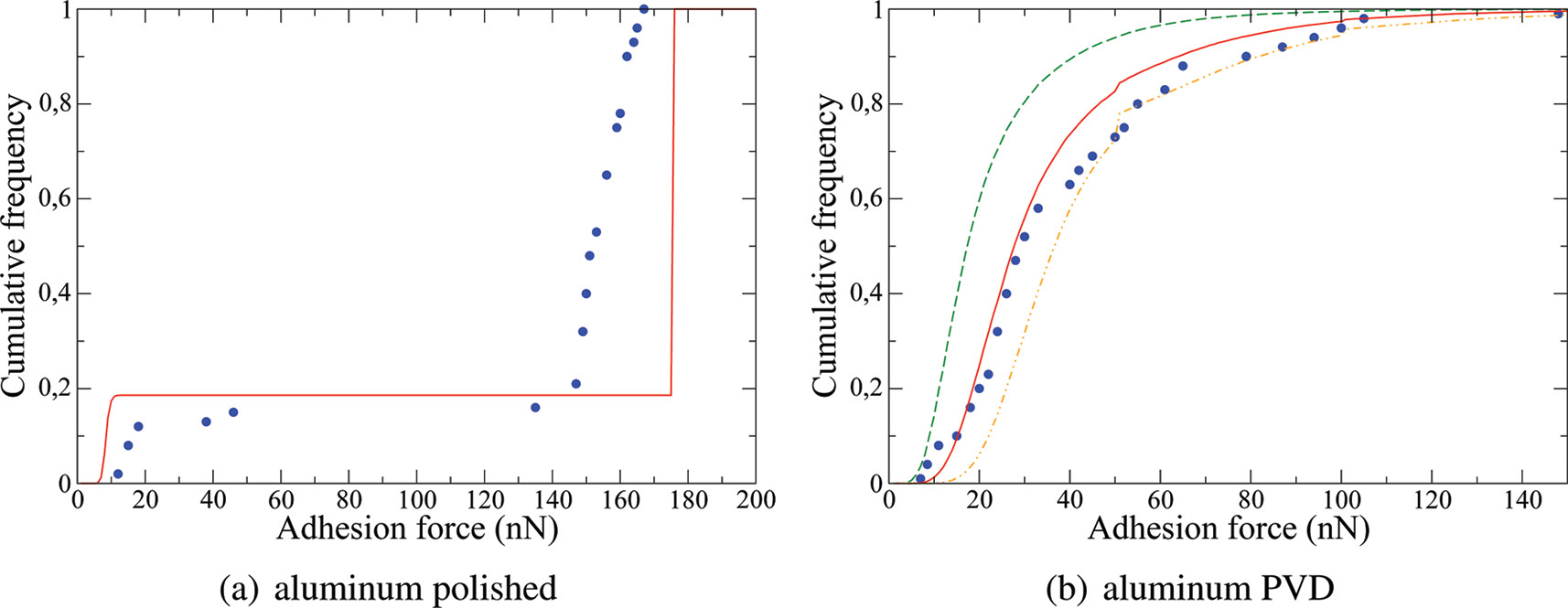}
   \caption{Adhesion force distribution between \SI{10}{\mu m} polystyrene particles and an aluminum polished substrate: comparison between experimental data (blue points) and numerical results (red line). Reprinted with permission from \cite{henry2012numerical}. Copyright 2012, American Chemical Society.}
   \label{fig:fig_henry_la_2012_valid_adh}
  \end{subfigure}
  \hspace{15pt}
  \begin{subfigure}{0.47 \linewidth}
   \centering
   \includegraphics[width=1.0\textwidth, trim=0cm 0.5cm 7.6cm 0cm, clip]{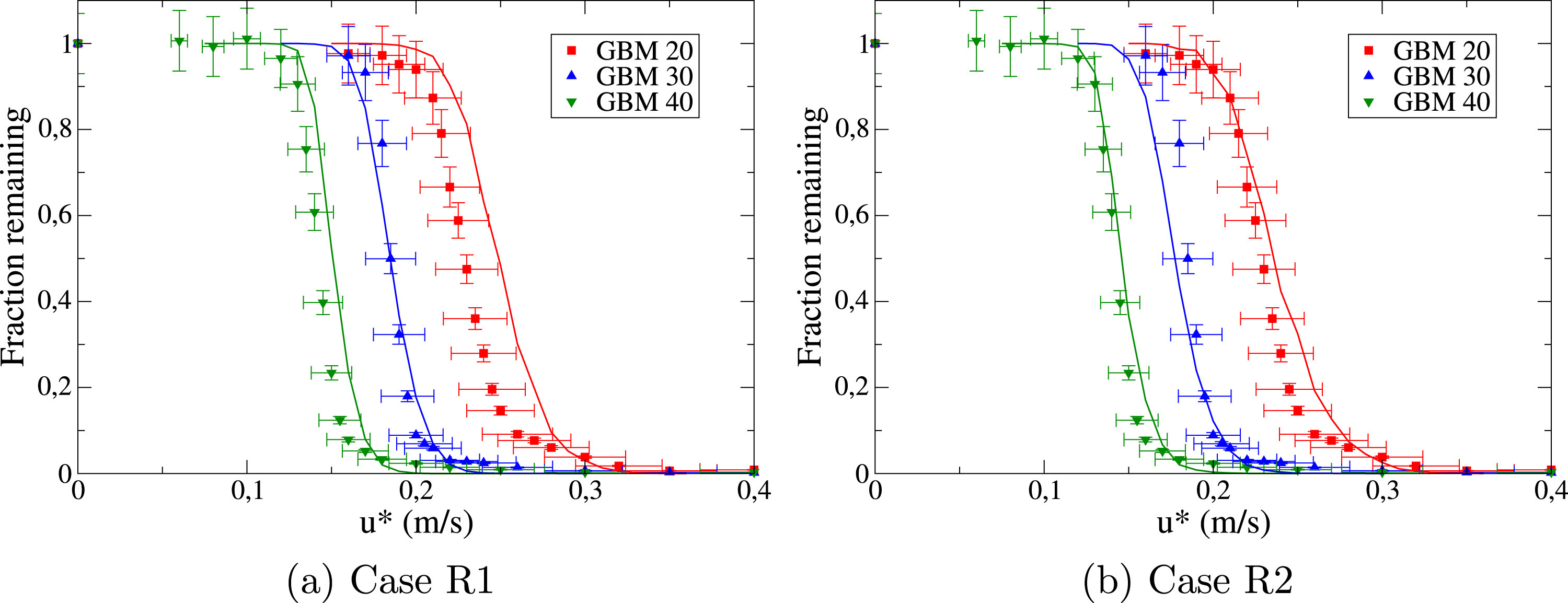}
   \caption{Fraction of glass microspheres remaining on a smooth glass substrate against the friction velocity: comparison between experimental data (symbols and error bars) and numerical results (lines). Reprinted with permission from \cite{henry2018colloidal}. Copyright 2018, Elsevier.}
   \label{fig:fig_henry_JAS_2018_valid_remain}
  \end{subfigure}
  \caption{Validation of a dynamic PDF approach showing the comparison between experimental data and numerical predictions for the adhesion force distribution (left) and remaining fraction on the surface (right). }
  \label{fig:fig_henry_JAS_2018_valid}
 \end{figure}
  
\subparagraph{Current use} 
 By construction, dynamic PDF approaches are consistent with PDF approaches used to simulate particle transport. They are thus usually coupled with RANS simulation of the fluid flow which provide one-point information on the two first moments of the velocity field, i.e., the mean velocity field and the covariance tensor of the fluctuating velocity components \cite{guingo2008new, henry2012numerical}. Special numerical schemes are used to integrate the governing SDEs \cite{peirano2006mean, henry2014stochastic} and statistics of interest are obtained by Monte Carlo methods.
 
 When coupled to proper models for the adhesion between rough surfaces (usually based on hemispherical asperities placed randomly on surface), these methods yield accurate predictions of both the adhesion force distribution (see Fig.~\ref{fig:fig_henry_la_2012_valid_adh}) and statistics on particle resuspension (like the remaining fraction displayed in Fig.~\ref{fig:fig_henry_JAS_2018_valid_remain}) to be obtained.
 
 More recently, dynamic PDF approaches have also been coupled to DNS simulations of an impinging jet flow \cite{camerlengo2018dns}. This allowed to investigate where particles are statistically more likely to be set in motion and to detach. As expected, the large majority of particle incipient motion and detachment events were found to occur in a region close to jet axis but less so as the distance from the jet axis increased (except just underneath the jet axis where the lateral velocity is too small to trigger such events).

 \subsubsection{Integrated kinetic approaches}
   \label{sec:models:approach:kin_PDF}
 
\subparagraph{General description} Integrated kinetic approaches consist in estimating each resuspension event without computing particle dynamics on surfaces. Hence, these approaches simplify the resuspension process to a kinetic one in which resuspension is assumed to be equivalent to particle detachment, thereby disregarding particle motion along walls. These models provide directly information on the probability of resuspension for each individual particle as well as on statistical quantities (like resuspension rates, threshold velocities and/or critical velocities), but without any account of particle motion prior to detachment. 

\subparagraph{Model formulation}
 Two main integrated approaches have been proposed in the literature. 
 \begin{itemize}
  \item Kinetic PDF approaches evaluate directly the probability for a particle to be resuspended from a monolayer deposit by considering only one specific scenario for particle resuspension. Drawing on the simple picture where resuspension results from a balance between motion-inducing and motion-preventing forces/torques, one can express the instantaneous resuspension rate $\mc{T}_r(t)$ by:
  \begin{multline}
   \label{eq:eq_kinPDF_generic}
   \mc{T}_r(t) = \lim_{\Delta t \to 0} \left[ 
   \int\int_{\rm resusp.\ cond.} p_{\rm ind.}(s; \mb{F}_{\rm ind.})\, p_{\rm prev.}(s;\mb{F}_{\rm prev.}) \dd\mb{F}_{\rm ind.} \dd\mb{F}_{\rm prev.} \right. \\
   \left. | \, \mb{X}_{\rm p}(t)\cdot\mb{e}_{\perp}=0 \, ; t \leq s \leq t+\Delta t \phantom{\int} \right]
  \end{multline}
where $p_{\rm ind.}$ (resp. $p_{\rm prev.}$) corresponds to the probability to have a certain value of the forces inducing motion $\mb{F}_{\rm ind.}$ (resp. of the forces preventing motion $\mb{F}_{\rm prev.}$). Analytic formulas for $\mc{T}_r(t)$ can be worked out provided that the distribution of forces acting on the particle is known and provided that a resuspension condition is expressed. 

  \begin{figure}[ht]
   \centering
   \begin{subfigure}[c]{0.47\textwidth}
    \centering
    \includegraphics[width=0.9\textwidth, trim=0cm 0.0cm 0cm 4cm, clip]{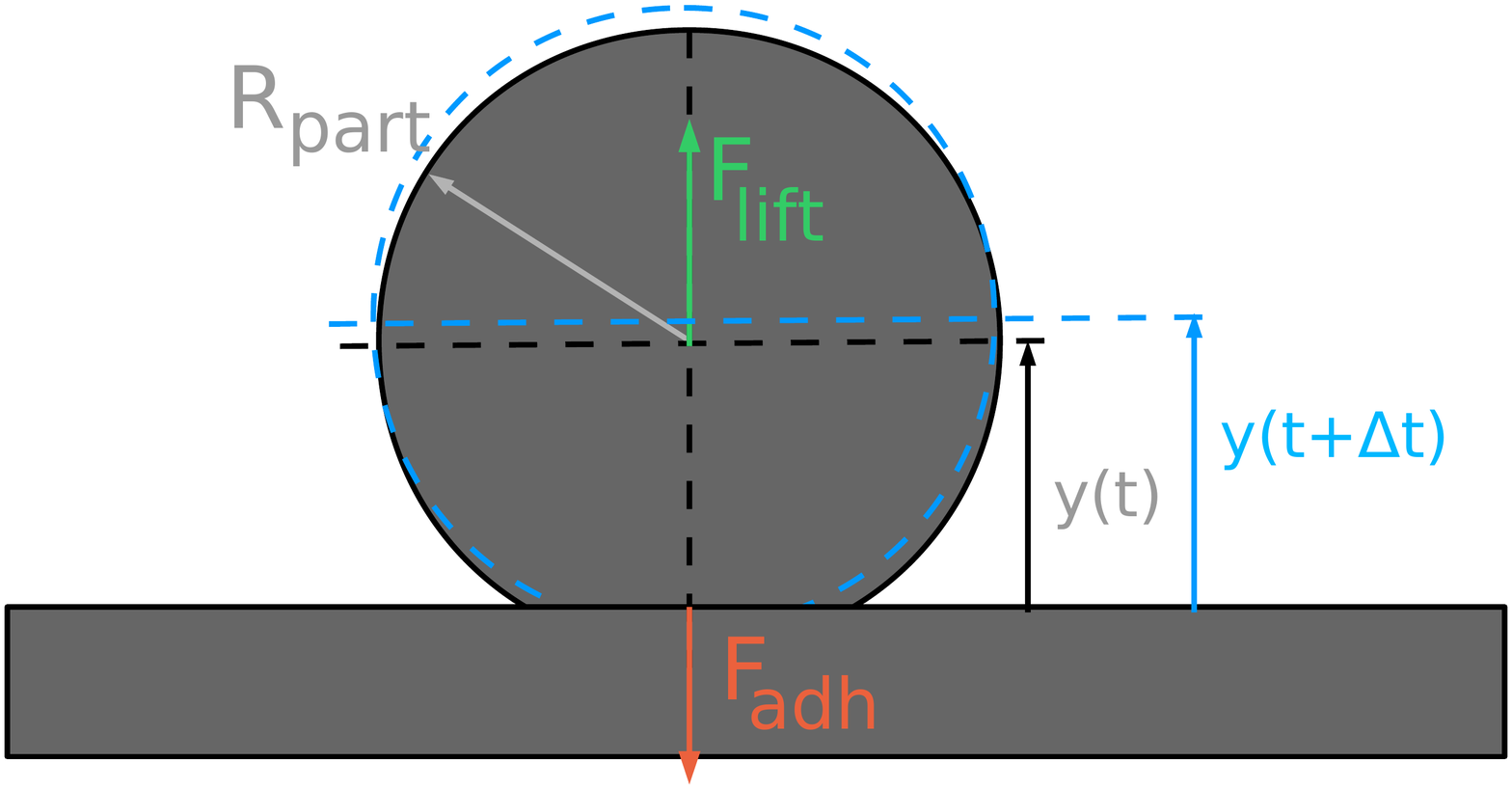}
    \caption{Sketch of the resuspension scenario used in the Reeks, Reed and Hall model: energy accumulates (until detachment) as a particle oscillates vertically due to turbulent fluctuations and to surface deformations. Reprinted with permission from \cite{henry2014progress}. Copyright 2014, Elsevier.}
    \label{fig:fig_henry_2014_RRH}
   \end{subfigure}
   \hspace{15pt}
   \begin{subfigure}[c]{0.47\textwidth}
    \centering
    \includegraphics[width=0.9\textwidth, trim=0cm 0cm 0cm 4cm, clip]{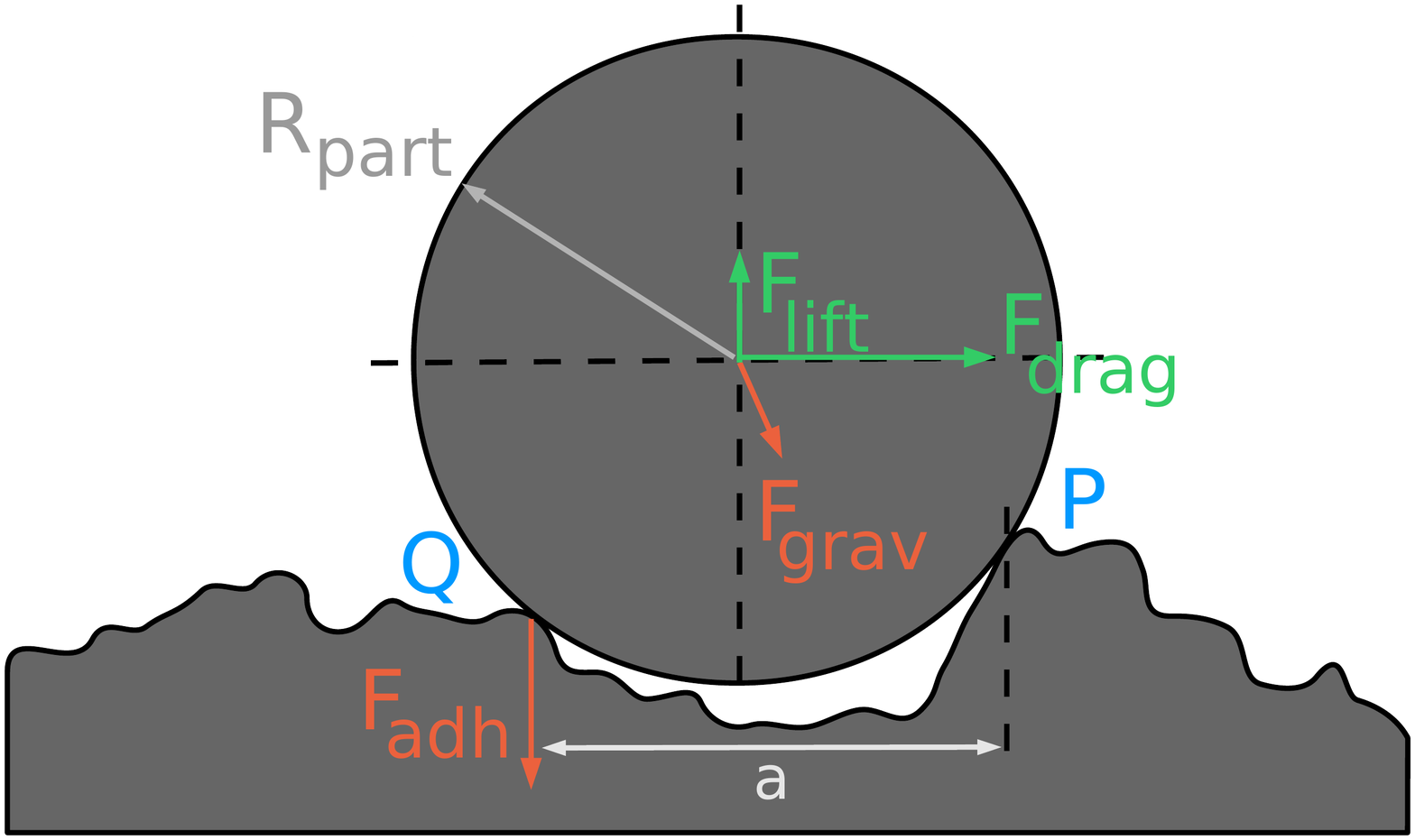}
    \caption{Sketch of the resuspension scenario used in the Rock'n'Roll model: a deposited particle oscillates around a pivot point due to turbulent fluctuations until the contact with the pivot point is broken. Reprinted with permission from \cite{henry2014progress}. Copyright 2014, Elsevier.}
    \label{fig:fig_henry_2014_RnR}
   \end{subfigure}
   \caption{Illustrations of the scenarios used in kinetic PDF approaches: Reeks Reed and Hall model (on the left) and Rock'n'Roll model (on the right).}
  \end{figure}
  
  To exemplify how this general formalism is put in practice and to illustrate such kinetic PDF approaches, it is instructive to briefly recall the main aspects of the RnR model (more details can be found in \cite{reeks2001kinetic, henry2014progress}). The RnR model assumes that particles are set in motion by their oscillations around a pivot point due to rolling motion, which are induced by the drag and lift forces due to the interaction with the near-wall turbulent flow. In that context, the angle of deflection $\theta$ for small oscillations around the pivot point P (see Fig.~\ref{fig:fig_henry_2014_RnR} is expressed by a damped oscillator model:
  \begin{equation}
   \frac{d^2\theta}{dt^2} + \beta_{RnR} \frac{d\theta}{dt} + \omega_{RnR} \theta = I_{\rm p}^{-1} \Gamma^{'}(t)
   \label{eq:eq_kinPDF_RnR0}
  \end{equation}
  with $I_{\rm p}$ the particle moment of inertia, $\omega_{RnR}$ the oscillation frequency, $\beta_{RnR}$ the damping and $\Gamma^{'}$ the fluctuating component of the torque $\Gamma$ exerted on particles. Assuming that the particle is exposed to drag, lift, gravity and adhesion forces, one can write the torque as \cite{henry2014progress}:
  \begin{equation}
   \Gamma = \frac{a}{2}\,F_{\rm lift} + r_{\rm p} F_{\rm drag} -  a\,F_{\rm adh} - \frac{a}{2}\, m\, \mb{g}\cdot \mb{n}_{\perp} + r_{\rm p} m\, \mb{g}\cdot \mb{n}_{||} .
   \label{eq:eq_kinPDF_RnR_torque}
  \end{equation}
  where $a$ is the lever-arm distance between the pivot point P and the point at which adhesion forces act. The resuspension rate can then be obtained by evaluating the number of particles (per unit time) which reach an extremum (maximum) of the potential energy when the wall-normal fluid velocity is positive (i.e., oriented out of the surface). Assuming a Gaussian distribution, this gives the following expression for the resuspension rate $\tau_r$ (per unit time) of a particle exposed to a given adhesion force:
  \begin{equation}
   k_r|_{F_{adh}} = \frac{\omega_{0}}{2\pi}\ \exp \left( -\frac{k(\lra{\, \Gamma\, |_{F_{adh}}\, })^2}{\langle f^2\rangle(1+\eta_{RRH})}\right) 
   \label{eq:eq_kinPDF_RnR1}
  \end{equation}
  with $\omega_0$ a frequency (related to the oscillation frequency $\omega_{RnR}$) and $\eta_{RnR}$ a contribution to the potential energy from the resonant energy transfer (linked with the oscillation frequency $\omega_{RnR}$ and the damping term $\eta_{RnR}$). The term $f$ appearing in Eq.~\eqref{eq:eq_kinPDF_RnR1} corresponds to the fluctuating component of the hydrodynamic contributions $F(t) = F_{\rm lift}/2 + r_{\rm p} \, F_{\rm drag}/a$. To obtain the unconditional resuspension rate $\tau_r$ (still expressed per unit time), one has to integrate this conditional rate over all possible adhesion forces (whose distribution is usually prescribed). 
  
  More generally, other scenarios have been considered: the model by Wen and Kasper \cite{wen1989kinetics} relies on a simple force-balance approach where a particle is detached from a surface as soon as the balance between hydrodynamic lift forces and adhesive forces is ruptured (i.e., $F_{\rm lift}>F_{\rm adh}$); the RRH model (proposed later by Reeks et al. \cite{reeks1988resuspension}) considers an energy accumulation as particles oscillate vertically due to the surface deformations induced by turbulent fluctuations, which can ultimately induces resuspension when enough vibrational energy has been accumulated to overcome the potential well due to adhesive force (see Fig.~\ref{fig:fig_henry_2014_RRH}). Regardless of the resuspension condition considered, the resuspension rate obtained with kinetic PDF approaches is usually expressed in the form of a Boltzmann distribution, i.e., with an exponential factor containing the ratio between the potential well preventing motion and the energy of forces/torques favoring particle incipient motion.

  \item Impulse criterion models have been developed to evaluate the probability that a particle escapes from a multilayer deposit. As illustrated in Fig.~\ref{fig:fig_pahtz_2020_impulse}, a particle deposited on top of a bed can start moving when the balance between motion-inducing and motion-preventing forces is ruptured. However, this condition is not sufficient for this particle to move freely from the deposit (i.e., either entrained by the fluid or migrating on the surface): this occurs only if the particle is able to escape from the pocket formed by the local deposit configuration. This means that resuspended particles have to acquire enough kinetic energy to overcome the potential barrier of the bed pocket. Hence, the following resuspension condition has been suggested: a particle is resuspended from a pocket bed when the forces acting on it are strong enough to induce motion and when these forces are acting over a long-enough duration. This translates into an impulse condition, where the impulse $I_f$ has to exceed a given threshold value $I_{fc}$ \cite{diplas2008role}:
  \begin{equation}
   I_f = \int_{t_0}^{t_0+T}\|\mb{F}_{\rm s\to p}(s)\|ds \ge I_{fc}.
   \label{eq:eq_impulse}
  \end{equation}
  The hydrodynamic forces entering this equation have to exceed a certain threshold $F_c$, so that particles are set in motion. Expressions for this threshold value $I_{fc}$ have been derived later considering the competition between drag forces, lift forces or gravity \cite{celik2010impulse, valyrakis2010role} and taking into account the slope of the river bed \cite{lamb2008critical}.

  \begin{figure}[ht]
   \centering
   \includegraphics[width=0.4\textwidth]{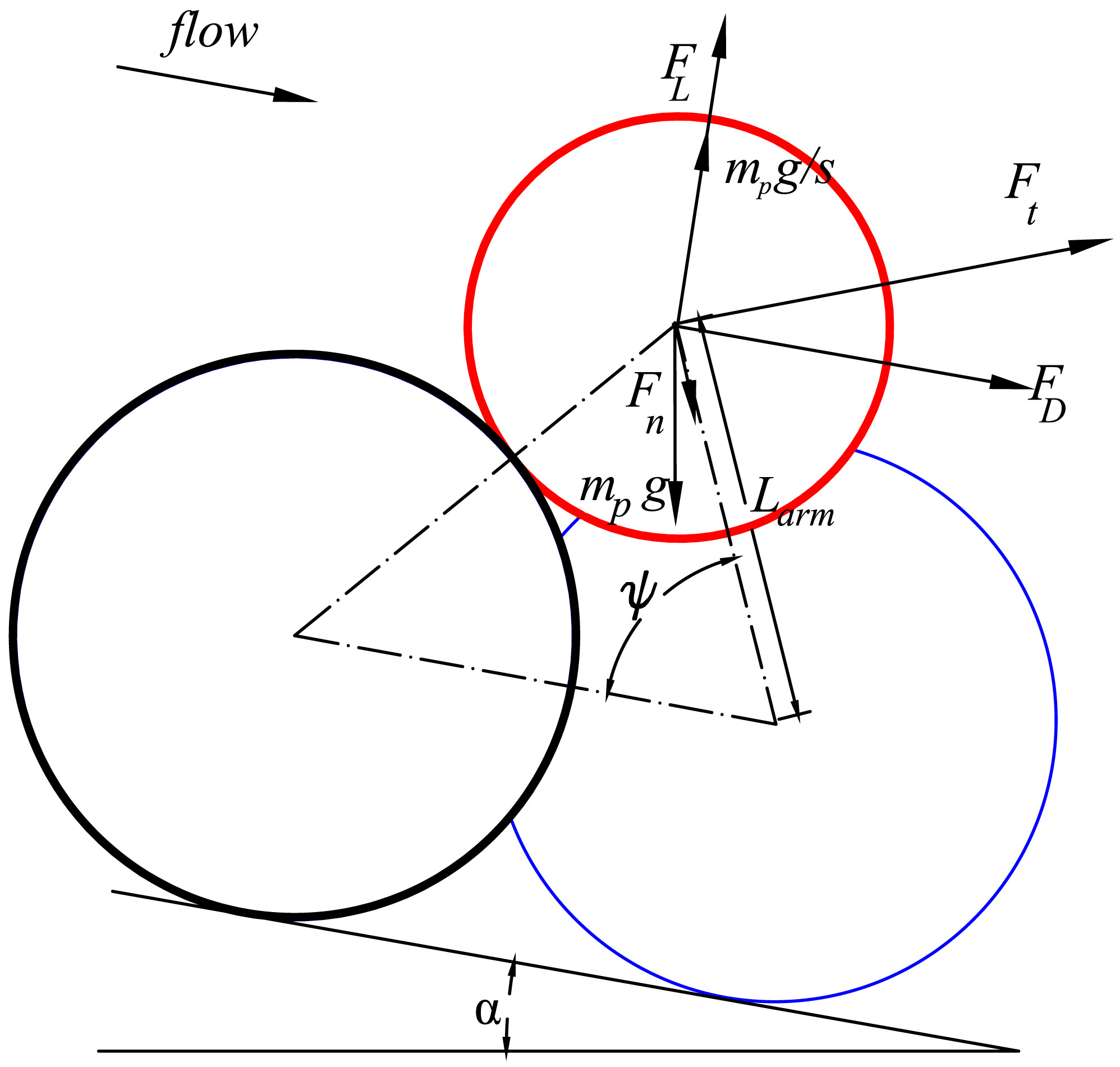}
   \caption{Sketch showing a pocket geometry: the red particle will be able to migrate on the surface only if the forces acting on it are strong enough to get the particle out of the pocket formed by other surrounding particles. Reprinted with permission from \cite{pahtz2020physics}. Copyright 2020, John Wiley and Sons.} 
   \label{fig:fig_pahtz_2020_impulse}
  \end{figure}

  More recently, models based on an impulse criterion have been replaced by models based on an energy criterion \cite{valyrakis2013entrainment}. The idea behind this change is that resuspension is more likely to happen when particles interact with high-intensity turbulent structures rather than with moderate-intensity ones. In fact, the energy transferred from the fluid to the particle is expected to be much higher when the driving hydrodynamic forces significantly exceeds the threshold force for incipient motion (i.e., $\|\mb{F}_{\rm s\to p}\|\gg F_{c}$) compared to the one obtained with near-critical forces (i.e., $\|\mb{F}_{\rm s\to p}\|\gtrsim F_{c}$). This has led to the introduction of a criterion based on the actual energy transferred to the particle \cite{valyrakis2013entrainment}:
  \begin{equation}
   C_{\rm eff}E_{f} = C_{\rm eff} \int_{t_0}^{t_0+T} P_{f}(s)ds \ge W_c.
   \label{eq:eq_impulse_energy}
  \end{equation}
  where $W_c$ is the minimal amount of work required for complete particle entrainment, $P_f(t)$ is the instantaneous flow power (proportional to the cube of the local flow velocity), and $C_{\rm eff}$ is the coefficient representing the efficiency of the energy transfer from the fluid to the particle. Again, expressions for the threshold values $W_c$ are available depending on the mode of motion (e.g., rolling or hopping \cite{valyrakis2013entrainment}).
 \end{itemize}

\subparagraph{Typical output and characteristics} As it follows from the above description, kinetic approaches yield estimates of the probability of a resuspension event. Yet, given that these models do not account for the migration dynamics, they can overpredict resuspension since some particles set in motion can nevertheless remain on the surface (and come to a stop at a downstream location \cite{wang1990effects}). In addition, kinetic PDF models have usually been designed considering only a single scenario (either sliding, rolling, lifting motion or even particle collisions). When resuspension occurs through a variety of processes (e.g., with particles rolling and sliding at the same time \cite{feuillebois2016three} or with mixed rolling/sliding and hopping motion as displayed in Fig.~\ref{fig:fig_kassab_2013_mechanisms}), these formulations can become more complicated as they require additional information on the relative importance between each event in the overall resuspension process. 

Another limitation is related to the input parameters required. In fact, these models require information on the forces/torques acting on the particle, which are often assumed to be described by normal or log-normal distributions (an extension to other distributions was given in \cite{zhang2013particle}). Nevertheless, additional complexity arise when complex multimodal distributions occur (as has been measured recently for the adhesive force between a particle and a surface exposed to outdoor contamination \cite{rush2018glass}). Another critical parameter of the RnR model is the lever-arm of the adhesion and lift forces. The only experimental estimation of such a parameter was based on centrifuge measurements in \cite{reeks2001kinetic} for a specific surface/particle pair and for only one particle diameter. The method allowed only for the determination of the mean value of the lever-arm and recent work has highlighted the RnR model sensitivity to the lever-arm parameter \cite{brambilla2020impact}. In addition, a basic hypothesis of the RnR model is that the two points of contact between the particle and the surface are symmetric with respect to the particle center. For an actual surface (e.g., like glass with nanometer roughness), a recent study also showed that the assumption that the points of contact are symmetric about the particle center is not always true, which implies that the lever-arm for the adhesion force may not be just twice that of the lift force \cite{brambilla2020impact}. 

\subparagraph{Current use}  Due to their easy use and relatively low computational costs, kinetic PDF approaches have been widely used. 
\begin{itemize}
  \item Kinetic PDF approaches:
  These approaches have been extensively used in the context of atmospheric resuspension and especially in nuclear safety \cite{stempniewicz2008model}. They provide quick access to the statistics on particle resuspension (including the remaining fraction or the time-averaged resuspension rate) while requiring information only on the distributions of adhesive and hydrodynamic forces. As a result, predictions of useful quantities for atmospheric resuspension can be obtained by measuring experimentally or computing numerically the mean and standard deviation of these forces. For instance, as displayed in Fig.~\ref{fig:fig_reeks_2001_fremain}, the Rock'n'Roll model was shown to reproduce the fraction of \SI{20}{\mu m} alumina particles remaining on a stainless steel substrate when the adhesion forces are assumed to follow a log-normal distribution (the best fit was obtained with a spread $\sigma_a = 10.4$) \cite{reeks2001kinetic}.
  \begin{figure}[ht]
   \centering
   \includegraphics[width = 0.57\textwidth]{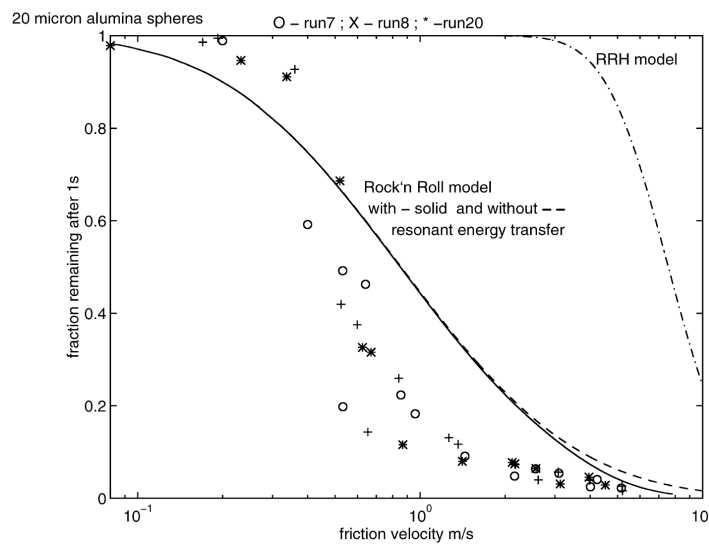}
   \caption{Comparison of experimental measurements and predictions with the Rock'n'Roll model for the remaining fraction of \SI{20}{\mu m} alumina spheres on a stainless steel substrate exposed to a turbulent flow for a duration of \SI{1}{s}. Reprinted with permission from \cite{reeks2001kinetic}. Copyright 2001, Elsevier.}
   \label{fig:fig_reeks_2001_fremain}
  \end{figure}
  
  Nevertheless, it should be remembered that characterizing distributions with information on their average value and their standard deviation is not always straightforward since these distributions can have very different shapes and do not necessarily belong to a unique class of PDF \cite{prokopovich2011adhesion, audry2009adhesion, rush2018glass}: recent studies have indeed highlighted that adhesion forces with outdoor surfaces (contaminated by tiny objects such as dust or sand) display multimodal distributions that are very different from those obtained with clean laboratory surfaces \cite{rush2018glass}. 
  
  In addition, since models are designed to capture either rolling, sliding or lifting motion, their range of validity is relatively well established. For instance, for a friction velocity of 0.1 \si{m/s}, the Rock'n'Roll model is applicable for particles smaller than \SI{700}{\mu m} (resp. \SI{50}{\mu m}) when they are suspended in the air (resp. in water). In ventilation ducts, the friction velocity is typically one order of magnitude larger, meaning that the limit of applicability of the RnR model drops by an order of magnitude to particles smaller than \SI{70}{\mu m}. In the context of nuclear safety, where LOCA accidents might occur in very high temperature gas-cooled reactors at friction velocities up to \SI{5}{m/s}, the RnR model will be applicable only to particles smaller than roughly \SI{15}{\mu m}. 
  
  \item Impulse criterion approaches:
  
  These models have been developed in the context of fluvial transport. As a result, they are widely used to predict the Shields parameter for the resuspension of sediments from rivers \cite{pahtz2020physics}. Indeed, applying an impulse- or energy-criterion is especially adapted to the turbulent flows encountered in marine systems (with high values of the Reynolds number). This is not the case for static force- or torque-balance approaches, which are more convenient to treat low Reynolds-number flows (such as Stokes flows \cite{deskos2018incipient}). For instance, such a method has been used to predict the Shields parameter using a work-based criterion where a particle has to escape a pocket geometry (as the one displayed in Fig.~\ref{fig:fig_pahtz_2020_impulse}) \cite{lee2012work}. 

 \end{itemize}

 \subsubsection{Force-balance/torque-balance approaches}
   \label{sec:models:approach:FBM}
 
 \subparagraph{General description} Force-balance, or torque-balance, approaches consist in evaluating directly the resuspension of each particle by assuming that resuspension corresponds to the rupture of balance between forces/torques acting on the particle. Hence, they are similar to integrated kinetic approaches since they do not consider the dynamics of particles on the surface and also assume that resuspension is synonym of particle detachment. In a sense, they can be regarded as formulations developed in physical space while integrated kinetic PDF models are formulated in sample space.

\subparagraph{Model formulation} Since the resuspension criterion simply amounts to a rupture of the static equilibrium between forces inducing particle motion and forces hindering it, this gives the following deterministic conditions for each of the three resuspension modes (considering only drag, lift, gravity and adhesion forces): 
 \begin{subequations}
  \label{eq:eq_FTB}
  \begin{eqnarray}
  \rm{Lift-off} & \mb{F}_{\rm lift} & > \mb{F}_{\rm grav} + \mb{F}_{\rm adh}, \\
  \rm{Sliding} & \mb{F}_{\rm drag} & > \mu_{\rm fr} \left(\mb{F}_{\rm adh} + \mb{F}_{\rm grav} - \mb{F}_{\rm lift} \right), \\
  \rm{Rolling} & \mb{M}_{\rm drag} + \mb{M}_{\rm lift}& > \mb{M}_{\rm adh} + \mb{M}_{\rm grav} 
  \end{eqnarray}
 \end{subequations} 
 with $\mu_{\rm fr}$ the static friction coefficient. A closed set of equations is then obtained by assuming that the forces entering Eqs.~\eqref{eq:eq_FTB} are given. For that purpose, one can either rely on measurement or on theoretical/empirical expressions for these forces (these expressions are described in Section~\ref{sec:models:forces}).

\subparagraph{Typical output and characteristics} As for integrated kinetic approaches, force/torque balance models provide information on the probability of resuspension for each individual particle as well as to resuspension statistics (like resuspension rates, threshold velocities and/or critical velocities). Yet, no information on particle velocities upon resuspension or migration on the surface is available with such models.

One of their drawbacks lies in the fact that a specific mode of resuspension is often chosen right from the onset. This means that the expression obtained are only applicable to a certain size of particles. In fact, experimental observations have shown that small particles embedded within the viscous sublayer are more prone to rolling motion while larger particles, which are protruding from the viscous sublayer, are more susceptible to interact with near-wall turbulent structure and undergo direct lift-off \cite{henry2014progress}. 

\subparagraph{Current use} These models have been widely used, especially in the early stages of resuspension studies. Their popularity is probably related to the observations of threshold velocities for incipient motion, which inspired the idea that resuspension corresponds to a rupture of balance between motion-inducing forces and motion-preventing ones. The exact expressions of these conditions have been applied to a range of configurations, most of them related to the resuspension from monolayer deposits involving spherical particles interacting with smooth substrates \cite{ibrahim2003microparticle} or rough substrates/beds \cite{ziskind1997adhesion}. These models were recently extended to compute the resuspension probability by accounting for the distributions in turbulent fluid forces or adhesive forces between rough surfaces. This has led to the use of such force/torque balance models in Monte Carlo approaches \cite{goldasteh2012model, benito2015monte, benito2016validation}. 

\begin{figure}[ht]
  \centering
  \includegraphics[width = 0.55\textwidth]{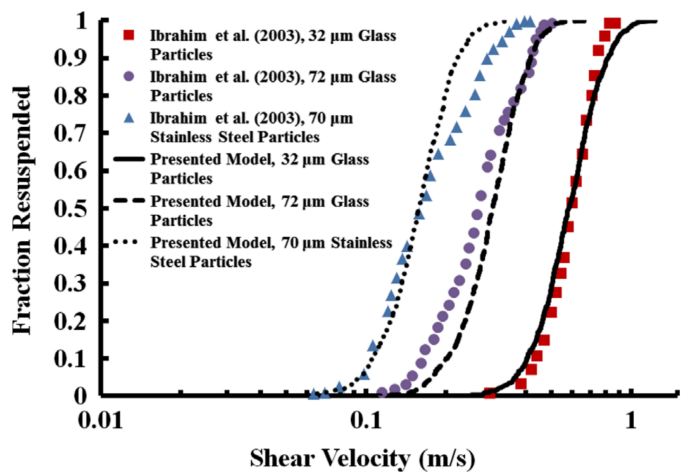}
  \caption{Comparison of experimental data from \cite{ibrahim2003microparticle} and predictions obtained with a Monte-Carlo model based on a torque-balance approach (for rolling motion). The fraction of glass particles (with three sizes) is plotted as a function of the shear velocity. Reprinted with permission from \cite{goldasteh2012model}. Copyright 2012, Elsevier.}
  \label{fig:fig_goldasteh_2012_FBM}
 \end{figure}

For instance, one model aims at capturing the effect of particle roughness and complex shape by assuming that uniform hemispherical bumps are present on the surface of spherical particles that are removed from the surface by rolling. The particle diameter, the bumps number, and their diameters are treated as random variables sampled from predefined distributions \cite{goldasteh2012model}. In particular, an experimentally-measured Gaussian distribution was used for the particle diameter, the number of bumps was assumed to follow a Poisson distribution, and the inverse of a roughness parameter was assumed to follow a log-normal distribution. The adhesion force was then computed using a JKR model (introduced in Section~\ref{sec:models:forces:contact}). Under these assumptions, both the adhesion and the aerodynamic forces become random variables. Hence, force-balance/moment-balance approaches can be used to describe single resuspension events for specific combinations of particle, surface, and fluid conditions but they can also capture statistical information on resuspension rates. By resorting to a Monte Carlo method based on torque-balance (for rolling motion), this formulation was then used to estimate the resuspension rate as function of the fluid shear velocity, as shown in Fig.~\ref{fig:fig_goldasteh_2012_FBM}.
 
 \subsubsection{System-scale models based on force/torque balance}
   \label{sec:models:approach:syst_FBM}
 
\subparagraph{General description} Another category of models consists in predicting directly information at the system scale, i.e., at a macroscopic scale. This means that, instead of solving the resuspension and motion of individual particles, an approximation of the quantity of interest is derived directly from a combination of theoretical arguments and empirical formulas. In that sense, system-scale models are similar to the force-balance approaches presented before. Yet, the main difference concerns the predicted quantity: instead of estimating the average resuspension rate, these models provide information on more macroscopic quantities (such as the threshold velocity for incipient motion or the Shields parameter).
 
\subparagraph{Model formulation} As indicated, system-scale models consist in proposing an analytical expression for the selected quantity of interest. For instance, a recent review \cite{dey2018advances} has shown that various approximations of the threshold Shields parameter were obtained in the literature. These expressions vary depending on the theoretical considerations made (e.g., propositions based on force-balance for single rolling, sliding, and/or lifting motion, type of forces involved) as well as on the expressions retained for each of the forces (e.g., hydrodynamic drag force with/without shielding effects). To estimate the Shields parameter, these methods frequently rely on the use of empirical laws which relate a shear stress (or a friction velocity) to the average velocity in a wind-tunnel. The resulting expressions are easy to use and depend on a number of variables (such as the bed slope, the particle density, the particle-based Reynolds number). For instance, considering the resuspension of particles from a bed through sliding motion due to drag \& lift \& gravity forces and using the statistical theory of turbulence, the threshold Shields parameter $\Theta_c$ has been expressed as \cite{ikeda1982incipient}:
 \begin{equation}
  \Theta_c = \frac{4}{3} \cdot \frac{\mu_{\rm fr}}{C_D + C_L\, \mu_{\rm fr}} \times \left\{ \frac{10.08}{R_{\star,c}^{10/3}}+\left[\frac{1}{\kappa}\text{ln}\left(1+\frac{4.5R_{\star,c}}{1+0.3R_{\star,c}}\right)\right]^{-10/3}\right\}^{0.6}
  \label{eq:eq_thetac_ikeda}
 \end{equation}
 with $\mu_{\rm fr}$ the friction coefficient, $C_D$ the drag coefficient, $C_L$ the lift coefficient (more details in Section~\ref{sec:models:forces}), $\kappa$ the von K\'{a}rm\'{a}n constant (usually set to $\kappa = 0.41$) and $R_{\star,c} = u_{\star} k_s / \nu_{\rm f}$ the Reynolds number based on the friction velocity $u_{\star}$ and $k_s$ the bed roughness height (proportional to the particle diameter).

\subparagraph{Typical output and characteristics} The main advantage of such expressions is that they provide quick and relatively reliable information on an observable of interest. For that reason, they are often used as a source term for the amount of resuspended material in numerical simulations of particle transport. However, these expressions should be handled carefully since the range of validity of each model depends on the assumptions made to derive the expression (and possibly on the empirical formulas used for the forces entering the force/torque balance model).

 \begin{figure}[ht]
  \centering
  \includegraphics[width=0.65\textwidth, trim=0cm 0cm 0cm 0cm, clip]{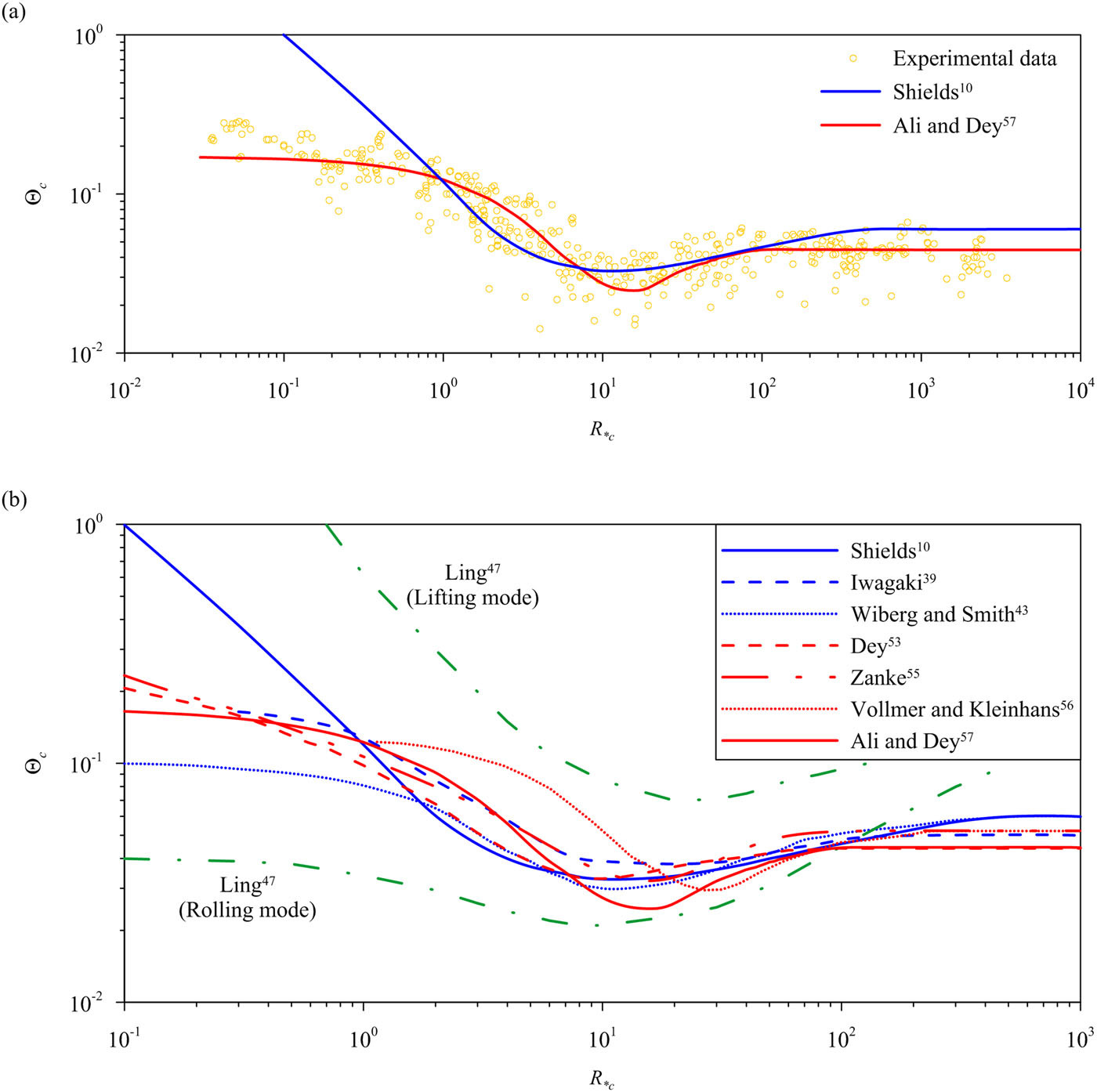}
  \caption{Critical Shields parameter $\Theta_c$ as a function of the Reynolds number based on particle size and friction velocity $R_{\star_c}$: a) Comparison between experimental data (yellow), the Shields curve (blue) and the model from Ali and Dey model (red); b) Comparison between various models. Reprinted with permission from \cite{dey2018advances}. Copyright 2018, AIP Publishing.}
  \label{fig:fig_dey_2018_comparison}
 \end{figure}
 
\subparagraph{Current use} These models are widely used in practical applications such as fluvial or aeolian transport. In fact, various formulations have been proposed to express the Shields parameter as a function of various variables \cite{dey2018advances} (such as the particle size, density, etc.). A comparison between the various predictions is displayed in Fig.~\ref{fig:fig_dey_2018_comparison}.
 
 Another example concerns river morphology where morphodynamic models often rely on the Exner equation \cite{ancey2020bedload1}, which expresses the conservation of mass between deposited sediments within a bed and suspended sediments. This equation relates the changes in the bed elevation to the amount of sediment that are resuspended (i.e., the divergence of the sediment flux) and is usually formulated as:
 \begin{equation}
  \frac{\partial z}{\partial t} = \frac{-1}{\rho_{\rm bulk}} \frac{\partial q}{\partial x}
  \label{eq:eq_exner}
 \end{equation}
 with $z$ the bed elevation, $\rho_{\rm bed}$ the bulk density of particles, and $q$ the sediment flux. This has led to the development of various expressions for this particle flux (interested readers are referred to dedicated reviews on this topic such as \cite{ancey2020bedload1, zhang2021aeolian}). Such formulations have also been used to predict the resuspension of sand due to wind around obstacle (see for instance \cite{tominaga2018wind}, which relies on RANS simulations of the fluid flow to estimate the local fluid velocity required as an input to these models).
 
 \subsubsection{Empirical formulas}
   \label{sec:models:approach:emp_form}
 
\subparagraph{General description} Empirical formulas correspond to models based on empirical observations instead of relationships based on theoretical descriptions of the phenomena at play. These models are usually obtained by careful analysis of data coming from experimental observations or numerical experiments. From this analysis, simple formulas are used to express a relationship between the desired quantity (often called the observable) and different input variables known to affect this observable. 

\subparagraph{Model formulation} When designing an empirical formula, the first step is to choose the observable of interest, and the second step is to select the input variables to be included in the formula. To be more specific, let us consider the case where we are interested in describing the time-averaged resuspension rate $\tau_r^{\Delta t}$ of aerosols in the atmosphere at short times (here less than \SI{24}{h}). We further assume that the empirical formula should capture the evolution of the time-averaged resuspension rate $\tau_r^{\Delta t}$ as a function of the friction velocity $u_{\star}$ and time $t$ only. The best fit obtained using experimental data from three experiments gives \cite{loosmore2003evaluation}:
 \begin{equation}
  \tau_r^{\Delta t} = 0.01 \frac{(u_{\star})^{1.43}}{t^{1.03}}.
  \label{eq:eq_empirical1}
 \end{equation}
 Another fit can be derived if we also include information on the particle diameter $d_p$, the particle density $\rho_{\rm p}$, and the average substrate roughness height $\langle r_{\rm a}\rangle$. This gives \cite{loosmore2003evaluation}:
 \begin{equation}
  \tau_r^{\Delta t} = 0.42 \frac{(u_{\star})^{2.13}\,d_{\rm p}^{0.17}}{t^{0.92}\,\langle d_{\rm a}\rangle^{0.32}\,\rho_{\rm p}^{0.76}}.
 \label{eq:eq_empirical2}
 \end{equation}
  Experiments (real or numerical) used to derive an empirical equation must therefore have a form of internal consistency regarding fluid flow conditions (laminar vs. turbulent), particle size (small vs. large with respect to the boundary layer), particle shape (quasi-spherical vs. one dominant dimension, like in rod-like particles), surface type (bare vs. with vegetation), surface roughness (small vs. large compared to the particle diameter), wettability (of both the particle and the surface). Note that it is possible to capture multiple regimes with empirical formula, for instance with asymptotic equations, but this is usually outside their scope. 
	
\subparagraph{Typical output and characteristics}  The main interest of empirical formulas is that they provide quick and relatively reliable information on an observable of interest. For that reason, empirical formulas are often used in practical applications (e.g., dispersion of pollen or allergens). Empirical formulas can also be used in numerical simulations of particle resuspension and transport, mostly as source terms for the amount of resuspended particles during a given amount of time knowing the local fluid velocity.
 
 As pointed out, their main drawback is their range of validity. In fact, these equations can only capture the physics and regimes that were present in the experiments used for model tuning, and they depend on the relative amount of data in each experiment as well as the estimation of model parameters that may have not been measured at the time of the experiment. Consequently, should we be interested in applying one such formula in a different context (where, for instance, the relative humidity of air plays a role), a new empirical formula would have to be designed to include the relative humidity as one of the input parameters (provided that information on this variable is available within the initial dataset). As a result, before applying an empirical formula to a specific problem, it is imperative to understand the range of applicability and the model limitations. For instance, the aforementioned empirical models fail to predict the short-term prompt resuspension which can account for the instantaneous aerosolization of a large portion of the deposit. This aspect is particularly important in the evaluation of health-effects.

\subparagraph{Current use}
 In the past, empirical formulas were based on dimensional analysis (see the review \cite{kim2010source} and references therein). This has led to a number of linear regression or power laws for the time-averaged resuspension rate \cite{loosmore2003evaluation, kim2016effects}. Nowadays, new machine learning (ML) type approaches have emerged to derive data-driven models. While the traditional black-box approach is not very appealing due to its lack of transparency, new methods allow for the derivation of equations and the imposition of physics-based constraints. For instance, symbolic regression is a type of regression analysis that combines mathematical building blocks like mathematical operators and functions to derive the simplest equations that best fit a data set. Compared to other regression techniques, the user does not have to know the functional relationship among the variables a priori and the method uncovers both the model structure and the model parameters. Prior knowledge can be incorporated by using shape-constrained symbolic regression, in which an expected behavior can be prescribed (e.g., monotonicity of the function with respect to certain inputs). In general, these functions will be more complex than those obtained by dimensional analysis and power laws and have the potential to reveal more descriptive equations of the underlying physical system.

\begin{figure}[ht]
  \centering
  \includegraphics[width=0.55\textwidth, trim=0cm 0cm 0cm 0cm, clip]{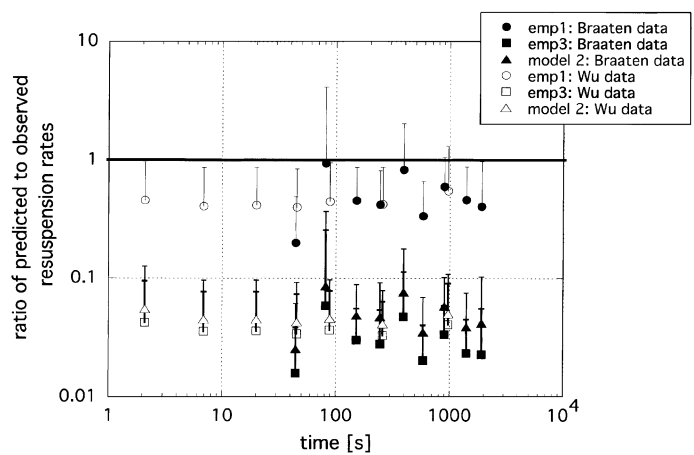}
  \caption{Evaluation of the performance of empirical models (see  Eqs.~\eqref{eq:eq_empirical1} and~\eqref{eq:eq_empirical2}) against the data from Wu and Braaten experiment on the resuspension of lycopodium spores. Reprinted with permission from \cite{loosmore2003evaluation}. Copyright 2003, Elsevier.}
  \label{fig:fig_loosmore_2003_empirical}
 \end{figure}

To illustrate the limitation of empirical approaches, various empirical models derived from various atmospheric resuspension measurements (given by Eqs.~\eqref{eq:eq_empirical1} and~\eqref{eq:eq_empirical2}) have been compared to additional experimental data on the resuspension of lycopodium spores. As illustrated in Fig.~\ref{fig:fig_loosmore_2003_empirical}, some of the models are not able to reproduce the additional measurements due to the lack of accurate considerations of surface properties (especially roughness) \cite{loosmore2003evaluation}.
 
 Nevertheless, due to their simplicity, these models have been extensively used and coupled to various approaches. For instance, empirical models are frequently applied to estimate particle resuspension from trees \cite{buccolieri2018review}: the principle is to express directly the fraction of deposit being resuspended as a function of the wind speed, the leaf area density, and particle concentration, which are then taken as source terms (i.e., new particles entering the physical domain) in CFD simulations of particle dispersion in atmospheric environments. Similar empirical expressions are introduced as source terms in CFD simulations of radionuclide dispersion (see for instance \cite{dovlete2018pre}). Expressions of the Shields parameter are also frequently used as source terms even in relatively high-level CFD simulations. For example, it has been used in combination to Large-Eddy Simulation (LES) of the fluid flow to study creep motion over a ripple \cite{chang2003entrainment} or even resuspension induced by helicopter hoover over a sandy surface \cite{wu2017particle}.

  \subsubsection{A hierarchical view on modeling approaches}
   \label{sec:models:approach:hierarchy}

\begin{figure}[ht]
 \centering
 \includegraphics[width=0.9\textwidth]{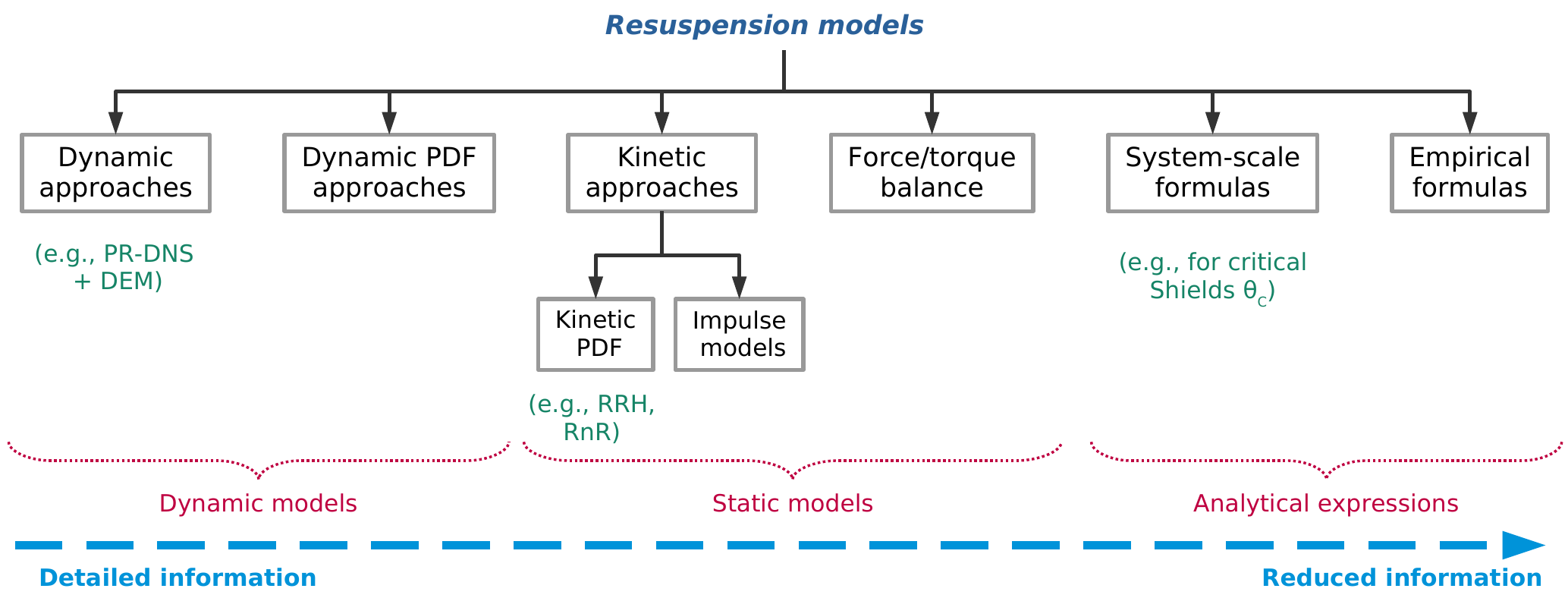}
 \caption{Sketch summarizing the various resuspension models and their level of description (arrow).}
 \label{fig:fig_sketch_model}
\end{figure}

The various modeling approaches outlined in this section can be organized as a function of their information content. This is shown in Fig.~\ref{fig:fig_sketch_model} in which the horizontal arrow indicates a reduction of information in the treatment of the particle resuspension process. A few remarks can be made: 
\begin{enumerate}[(a)]
\item Only two methods, namely the exact dynamical and the dynamical PDF, consider the complete resuspension process in the sense that they distinguish between detachment and actual resuspension. Both introduce specific effects for deposited particles, such as adhesive forces, but retain the same formulation as for particles transported by the fluid flow away from walls. Consequently, they are naturally consistent with corresponding simulations of particle transport in the fluid near-wall boundary layers. This makes them attractive candidates if one wishes to address the overall process in which deposition needs also to be considered (for instance, for saltating particles);
\item In contrast, the kinetic and force-balance approaches appear as static methods since they assume that resuspension is equivalent to detachment and, therefore, to the rupture of an initial equilibrium, or static, state of deposited particles. Regardless of the differences in how adhesive and fluid entrainment forces are handled, these two models can be seen as equivalent, with one (the kinetic approach) being formulated in sample space and the second (the force-balance approach) written directly in physical space;
\item The last two methods, the system-scale model and empirical formulas, bypass the treatment of the physical process at play in particle resuspension altogether and attempt directly to provide analytical expressions for either some key parameter or a time-averaged particle resuspension rate. 
\end{enumerate}
 
 \section{The art of modeling: new insights into existing models}
 \label{sec:art_model}

Two remarks can be drawn from the description of the physics involved in particle resuspension and current attempts at capturing its key aspects. First, there is no first-principle formulation acting as the reference theory and from which simplified versions would be derived. Second, there is no lack of models but rather of profusion of them. What is however puzzling is that not only do these models operate at different levels of description but they involve sometimes completely different treatments of the same physical phenomena. For instance, adhesion forces are regarded as input parameters in some models while they are considered as characteristic features to be predicted in others. This situation indicates that very different conceptions of a modeling approach are present but without being explicitly stated. To chart the modeling landscape, it is therefore useful to step back and address specifically what constitutes a model. 

In practice, the formulation of a model implies building a consistent theoretical framework under known hypotheses which have direct impact on the quantities being predicted and the range of applicability. Some of these choices follow from theoretical considerations (like choosing the relevant mechanisms and forces at play), some from the applications which are studied (with the expected range of validity and applicability), while others refer to practical issues (e.g., the tractability of a model in real-life situations or its desired efficiency). Thus, modeling requires to combine and balance widely different aspects often manifested as constraints to meet. In that sense, developing a model is akin to an art and it is henceforth referred to as the ``art of modeling''. 

The objective of this section is to introduce, or recall, the underlying principles which makes up the reading grid with which the various modeling approaches presented in Section~\ref{sec:models:approach} will be assessed in Section~\ref{sec:analysis}. To do so, we consider the following questions:
\begin{enumerate}[(1)]
 \item What is a model?
 \item How is a model conceived?
 \item How is a model used and evaluated?
\end{enumerate}

 \subsection{Defining a model}
 \label{sec:art_model:principle:definition}
 
Etymologically, the word model is derived from Latin \textit{modulus} meaning small measure. In the field of physical sciences, a model corresponds to ``a system of postulates, data, and inferences presented as a mathematical description of an entity or state of affairs''. By extension, it can also refer to ``a computer simulation based on such a system''\footnote{Merriam-Webster. (n.d.). Model. In Merriam-Webster.com dictionary. Retrieved in June, 2022, from \href{https://www.merriam-webster.com/dictionary/model}{https://www.merriam-webster.com/dictionary/model} (see sense 3 and etymology)}. This definition shows that there are several notions that need to be addressed separately: 
\begin{itemize}
 \item The model, which corresponds to a simplified representation used to explain the behavior of a physical system or the workings of a physical process. 
 \item Its implementation in a software, which involves the development/use of dedicated numerical algorithms to solve the set of equations defining the model (or provide accurate approximations). 
 \item The realization of one or several numerical simulations using the model implemented in a software together with the analysis of the results obtained.
\end{itemize}

In this article, we essentially focus our attention on the first notion, i.e., the conception of a model as a description aiming at reproducing some key features of a physical system of interest. 
The numerical implementation of such models is left out of the scope of the present paper because it relates more to the fields of computational physics, numerics, and computer sciences (interested readers are referred to publications dedicated to this topic, such as \cite{press2007numerical, prosperetti2009computational}). This does not mean that the development of accurate and efficient algorithms should be overlooked, as it has direct consequences on the precision of the results obtained and on their usefulness to the scientific community. 

 \subsection{Conceiving a model}
 \label{sec:art_model:principle:conception}
 
Modeling refers to the action of building models. However, beneath this deceitfully trivial definition lies a much richer and complex endeavor. For instance, Bonate mentioned in his introduction to the book entitled \textit{The art of modeling} \cite{bonate2011art}: ``there is more to modeling than the act of modeling. To be an effective modeler requires understanding the different types of models and when one type of model is more appropriate than another, how to select a model from a family of similar models, how to evaluate a model's goodness of fit, how to present modeling results both verbally and graphically to others, and to do all these tasks within an ethical framework''. 

\begin{figure}[ht]
 \centering
 \includegraphics[width=0.75\textwidth]{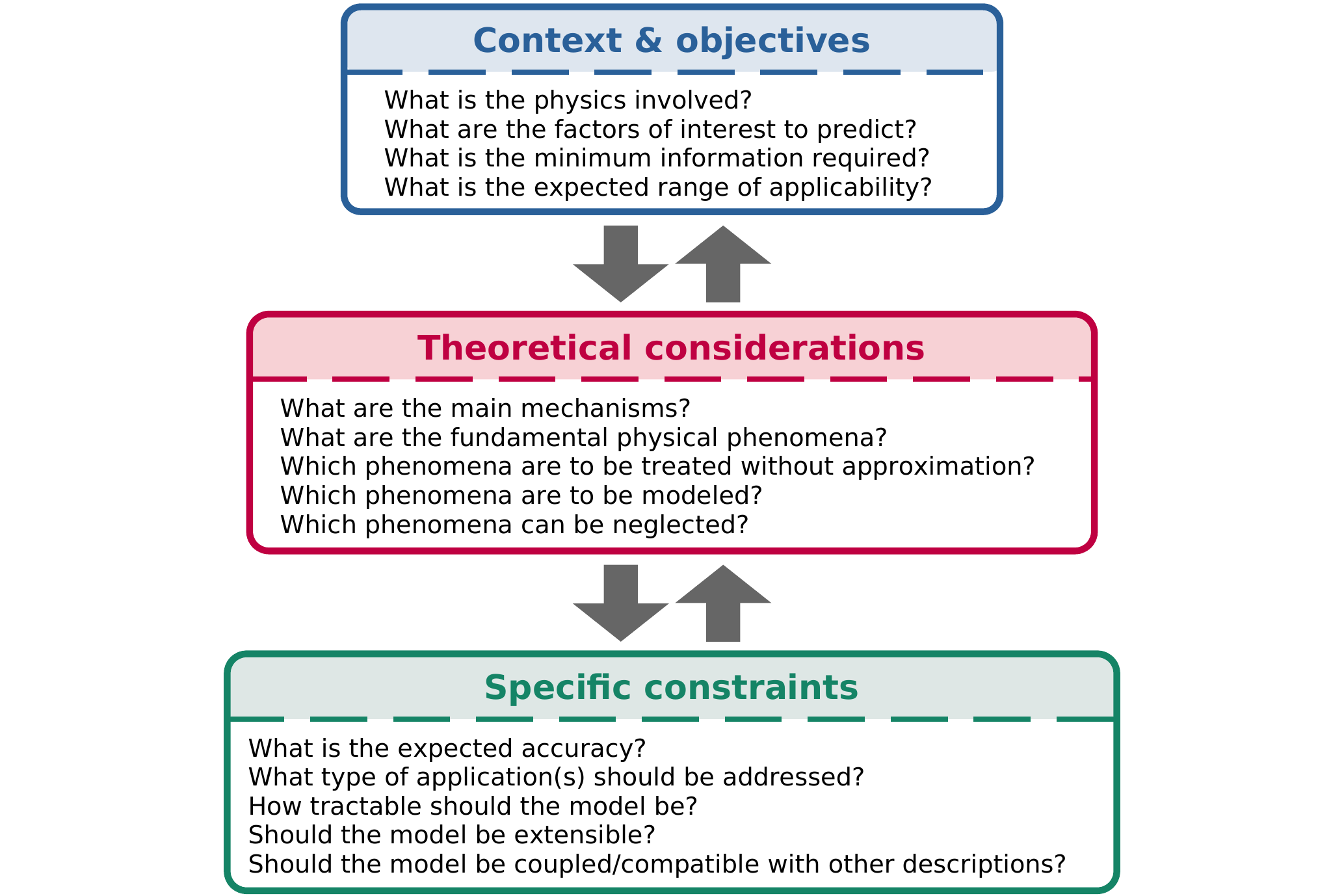}
 \caption{Sketch summarizing the various notions involved when conceiving a model including: aspects related to the context, theoretical considerations and specific constraints.}
 \label{fig:sketch_model_requirement}
\end{figure}

To put it differently, we suggest here that modeling requires to address the following aspects, which are indicated in Fig.~\ref{fig:sketch_model_requirement} (some of these notions are discussed in \citep[Section 3]{minier2016statistical}). 

\begin{itemize}
 \item The context and objectives of a study are one of the main factors influencing the choice of a model. Indeed, once the specific problem at hand is outlined, one has to start by identifying the quantities of interest that are relevant to fulfill the selected objectives. Given that it is unfortunately often overlooked, it is worth emphasizing that this step is an essential one in the construction of a model.

For instance, in the context of multiphase flows, one could be interested in various aspects, including: identifying where particles are coming from (especially to distinguish between natural and anthropogenic sources), characterizing these particles (particularly their nature and toxicity for living organisms), measuring their dynamics (with fluxes and rates of transport, deposition and/or resuspension), or examining their fate (such as the location where they accumulate in the environment or if they can be inhaled by humans). The choice of a quantity of interest is directly derived from such questions: chemical analysis of the particles is paramount to characterize their composition, while measurements of rates/fluxes is more relevant to investigate the dynamics of particles in a flow or their fate (e.g. by placing sensors at specific locations, such as the average human breathing height in medical environments). 
 
Once the factors of interest are determined, the minimum information required to derive these factors of interest needs to be clarified. For example, in the frame of statistical formulations of turbulent reactive single-phase flows, the objective is usually to obtain the average value of chemical source terms. As a result, the minimum information corresponds to the one-point one-time PDF of the concentration of scalars involved in the chemical reactions. This example is developed in \citep[Section 3.2]{minier2016statistical} in more details. What should be remembered is that the models thereby developed should provide at least this minimum amount of information. Of course, more complete models providing additional information are still viable (strictly speaking, this extra information is not required with respect to the objectives).
 
 Another item to specify is the range of applicability. For instance, a model designed to capture the transport of sand in the atmospheric boundary layer (where canopy effects are important) cannot usually be directly applied to reproduce the motion of plankton in the ocean. In fact, it requires several adaptations: first, the model has to be adjusted/customized to capture the key features of the fluid flow oceans (possibly with stratification effects as those described in \cite{boegman2009flow}); second, the model needs to be extended to treat the motion of plankton (which can be self-propelled and even display collective motion \cite{koch2011collective}). It is therefore important to define right from the outset whether a model is intended only for one specific application or for a range of situations.
 
 \item Having defined the context and objectives, the next step consists in analyzing the physics involved to identify the key mechanisms and the fundamental physical phenomena at play. For particle resuspension, these issues were already touched upon in Section~\ref{sec:physics:phenomenology}. As indicated in Fig.~\ref{fig:sketch_model_requirement}, this provides answers to the following questions: What are the essential phenomena? What are the phenomena that can be neglected? And what are the phenomena that can be approximately accounted for but without too many details? These answers can be used as guidelines for a model construction since we look for approaches that allow to treat the essential phenomena without approximation while still containing, or resolving, enough information to model those phenomena that can be approximated.

 For instance, if someone is interested in the motion of large suspended particles, it is reasonable to neglect Brownian motion while gravity must be properly accounted for. However, this means that the model cannot be directly applied to study the motion of small and light colloid particles for which Brownian motion is paramount while gravity can be neglected. Depending on which type of particles is considered, the same physical force is then regarded as either essential or negligible. Note, however, that if the objective is to treat any group of particles with sizes ranging from colloids to sediments, both Brownian motion and gravity become essential phenomena and, therefore, must be duly included (even though they can have little impact for a specific subsets of particles). This is in line with the remark above on the chosen range of applicability of the model to develop/select. In a similar situation, if we consider particles embedded and carried by turbulent flows (think about aerosols emitted from a chimney), particle transport is obviously a key phenomenon. If, furthermore, we wish to handle without loss of accuracy any set of particles having different diameters or even chemical properties, then this clearly tips the scales in favor of particle tracking methods. More detailed discussions can be found in \cite{minier2001pdf,minier2016statistical} on distinctions between slow variables representing degrees of freedom that need to be treated without approximation and fast variables representing other degrees of freedom whose effects can be mimicked with stochastic models. 
 
 \item The conception of a model can be further restrained by specific constraints, such as those related to the model precision, flexibility, tractability, and/or modularity. More specifically, let us consider again the case of particle transport by turbulent flows for which, as pointed out above, particle tracking approaches are attractive candidates. Yet, depending on the level of precision required, the choice of the expressions for the hydrodynamic forces and surface forces can vary. For instance, if particles are much larger than the Kolmogorov scale, one can choose between very fine computations of hydrodynamic forces (based on PR-DNS simulations) or approximate expressions (using only the fluid velocity from PP-DNS methods, see Section~\ref{sec:models:turb_rough:turb}). Of course, this has consequences in terms of the level of information contained in each formulation. Yet, it also implies that the computational costs associated with these two models are not comparable since the level of information is not the same. If the first approach is chosen, this limits ourselves to academic situations involving simple geometries and low Reynolds-number flows whereas the second one is more tractable and applicable to a much wider range of practical and real-life situations (see a similar discussion in \citep[section 2.4]{minier2016statistical}). 

 A complementary issue is related to the extensibility of the model, which concerns how easily a model can be modified to account for additional phenomena and mechanisms. The issue is to ascertain whether a chosen formulation is able to accommodate future developments when more complex physics is considered. For instance, we may wish to have formulations that can be extended in a straightforward manner by adding extra variables attached to each particle (representing additional degrees of freedom, such as rotation or orientation, etc.) without having to revisit the whole construction when we go, for example, from spherical particles to particles with more complex shapes. This is indeed one of the main issue to be addressed later on in Section~\ref{sec:next_model}. 

\end{itemize}

In short, it is seen that modeling requires a deep understanding of the issues at stake, the specific objectives, as well as the limitations behind current fundamentals/theories. To rephrase Bonate, one could say that a modeler has to contextualize all the notions required. This contextualization is a long-lasting process which requires patience and efforts to avoid over-simplifying or overlooking some of the information. 

 \subsection{Evaluating a model}
 \label{sec:art_model:principle:evaluation}
 
\begin{figure}[ht]
 \centering
 \includegraphics[width=0.75\textwidth]{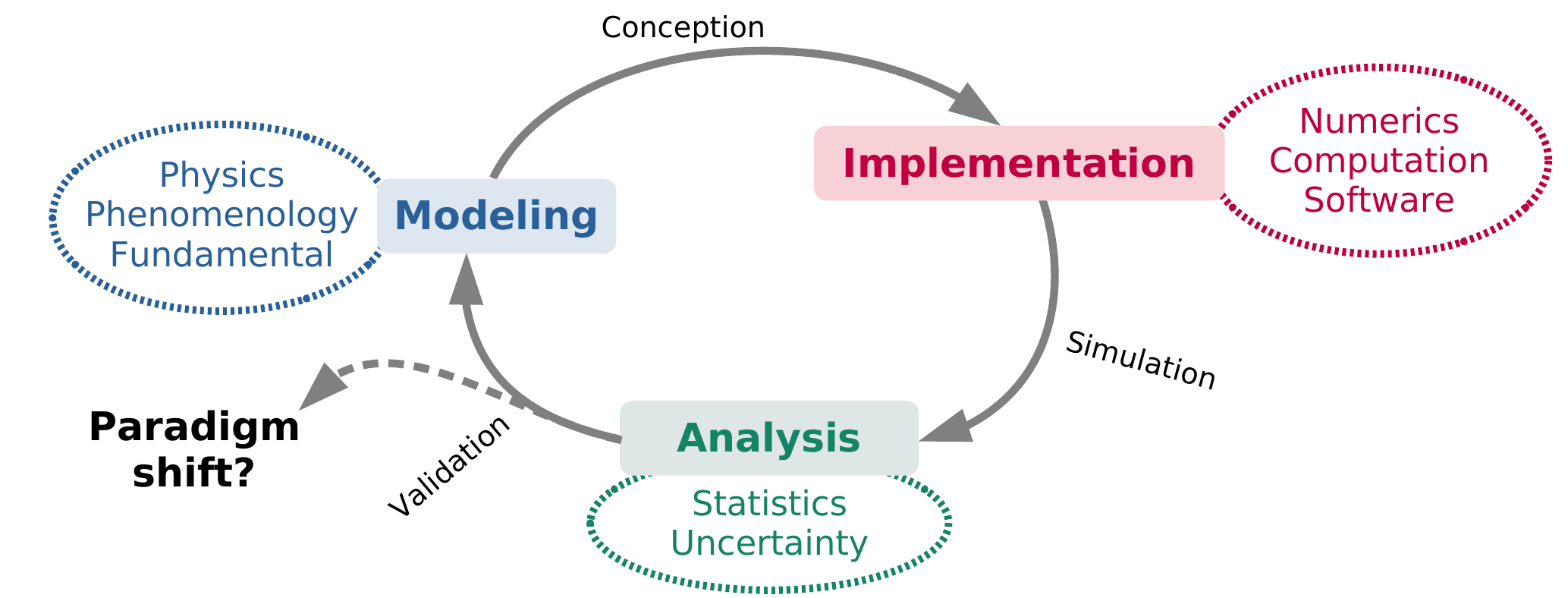}
 \caption{Illustration of the life cycle of scientific models: models constantly evolve as they are further tested in new configurations/applications and, on singular occasions, a sudden shift in the paradigm can lead to the development of a whole new field/category of models.}
 \label{fig:sketch_model_cycle}
\end{figure}

Once a model is formulated, its accuracy has to be assessed: this corresponds to the so-called validation phase of a model. For that purpose, the model is usually implemented in a software. Then, either analytic formulas derived from this model or numerical simulations obtained by running the corresponding software are compared with available experimental data. This three-step process is illustrated in Fig.~\ref{fig:sketch_model_cycle}, which shows that there is actually a cycle between this triptych (modeling, implementation, analysis). In fact, comparison of a model predictions to new experimental data can bring out that some features are not properly captured. This can be related either to new properties that have been unveiled by more recent observations or to properties already known but that were previously deemed unimportant. Regardless of the origin of such discrepancies between observations and model predictions, the result of this cycle is that models are not frozen objects but are constantly evolving. 

For instance, in the community of particle-laden flows, accurate predictions of the transport and deposition of particles have been a long-standing topic \cite{henry2012towards}. Yet, even a cursory look at the evolution of these models reveals how they have been continuously enriched. In fact, numerical simulations first took into account particle transport by a flow (often turbulent) while particle-wall interactions were considerably simplified or downright neglected. Indeed, particles reaching the wall were often considered either as destroyed (numerically speaking) or as irreversibly deposited on the surface \cite{henry2012towards}. This gave first practical evaluations of the amount of particles deposited in a pipe as a function of time. Later, new experimental measurements provided more precise information on the morphology of the multilayer deposits thus formed, as well as on their effect on the flow (up to the complete clogging of the pipe). In order to reproduce these phenomena, questions related to the modeling of particle-wall interactions quickly came up since they play a key role in the morphology and cohesion of the deposits formed. Deposition models were therefore extended with new descriptions taking into account these particle-wall (and similarly particle-particle) interactions. At the same time, progress in the field of surface science (especially with AFM techniques) highlighted the role played by surface roughness on inter-surface forces. As a result, models for surface forces were also further enriched to account for the role of surface roughness.

 \subsection{Summary} 

In this section, we have introduced the ``art of modeling'', which corresponds to a set of guidelines to rely on when establishing a model. The list of criteria used in this analysis does not pretend to be the only one, since additional criteria could be selected and their ordering changed. We believe, however, that these criteria already constitute a detailed enough framework allowing for a comprehensive investigation. Applied to the resuspension models, they turn out to be helpful in revealing the complexity of the models themselves and how different physical notions are deeply entwined (this is detailed in the enxt Section~\ref{sec:analysis}).

These guidelines are intended to help scientists and engineers alike in developing future models or in choosing the appropriate model with respect to the context of their study. This is indeed a task that should not be overlooked, especially due to the richness of the phenomena at play in particle resuspension, which involve various disciplines and a range of spatio-temporal scales. As it follows from the previous discussions, the choice of a model requires to weigh the desired range of validity and applicability of the model as well as how accurately the key phenomena should be described (e.g., mechanisms and forces at play). Therefore, as also claimed by Ancey \cite{ancey2020bedload2}, modeling ``does not rely solely on explicit knowledge - that is a collection of equations, rules, diagrams - but also on the implicit knowledge gained through personal experience and interaction with highly practiced peers''.

\section{Analyzing resuspension models}
 \label{sec:analysis}

Drawing on the guiding principles presented in Section~\ref{sec:art_model}, we propose now to shed new light on existing resuspension models. This requires to adopt a leading thread. Indeed, the steps sketched in Fig.~\ref{fig:sketch_model_requirement} should not be regarded as marking a one-way street leading from one context to the details of the models. Most of the time, the construction of a model implies to address the various aspects discussed in Section~\ref{sec:art_model:principle:conception} in a different order than the one sketched in Fig.~\ref{fig:sketch_model_requirement}. 

\begin{figure}[ht]
 \centering
 \includegraphics[width=0.85\textwidth]{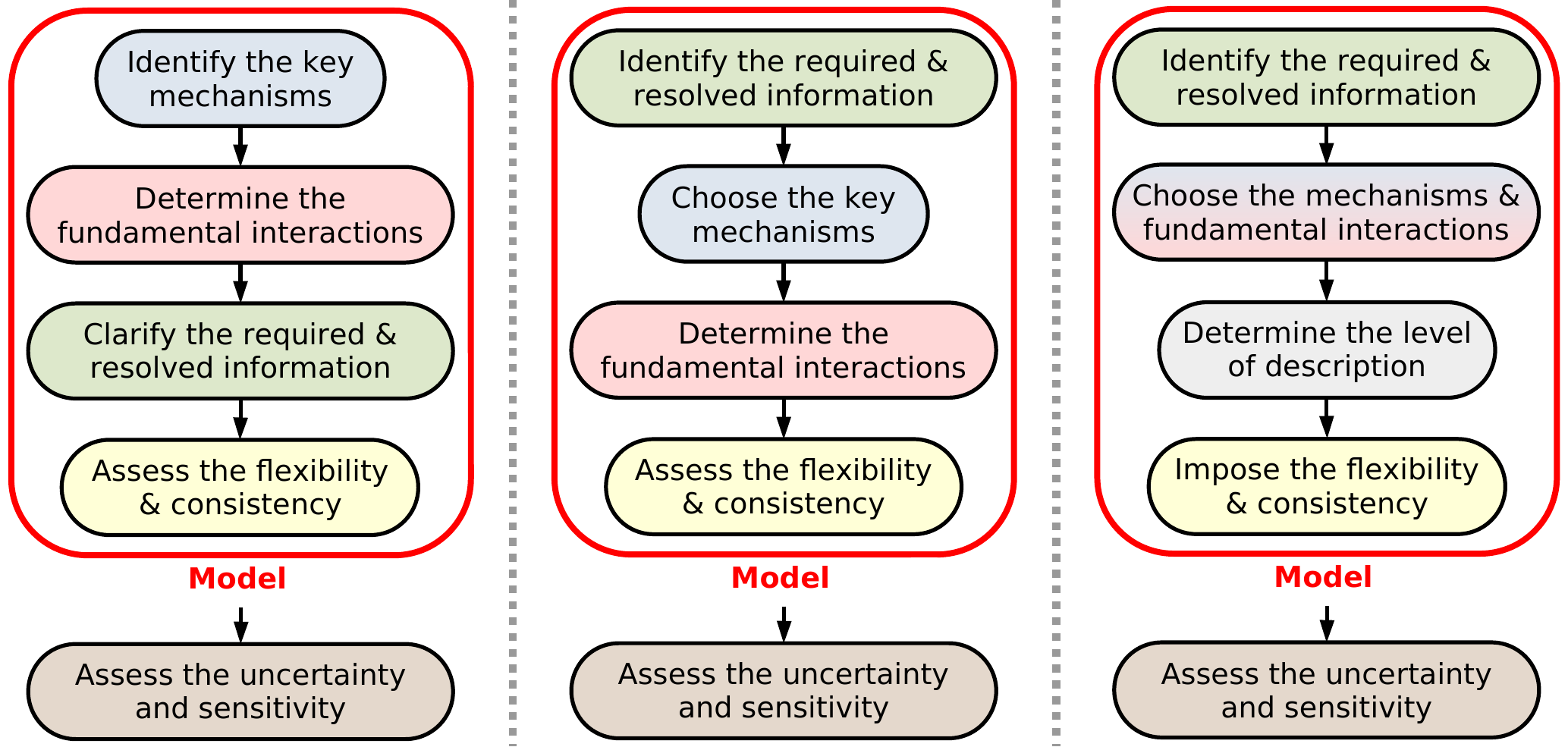}
 \caption{Illustration of 3 different reading grids to analyze resuspension models. The grids differ mostly in the order through which the criteria are treated (with slight variations in the content of each criterion). The left one is used throughout Section~\ref{sec:analysis}.}
 \label{fig:sketch_criteria_intro}
\end{figure}

To avoid a cumbersome presentation with too many entries, we choose here to concentrate on the physics involved (the mechanisms, fundamental interactions, forces, etc.) and to introduce considerations about various contexts in the course of the discussion. This can lead to a variety of reading grids depicted in Fig.~\ref{fig:sketch_criteria_intro}: they differ by the order of the criteria used to assess modeling approaches. In the following, we rely on the one displayed on the left, with the different criteria ordered as:

\begin{enumerate}[6.1 -]
 \item Identifying, selecting and modeling the relevant mechanisms and phenomena considered;
 
 \item Determining the fundamental interactions at play and choosing how they are described;
 
 \item Clarifying the required and resolved information (including its physical meaning and level of description);
 
 \item Determining if the model is coupled to other approaches, and assessing the consistency / coherency between these descriptions (including the level of information);
 
 \item Evaluating the uncertainty and sensitivity of the outputs, thereby claryfying the range of applicability and precision of the model.
\end{enumerate}

 \subsection{Accounting for resuspension mechanisms}
  \label{sec:analysis:mechanisms}

The first criterion is related to the identification and selection of the relevant mechanisms. In particular, the objective is to answer the following questions: What are the main physical mechanisms? Which ones should be treated without approximation? Which ones can be approximated or even neglected?

\subparagraph{Selection of the resuspension scenario} Various mechanisms have been highlighted in Section~\ref{sec:physics:phenomenology:mechanisms}, including: rolling, sliding and lifting motion as well as collision-induced events. These mechanisms are not necessarily exclusive since a particle can move through a combination of several of them (see Fig.~\ref{fig:fig_kassab_2013_mechanisms}). 

It is then tempting to sort existing models with respect to the mechanisms that are accounted for. However, this would not reflect the assumptions that have motivated such choices. For instance, it appears reasonable to neglect inter-particle collisions in the case of sparse monolayer deposits, i.e., with only a few particles on a large surface.
Yet, this assumption is erroneous when applied to dense monolayer deposits (i.e., where the inter-particle distance is smaller or equal to the particle size) or to multilayer deposits. Hence, it is not surprising to find out that models established in the context of fluvial and aeolian transport account for inter-particle collisions \cite{pahtz2020physics} (e.g., impulse criterion and some system-scale models) since they are by essence designed to be applied to complex multilayer beds. Meanwhile, many models developed in the multiphase flow community aimed at reproducing the measured resuspension from sparse monolayer deposits. As a result, these models often neglect inter-particle collisions while focusing on rolling/sliding/lifting motion. Yet, these models often focus on a single mechanism: some account only for rolling motion (e.g., the dynamic PDF approach \cite{guingo2008new} or the Rock'n'Roll model \cite{reeks2001kinetic}), others only for lifting motion (like the RRH model \cite{reeks1988resuspension}), or even a combination of rolling/sliding/lifting (as in force- and torque-balance approaches \cite{ibrahim2003microparticle}). Another noteworthy consequence of neglecting inter-particle collisions is that it allows to treat resuspension events as independent from one another. 

\begin{figure}[ht]
 \centering
 \includegraphics[width=0.85\textwidth]{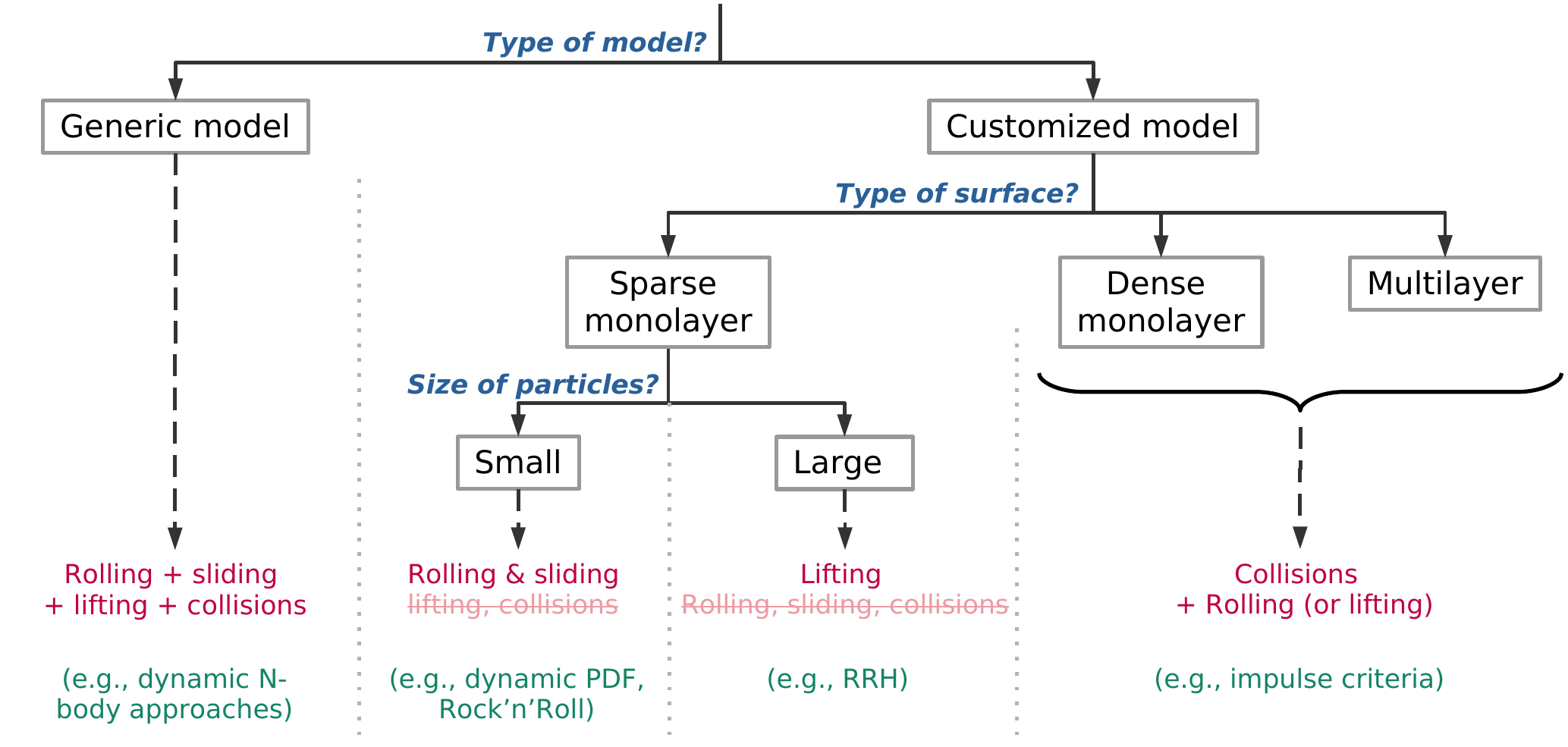}
 \caption{Sketch summarizing the various notions involved when choosing the relevant mechanisms at play in a model for particle resuspension. They are sorted according to the type of model (generic or customized), the type of surface (sparse/dense monolayer or multilayer) and the type of particles. The consequences in terms of the key mechanisms at play are highlighted (red) together with the possible models that can be used (dark green).}
 \label{fig:sketch_criteria_mechanisms}
\end{figure}

From this discussion (see Fig.~\ref{fig:sketch_criteria_mechanisms}), it appears that the main issue is whether the model should be universal or specialized/customized for a given task/application. This choice is not trivial and has many consequences in terms of the range of applicability and validity of a given model. For instance, general-purpose resuspension models present the advantage of working for any size of particles and different type of deposits (i.e., monolayer or multilayer). This is the case for particle tracking formulations in general, and especially for the recent dynamic N-body approach since it reproduces any type of motion for spherical particles within a complex sediment bed \cite{vowinckel2016entrainment}. On the contrary, customized models are not necessarily applicable to other context/applications. This is obvious for empirical formulations, which can be viewed as ultimate customized models: empirical approaches can indeed only be used in a context that is representative of the data used to generate the empirical equation. 

\subparagraph{Description of the mechanisms} Having chosen the mechanisms to be included in the model, the next step consists in selecting the appropriate level of description. For the sake of clarity, let us consider the case where a customized model for the rolling motion of colloidal particles is developed. One has now to decide how to treat rolling motion in the model. In line with the discussions in Section~\ref{sec:art_model}, there are three options. First, rolling can be treated without approximation by resorting to a dynamic model, which explicitly tracks the motion of particles as they roll on the surface (like in the dynamic PDF approach \cite{guingo2008new}). Second, rolling can be modeled to capture only its effect on the resuspension by resorting, for example, to impulse criterion \cite{valyrakis2013entrainment} or to kinetic PDF approaches \cite{reeks2001kinetic}. Third, rolling can be simplified to a rupture of balance for the incipient motion, as done in static torque-balance approaches. 

\subparagraph{Modularity of a model} Another noteworthy aspect is the flexibility of a model. In the context of the mechanisms at play in particle resuspension, a model is considered flexible if the mechanisms can be changed without having to rework on all the foundations of the model. In practice, adding/removing/changing one of the mechanisms considered is a relatively straightforward operation when using Lagrangian tracking formulations (either dynamic ones or static ones like force- and torque-balance models). It basically amounts to adding/removing/changing one of the equation of motion of the particles. It can turn out to be a little bit more complicated when using integrated kinetic approaches, since a specific formula is obtained considering a set of predefined mechanisms. Hence, a new formula can still be obtained when the set of mechanisms is changed, but it has to be rederived (if it has not yet been done). Similarly, formulas obtained with system-scale approaches can also be extended to account for a different set of mechanisms since they are based on force- and torque-balance approaches. Empirical models are the only category that does not allow any flexibility on the mechanisms involved since each expression is fully determined by the data set and the corresponding conditions.

 \subsection{Fundamental interactions at play}
  \label{sec:analysis:fund_inter}

The second criterion is related to the selection of the relevant interactions. More precisely, the aim is to provide answers to the following questions: What are the fundamental interactions in particle resuspension? Which ones should be treated explicitly? Which ones are to be modeled? Which ones can be neglected?
 
To address these issues, we shed light on the reasoning that drives the choice of the relevant fundamental interactions and on the subsequent selection of an appropriate description for each force.

\subparagraph{Selection of the fundamental interactions} Various forces can play a role in resuspension. As detailed previously in Section~\ref{sec:models:forces}, this includes contributions from: interactions with external fields (like gravity), hydrodynamic forces (due to particle-fluid interactions), contact and non-contact forces (related to particle-surface or particle-particle forces).

As for the resuspension mechanisms, the issue is whether one aims at obtaining a general-purpose or a customized model. To be more specific, let us imagine that we wish to develop a general-purpose model for the resuspension of hard spherical particles with arbitrary properties (size, density, and composition), for any type of deposit (i.e., monolayer or multilayer), and exposed to any kind of flow (such as gases or fluids, laminar or turbulent). In that case, it is natural to take into account most (if not all) of the forces listed in Section~\ref{sec:models:forces}. On the contrary, customized models allow to neglect some of the forces. For example, in the case of sparse monolayer deposits, inter-particle collisions effects as well as particle-particle interactions can be neglected compared to hydrodynamic effects. Hence, when choosing whether to include a given force within the equations of motion, the issue is related to the range of applicability of the model. Drawing on this notion, we propose to review in the following the motivations that allow to neglect some of the forces, instead of comparing each model separately. 
\begin{itemize}
 \item The most common external force included in resuspension models is gravity (since particles studied are usually deposited on surfaces within a given planet). The case of electromagnetic forces is easier to discern, since these forces are typically included in a model only when particles having electro-magnetic properties are actually exposed to an electromagnetic field.
 
 \item Among the various hydrodynamic forces, the most important ones are the drag and lift forces. These forces are always occurring for particles deposited on a surface due to the strong shear flow in the vicinity of the surface. Yet, it is possible to neglect one of the two forces depending on the case studied. In fact, models for monolayer deposits are often customized to capture either the resuspension of small colloidal particles through rolling motion (which is induced by drag forces) \cite{henry2018colloidal} or of large sediments through direct lift-off (with negligible drag forces) \cite{reeks1988resuspension}. 
 Apart from these specific cases, other hydrodynamic forces can also enter the particle equation of motion, such as: the added-mass force (which is negligible when particles are much heavier than the fluid) and Basset history forces (which are often neglected since precise expressions are rather complex). As already discussed, Brownian motion can also be included when studying the motion of small colloidal particles.
 
 \item Surface forces are often simplified to adhesive and friction forces. Indeed, other surface forces only occur in specific situations, making it relatively easy to know if these forces should be accounted for. To be more specific, capillary forces are expected to occur when dealing with hydrophilic particles exposed to an airflow with a high relative humidity and for low surface roughness. An exception may be when dealing with organic particles, like spores, whose outer layer can change properties in humid environments. Hence, most situations are handled with only two major contributions: adhesion forces and friction forces. Since friction is a force that resists the translational motion between two surfaces in contact, it has to be included when sliding is a dominant mechanism (and can be neglected if sliding is negligible). Meanwhile, adhesive forces are usually preventing the motion of small particles (like colloids) and have to be included in such cases. However, when dealing with large and heavy particles (such as gravels in rivers \cite{valyrakis2013entrainment}), adhesion forces can be neglected since they are several orders of magnitude smaller than gravity.
 
\end{itemize}

At this point, it is also worth noting that the previous discussion applies to all resuspension models. While this seems obvious for dynamic models (which include a number of forces in the equation of particle motion), it remains valid for empirical formulas too. In fact, empirical formulas can be adapted to describe the evolution of a quantity of interest as a function of the forces involved (similarly to system-scale models, see \cite{dey2018advances}). This would lead to empirical expressions of the form: $\tau_r^{\Delta t} = \mathcal{G}(F_{\rm hydro}, F_{\rm surf}, F_{\rm ext})$. Yet, such empirical formulas would be more intricate to use since they would require information on the forces rather than (or complementary to) information on the fluid/particle/surface properties. Hence, in terms of applicability, it is harder to keep track of the range of validity of such formulas. A more complete discussion on the required/resolved information from each model will be given later in Section~\ref{sec:analysis:information}.

\begin{figure}[ht]
 \centering
 \includegraphics[width=0.95\textwidth]{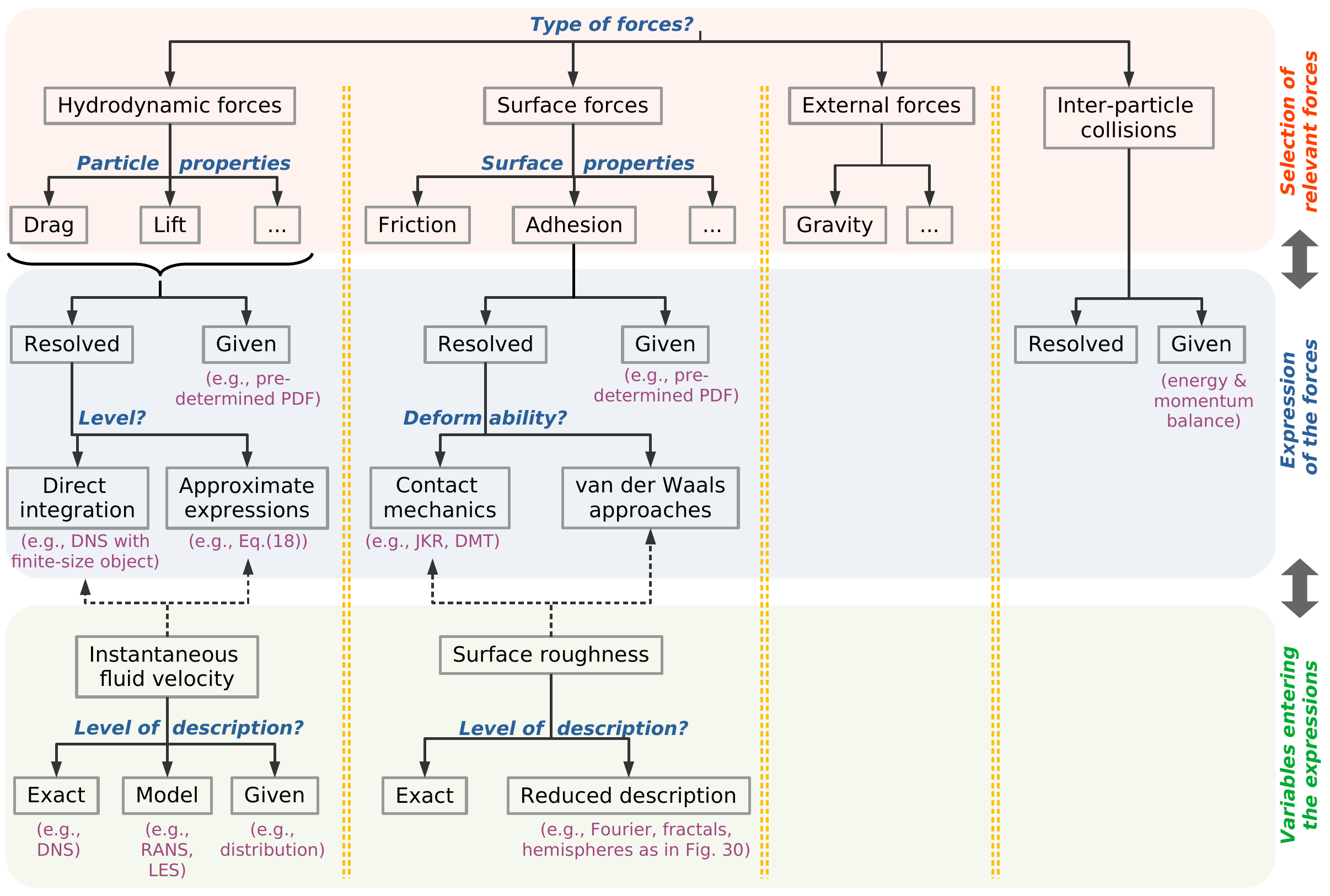}
 \caption{Sketch summarizing the various notions involved when choosing the fundamental forces in models for particle resuspension. It is presented in terms of the various forces involved (hydrodynamic, surface, external, collisions) together with the expressions of these forces and the information required on additional variables that enter these expressions (when relevant).}
 \label{fig:sketch_criteria_force}
\end{figure}

\subparagraph{Description of the selected interactions} Once the relevant fundamental interactions are selected, one has to decide how to treat each of these forces in the model. In practice, two choices can be made: a force can be treated without approximation or it can be modeled. In the following, we illustrate these issues for the hydrodynamic and surface forces, which pose the greatest modeling challenge (see also Fig.~\ref{fig:sketch_criteria_force}).
\begin{itemize}
 \item Since hydrodynamic drag and lift forces are proportional to the local instantaneous fluid velocity around each particle, they fluctuate due to near-wall turbulence (see \cite{frisch1995turbulence, pope2000turbulent}). As a result, models for drag and lift forces involve two questions: how is the force evaluated and how is the relative velocity estimated? As illustrated in Fig.~\ref{fig:sketch_criteria_force}, hydrodynamic forces can be obtained with two approaches. First, simple models consider that the distribution of hydrodynamic forces is predetermined right from the onset (e.g., using Gaussian distributions as in kinetic PDF approaches \cite{reeks2001kinetic}). The second approach relies on the use of dedicated models to obtain the distribution of hydrodynamic forces as a result. This includes both fine-scale descriptions (based on surface integration) or approximate expressions (see Section~\ref{sec:models:forces:hydro}). Yet, when approximate expressions are chosen, information on the instantaneous fluid velocity at the particle position is needed. As described in Section~\ref{sec:models:turb_rough:turb}, this instantaneous flow velocity can be accessed using three methods: it can be computed directly (using DNS), it can be approximated using turbulence models (such as RANS/PDF models), or one can consider a predefined distribution of these velocities. 
 
 \item Similarly, adhesive forces can fluctuate very rapidly due to surface roughness (see Section~\ref{sec:models:turb_rough:rough}). Information on adhesion forces can be obtained by solving numerically specific models or by assuming predetermined distributions (see also Fig.~\ref{fig:sketch_criteria_force}): 
 \begin{itemize}
  \item There are two main difficulties when trying to come up with a model for surface forces between rough surfaces: (1) how to compute the forces? and, (2) how to account for the presence of roughness? As mentioned in the analysis of existing models for contact forces (see Section~\ref{sec:models:forces:contact}), these forces can be evaluated either with contact mechanics theories (that include deformation effects) or with a Hamaker approach for van der Waals forces (i.e., neglecting deformations). Hence, the choice between these two descriptions of surface forces is driven by the particle properties (here, their ability to deform). Yet, rough surfaces are very chaotic in nature and they display a range of roughness sizes and non-trivial distributions on the surface (see Fig.~\ref{fig:surf_roughness}). Nevertheless, methods for the precise calculation of the force between rough surfaces are available when information on the exact surface topology is accessible. 
  When models are expected to remain tractable in complex realistic situations, roughness is approximated with known functions (e.g. resorting to Fourier transforms, fractal representations, or simple hemispherical asperities placed on a smooth plane, see Fig.~\ref{fig:sketch_represent_roughness}). This choice results therefore from a careful balance between the precision required, the minimum amount of information required/available, and the computational costs associated to the method. 
  \item Models can be even further simplified by imposing the distribution of adhesion forces right from the onset (as in kinetic PDF approaches \cite{reeks2001kinetic}). The interest of such models is that they do not require any sort of complex computation. But one of their main drawbacks is that information on adhesion force distributions has to be accessible and provided beforehand, either from experimental measurements or fine numerical simulations. This is an important point, which questions the universal nature of such formulations (see the discussion in Section~\ref{sec:analysis:information}).
 \end{itemize}
\end{itemize}

\subparagraph{Flexibility of the model} In the context of fundamental interactions, model flexibility is related to inter-operability and modularity. Inter-operability means that a force can be replaced by a similar force without having to rework on all the foundations of the model. Modularity means that the force can be extended by another formulation (including possibly extra effects). As for the mechanisms, model flexibility is inherent to particle tracking approaches, which allows to change not only any of the fundamental interactions entering the equations of motion but also their expressions. Meanwhile, integrated kinetic approaches can be adapted to account for additional forces but also for different expressions of the forces entering these equations. 
The only category of models that suffer from a lack of flexibility is the empirical models (since they rely on a predetermined set of data).

 \subsection{Nature of resolved/required information}
  \label{sec:analysis:information}

The third criterion consists in clarifying the required and resolved information within each model (including its physical meaning and its level of information). More precisely, this step aims at answering the following questions: What are the input variables required by a model? What are the quantities expected to be predicted by a model? What is the physical nature of these variables? What is the minimum level of information required for these variables?

\begin{figure}[ht]
 \centering
 \includegraphics[width=0.85\textwidth]{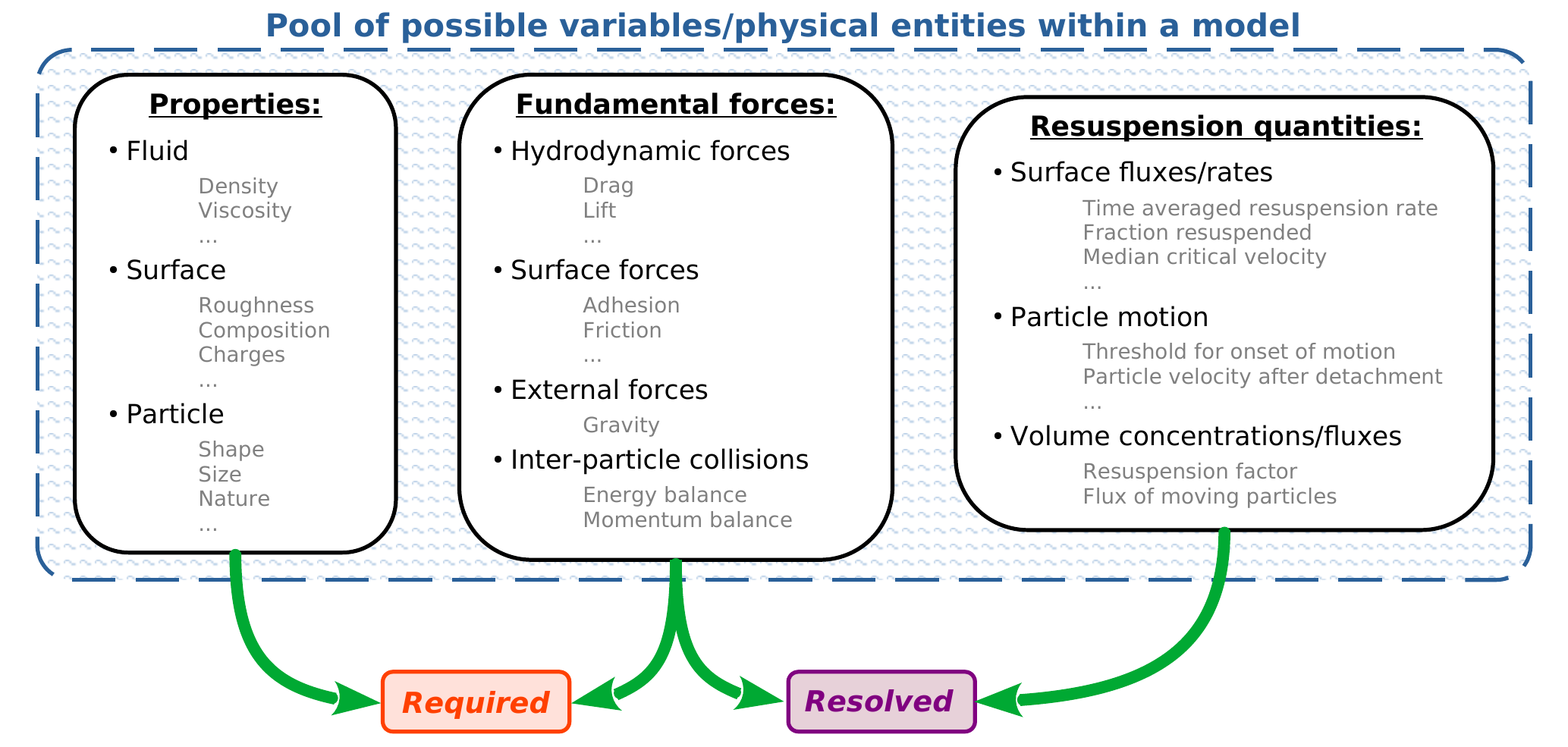}
 \caption{Illustration of how the various variables can be used either as required information or resolved information. These variables/physical entities are sorted according to three categories (properties related to fluid/particle/surface, fundamental forces or resuspension quantities).}
 \label{fig:sketch_criteria_info}
\end{figure}

By the term ``variable'', we mean here a number of possible physical entities that enter the formulation of a model. As displayed in Fig.~\ref{fig:sketch_criteria_info}, this covers three types of variables: the properties related to fluid/surface/particle (e.g., the fluid density or the surface roughness), the values of the fundamental forces at play (like the drag force acting on a particle), and the quantities describing particle resuspension (such as the fraction remaining).

As illustrated in Fig.~\ref{fig:sketch_criteria_info}, all these variables can be viewed as forming a sort of ``pool of possible variables'' and, depending on the context of the study, one may decide which are the required/resolved quantities. Here, we wish to emphasize that the choices made during the first two steps (namely the mechanisms and the fundamental forces) have profound consequences on the information predicted by a model and the information required to use it. To be more specific, let us consider the case of a dynamic model coupled to a PR-DNS (as in \cite{vowinckel2016entrainment}). The use of an individual particle tracking approach implies that detailed information can be obtained about resuspension quantities (such as velocities of individual particles, surface fluxes/rates like the Shields parameter or even volume concentrations). In addition, such a model yields predictions on the forces at play, especially on hydrodynamic drag and lift forces which are here explicitly treated (note that no information is available on adhesion forces since they are neglected for such large particles but they can easily be included if the model is extended to treat small particles). Meanwhile, this model requires detailed information on particle properties (e.g., shape and size) and on fluid properties (density, viscosity, velocity). 

As it transpires from Fig.~\ref{fig:sketch_criteria_info}, values of the fundamental interactions at play can also appear as required information in other models. One example is the kinetic PDF approaches (like the Rock'n'Roll model \cite{reeks2001kinetic, you2017statistical}), which computes information on resuspension rates using predetermined distributions for adhesion and hydrodynamic forces. In practice, the RnR model usually relies on log-normal and Gaussian distributions for adhesion and hydrodynamic forces \cite{reeks2001kinetic}, respectively. More recently, the model was extended to account for other distributions (like a Weibull distribution for adhesion forces, as in \cite{you2017statistical}). 
However, recent measurements have shown that adhesion forces can display complex multimodal distributions when using realistic outdoor surfaces contaminated by dirt instead of clean laboratory surfaces \cite{brambilla2018glass}. Not only can the distribution of adhesion forces take complex shapes, but it is also highly dependent upon the characteristics of each surface. In other words, unless we limit ourselves to the same well-characterized surfaces, the distribution of adhesion forces cannot be regarded as following a universal form. This means that, if we expect to apply the same model to any kind of particle-surface situations, adhesion forces should be resolved by the chosen modeling approach. In that sense, a universal formulation based on kinetic PDF can only be obtained provided that the distribution of adhesion forces are computed (and not prescribed). In fact, the same remark holds for hydrodynamic forces in kinetic PDF approaches.

Another important notion is related to the level of information contained in a model inputs/outputs. In fact, the level of information contained in the required variables has consequences on the level of information which can be resolved for the predicted quantities. This is especially the case when treating the adhesion forces between rough surfaces (provided that these are relevant interactions). If surface roughness is described using reduced statistical information (like the rms roughness size and the average peak-to-peak distance), only statistical information on the adhesion forces can be predicted by the model. Hence, unlike with detailed representations of surface roughness, the adhesive force of an individual particle at a precise location over the surface cannot be accessed. While this seems obvious for the computation of adhesion forces, it can be more ambiguous in the overall resuspension models (which involves descriptions of roughness, hydrodynamic forces, etc.). For instance, even if hydrodynamic forces are treated without approximation, the use of such a statistical approach for roughness implies that only statistical information on resuspension quantities can be predicted. This aspect resurfaces in Section~\ref{sec:analysis:consistency} about consistency issues. 

 \subsection{Consistency issues}
 \label{sec:analysis:consistency}
 
The fourth criterion consists in assessing the consistency of the overall resuspension model, especially when it is composed of multiple sub-models (e.g., with description of hydrodynamic forces and the corresponding turbulent flow together with description of adhesion forces and the corresponding surface roughness). In particular, the objective is to evaluate the coherency and compatibility of the expressions entering the model with respect to the following issues: What is the application domain considered? What is the expected range of applicability? What is the minimum level of information expected/required?

Due to the multidisciplinary and multiscale aspects involved in particle resuspension, it is hard to keep track of all the assumptions involved in each formulation. However, checking consistency between these formulations is paramount in order to assess the range of validity and applicability of a model. In fact, the precision of a model is fully determined by the minimum level of information contained across all the variables entering a model (see the previous section). We illustrate these issues with a few examples:
\begin{itemize}
 \item First, resuspension models can be inconsistent in terms of the descriptions of the variables entering the model. For instance, let us consider the case of a model based on a direct calculation of the adhesion force using information on the exact surface topology. We further assume that this formulation is combined with approximate expressions for hydrodynamic forces (like Eq.~\eqref{eq:Fdrag}) where the fluid velocity is not fully resolved in the near-wall region (e.g., through RANS simulations based on so-called wall functions or, in other words, with no down-to-the-wall integration). This would lead to a discrepancy between the level of description of the adhesion and hydrodynamic forces. In fact, such a model would amount to solving exactly physical aspects at the nanoscopic scale (where adhesion forces matter) while simplifying other effects acting at the hydrodynamic scale (i.e., over millimeter-size scales). Hence, the error made at the hydrodynamic scale would completely offset the precision of the methods for adhesion forces. Strictly speaking, such adhesion models would only be consistent with fully-resolved information on the near-wall instantaneous fluid velocity (e.g., using PR-DNS as in \cite{vowinckel2016entrainment}). 
 
 While such inconsistencies are relatively obvious since adhesion forces act at scales much smaller than hydrodynamic forces, the opposite is not necessarily straightforward. In fact, due to the wide separation of scales, it is tempting to simplify or even neglect adhesion forces compared to hydrodynamic forces. This would be reasonable when studying large suspended particles for which adhesion forces are much smaller than the gravity force. However, when studying smaller particles, it is preferable to have a coherent level of information for both forces, as well as for their corresponding required physical entities (near-wall fluid velocity and surface roughness). Such issues have been largely explored in the context of particle transport and deposition \cite{henry2012towards}. 
 
  \begin{figure}[ht]
  \centering
  \includegraphics[width = 0.9\textwidth]{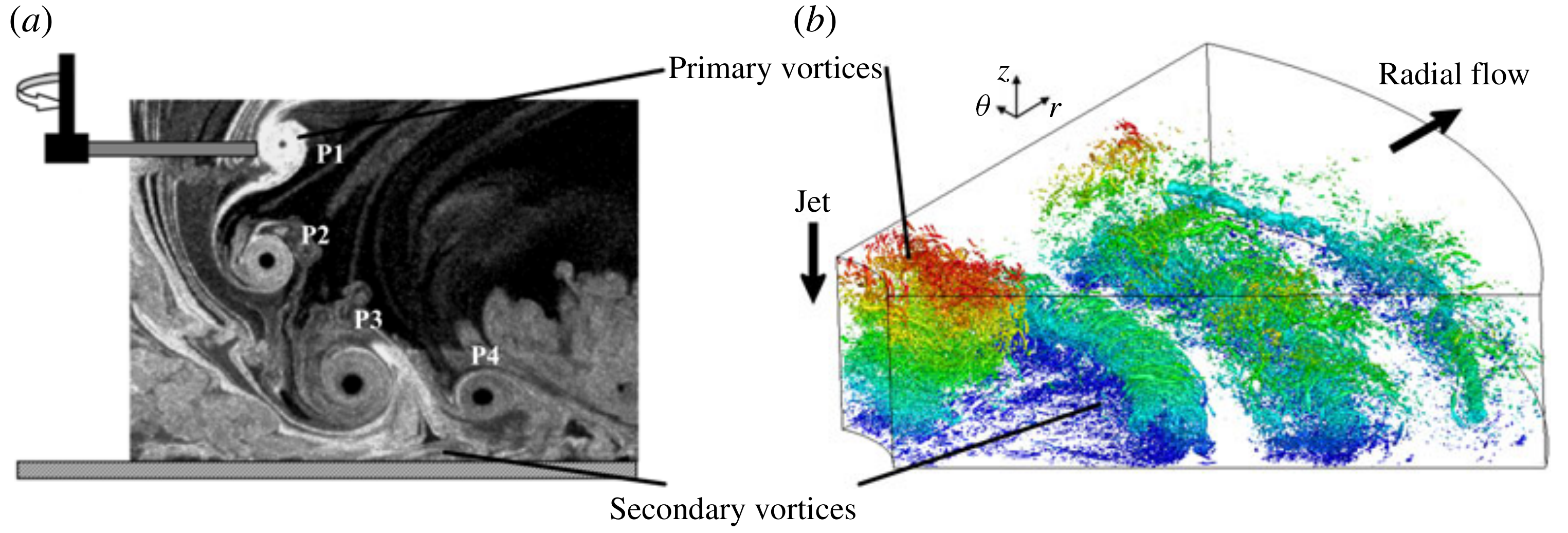}
  \caption{(a) Visualization of the wake of a rotor in ground effect. Reprinted with permission from \cite{lee2010fluid}. Copyright 2010, Vertical Flight Society.
  (b) Instantaneous flow structures in an excited, round impinging jet with embedded azimuthal vortices. Isosurfaces of the second invariant of the velocity-gradient tensor are colored according to their distance from the bottom surface. This shows the complex flow induced by a rotor which directly affects re-entrainment of detached particles in the near-wall region. Reprinted with permission from \cite{wu2017particle}. Copyright 2017, Cambridge University Press.}
  \label{fig:fig_wu_2017_rotor}
 \end{figure}

 \item Second, resuspension models are often coupled to other approaches to compute the fluid and particle dynamics.
 For instance, it would be consistent and coherent to couple a dynamic resuspension model to a PR-DNS combined to N-particle tracking (as in \cite{vowinckel2016entrainment}). In fact, this amounts to using a N-body Lagrangian formulation to describe both dynamics and resuspension. In contrast, associating a dynamic resuspension models to a reduced description of a two-phase flow (like RANS) would be inconsistent, since the only information provided by the flow solver is the average one-point velocity field and not the instantaneous velocity field (this point is detailed in Section~\ref{sec:analysis:cross}). The reverse also holds: resuspension models that require information on the fluid velocity should be coherent with the level of information available in the chosen approach for the fluid phase. For example, an empirical expression for the resuspension rate can be coupled to fine simulations of a turbulent flow (like LES) and to Lagrangian tracking of detached particles (as in \cite{wu2017particle}). As displayed in Fig.~\ref{fig:fig_wu_2017_rotor}, this has allowed the authors to investigate the role of near-wall turbulent structures induced by helicopter rotors on particle dynamics in the near-wall region. While this does not pose any difficulty in terms of numerics, it can be questionable in terms of physical descriptions. In fact, the choice of an empirical formula implies that the average number of resuspended particles is evaluated using only a one-point one-time information on the instantaneous velocity. Yet, LES does provides more detailed information about the instantaneous but filtered velocities, which is not used for the resuspension model (i.e. attached particles) but is taken into account as soon as they get detached from the surface.

\end{itemize}

 \subsubsection{Avoiding typical statistical pitfalls}
  \label{sec:analysis:cross}
	
Coupling inconsistent descriptions of the fluid and particle phases does not just limit the overall precision but can also result in deeply flawed formulations. This typically occurs when the nature of the required statistical information is not properly analyzed and when what can - or cannot - be extracted from a given statistical model is not well understood. 

\begin{figure}[ht]
 \centering
 \includegraphics[width = 0.95\textwidth]{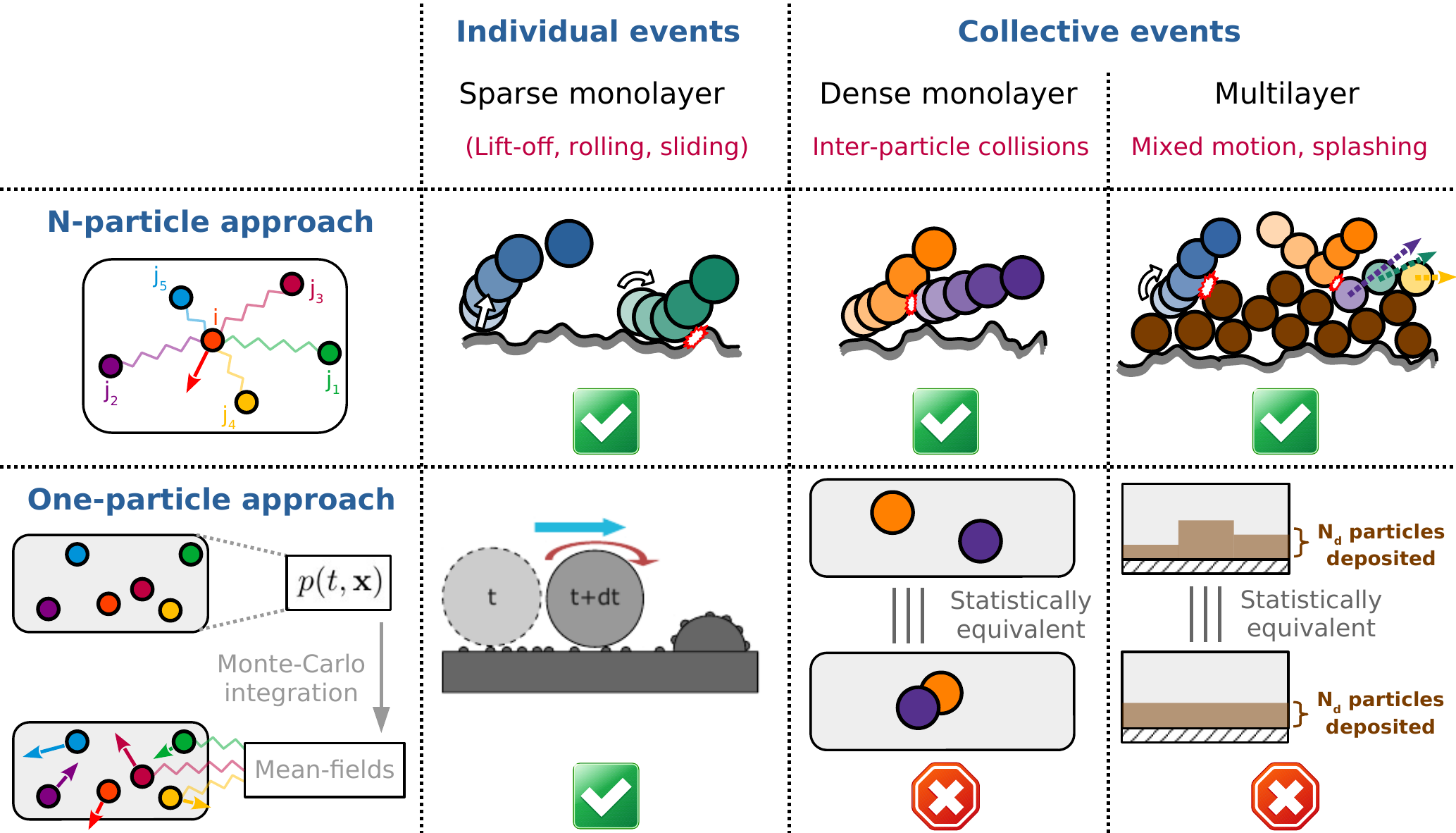}
 \caption{Illustration of consistency issues, with the amount of information available in N-point PDF formulations (middle) and one-point PDF approaches (bottom).}
 \label{fig:sketch_consistency}
\end{figure}

To exemplify such pitfalls, let us consider a small volume, say $\delta \mc{V}(\mb{x})$, around a given point $\mb{x}$ in which ${\rm N}(\mb{x})$ particles are located at a given time $t$. In the spirit of N-body approaches (like DEM), we may wish to determine all the possible particle-particle interactions between these ${\rm N}(\mb{x})$ particles, including inter-particle collisions or whether one particle is rolling on another one (see Fig.~\ref{fig:sketch_consistency}). Such a direct treatment of particle-particle interactions requires, however, to know the relative position of each particle within the small volume. In a statistical description, this means that we must have access to the PDF $p(t;\mb{x}_{\rm p}^{(1)}(t), \mb{x}_{\rm p}^{(2)}(t), \dots, \mb{x}_{\rm p}^{({\rm N}(\mb{x}))}(t))$ of the ${\rm N}(\mb{x})$ particle locations, $\mb{x}_{\rm p}^{(i)}(t)$ being the position at time $t$ of the particle labeled $(i)$ with $i=1,\dots,{\rm N}(\mb{x})$ such that $\mb{x}_{\rm p}^{(i)}(t) \in \delta \mc{V}(\mb{x})$. This information is only accessible from a N-particle PDF method.

However, if we are using a one-particle Lagrangian PDF approach, it is essential to be aware that we have only access to the one-particle PDF of particle locations, that is $p(t;\mb{x}^{(i)}(t))$ ($\forall i=1,\dots,{\rm N}$), but not to the ${\rm N}(\mb{x})$-PDF. Actually, by resorting to a locally homogeneity hypothesis, through which probabilistic expectation at point $\mb{x}$ is replaced by an ensemble averaging over the ${\rm N}(\mb{x})$ particles located in the small volume $\delta \mc{V}(\mb{x})$, these ${\rm N}(\mb{x})$ particles correspond to ${\rm N}(\mb{x})$ samples of the one-particle PDF $p(t;\mb{x})$. Since such one-particle PDF methods are developed in the spirit of mean-field formulations, this means that the only information that can be extracted corresponds to one-point statistics (like the particle volume fraction $\alpha_{\rm p}(t,\mb{x})$ or the mean particle velocity $\lra{\mb{U}_{\rm p}}(t,\mb{x})$). In a weak sense, all we can write for the particle position PDF is
\begin{equation}
p(t,\mb{x}) \simeq \frac{1}{{\rm N}} \sum_{i=1}^{{\rm N}} \delta (\mb{x} - \mb{x}^{(i)}(t))~,
\end{equation}
which, using the local homogeneity assumption mentioned above, translates into $p(t,\mb{x})\delta \mc{V}(\mb{x}) \simeq {\rm N}(\mb{x})/{\rm N}$. Therefore, all the possible ways through which the ${\rm N}(\mb{x})$ particle locations are distributed within the small volume $\delta \mc{V}(\mb{x})$, such as the two ones displayed in Fig.~\ref{fig:sketch_consistency}, cannot be distinguished. In the context of one-particle PDF methods, they are all equivalent since they yield the same weak approximation of the one-particle PDF, and correspondingly of the one-point PDF (e.g., mean concentration, mean velocity). Consequently, all attempts at calculating direct particle-particle interactions between the subset of ${\rm N}(\mb{x})$ particles are bound to be meaningless. 

Note that the same reasoning indicates that in one-particle method, even if the number of particles making up a deposit is correctly predicted, we still cannot deduce exactly from the simulations the resulting morphology or structure of the deposit. This means that the different ways according to which deposited particles are organized (see, for instance, the left panel in Fig.~\ref{fig:sketch_morphology} in 
Section~\ref{sec:physics:phenomenology:mechanisms}) cannot be distinguished. Capturing deposit morphology is, of course, possible but requires additional hypothesis to be introduced and a specific model to be developed.

A similar pitfall appears if we try to account directly for actions at distance between particles when information is incomplete. To emphasize this point, let us now consider two small volumes, $\delta \mc{V}(\mb{x}_1)$ around point $\mb{x}_1$ and $\delta \mc{V}(\mb{x}_2)$ around point $\mb{x}_2$ which contain ${\rm N}(\mb{x}_1)$ and ${\rm N}(\mb{x}_2)$ particles, respectively. If we wish, for instance, to extract the spatial correlation between particle velocities (so as to deduce the length correlation or the energy spectrum) at points $\mb{x}_1$ and $\mb{x}_2$, we are then considering
\begin{equation}
\label{eq: mean spatial correlation}
\lra{ \mb{U}_{\rm p}^{(1)}\, \mb{U}_{\rm p}^{(2)} \,|\, (\mb{x}_{\rm p}^{(1)} \in \delta \mc{V}(\mb{x}_1) \, ; \, \mb{x}_{\rm p}^{(2)} \in \delta \mc{V}(\mb{x}_2) } = \int \mb{V}_{\rm p}^{(1)} \mb{V}_{\rm p}^{(2)} \, p(t; \mb{x}_1, \mb{V}_{\rm p}^{(1)}, \mb{x}_2, \mb{V}_{\rm p}^{(2)}) \, \dd \mb{V}_{\rm p}^{(1)} \, \dd\mb{V}_{\rm p}^{(2)}~.
\end{equation}
Similarly, if we consider two-body particle interactions deriving from a potential $U_{\rm p}^{\rm int}(\mb{r})$, then the total force acting on a particle located at $\mb{x}_1$ is 
\begin{equation}
\label{eq: mean two-body integration}
\mb{F}_{\rm p}(\mb{x}_1)= \int \nabla U_{\rm p}^{\rm int}(\mb{x}_{\rm p}^{(2)} - \mb{x}_1) \, p(t; \mb{x}_1, \mb{x}_{\rm p}^{(2)}) \, \dd \mb{x}_{\rm p}^{(2)}~,
\end{equation}
which involves the two-particle location PDF (obtained as the marginal of the two-particle position-velocity PDF in Eq.~\eqref{eq: mean spatial correlation}).  Using the principles introduced in Section~\ref{sec:art_model:principle:conception}, this means that, if our objective is to handle without approximation quantities such as the ones in Eqs.~\eqref{eq: mean spatial correlation}-\eqref{eq: mean two-body integration}, then the minimal information required is the two-particle PDF. However, if we have first selected a one-particle PDF model, this information has been averaged out and we have only access to the one-particle marginals. In practice, this means that we have ${\rm N}(\mb{x}_1)$ and ${\rm N}(\mb{x}_2)$ particles in the two small volumes $\delta \mc{V}(\mb{x}_1)$ and $\delta \mc{V}(\mb{x}_2)$, but for each particle labeled $(i)$ in the first volume (with $i=1,\dots,{\rm N}(\mb{x}_1)$), we cannot determine anymore its corresponding partner $(j)$ in the second volume (with $j=1,\dots,{\rm N}(\mb{x}_2)$). The ${\rm N}(\mb{x}_1)$ samples can be used to derive mean quantities at point $\mb{x}_1$ by Monte Carlo estimates while the ${\rm N}(\mb{x}_2)$ samples can be used to derive mean quantities at point $\mb{x}_2$ by the same procedure, but no correlation between the two particle sets can be extracted from the one-particle PDF simulation. To be able to do so implies to track not a large number of particles but a large number of particle pairs.

The conclusion is that, if the objective is to treat without approximation quantities involving particle-particle interactions at a distance or at contact, we must be using a N-particle PDF approach. This requires that the velocity field of the carrier fluid (the continuous phase) be exactly known. In consequence, this rules out typical turbulence models, such as the RANS and even the LES ones, since they correspond to one-point statistical formulations. Applying a DEM approach for the particle phase is therefore only consistent with performing a DNS for the fluid phase. Phrased differently, coupling directly a DEM formulation to a one-point turbulence model leads to a deeply flawed and ill-based overall formulation. This is, unfortunately, a rather common mistake due to a poor analysis of the statistical requirement in terms of available information.

 \subsection{Uncertainty quantification}
 \label{sec:analysis:uncertainty}

The fifth (optional) criteria is related to the statistical analysis of the model. The aim is to provide answers to the following queries: what are the overall performances of a model? In particular, how sensitive are the results to the variables entering a model? What are the uncertainties on the required variables and their effect on the resolved variables?

\begin{figure}[ht]
 \centering
 \includegraphics[width = 0.8\textwidth]{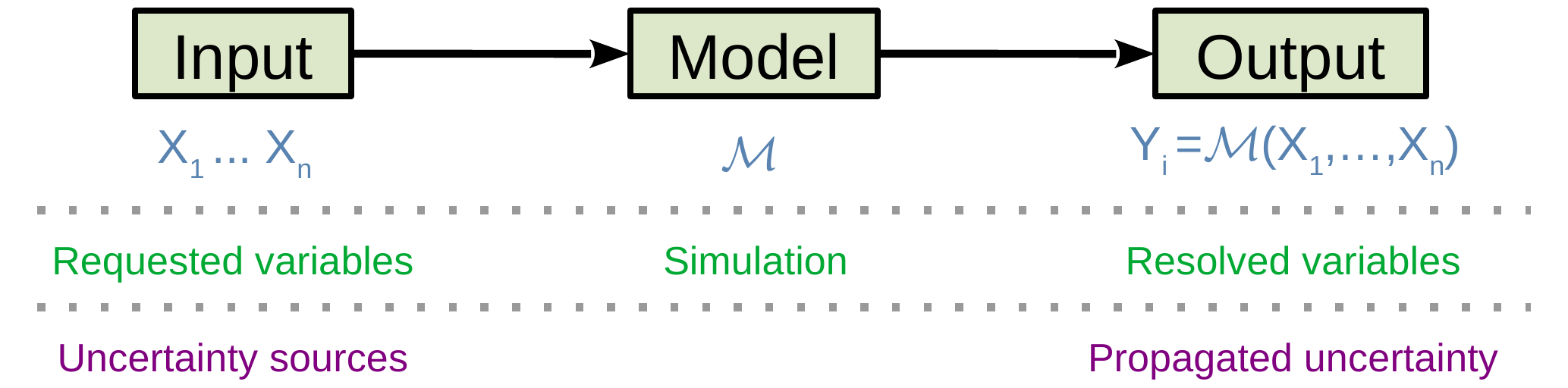}
  \caption{Sketch summarizing the key features in the analysis of results obtained from simulations: by identifying the requested variables (together with the associated uncertainty) and the resolved variables entering a model, one can evaluate the uncertainty propagation.}
 \label{fig:sketch_criteria_UQSA}
\end{figure}
As displayed in Fig.~\ref{fig:sketch_criteria_UQSA}, this step requires to have clarified all the notions introduced in the previous sections (see also Fig.~\ref{fig:sketch_criteria_intro}), especially the identification of the required/resolved variables. Hence, if performed, this analysis should be carried out after choosing/establishing a model. It usually relies on existing statistical tools, which allow two kinds of examinations:
\begin{itemize}
 \item Sensitivity analysis (SA):
 
 Sensitivity analysis consists in assessing how the outputs of a given model respond to variations in its inputs. In practice, sensitivity analysis ranks the inputs with respect to their importance in the variability of the output. Hence, this allows to distinguish the requested variables that mostly influence the resolved quantities (often in non-linear/exponential ways) and those that are less relevant. In addition, sensitivity analysis can reveal interactions among input variables and correlations among output quantities. 
    
 \item Uncertainty quantification (UQ):
 
 Uncertainty quantification is the science of quantitative characterization and reduction of uncertainties in models (and more generally any computational or practical application). It tries to determine how likely certain outcomes are if some aspects of the system are not exactly known. For that purpose, one usually relies on uncertainty propagation to estimate how uncertainty in the input variables is propagated through a model to the resolved variables. 
 
\end{itemize}

From the previous definitions, it appears that uncertainty analysis and sensitivity analysis are closely related to each other. In fact, a good practice is to run both of them in tandem.
 
In the following, we would like to insist on the interest for scientists to rely more on such statistical analyses, which have been seldom used in the context of particle resuspension \cite{raffaele2016windblown, soepyan2016threshold, raffaele2018uncertainty}. For that purpose, we do not delve into the existing statistical tools developed for SA and UQ techniques (interested readers are referred to dedicated literature on this topic, such as \cite{smith2013uncertainty, iooss2015review}). Instead, we wish to underline what can be gained from such analyses. In particular, these analyses can prove very helpful in:
\begin{itemize}
 \item Model calibration: Since a large number of variables enter resuspension models, sensitivity analysis can facilitate the calibration stage by focusing on the most sensitive parameters. Without such knowledge on the ranking of input variables, the risk is to lose time calibrating non-sensitive parameters.
 \item Model simplification: SA and UQ techniques allow to identify redundant parts within a model or inputs that have no effect on the desired output quantity. Hence, such information is helpful to potentially remove such parts/inputs from existing models in order to simplify them.
 \item Model improvement: SA and UQ techniques can reveal previously unsuspected connections/correlations between experimental measures, model inputs, and model predictions \cite{raffaele2016windblown}. This helps designing updated or higher-order models in which the depiction of the observed phenomena is improved. Such methods can even help inching towards unified models that actually bridge the gap between monolayer and multilayer deposits as well as between aeolian and fluvial resuspension (see also \cite{pahtz2020unification, pahtz2021unified}).
\end{itemize}
At this stage, it is worth pointing out that gaining such precious knowledge is not free. In fact, one of the drawbacks of statistical analyses is that they require extensive amounts of data, each dataset corresponding to a slight variation in the input variables. In addition, replicates of the same input conditions are needed to assess the intrinsic stochasticity on a given output (like the time-averaged resuspension rate $\tau_r^{\Delta t}$). Hence, SA and UQ are delicate and time-consuming tasks, especially when relying on long experimental measurements and/or on computationally expensive models \cite{raffaele2018uncertainty}.

This points to the importance of understanding first these tools carefully in order to make the best out of them, before actually applying them in practical situations. This relates to one comment in Gigerenzer's book entitled \textit{Reckoning with risk: learning to live with uncertainty} \cite{gigerenzer2003reckoning}, where he criticized that ``data analysis is typically taught as a set of statistical rituals rather than a set of methods for statistical thinking''.

 \subsection{Summary}
  \label{sec:analysis:summary}
	
At first sight, the presentation of the guiding principles related to the `art of modeling' in Section~\ref{sec:art_model} may appear like a step into a domain more akin to the philosophy of sciences than physics itself. However, not only is it sometimes sound to revisit the basics behind modeling efforts, but addressing the complex maze of the modeling proposals introduced in Section~\ref{sec:models} from such a standpoint has proven fruitful in several aspects. First, this has allowed to bring out a reading grid according to which these model proposals have been organized and assessed. Second, this has provided us with a general framework that is helpful when attempting to go beyond present model limitations, in particular, by keeping us clear of two false impressions. 

Indeed, new researchers in particle resuspension might derive the impression that models are bound to be incomplete, one way or another. Yet, once present shortcomings are clearly brought out, this incompleteness can be regarded as an incentive to pursue research work so as to push back present limits. Another possible impression from the profusion of available models is the misleading belief that particle resuspension has been thoroughly investigated, leaving no areas on which to shed new light. As recalled in Section~\ref{sec:art_model}, models are, however, regularly evolving through the triptych cycle (modeling/implementation/analysis). Therefore, thanks to this analysis, the stage for a proper assessment of model formulation is set, thereby paving the way for future extensions/developments of these models. This is addressed in the following Section~\ref{sec:next_model}.

\section{Paving the way for new models and experiments}
 \label{sec:next_model}

The objective of this section is to introduce future works needed to improve our understanding of particle resuspension as well as its modeling. Although still relatively unexplored at the moment, we believe that two directions in particular have the potential to become avenues of progress. They concern: 
\begin{enumerate}[i.]
 \item More complete and realistic models accounting for particles with complex shapes  (e.g., non-spherical, deformable, rough particles) and intricate surface forces (related to surface heterogeneities); 
 \item More general-purpose and unified approaches bridging the gap between two separate existing class of models (dedicated either to monolayer or multilayer resuspension) and covering a larger range of applications (i.e., not necessarily customized for either fluvial or aeolian resuspension).
\end{enumerate}

In the following, these two roads are described in more details by addressing, at the same time, new experimental measures and refined models accounting for these new phenomena.

 \subsection{Beyond spherical particles: towards more realistic descriptions}
  \label{sec:next_model:complete}

From the discussions on existing models in Sections~\ref{sec:analysis:mechanisms} and~\ref{sec:analysis:fund_inter}, it appears that a complete description of the resuspension process relies not only on the knowledge of the mechanisms at play but also on the specific forces entering these formulations. In turn, information on particle properties (e.g., size, density) as well as on surface characteristics (like roughness, composition) is required for adhesion forces (when relevant), while information on the fluid flow is needed for hydrodynamic forces. This implies that more complete descriptions of particle resuspension in realistic situations require further investigations of two aspects detailed below: the properties of particles and those of surfaces.

 \subsubsection{Resuspension of particles of complex shapes}
  \label{sec:next_model:complete:part} 

Up to now, a large majority of particle-tracking approaches have focused on the case of spherical particles. This might appear as a somewhat strong assumption since many particles are not spherical (like pollen, fibers, or sand). Nevertheless, it has allowed to analyze precisely the physical phenomena involved and to develop satisfactory models that accurately reproduce the key resuspension features of spherical particles. However, we need now to go beyond this first approximation and consider more realistic particle shapes. In particular, two categories of particles are worth exploring:

\begin{itemize}
 \item Non-spherical rigid particles:
 
 \textbf{Physical origin:} Particles in the environment often display complex shapes. For example, many biological agents have rod-shaped geometries \cite{brambilla2017adhesion} (like bacillus spores shown in Fig.~\ref{fig:illustr_nonsphere_spore}), while fibers are usually thin and elongated objects \cite{lundell2011fluid} (as seen in Fig.~\ref{fig:illustr_nonsphere_fiber}). Other particles have irregular surface features \cite{chen2019experimental}, like sand grains shown in Fig.~\ref{fig:illustr_nonsphere_sand} or pollens with spiky surfaces  visible in Fig.~\ref{fig:part_roughness_pollen}. 
 
 \begin{figure}[h]
  \centering
  \captionsetup[subfigure]{justification=centering}
  \begin{subfigure}{0.32 \linewidth}
   \centering
   \includegraphics[width=0.76\textwidth, trim = 12cm 10cm 14cm 13cm, clip]{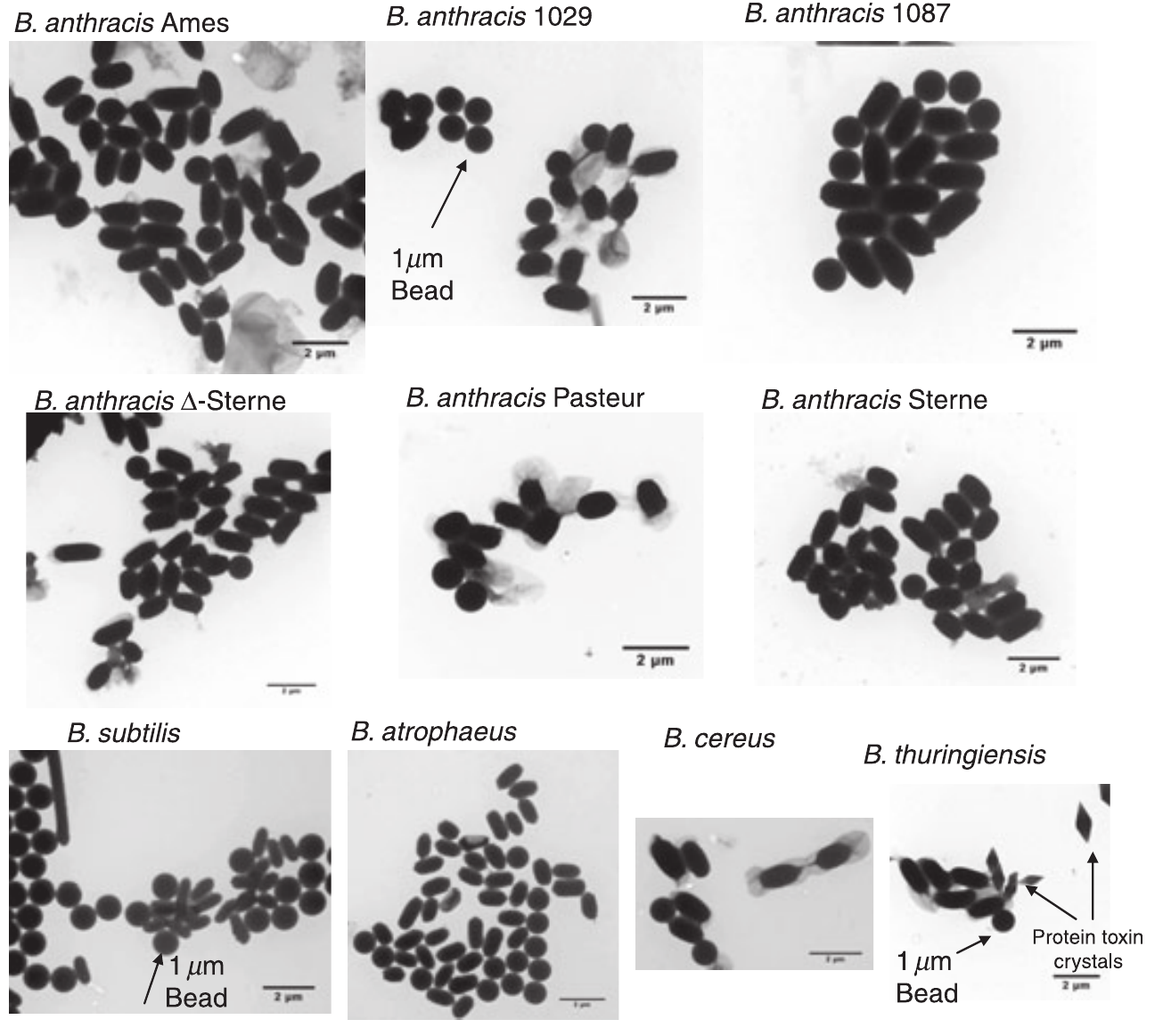}
   \caption{Electron micrographic image of \textit{Bacillus anthracis (Pasteur)} spores. Reprinted with permission from \cite{carrera2007difference}. Copyright 2006, John Wiley \& Sons.}
   \label{fig:illustr_nonsphere_spore}
  \end{subfigure}
  \hspace{3pt}
  \begin{subfigure}{0.32 \linewidth}
   \centering
   \includegraphics[width=0.76\textwidth, trim = 1.5cm 1cm 29cm 14.5cm, clip]{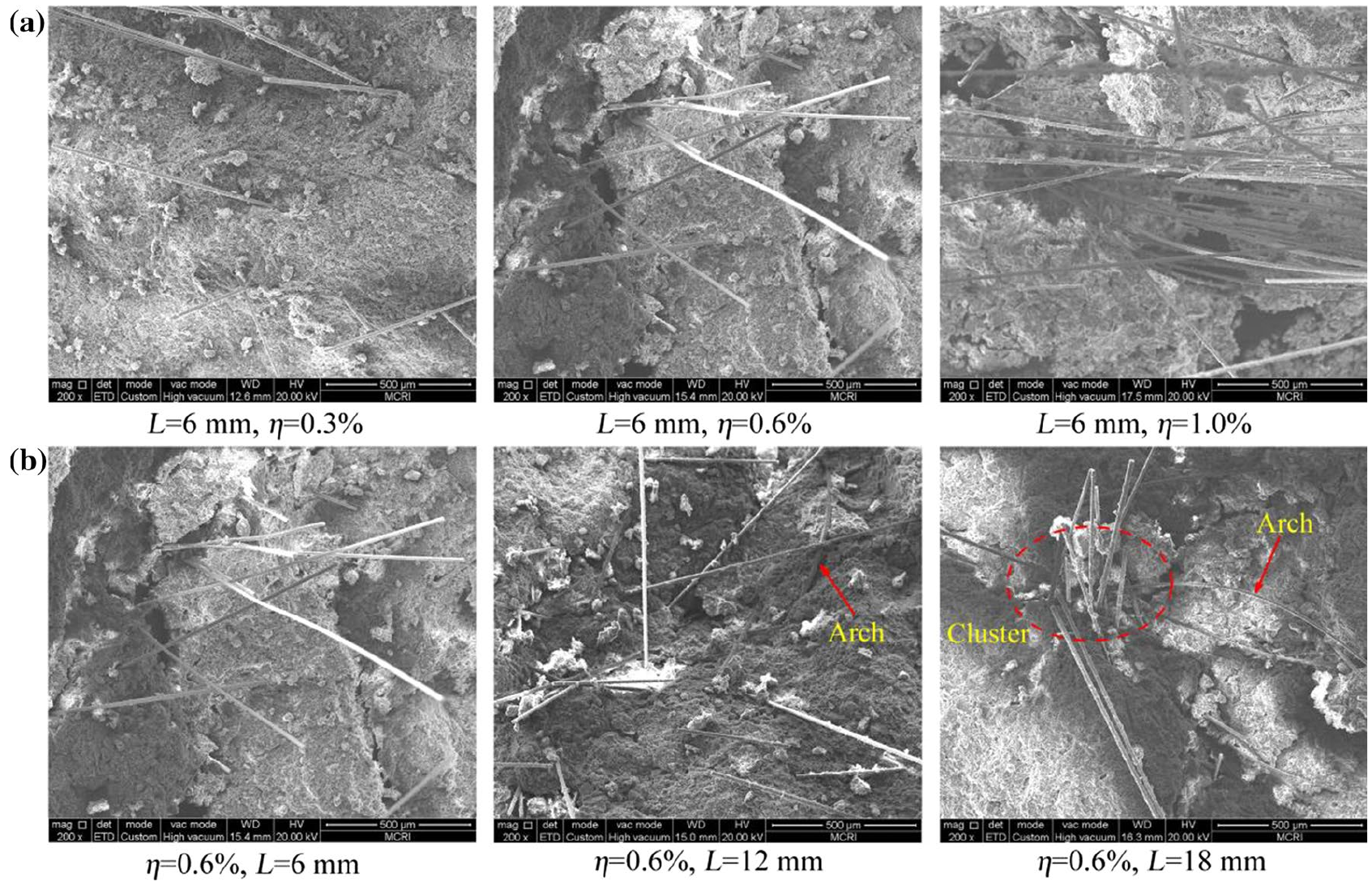}
   \caption{Electron micrographic image of \SI{6}{mm} long Basalt-fiber reinforcing a soil. Reprinted with permission from \cite{xu2021study}. Copyright 2021, Springer Nature.}
   \label{fig:illustr_nonsphere_fiber}
  \end{subfigure}
  \hspace{3pt}
  \begin{subfigure}{0.32 \linewidth}
   \centering
   \includegraphics[width=0.76\textwidth, trim=8.2cm 1cm 0cm 0cm, clip]{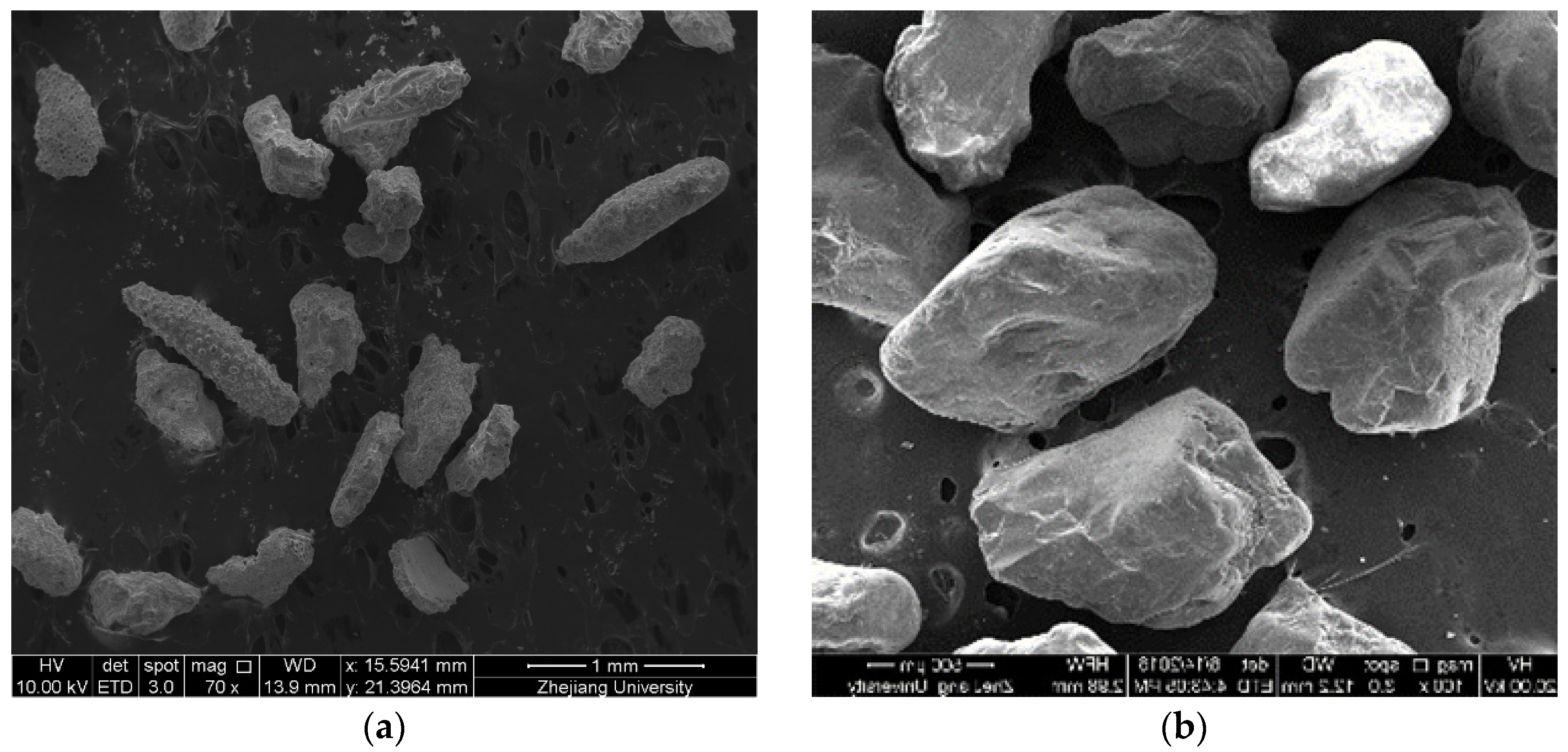}
   \caption{Electron micrographic image of quartz sand. Reprinted with permission from \cite{he2021effect}. Copyright 2021, Multidisciplinary Digital Publishing Institute.}
   \label{fig:illustr_nonsphere_sand}
  \end{subfigure}
  \caption{Images of particles with a non-trivial geometry, ranging from rod-like shapes (left), thin elongated objects( middle) to multifaceted particles (right).}
  \label{fig:illustr_nonsphere}
 \end{figure}
 
 \textbf{Role in resuspension:} Particle shape anisotropy can impact resuspension in the following ways:
 \begin{itemize}
  \item First, hydrodynamic and adhesion forces differ from the case of spherical particles. In particular, the strength of both forces depend on the orientation of the particle with respect to the underlying surface and to the fluid streamlines. For instance, as depicted in Fig~\ref{fig:sketch_spheroid_forces}, elongated particles undergo much lower adhesion forces when only the tip of the particle is in contact with the surface (left figure) compared to the adhesion force obtained when the particle has its major axis oriented parallel to the surface (middle figure). Meanwhile, hydrodynamic drag and lift forces are much lower in the case of a streamwise orientation (middle figure) than a wall-normal or tangential orientation (left or right figures). This comes from Eq.~\eqref{eq:Fdrag}, which shows that hydrodynamic forces are directly proportional to the area exposed to the fluid flow. 
  
  \begin{figure}[ht]
   \centering
   \includegraphics[width=0.9\textwidth]{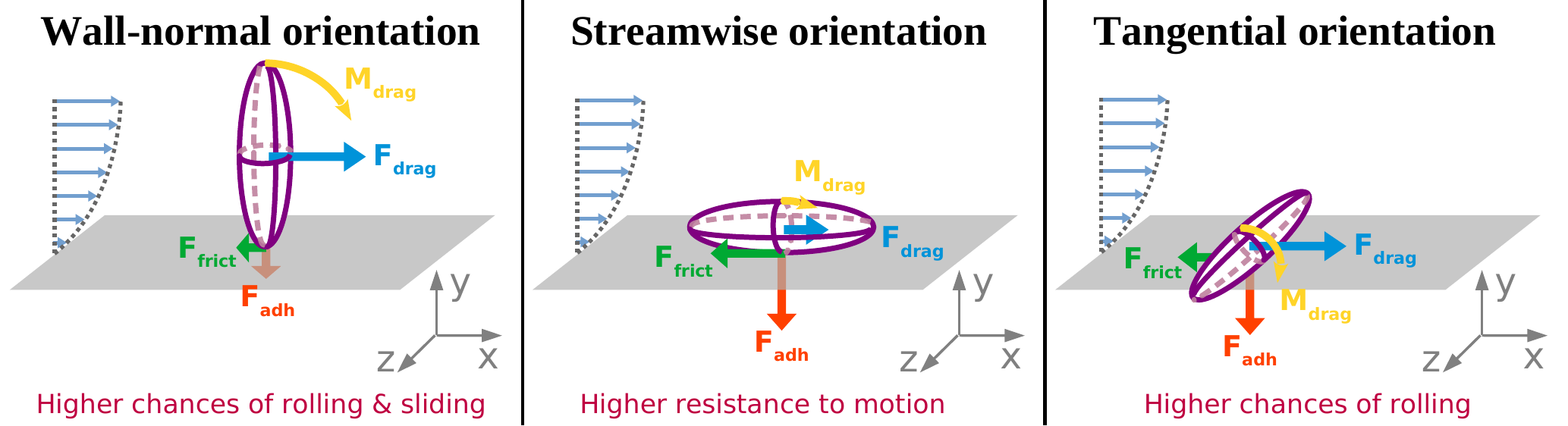}
   \caption{Illustration of how forces and probable motion change with respect to a spheroid orientation in contact with a surface.}
   \label{fig:sketch_spheroid_forces}
  \end{figure}
  
  \item Second, as a result of these forces, incipient motion depends also on the orientation of the particle with respect to the surface and to the direction of the flow. In particular, the relative importance between the resuspension modes (namely sliding, rolling, and lifting) varies with the particle orientation. In particular, it has been shown that sliding occurs sooner than rolling for hemispherical or irregular particles while rolling usually occurs sooner than sliding for spheres \cite{stevenson2002incipient}. 
  To give another example, let us consider the case of a spheroid (as the one displayed in Fig~\ref{fig:sketch_spheroid_forces}). When the spheroid is in the tangential orientation case (i.e. when the flow is perpendicular to the particle rotation axis), the rod is expected to have a higher chance of rolling on the surface compared to a sphere with a radius equal to the spheroid second axis. On the contrary, when the flow is perfectly aligned with the particle major axis (streamwise orientation), the rod is expected to be harder to move than a sphere with a radius equal to the rod thickness due to lower drag force and the fact that this orientation somewhat impedes rolling. 
	
 \end{itemize}
 
\textbf{Experimental needs:} This brings out new challenges in terms of measurements. In particular, new experiments are needed to determine the type of motion of such non-spherical particles, both at the onset of motion and after detachment. For that purpose, one might consider techniques that have recently proven useful to measure the rolling motion of large millimeter-size particles \cite{agudo2017detection}: it consists in using transparent particles with tiny black spots drawn on their surfaces while monitoring their motion using optical detection methods (see Fig.~\ref{fig:fig_agudo_2017_roll_technique}). This technique provides access to both translational and rotational motion of particles rolling on the surface (see Fig.~\ref{fig:fig_agudo_2017_roll_trans_vel}). Hence, it can deepen our understanding of the resuspension modes for particles with complex shapes while providing valuable information for the development and validation of models for such particles. A parameter to carefully control is the boundary layer depth, which can be modified by the choice of fluid (e.g., air vs. water), to capture both particles immersed in the boundary layer and particles sticking out of it.

\begin{figure}[ht]
 \centering
 \captionsetup[subfigure]{justification=centering}
 \begin{subfigure}{0.45 \linewidth}
  \centering
  \includegraphics[width=\textwidth, trim = 0cm 0cm 0cm 0cm, clip]{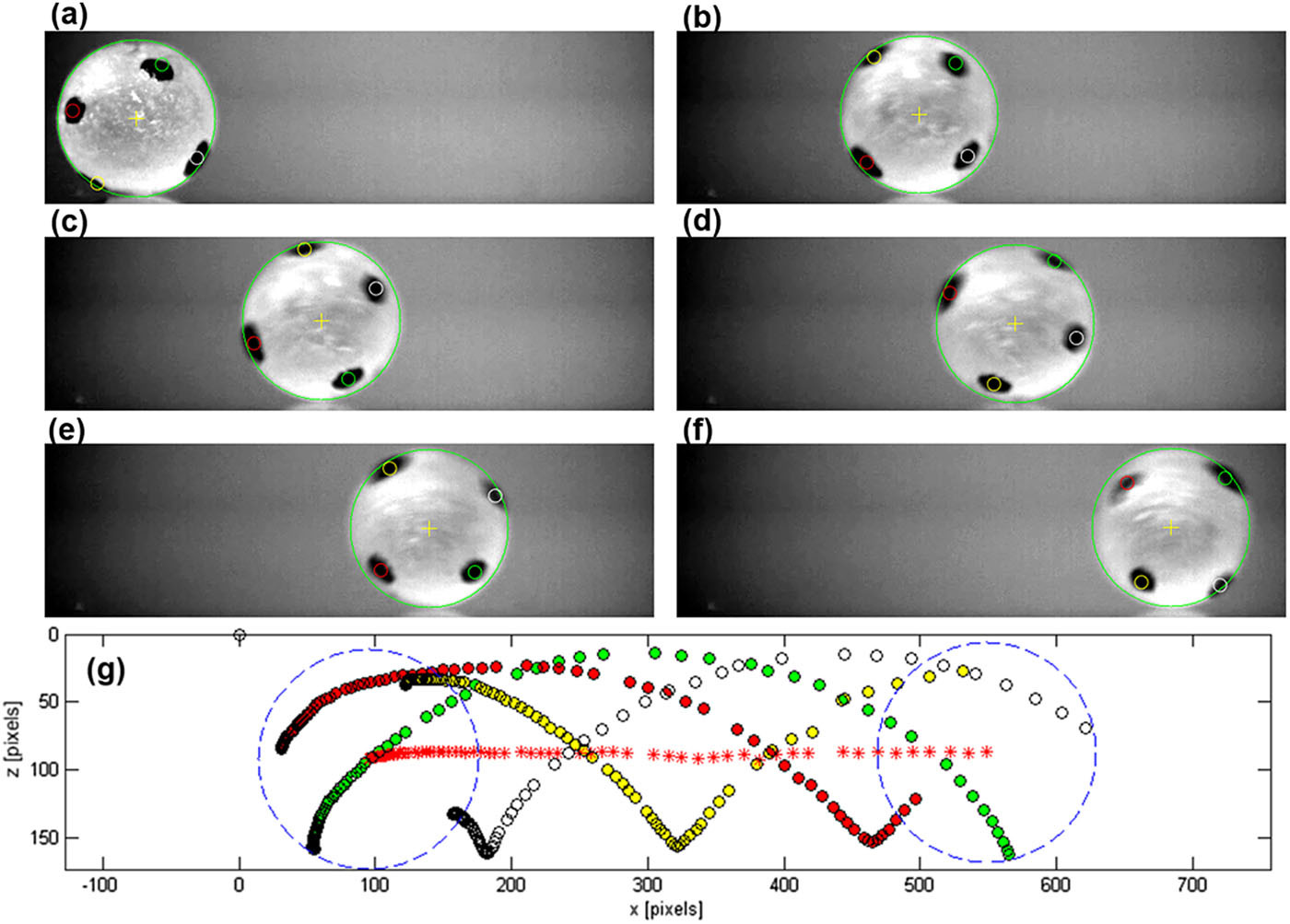}
  \caption{Snapshots showing the rolling motion of a single \SI{1.94}{mm} polystyrene bead on an inclined Teflon surface. The rolling motion is extracted thanks to the four dots on the surface (see the corresponding yellow, green, red and white circles on the bottom figure). Reprinted with permission from \cite{agudo2017detection}. Copyright 2017, AIP Publishing.}
  \label{fig:fig_agudo_2017_roll_technique}
 \end{subfigure}
 \hspace{10pt} 
 \begin{subfigure}{0.45 \linewidth}
  \centering
  \includegraphics[width=\textwidth, trim=0cm 0cm 0cm 4.3cm, clip]{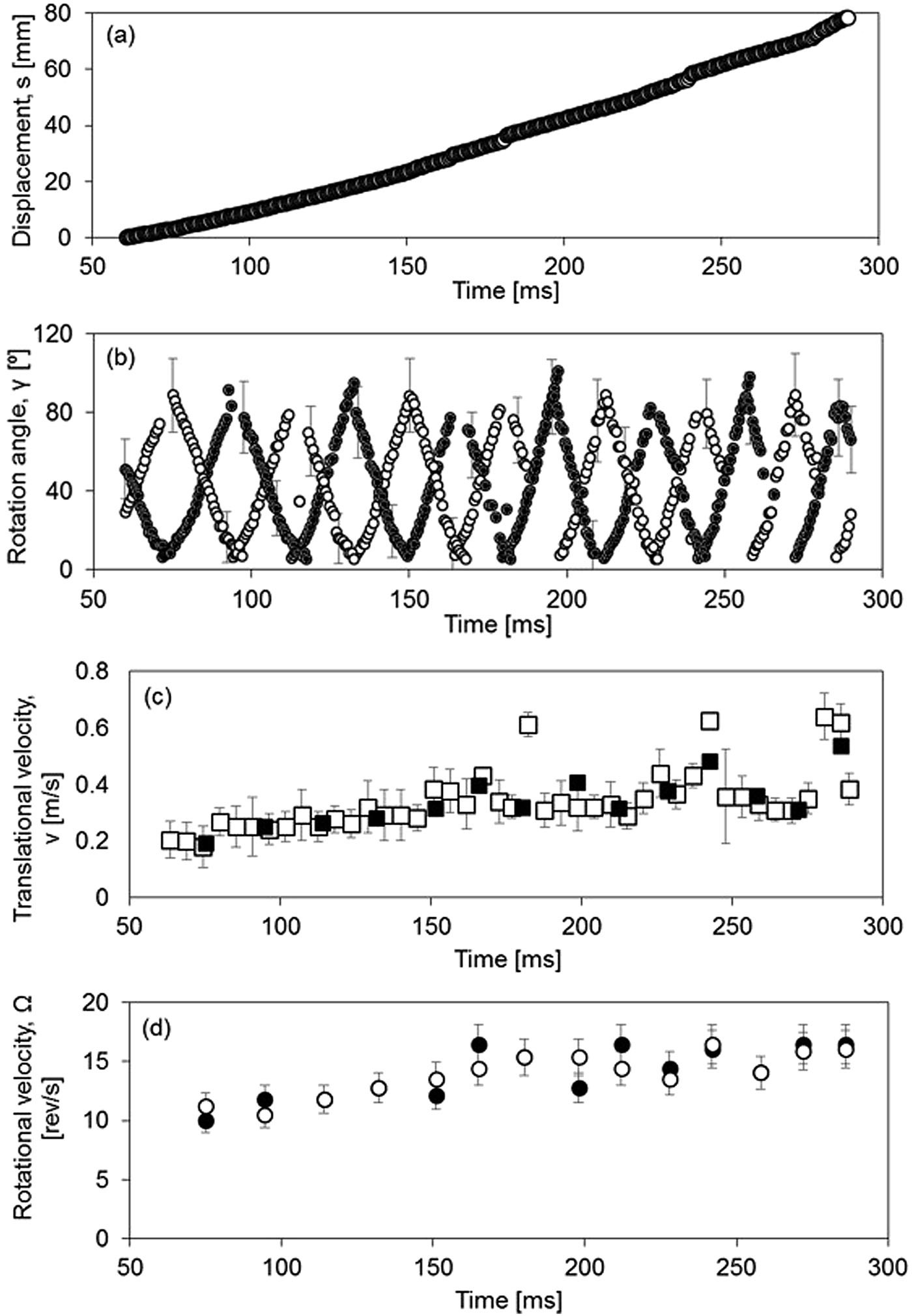}
  \caption{Bead translational velocity (top), and angular velocity (bottom) as a function of time extracted using the detection algorithm (open and solid symbols indicate results obtained from two different marks on the surface). Reprinted with permission from \cite{agudo2017detection}. Copyright 2017, AIP Publishing.}
  \label{fig:fig_agudo_2017_roll_trans_vel}
 \end{subfigure}
 \caption{Images showing how markers on the surface of a transparent particle can be used to extract information on the rolling motion and, as a result, on the translational and rotational velocities.}
 \label{fig:fig_agudo_2017_roll}
\end{figure}

\textbf{Modeling challenges:} Following the methodology introduced in the description of current models in Section~\ref{sec:models:approach}, the first step consists in selecting the degrees of freedom that characterize non-spherical particles as mechanical objects. To illustrate the additional modeling challenges compared to the spherical-particle situation, it is interesting to start with what seems a simple case of anisotropic particles, namely rigid ellipsoids. Indeed, ellipsoidal particles correspond to basic geometrical forms obtained by deforming a sphere by means of affine transformations. They are characterized by their extension ($r_1, r_2, r_3$) in each of the three directions, leaving a number of possible shapes such as: spheroids (i.e., ellipsoids of revolution $r_2=r_3$), prolate shapes (extended ellipsoids like rods or pencils $r_1 > \{r_2, r_3\}$), oblate shapes (flattened ellipsoids like disks $r_1 < \{r_2, r_3\}$), and Zingg ellipsoids ($r_2/r_1=2/3$ and $r_3/r_2=2/3$, representative of sediment shapes \cite{jain2019collision}). Let us even consider the specific case of rod-shaped particles, which corresponds to a prolate spheroid (as shown in Figs.~\ref{fig:sketch_spheroid_forces} or~\ref{fig:sketch_fibre_model}). Such rod-shaped particles can be characterized by a single parameter, namely the spheroid aspect ratio $\lambda = r_{\rm p,L}/r_{\rm p,S}$, which measures the ratio between the semi-major length $r_{\rm p,L}$ (i.e. along the symmetry axis) and the semi-minor length $r_{\rm p,S}$. From the sketches in Fig.~\ref{fig:sketch_spheroid_forces}, it is however clear that the basic forces acting on a rod-shaped particle depend on its orientation since the projected areas are now function of the rod orientation with respect to the flow streamlines. In a kinetic description (the extension to non-fully resolved flow fields is addressed later), this means that the particle state vector retained for spheres, $\mb{Z}_{\rm p}=(\mb{X}_{\rm p}, \mb{U}_{\rm p})$, must be extended to account, at least, for the particle orientation $\mb{P}_{\rm p}$, so that the minimum choice is to have $\mb{Z}_{\rm p}=(\mb{X}_{\rm p}, \mb{U}_{\rm p},\mb{P}_{\rm p})$. The rod orientation evolves as a function of the particle rotational velocity $\mb{\Omega}_{\rm p}$, since we have
\begin{equation}
\frac{\dd \mb{P}_{\rm p}}{\dd t}= \mb{\Omega}_{\rm p} \times \mb{P}_{\rm p}~.
\end{equation}
It is worth pointing out that keeping only the rod orientation implies that some (rotational) information is lost. Indeed, one can capture tumbling motions (i.e., rotation around directions orthogonal to $\mb{P}_{\rm p}$), but not spinning motion (i.e., rotation around the direction aligned with $\mb{P}_{\rm p}$). For this reason, a slightly extended particle state vector is obtained by retaining the particle rotational velocity, which means that we are now considering $\mb{Z}_{\rm p}=(\mb{X}_{\rm p}, \mb{U}_{\rm p}, \mb{\Omega}_{\rm p})$. 
For this state vector, the evolution equations which govern the dynamics of rods are given by \cite{voth2017anisotropic}: 
\begin{subequations}
 \label{eq:eq_Jeffery}
 \begin{align}
  \frac{d\mb{X}_{\rm p}}{dt} &= \mb{U}_{\rm p}~, \label{eq:eq_Jeffery_Xp}\\
  \frac{d\mb{U}_{\rm p}}{dt} &= \frac{\nu_{\rm f} \ \rho_{\rm f}}{m_{\rm p}} \, \mathbb{R}^{-1}  \mathbb{\widehat{K}} \ \mathbb{R} \cdot \left(\mb{U}_{\rm s} - \mb{U}_{\rm p}\right) + \mb{F}_{\rm non-drag}~,\label{eq:eq_Jeffery_Up}\\
  \frac{d(\mathbb{I}_{\rm p} \cdot \mb{\widehat{\Omega}}_{\rm p})}{dt} &= -\mb{\widehat{\Omega}}_{\rm p} \times \left(\mb{\mathbb{I}_{\rm p} \cdot \widehat{\Omega}}_{\rm p}\right) + \mb{\widehat{T}}~. \label{eq:eq_Jeffery_Op}
 \end{align}
\end{subequations}
In the equation for the translational velocity, Eq.~\eqref{eq:eq_Jeffery_Up}, the drag force has been singled out in the first term on the rhs while $\mb{F}_{\rm non-drag}$ refers to all other relevant forces (lift, adhesion, etc.). It is seen that the expression of the drag force for rod-shaped particles involves a non-isotropic resistance tensor $\mathbb{K}$ since this drag force depends on the spheroid orientation with respect to the fluid. This resistance tensor can be expressed by a purely diagonal tensor $\mathbb{\widehat{K}}$ in the local coordinate system attached to each spheroid ($\widehat{x},\widehat{y},\widehat{z}$) (see Fig.~\ref{fig:sketch_fibre_model}). The tensor $\mathbb{R}$ hence corresponds to the rotation tensor that allows to go from the global frame of reference to the local one. The exact expressions of this resistance tensor can be expressed in terms of the Euler parameters and are detailed elsewhere (as in \cite{mortensen2008dynamics, siewert2014orientation}). Meanwhile, the equation for the rotational velocity involves the rotational inertia tensor $\mathbb{I}_{\rm p}$ and the torque $\mb{\widehat{T}}$ of all forces acting on a spheroid (more details in the pioneering work of Jeffery \cite{jeffery1922motion}). Note that, in order to keep the rotational inertia tensor constant in time, Eq.~\eqref{eq:eq_Jeffery_Op} is not written in the original inertial frame of reference, as in Eq.~\eqref{eq:eq_Jeffery_Up}, but in the particle-attached frame of reference (this is represented, for instance, by the notation $\mb{\widehat{M}}$ which indicates that the torque $\mb{M}$ is to be expressed in the frame of reference shown in the right panel of Fig.~\ref{fig:sketch_fibre_model}). Regardless of the details of the Jeffery model, it is important to point out that $\mb{\widehat{M}}$ is function of the fluid velocity gradients at the rod center of mass, which we can express by $\mb{\widehat{M}}=\mc{F}(\mathbb{G}_{ij})$ where $\mathbb{G}_{ij}=\partial U_{\rm f,i}/\partial x_j$ is the fluid velocity gradient. Therefore, the Jeffery description relies on both the fluid velocity seen $\mb{U}_{\rm s}$ and the velocity gradient seen $\mathbb{G}_{ij}$ (note that while $\mb{U}_{\rm s}$ is a vector, $\mathbb{G}$ is a second-order tensor). 

As it transpires from Eqs.~\eqref{eq:eq_Jeffery}, models for the dynamics of rigid spheroids require to have proper expressions for the forces acting on the particles, among which $\mb{F}_{\rm non-drag}$ or expressions of the drag force that go beyond the limit of validity of the Stokes regime (low particle Reynolds numbers).
\begin{enumerate}[i -]
 \item \textit{Hydrodynamic drag and lift forces}: As mentioned in Section~\ref{sec:models:forces:hydro}, hydrodynamic drag and lift forces can be computed with microscopic descriptions, based on a direct integration of the extra fluid stress tensor along the entire surface of the particle. However, when developing models that remain tractable in realistic applications, one has to rely on approximate expressions. In the case of rigid spheroids, the drag force can be correlated to the angle $\phi$ between the direction of the fluid and the spheroid main axis \cite{zastawny2012derivation}:
 \begin{align}
  \label{eq:eq_CD_spheroid}
  \begin{aligned}
   C_D & = C_{D,\phi=0\si{\degree}} + \left( C_{D,\phi=90\si{\degree}} - C_{D,\phi=0\si{\degree}}\right) \, {\rm sin}^{a_0}\phi \\
   {\rm with} & \quad C_{D,\phi=0\si{\degree}} =  \frac{a_1}{\Rep^{a_2}} - \frac{a_3}{\Rep^{a_4}} \quad \text{and} \quad C_{D,\phi=90\si{\degree}} =  \frac{a_5}{\Rep^{a_6}} - \frac{a_7}{\Rep^{a_8}}.
  \end{aligned}
 \end{align}
 where $a_{i\in[0...8]}$ are nine shape-dependent coefficients fitted to PR-DNS integrations of the drag force on spheroids (see \citep[table 2]{zastawny2012derivation}). 
 
 Similar expressions have been suggested for the lift force:
 \begin{equation}
  \label{eq:eq_CL_spheroid}
  C_L = \left( \frac{b_1}{\Rep^{b_2}} + \frac{b_3}{\Rep^{b_4}} \right) \, \left( {\rm sin} \phi \right)^{b_5+b_6\ \Rep^{b_7}} \, \left( {\rm cos} \phi \right)^{b_8+b_9\ \Rep^{b_10}}
 \end{equation}
 with $b_{i\in[1...10]}$ ten shape-dependent coefficients fitted to PR-DNS integrations of the drag force on spheroids (see \citep[table 3]{zastawny2012derivation}). 
 
 These expressions were tested and validated for different particle Reynolds numbers $1\le\Rep\le300$, several aspect ratios $\lambda=(1.25,\ 2.5,\ 5)$ and various angles $0\le\phi\le90$. To the authors knowledge, no generic expressions valid for a larger range of shapes/Reynolds number have been proposed.
 
 \item \textit{Adhesion forces}: As mentioned in Section~\ref{sec:models:forces:contact}, the adhesion between rigid bodies is well described by van der Waals forces. Hence, it is possible to use a microscopic description based on SEI methods (to integrate molecular-molecular forces over the exact geometry). Yet, one can also resort to analytical formulas using the Derjaguin approximation (for short separation distances). This gives a formula for VDW potential energy between a spheroid and a sphere which is similar to Eq.~\eqref{eq:eq_VDW_sphsph} for two spheres, but with a different geometrical prefactor \cite{brambilla2017adhesion}:
 \begin{equation}
  E_{\rm VDW, Spheroid1-Sph2} =  -\frac{A_{\rm Ham}}{6h} \frac{r_{\rm p1,L}\,r_{\rm p,2}}{\sqrt{\left(\frac{r_{\rm p1,L}^2}{r_{\rm p1,S}}+r_{\rm p,2}\right)\left(r_{\rm p1,S}+r_{\rm p,2}\right)}}
 \end{equation}
 Using the same arguments on the prefactor, the formulas that include the effect of surface roughness can also be generalized for spheroidal particles (see \cite{brambilla2017adhesion}).

 \item \textit{Other forces:} To the authors' knowledge, no general formulation has been proposed for the capillary force or for the electrostatic forces between a spheroid and a plate \cite{brambilla2017adhesion}.
 
\end{enumerate}

At present, models for rigid non-spherical particles are usually limited to ellipsoids and follow the Jeffery formulation, Eqs.~\eqref{eq:eq_Jeffery}. These geometries are indeed easy to describe with mathematical expressions, making them compatible with fine simulations of the underlying fluid flow. For instance, PR-DNS \cite{jain2021impact} or LES \cite{zhang2020numerical} were performed in the context of large spheroids in water. These simulations have already confirmed the fact that ellipsoidal particles are more prone to sliding motion (or mixed combined motion) than purely rolling motion \cite{zhang2020numerical}. With respect to the models considered in Section~\ref{sec:models:approach}, this means that only the first category, namely the exact dynamical formulation, has started to be extended to address rigid non-spherical particles. In this framework, complete information on the fluid flow is accessible and the particle state vector is indeed limited to $\mb{Z}_{\rm p}=(\mb{X}_{\rm p}, \mb{U}_{\rm p}, \mb{\Omega}_{\rm p})$. In contrast, the formulation of extended dynamic PDF models based on the Jeffery description raises considerable challenges since the particle state vector is likely to be considerably augmented to include both the fluid velocity as well as the velocity gradients seen by rod-shaped particles. First suggestions have been made \cite[Section 10.2]{minier2016statistical} but work remains to be done to clarify the minimum amount of information on the fluid velocity gradients which is needed and to formulate new stochastic models able to capture their key statistical characteristics for rod dynamics. At the other side of the model spectrum, empirical models require ad-hoc experiments and parameters describing non-sphericity will become input variables. For intermediate formulations like kinetic PDF models, non-sphericity could be captured in the input PDF of adhesion forces. The issue is then to prescribe the resuspension mechanism, or their relative importance, depending on the amount of particles for each orientation (if not all equally possible). 

\textbf{Summary:} The key challenge for non-spherical rigid particles is to come up with general-purpose models that can be applied regardless of particle size. Indeed, particle anisotropy has always an impact on hydrodynamic and adhesion forces, however different for low- or large-inertia particles. Tracer particles (i.e., with a small inertia) were recently shown to follow the fluid streamlines and align their orientation with the velocity gradient \cite{allende2018stretching}. As a result, their anisotropy might not severely change their resuspension. The same does not hold for inertial particles, whose dynamics is not aligned with the flow streamlines. This suggests that a universal model can only be achieved provided that, at the very least, the particle orientation is included in the description.
 
 \item Deformable and/or flexible objects
 
 \textbf{Physical origin:} Apart from having complex shapes, particles can also deform. For example, plastic debris can undergo plastic deformations when trapped in a sediment bed, leading to changes in shape near the contact area. Meanwhile, elongated particles (like fibers \cite{lundell2011fluid} or polymers) as well as quasi-2D objects (such as leaves) can deform under the action of external forces, leading to stretching, bending or twisting (as illustrated in Fig.~\ref{fig:illustr_deform}). Another example is related to biological particles, which can change their size depending on the local environment (like spores shrinking in low relative humidity \cite{qian2014walking}).
 
 \begin{figure}[ht]
  \centering
  \captionsetup[subfigure]{justification=centering}
  \begin{subfigure}{0.45 \linewidth}
   \centering
   \includegraphics[width=0.7\textwidth, trim = 1.8cm 0cm 1.8cm 0cm, clip]{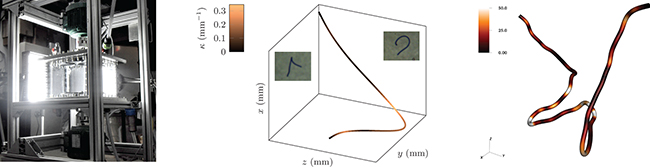}
   \caption{Example of a 3D reconstruction of a fiber conformation from experimental data (inset showing the side and top view, the color reflects the curvature). Reprinted with permission from \cite{gay2018characterisation}. Copyright 2018, IOP Science.}
   \label{fig:fig_gay_2019_fiber}
  \end{subfigure}
  \hspace{5pt} 
  \begin{subfigure}{0.45 \linewidth}
   \centering
   \includegraphics[width=0.8\textwidth, trim = 0cm 0cm 0cm 0cm, clip]{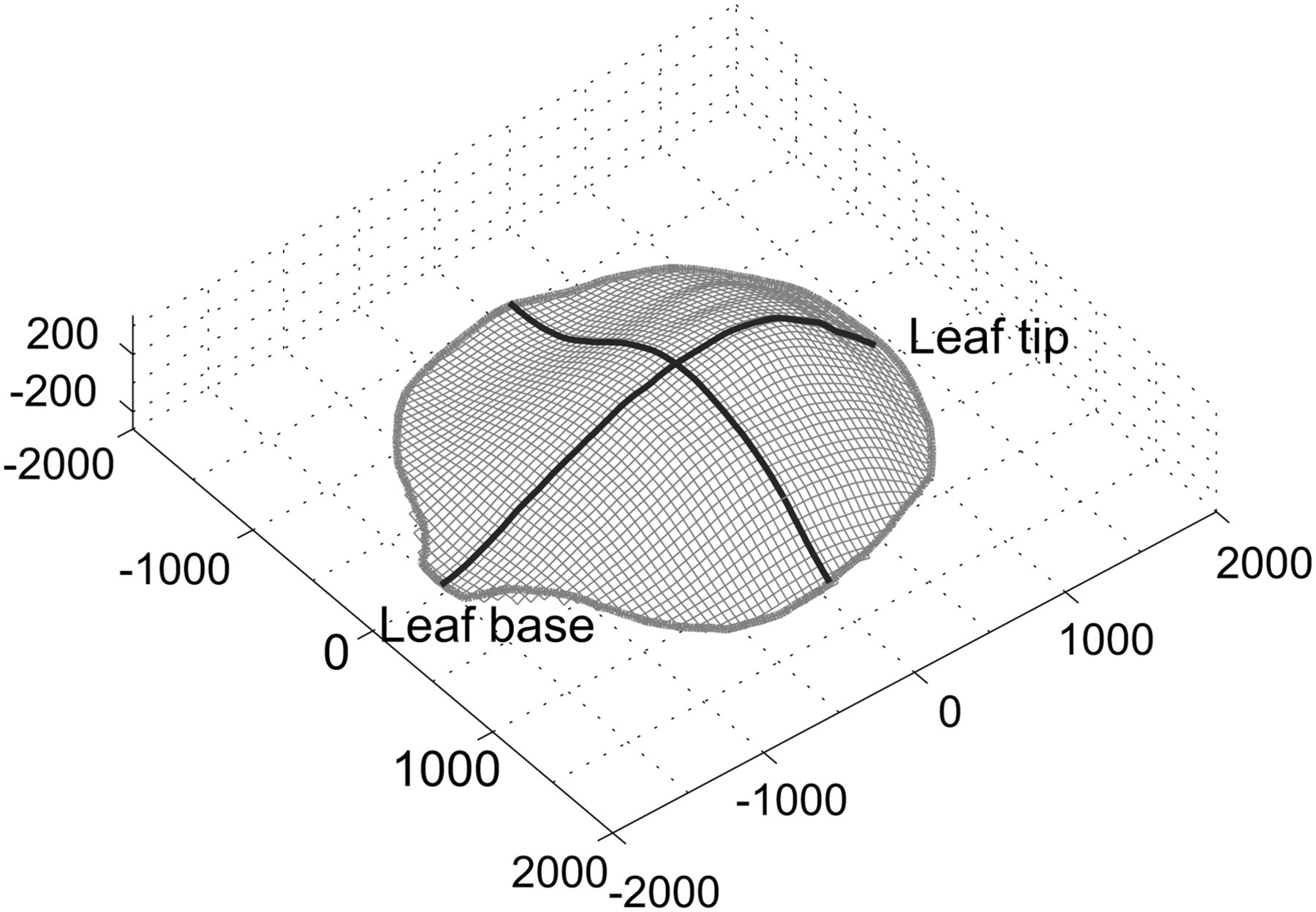}
   \caption{3D reconstruction of a DAS10 leaf with digital longitudinal (i.e., from leaf base to leaf tip) and transverse cross sections shown as black lines (units are in \SI{}{\mu m}). Reprinted with permission from \cite{rolland2014quantifying}. Copyright 2014, Oxford University Press.}
   \label{fig:fig_rolland_2014_leaf}
  \end{subfigure}
  \caption{Images of deformable particles (such as elongated 1D fibers or flat 2D surfaces).}
  \label{fig:illustr_deform}
 \end{figure}

 \textbf{Role in resuspension:} Resuspension of deformable particles can differ significantly from the one of rigid particles. This is due to a combination of two effects: first, plastic deformations of the surfaces in contact can make adhesive forces much more complex; second, the flexibility of elongated particles (like fibers or polymers) can lead to a partial contact between the fiber and the surface (the remaining parts forming arches above the surface, as shown also Fig.~\ref{fig:adhesion_fiber_grebikova1}).
 
 \textbf{Experimental needs:} New measurements of the adhesion between flexible elongated particles and surfaces are required to characterize how adhesion between them depends on the fiber orientation, the fiber properties (flexibility), the contact configuration (i.e., the the size of the regions actually in contact) and the pulling angle. In fact, a flexible polymer has a certain orientation on the surface which induces a higher resistance to lateral sliding/rolling motion along the major axis. By grafting the end of a polymer to an AFM and pulling it away from the surface \cite{grebikova2017pulling}, recent experimental measurements have confirmed that desorption changes according to the pulling angle (see Fig.~\ref{fig:adhesion_fiber_grebikova2}). Ideally, these techniques should be coupled with real-time optical measurements of the particle shape to record the contact configuration between the flexible particle and the surface. They should help designing fine models to capture such intricate adhesion regimes, using for instance discretization techniques where a flexible particle is represented as a chain of ellipsoids or other tractable shapes \cite{fan1997flow}.
  
  \begin{figure}[ht]
  \centering
   \begin{subfigure}{0.45 \linewidth}
    \centering
    \includegraphics[width=0.65\textwidth,trim = 0cm 0.0cm 4.5cm 0cm,clip]{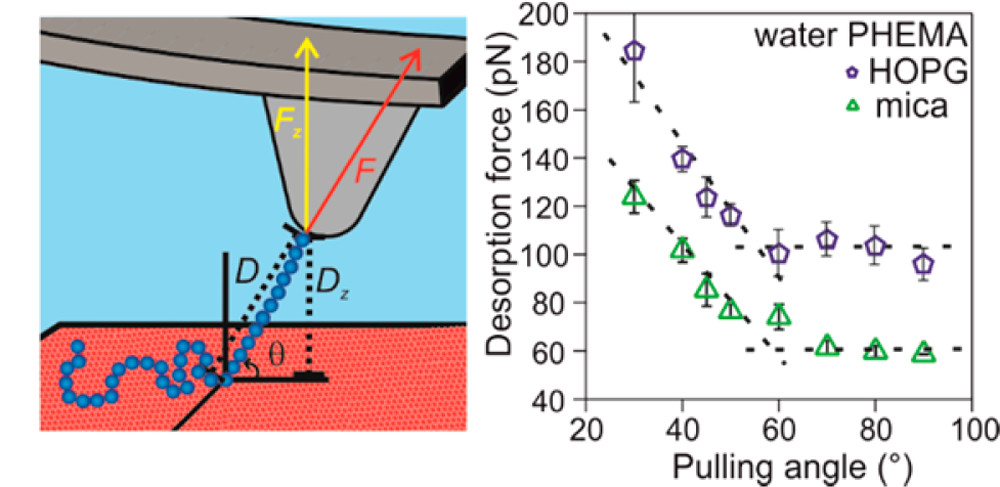}
    \caption{Schematic of the experiment, with a polymer partially adsorbed on the surface and pulled by AFM tip.}
    \label{fig:adhesion_fiber_grebikova1}
   \end{subfigure}
   \hspace{10pt}
   \begin{subfigure}{0.45 \linewidth}
    \centering
    \includegraphics[width=0.65\textwidth,trim = 4.4cm 0cm 0cm 0.0cm,clip]{adhesion-fiber-grebikova2}
    \caption{Average desorption force values plotted as a function of the pulling angle.}
    \label{fig:adhesion_fiber_grebikova2}
   \end{subfigure}
   \caption{Adhesion of a deformable polymer on a surface. Reprinted with permission from \cite{grebikova2018angle}. Copyright 2018, American Chemical Society. }
   \label{fig:adhesion_fiber_grebikova}
  \end{figure}
 
 \textbf{Modeling challenges:} Apart from the extension of adhesion models to account for plastic deformations (briefly addressed in Section~\ref{sec:models:forces:contact}), the key challenge is to develop new models for deformable 3D particles. Following again the methodology introduced in the description of current models in Section~\ref{sec:models:approach}, a two-step process if used: (a) the first step consists in defining a set of additional degrees of freedom to capture all the features/phenomena observed experimentally; (b) the second step consists in developing the corresponding evolution laws for these extra particle-attached variables. To illustrate the additional modeling challenges compared to rigid anisotropic particles, it is interesting to start with what seems to be a simple case of 3D deformable objects, namely flexible elongated fibers. Compared to rod-shaped particles, a flexible fiber is a thin and elongated object that can hardly be described by the sole end-to-end distance (as displayed in Fig.~\ref{fig:sketch_fibre_model}). Instead, depending on the forces acting on the fiber, it can form complex sinuous shapes (somewhat like a snake). Describing such objects requires therefore to handle additional degrees of freedom compared to rigid rod-shaped particles. In the frame of particle tracking approaches, this would be directly implemented by extending the state vector and adding the corresponding equations. In practice, one can resort to three levels of description shown in Fig.~\ref{fig:sketch_fibre_model}:
 
 \begin{figure}[ht]
  \centering
  \includegraphics[width=0.9\textwidth]{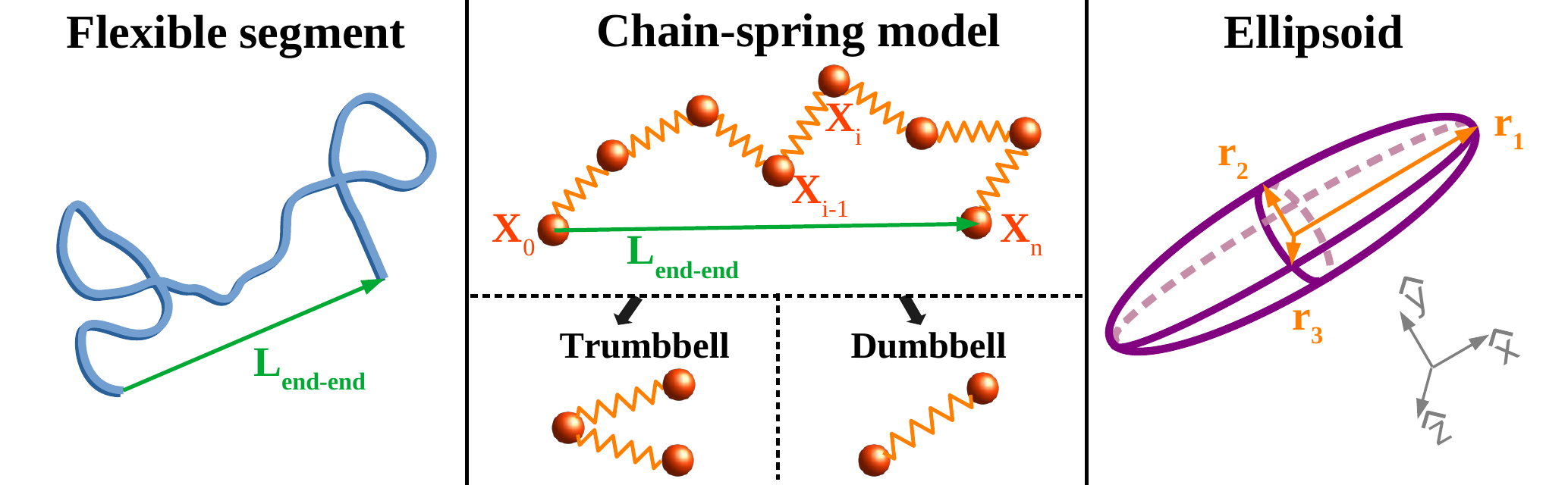}
  \caption{Illustration of the possible descriptions of flexible elongated particles, from the most detailed one (on the left) to the reduced models (like chain-spring, dumbbell or ellipsoids, on the right).}
  \label{fig:sketch_fibre_model}
 \end{figure}
 \begin{enumerate}[i -]
  \item \textit{Slender body theories} consist in describing a fiber as a slender elastic filament whose curvature is driven by the stresses acting on the fiber. In that case, its deformability can be described using beam theory, i.e. as an elongated cylindrical object whose deformation is given by the Euler-Bernoulli theory \cite{shelley2000stokesian}. In the case of an inextensible and inertialess fiber in a viscous flow, the dynamics of a fiber of length $\ell$ and circular cross-section $a$ can be approximated using the following equation \cite{lindner2015elastic, tornberg2004simulating, tornberg2006numerical}:
  \begin{eqnarray}
   &&\partial_t \mathbf{X} = \mathbf{U}_{\rm f}(\mathbf{X},t)+ \frac{1}{\alpha}\,\mathbb{D}\, \left[\partial_s(T\,\partial_s \mathbf{X}) - F\,\partial_s^4\mathbf{X}\right], \quad  \mbox{with} \quad |\partial_s\mathbf{X}|^2 = 1  \label{eq:eq_SBT} \\ 
   && \mbox{where} \quad \alpha =\frac{8\pi \,\rho_{\rm f}\,\nu_{\rm f}}{b} \quad \mbox{and}\quad \mathbb{D} = \mathbb{I} + \partial_s\mathbf{X}\,\partial_s \mathbf{X}^{\mathsf{T}},
     \nonumber
  \end{eqnarray}
  with $s\mapsto \mathbf{X}(s,t)$ the curvilinear axis (parametrized by the arc-length coordinate $s\in[-\ell/2,\ell/2]$), $F$ the fiber's bending modulus (also called flexural rigidity, defined as $F=EI$ where $E$ is the Young modulus and $I$ the fiber moment of inertia), $b = -\rm ln(\ell^2/a^2 \textit{e})$ the shape parameter and $\mathbb{D}$ the mobility matrix (whose form arises from the anisotropic drag exerted by the fluid on the slender fiber). This equation has to be complemented with boundary conditions at both ends of the fiber (free-end or fixed-end). This equation shows that such fibers deform due to the action of a velocity gradient along the fiber's length. Solving such equations numerically requires to discretize the curvilinear axis as an ensemble of inextensible and undeformable segments, i.e. with a length that is much smaller than the flexibility length (which measures the minimum length at which a fiber deforms under a given stress).
  
  \item \textit{Chain-spring models} are more simplified models, where an elongated deformable fiber is represented by a chain of beads connected together through spring-like segments, or through chain-rod models that use rigid rods \cite{minier2016statistical, ottinger2012stochastic, venerus2018modern}. Hence, for each bead $\mb{X}_i$ composing the fiber, one has to solve the following equation:
  \begin{equation}
   m_i\, \frac{d^2 \mb{X}_i}{dt^2} = -\zeta \left[\frac{d \mb{X}_i}{dt} -\mb{U}_{\rm f}(\mb{X}_i,t)\right] +\lambda_{i}\,(\mb{X}_i - \mb{X}_{i-1}) - \lambda_{i+1}\,(\mb{X}_{i+1}-\mb{X}_i),
  \label{eq:eq_chainspring}
  \end{equation}
  where $\zeta$ denotes the individual drag coefficient of each bead $i$ and $\lambda_i$ is the tension between the $i$-th and the $(i-1)$-th beads. This tension can be described using simple spring models or fixed distance (as in Kramer chains). Such formulations allow to capture bending and twisting of elongated flexible particles with reduced costs. Even more reduced descriptions can be obtained using three beads connected with spring-like segments (trumbbells) or two connected beads (dumbbells).
  
  \item \textit{Ellipsoids} can also be used as an extermely simplified representation of 3D elongated and flexible fibers. When dealing with inertia-less fibers with a size smaller than the Kolmogorov scale, it has indeed been shown recently that the dynamics of such objects is similar to the one of inertia-less ellipsoids. This is due to the fact that they follow the fluid streamlines and fibers are always in a stretched state, except when interacting with rare but strong turbulent vortices \cite{allende2018stretching}. Yet, by resorting to such reduced descriptions, details about the exact curvature of an elongated object are definitely lost.
 \end{enumerate}

Coming up with a reduced formulation for the orientation and deformation of a flexible fiber requires to clarify first the resolved information that is required. For instance, when one is interested in accurately reproducing the possible entanglement of fibers deposited on the wall, resorting to reduced descriptions based on ellipsoids or dumbbells is irrelevant since they only provide information on an end-to-end length and not on the curvature. In such cases, models should rely at least on formulations based on three or more degrees of freedom (i.e., trumbbells or any N-bbells with $N\ge3$). 
 
 Things become even more complex when dealing with quasi-2D surfaces like leaves. In that case, few studies have explored their deformation due to the complexity of the fluid-structure interactions involved and the fact that each result can hardly be generalized to other geometries. For example, a recent study on 2D sheets of papers has shown that bending such surfaces can induce stronger stiffness than unbent sheets of paper \cite{pini2016two}. In turn, material stiffness can affect particle resuspension. For instance, similarly to pole vault, a semi-flexible fiber can have only one of its ends in contact with the surface and be bent due to local shear. As the fiber resuspend, the energy that has built-up through elastic deformation can be converted into a wall-normal velocity, thereby leading to its motion away from the surface. In addition, tractable models require developing new reduced descriptions for the dynamics and deformation of such complex 2D surfaces, if relevant.

 \textbf{Summary:} The key challenges for deformable 3D particles is to come up with general-purpose models that can be applied regardless of the particle shape and mechanical properties (flexibility, extensibility). This poses difficulties both in terms of experimental characterizations and numerical descriptions of the deformation of such particles across their whole surface (which can be a simple 2D line as in fibers or more complex 3D surfaces like leaves). Coming up with reduced formulations for the orientation and deformation of such complex 3D objects is thus paramount but requires first to clarify the resolved information that is needed.
 
\end{itemize}
  
 \subsubsection{Resuspension from complex surfaces}
 \label{sec:next_model:complete:surf} 
 
Current resuspension models already account for surface roughness. Nevertheless, several issues remain to be addressed and are described below.

\textbf{Experimental needs:} Current measurement techniques do not allow for simultaneous measurement of both adhesion forces and particle resuspension. This is due to the procedures currently used to measure the adhesion force: all existing methodologies rely on detecting the moment at which the contact between the particle and the surface is ruptured. In practice, various methods can be used to measure adhesion forces, which are summarized below and in Fig.~\ref{fig:sketch_adhesion_exp}:
  
\begin{figure}[ht]
 \centering
 \includegraphics[width=0.8\textwidth, trim = 0cm 0cm 0cm 1.6cm, clip]{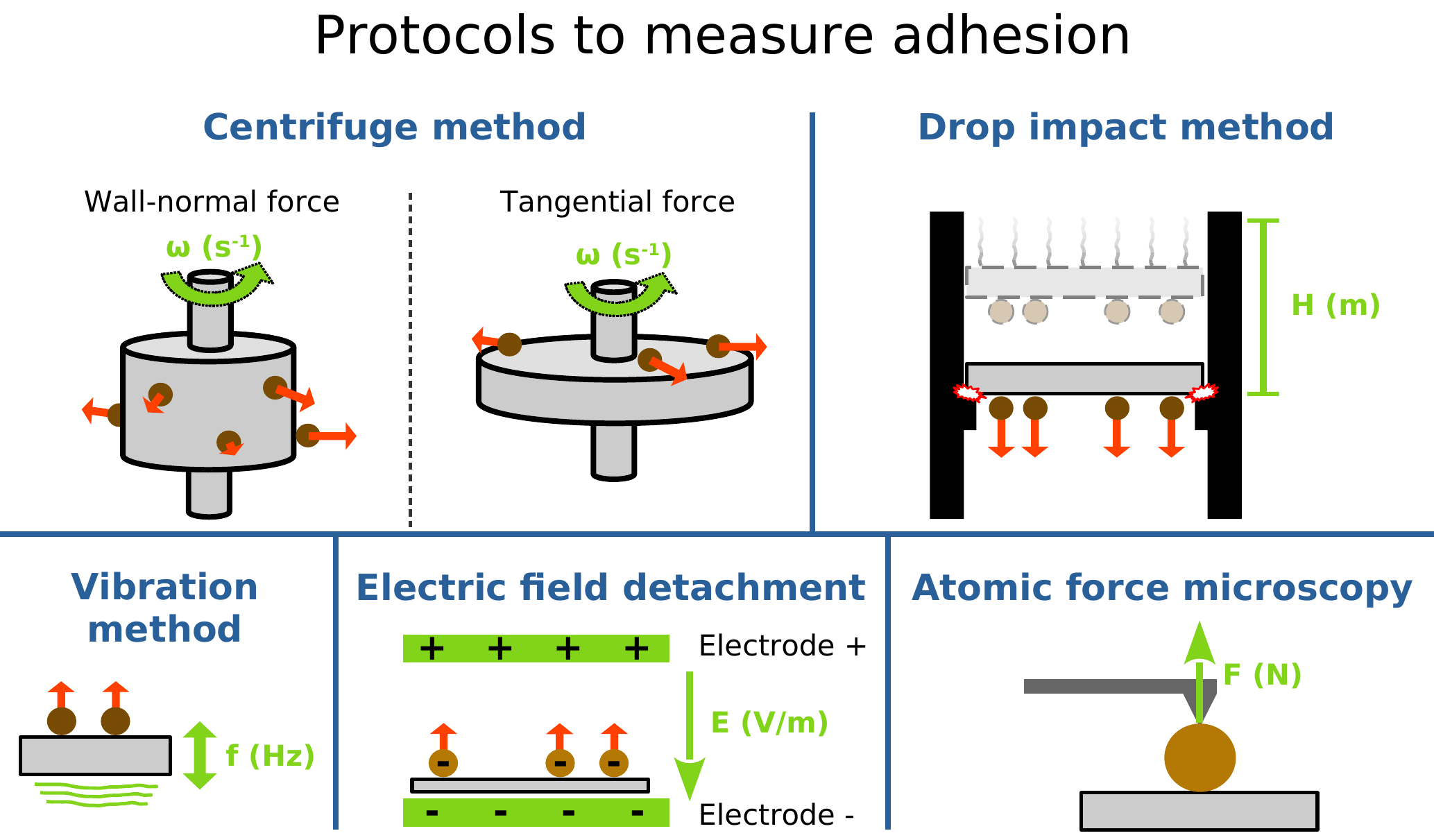}
 \caption{Illustration of the existing techniques to measure adhesion forces. They are based on measuring particle detachment through various methods (such as centrifuge, drop impact, vibration, electric field detachment or AFM).}
 \label{fig:sketch_adhesion_exp}
\end{figure}
 
\begin{itemize}
 \item The \textit{centrifuge method} consists in placing a surface with particles deposited on it in a centrifuge and measuring the velocity at which the particles are detached. Provided that the properties of each particle are known (size, density), one can reconstruct the detachment force using information on the velocity at which detachment occurs \cite{lam1992influence}. This can be used to measure either the wall-normal or tangential forces required to set particles in motion \cite{moeller2017adhesion}. In addition, recent advancements have allowed to track the motion of particles as they move close to the surface (using continuous image acquisition \cite{knoll2017integration}).
   
 One of the advantages of this technique is that it can easily accommodate particles immersed in liquids. Hence, this method is popular in the pharmaceutical and food industry for evaluating the adhesive properties of polydisperse powders. Yet, its drawback is that it requires information on particle properties (mass, volume) and any uncertainty directly impacts the precision of the measurement.
   
 \item The \textit{drop impact test method} consists in attaching particles upside down to a support stub inside a tube. The stub is then dropped from a given height and comes to a sudden stop when it hits a stopper. Particle detaching from the surface fall through a hole in the stopper. The number of particles detached is determined by microscope image analysis of the surface before and after the fall or by weighting the sample. Video analysis of the impact provides impact velocity and contact time, from which adhesion force can be computed. By dropping the surface from different height, the fraction detached and the corresponding adhesion force can be determined. 
   
 This method has been used for instance in the food industry, to determine the adhesiveness of seasoning to tortilla chips \cite{ermis2009direct}. This method has the same disadvantages as the centrifuge one.

 \item The \textit{vibration method} consists in inducing vibrations at a given frequency and intensity until the particles deposited on a surface are detached. Particle detachment events are continuously monitored by image analysis and correlated with acceleration and surface displacement, which can be measured via a laser-scanning-vibrometer \cite{hein2002analysis}. 
  
 This method provides information on adhesion force distributions (see for instance \cite{ripperger2005measurement}). Another advantage is that the apparatus can be enclosed in a humidity chamber to control relative humidity and, therefore, study capillary forces \cite{hubbard2012experimental}. However, this method has drawbacks similar to the centrifuge one: the particle mass needs to be known to infer adhesion. In addition, the use of optical acquisition techniques restricts the applicability of this method to particles with a size close to \SI{1}{\mu m} \cite{hein2002analysis, hubbard2012experimental}. Another important issue is that it is restricted to mechanically stable substrates \cite{almeida2021adhesion}. 
  
 \item The \textit{electric field detachment method} characterizes particle adhesion by using an electric field to remove particles from a surface placed between two parallel planar electrodes \cite{takeuchi2006adhesion}. This method requires to charge particles using induction up to the saturation charge \cite{hu2008measurements}. Once DC voltage is applied, particles move from the bottom electrode to the top electrode. Particle detachment is observed by monitoring the current flowing between the electrodes. The adhesion force is computed from the particle charge and the electrostatic field strength. 
  
 This technique allows to obtain the adhesion force distribution by using different applied voltages (as in \cite{takeuchi2006adhesion}). In addition, the adhesion of particles of various sizes can be measured simultaneously. However, its main limitation is that it can only measure adhesion of conducting particles. In addition, the particle diameter is needed to compute the saturation charge. 
  
 \item The \textit{Atomic Force Microscopy} (AFM) is a form of scanning probe microscope that can be operated in various medium (air, gases, vacuum, or liquid) \cite{whitehouse2002surfaces}. It can be used directly to measure surface roughness (with a resolution of a couple of \SI{}{nm} in the transverse direction and \SI{0.1}{nm} in the wall-normal direction). Alternatively, a particle can be glued to the tip of a cantilever and, by bringing the probe particle in contact with a fixed surface and then retracting the probe, one can measure adhesion forces (simply by recording the deflection of the cantilever due to adhesion). To obtain statistical information on adhesion force distributions, the same procedure is repeated at different locations by moving the probe over the surface in a raster pattern. 
  
 The key advantage of AFM is that it allows to measure adhesion forces as small as $10^{-18}$ N \cite{nguyen2010centrifuge}. In addition, the force is determined without any assumption concerning the particle size or shape. However, one of the limitations is that the technique can only determine the adhesion of one particle at a time, hence, it could be time-consuming for polydisperse particle collections or irregular particles. Another limitation is the fact that the direction of contact is predetermined by the way the particle is attached to the cantilever and the particle cannot adjust freely on the surface, like when real powders are dispersed on a surface.

\end{itemize}
  
This overview shows that, in a sense, measuring adhesion seems like measuring detachment events, except that the forces used to measure adhesion may be different from those that induce resuspension in the system of interest. For instance, surface vibration can be used to determine the adhesion between sand grains while, in practical applications, detachment is mostly induced by the wind. In addition, no information is provided on the re-entrainment of the particles once they detach from the surface. As a result, the two phenomena of detachment and resuspension are not measured concurrently. New experimental protocols or techniques might allow for simultaneous measurements in the near future. However, this might prove to be a very intricate tasks especially if plastic deformations occur (since the surface topology will be definitely changed after each contact/detachment series, making subsequent measurements unrelated to the previous one). 

\medskip
\textbf{Modeling challenges:} The absence of simultaneous measures of adhesion forces and resuspension events has profound consequences for the models of small colloidal particles. In fact, these models were validated by comparing only statistical information, like the time-averaged resuspension rate $\tau_r^{\Delta t}$, and not information on individual events, like the instantaneous resuspension rate of a single particle and its velocity. While models with a reduced description have proven useful to predict such averaged quantities (especially the median critical velocity $u_{r=50\%}$), more advanced formulations are needed to better capture rare events like resuspension at very low/high velocities \cite{chkhetiani2012dust} and the influence of seasonal variations \cite{kinase2018seasonal}. This brings out two issues for reduced models:
  
 \begin{itemize}
  \item \textit{Refined models for rough surfaces:} To capture both short- and long-time resuspension rates, one cannot rely anymore on simplified descriptions of surface roughness (or adhesion forces). In fact, predictions of rare events (like resuspension at low or high velocities) can only be obtained provided that the tails in the distribution of adhesion and hydrodynamic forces are properly captured. This has implications in the current models for adhesion forces or surface roughness.
  
  When the adhesion force is computed directly from a reduced description of surface roughness (see Section~\ref{sec:models:turb_rough:rough}), roughness is often described with a statistical approach (like fractals, wavy sinusoids or random hemispherical asperities on a plate). In particular, approximate formulas for the adhesion force often rely on simple average parameters that are extracted from surface topologies (namely, the average roughness, the rms roughness, the peak-to-peak distance, or the highest/lowest peaks displayed in Fig.~\ref{fig:sketch_roughness_Ra}). Yet, these parameters do not contain enough information on surface features, especially on the whole range of sizes and their corresponding density on the surface \cite{henry2018colloidal}. Since large roughness features are expected to induce higher adhesion forces than smaller ones and since they can provide shelter for particles in their wake, such large features can be responsible for the presence of particles that are very hard to resuspend (and only do so at high velocities). Hence, to go beyond current limitations, more systematic measurements of surface roughness are needed to describe the whole distribution of roughness sizes and the corresponding density on the surface.
  
  When the adhesion forces are presumed (as in kinetic PDF approaches), the main difficulty comes from the use of prescribed laws (like Gaussian or log-normal distributions) to fit the measured distribution. In fact, such fit are expected to filter the information on the tails of the distributions, possibly leading to erroneous predictions of rare events. This can be easily circumvented either by using more complex fit (e.g., based on multi-peak distributions, as shown recently in Fig.~\ref{fig:sketch_roughness_Fadh}) or using directly the distribution of forces measured experimentally (provided it is detailed enough to contain relevant statistics on the extreme values of the adhesion forces).
  
 \begin{figure}[ht]
 \centering
  \begin{subfigure}{0.45 \linewidth}
   \includegraphics[width=0.99\textwidth,trim = 0cm 0cm 0cm 0cm,clip]{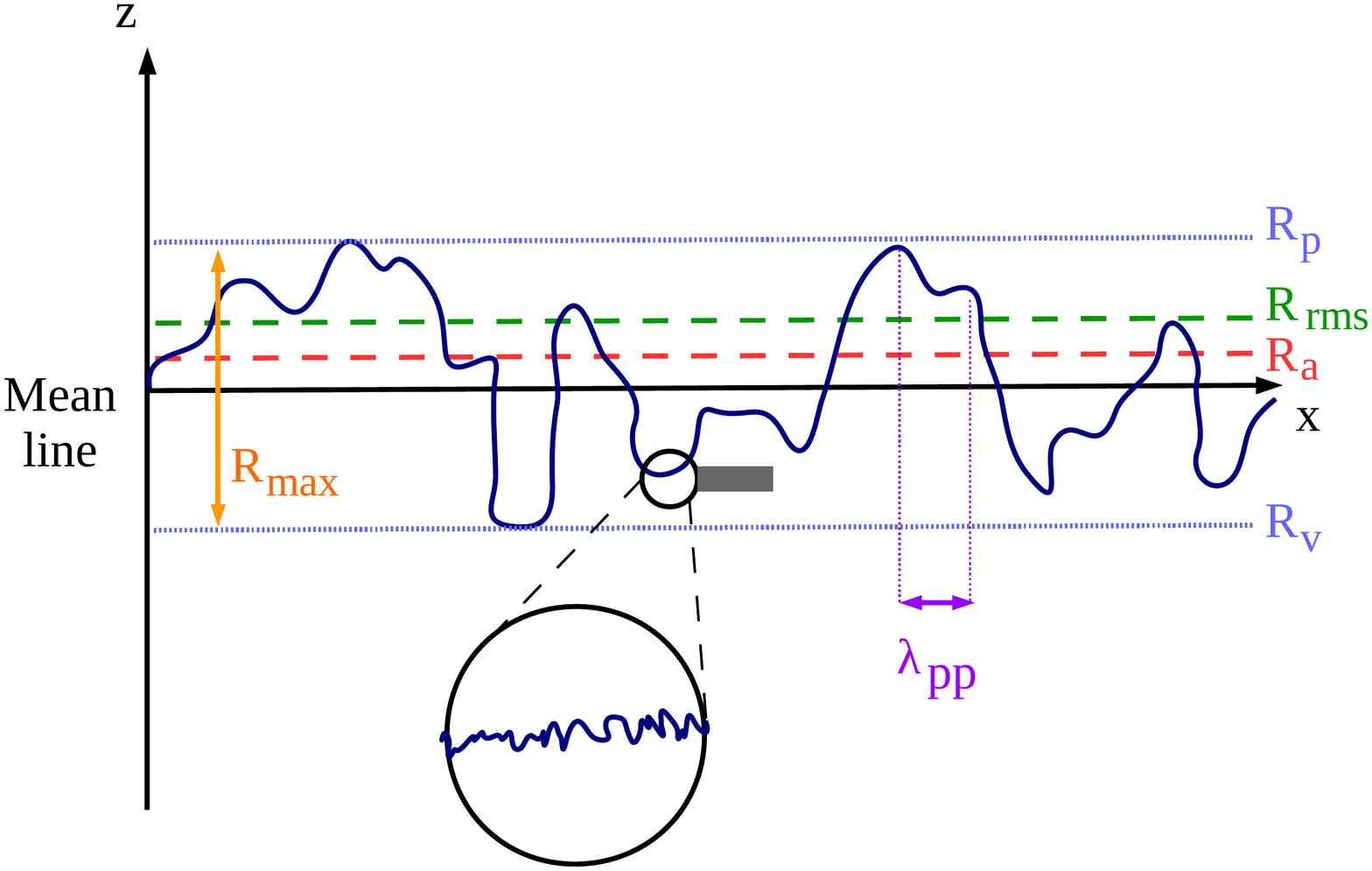}
   \caption{Sketch of a rough surface and the corresponding roughness parameters ($R_{\rm a}$ and $R_{\rm rms}$). Reprinted with permission from \cite{henry2018colloidal}. Copyright 2018, Elsevier.}
   \label{fig:sketch_roughness_Ra}
  \end{subfigure}
  \hspace{10pt}
  \begin{subfigure}{0.45 \linewidth}
   \includegraphics[width=0.85\textwidth,trim = 10.2cm 0cm 0cm 0.0cm,clip]{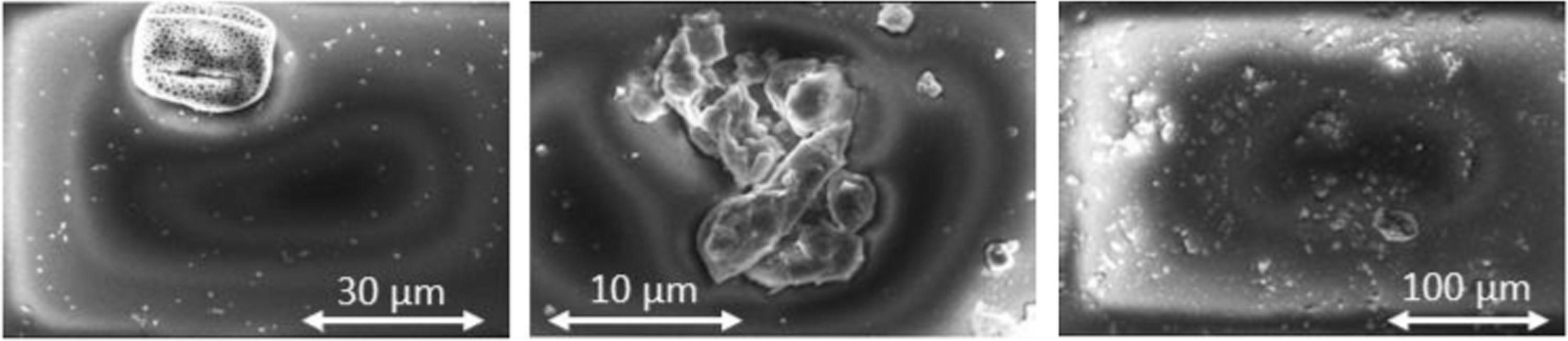}
   \caption{SEM images of a glass substrate contaminated by inorganic debris. Reprinted with permission from \cite{rush2018glass}. Copyright 2018, Elsevier.}
   \label{fig:fig_rush_2018_contaminated}
  \end{subfigure}
  \caption{Illustrations of the complexity in describing rough surfaces: (a) average roughness parameters only provide limited information on roughness; (b) substrates can display very different roughness characteristics due to local inhomogeneities (e.g., when contaminated particles are present on the substrate).}
  \label{fig:illustr_rough}
 \end{figure}
  
 \item \textit{Models for inhomogeneous rough surfaces:} An additional difficulty to model real surfaces is that they can become contaminated by other objects that can bound to the surface. Such a case is displayed in Fig.~\ref{fig:fig_rush_2018_contaminated}, which shows inorganic debris contaminating a glass substrate left outdoors for a certain duration \cite{rush2018glass}. Similar contamination can occur when studying the resuspension of biological particles from surfaces in outdoor environments \cite{qian2014walking}. The consequence of contamination is that surface roughness and adhesion forces can significantly change when a particle is in contact with regions of the surface that are not contaminated compared to regions that are contaminated. As a result, the adhesion force can display very complex multimodal distributions \cite{rush2018glass}. In addition, if contaminants are present on either the particle or the surface, more measurements are needed to characterize the spectrum of particle/surface interactions \cite{rush2018glass}. This implies more rigorous and time-consuming measurements so that the surface area used to ``scan'' the surface roughness and/or adhesion forces is large enough to be representative. To describe such surfaces, one can consider describing independently each region with uniform roughness properties, leading to the identification of small patches (i.e., isolated regions with irregular roughness features).
      
 \item \textit{Models for other surface heterogeneities:} Drawing on the previous comment, more realistic descriptions of particle resuspension should also include information on any type of surface heterogeneities. In the present paper, we have extensively discussed the role played by surface roughness, which corresponds to heterogeneities in the topology. However, surfaces can also display heterogeneities in their composition (as for composite materials like concrete) as well as in their chemical properties (e.g., charge). This can result in significant local variations in the resuspension of particles attached to such complex materials. To address such issues, the question is whether one is interested in capturing the variations of the resuspension rate on each local region (depending on its properties) or if only macroscopic information has to be resolved (e.g., on the average resuspension over the whole surface).
  
\end{itemize}
 
\textbf{Summary:} The limitations discussed here illustrate that improved understanding of the adhesion and resuspension processes are expected to emerge only from tight coupling between new experiments and fine-scale simulations. This can then lead to more advanced models that are based on reduced descriptions of these phenomena.

 \subsection{Beyond individual resuspension events and collection motion: towards unified descriptions}
  \label{sec:next_model:unify}

Broadly speaking, most of the models presented in Section~\ref{sec:models:approach} fall into two main but separate categories. On the one hand, many models have been designed to capture resuspension from sparse monolayer deposits where particles are hardly affected by the presence of other particles. On the other hand, some models have been formulated to address directly multilayer resuspension, therefore considering that a bed of particles is initially present on the surface and accounting for collective particles effects. Apart from the extensions related to more complex individual particle characteristics discussed in Section~\ref{sec:next_model:complete}, it is important to address what takes place in between these two asymptotic cases and, more precisely, the new phenomena induced when particle-particle interactions interplay with particle-fluid and particle-surface forces. The development of unified approaches (i.e., descriptions that can span all the range of possible applications and types of deposits) hence requires to investigate two important aspects detailed below: the transition between individual/collective events and the role of collective motion on resuspension.
 
 \subsubsection{Bridging the gap between monolayer and multilayer deposits}
  \label{sec:next_model:unify:mechanism}
  
Thanks to the use of dense monolayer deposits (i.e., deposits where the initial surface concentration of particles is high enough so that the average inter-particle distance becomes comparable to their diameter), studies on the influence of particle-particle interactions on resuspension have recently emerged  \cite{matsusaka2015high, rondeau2021evidence, banari2021evidence}. These studies have revealed several phenomena that can help bridging the gap between the descriptions of sparse monolayer and multilayer resuspension:

\begin{itemize}
 \item Collision propagation:
 
 As displayed in Fig.~\ref{fig:fig_banari_2021_coll} in Section~\ref{sec:physics:phenomenology:mechanisms}, when the initial distance between deposited particles is small enough, collision propagation can occur due to the motion induced by a particle colliding with one of its neighbors. As a result, when dealing with dense monolayers, particle collision effects cannot be neglected as usually done in models for sparse monolayer deposits.
 
 The role of particle collision in dense monolayers is two-fold:
 \begin{itemize}
  \item First, the collision frequency is directly driven by the initial particle layout on the surface. Indeed, it is obvious that two particles located next to each other collide if upstream particles are set in motion due to hydrodynamic actions. Hence, the initial presence of clusters of particles on the surface can lead to a large number of particles being set in motion, especially if one of the upstream particles starts moving, which is more probable since upstream particles are more prone to interact with a flow that is unaffected by the presence of other particles. On the contrary, a particle initially deposited further downstream is going to be hit by a moving particle only if it is within its actual path and provided that it is not too far (at which point, the moving particle might be too distant from the surface to interact with other deposited particles, see Fig.~\ref{fig:sketch_coll_prop}). 
  \begin{figure}[ht]
   \centering
   \includegraphics[width=0.8\textwidth, trim = 0cm 0cm 0cm 0cm, clip]{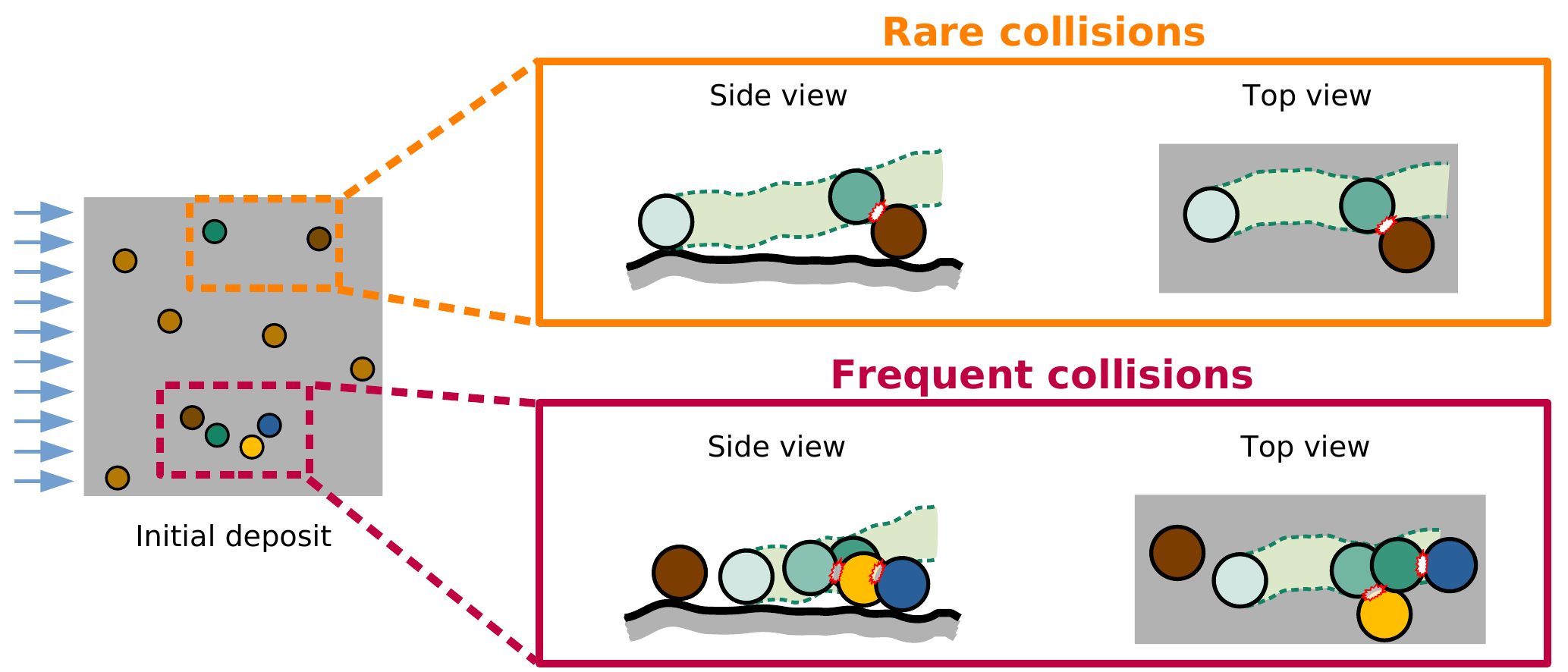}
   \caption{Sketch showing how inter-particle collisions can lead to multiple resuspension events depending on the initial layout of the particles: only few collisions are expected in regions with low surface concentrations while numerous collisions are expected in regions with high surface concentration (e.g., due to clustering).}
   \label{fig:sketch_coll_prop}
  \end{figure}
  
  This means that new experimental studies should provide detailed information on the initial layout of particles on the surface, their gaps and patches, especially to identify the regions where the mean particle distance is lower than the average value derived by assuming a uniform particle distribution over the surface (see also Fig.~\ref{fig:sketch_coll_prop}). For that purpose, two types of measurements are relevant depending on the level of information required: first, the exact location of each particle (using optical techniques and algorithm detecting the presence and position of each particles); second, reduced statistical information, for example by resorting to radial distribution functions (which can prove very helpful to characterize each region where particles are homogeneously distributed). These experimental data can then be used either in deterministic particle-tracking approaches (where each particle is placed according to the exact positions measured) or in PDF approaches (where only statistical information on the initial particle placement is required). Regardless of the level of description chosen, the corresponding resuspension models should be extended to include specific algorithms to detect or to account for inter-particle collision events: collisions can naturally be detected in N-body tracking approaches whereas statistical models have to be devised in PDF formulations. These statistical models can be built upon the notion of a ``residence time'' \cite{banari2021evidence} which corresponds to the amount of time spent by a moving particle in the vicinity of the surface, where it can interact with its neighbors. As displayed in Fig.~\ref{fig:sketch_coll_prop}, this allows to determine the average area covered by a particle as it moves near the surface and, using the statistics on the particle layout, to estimate the collision probability with other particles within this area. 
  
  \item Second, information on the outcome of each collision is needed. In fact, as displayed in Fig.~\ref{fig:sketch_coll_outcome}, four scenarios can take place for the outcome of a binary collision depending on the motion of each particle after the collision: (i) only the impacting particle moves; (ii) only the initially fixed particle moves; (iii) both particles move; (iv) both particles are immobile. Drawing on existing results for the collision of particles embedded in a fluid \cite{elimelech2013particle}, one can expect that the velocity and direction of motion of each particle after the collision depend on a number of physical variables, including the impact angle, the impact velocity, the particle properties (deformation) and the adhesive forces (especially those acting on the fixed particle which prevent motion). 
  
  New experiments should be carried out to investigate the outcome of inter-particle collisions, especially to identify the energy and momentum exchanges during such events. In turn, new models are expected to be developed in the near future to account for these possible outcomes (possibly including theories for plastic deformations which can lead to significant dissipation of energy during such collisions, when relevant).
  \begin{figure}[ht]
   \centering
   \includegraphics[width=0.8\textwidth, trim = 0cm 0cm 0cm 0.8cm, clip]{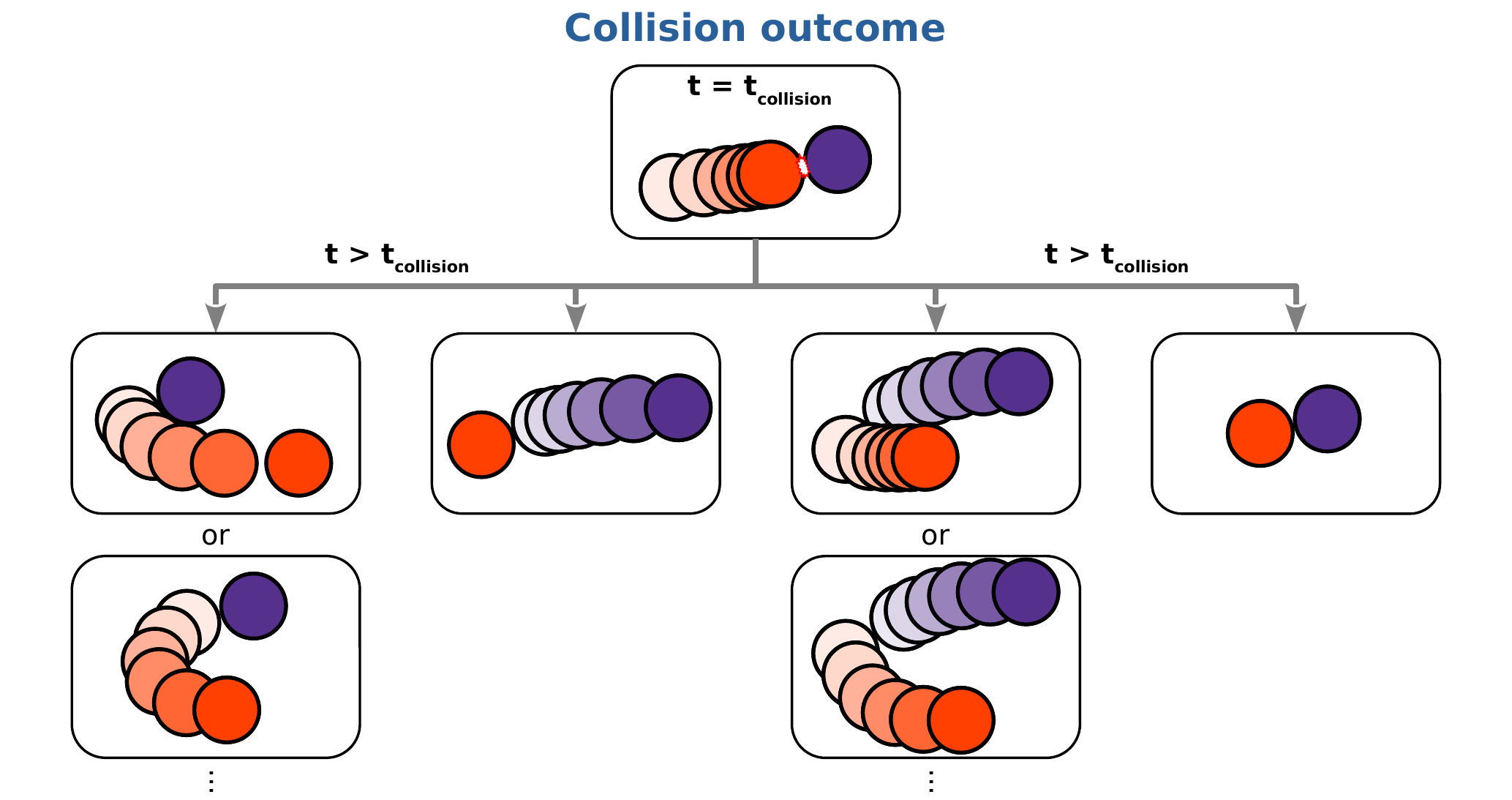}
   \caption{Sketch showing the possible outcomes of each inter-particle collision event: the moving particle can either come to a halt or continue its motion, while the initially fixed particle can remain deposit or start moving. This leads to 4 possible outcomes (with variations in each one due to velocity differences for example).}
   \label{fig:sketch_coll_outcome}
  \end{figure}
  
 \end{itemize}

 In the previous paragraph, we have insisted on the role of collision propagation assuming that only binary collisions take place. However, when the initial surface concentration of particles is high enough, more complex collisions involving three (ternary) or more particles can occur more frequently. Therefore, future studies should investigate whether such complex collisions have an impact on collision propagation and, when relevant, develop accurate models to reproduce their effect. For that purpose, it might prove useful to get inspiration from existing works on avalanche \cite{ge2021mechanical} and more generally granular media \cite{forterre2008flows}. We will come back to the role of such collective motion later in Section~\ref{sec:next_model:unify:collective}.
 
 \item Complex surface motion:
 
 Studies on dense monolayer deposits can also shed light on more complex surface motion. As we have seen, descriptions of sparse monolayer deposits are typically based on three resuspension modes (namely rolling, sliding and lifting). However, as particles collide with each other, some of them can stop moving (especially when the impacted particle is tightly bound to the surface) \cite{charru2004erosion, charru2007motion}. In multilayer resuspension, the cessation of motion has been studied due to its important role in sediment motion in rivers \cite{pahtz2018cessation}. It was shown that, once particles start moving, the fluid velocity can be decreased below the threshold of incipient motion without resulting in an immediate halt of the particle. This is related to the inertia of sediments, which can continue their motion for a certain amount of time even when the flow stops. 
 
 Note that such complex motions with possible arrest are also possible in sparse monolayer resuspension. For example, rolling/sliding colloids can encounter varying adhesion forces as they migrate on the surface. As a result, if they reach a region with high enough adhesion forces/torques which lasts for a sufficient duration, these particles can eventually stop moving as displayed in the right panel of Fig.~\ref{fig:sketch_complex_motion} (see for instance \cite{agudo2014neighbors, duru2015three, kalasin2015engineering}). Unfortunately, there is at the moment little information available on the frequency and importance of such events on resuspension. There is however a difference in the physical interactions involved. In sparse monolayer deposits, complex particle motion along a wall surface is only due to the competition between hydrodynamic and adhesion forces. In denser monolayer deposits, particle-particle collisions start to play a role and this leads to a much more intricate interplay between hydrodynamic, adhesion, and collision effects.
 
 As a result, more systematic characterization of particle dynamics near the surface is needed, including the distinction between their mode of motion (rolling/sliding) and potential arrest of the particles. This brings out interesting questions about migrating particles that come to a stop: can these particles start moving again later? Or do they remain immobile due to the local strong forces preventing their motion? Such information can be key to model long-term resuspension. In fact, as depicted in Fig.~\ref{fig:sketch_complex_motion}, if particles come to rest and reattach on the surface in regions with very strong adhesion forces (and/or possibly low hydrodynamic forces due to sheltering/shielding effects), these particles are likely to remain on the surface for a long time.

 \begin{figure}[ht]
  \centering
  \includegraphics[width=0.8\textwidth]{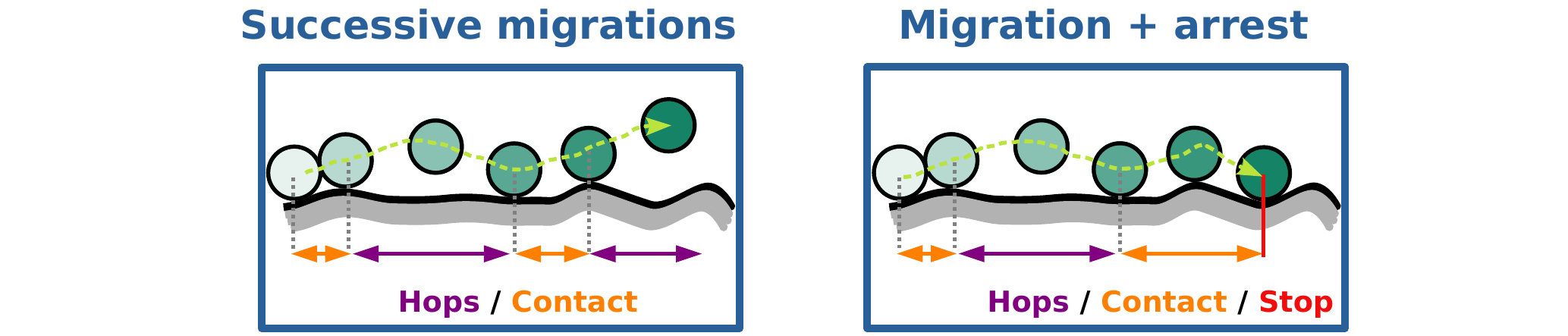}
  \caption{Illustration of the complex motion of particle migrating on a surface.}
  \label{fig:sketch_complex_motion}
 \end{figure}

Hence, a noteworthy consequence of such complex motion is that it can lead to more pronounced extreme events: adhesion measurements might show the rare occurrence of strong adhesive forces but, due to this complex dynamics, many more particles can end up trapped in these regions. This is all the more important when dealing with dense monolayers where, once one particle is trapped in such a region, a cluster can quickly be formed due to particles colliding with this already deposited particle (that can act as an attractor for this region). Consequently, the need for more detailed investigations of adhesion forces on rough surfaces (mentioned previously in Section~\ref{sec:next_model:complete:surf}) should be supplemented with careful analysis of particle dynamics. This implies that previous measurements of particle resuspension by placing particles and simply counting the number of particles remaining on consecutive images after being exposed to a given flow are not sufficient anymore.
 
 We believe that precise measurements of such particle dynamics are now within our grasp thanks to recent advances in terms of experimental techniques (especially to measure rolling motion \cite{agudo2017detection}, as displayed in Fig.~\ref{fig:fig_agudo_2017_roll}). In fact, some insights have already been provided regarding the rotational/translational velocity of migrating particles as well as the time/distance traveled between their incipient motion and their detachment. For instance, micrometer-size particles were shown to be able to travel a distance roughly equal to hundred times their diameter before detaching from engineered surfaces \cite{kalasin2015engineering}. Another noteworthy issue is that the distinction between detached particles and particles moving on a surface raises new questions regarding the accuracy of previous measurements of particle resuspension. Actually, previous experiments often relied on counting the number of particles remaining on a certain region of interest. Yet, there is no distinction made between detached particles and particles rolling/sliding out of the observation area. This would be especially true if the mean distance traveled by moving particles before actual detachment is large. 
  
 \begin{figure}[ht]
  \centering
  \includegraphics[width = 0.45\textwidth, trim=0cm 0.5cm 0cm 0.3cm, clip]{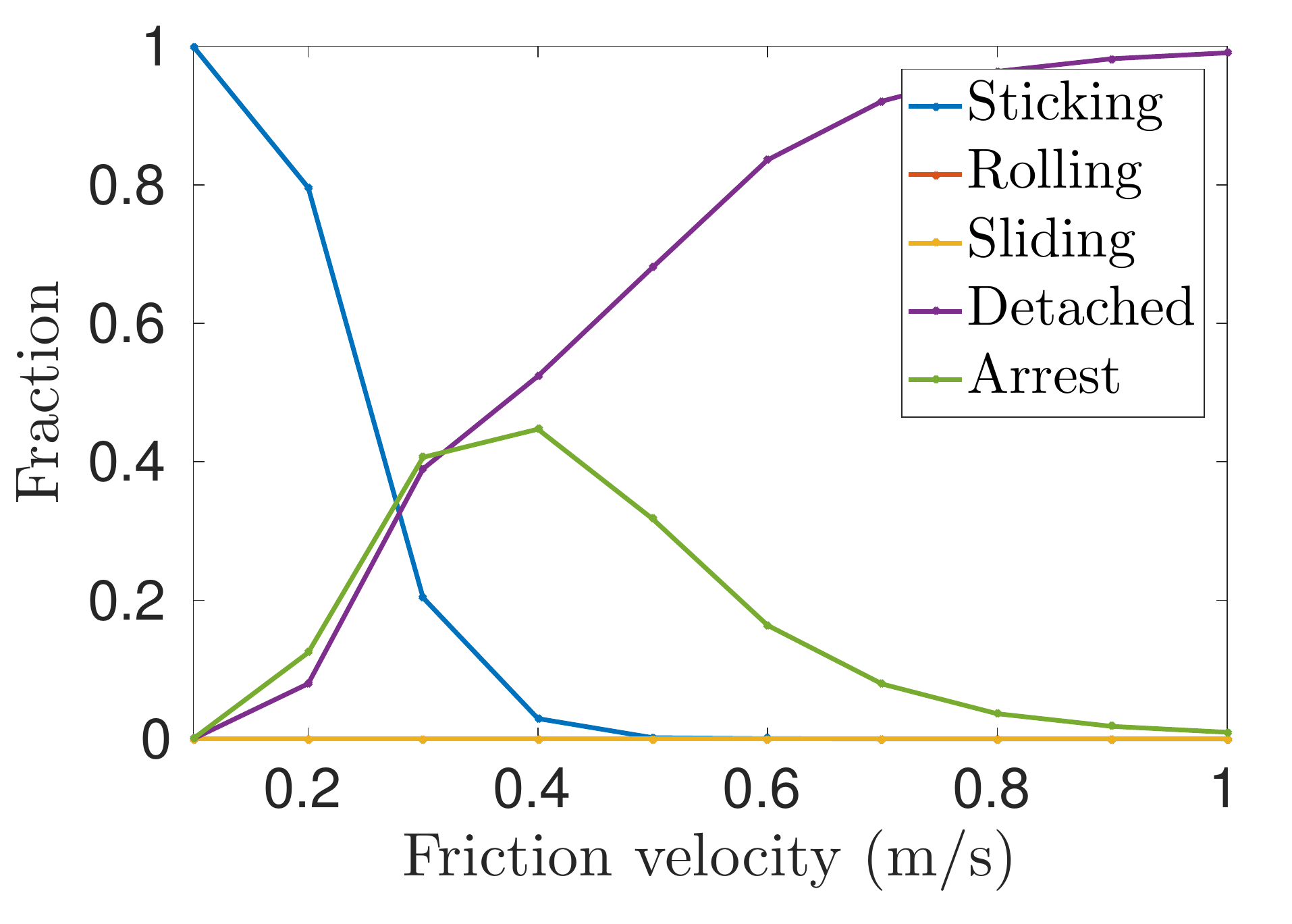}
  \caption{Simulations showing the relative importance of the migration (i.e., rolling, sliding), detachment and arrest as a function of the friction velocity. Simulations have been carried out using a recent dynamic PDF model \cite{henry2014stochastic} and considering the case of \SI{40}{\mu m} particles on a rough substrate (\SI{25}{nm} asperities covering \SI{2}{\%} of the surface) exposed to an airflow.}
  \label{fig:plot_arrest_relative_role}
 \end{figure}
 
As far as modeling approaches are concerned, such complex motion would be naturally accounted for by particle tracking approaches. To illustrate this aspect, a numerical simulation has been performed using the recent dynamic PDF approach for \SI{40}{\mu m} spherical particles \cite{henry2014stochastic}. The model has been extended to account for sliding and lifting motion (arrest is captured when the adhesion forces/torques exceed the hydrodynamic forces). As seen in Fig.~\ref{fig:plot_arrest_relative_role}, the relative importance of these mechanisms varies with the fluid velocity. Migration arrest happens to reach its highest value at intermediate velocities. This can be interpreted as follows: the number of particles reaching halting regions initially increases with the fluid velocity since particles are able to travel longer distance (thereby increasing their probability to reach such regions); yet, as the fluid velocity goes beyond a certain threshold, the adhesion forces in these regions becomes too small to prevent motion, leading to less and less particles coming to a stop. More studies of the relative importance between these mechanisms would help to understand the role of each particle/fluid/surface/deposit parameters (especially with recent measurements showing the importance of rolling even for large millimeter-size grains exposed to low Reynolds flows \cite{kudrolli2016critical, seil2018onset})

\end{itemize}

  \subsubsection{The role of collective motion}
  \label{sec:next_model:unify:collective}

As soon as a significant number of particles are deposited on a surface, they can form isolated clusters or even complex multilayer structures. Despite considerable progress in recent years (see for instance \cite{pahtz2020physics}), there are still several open issues related to particle collective motions:
\begin{itemize}
 \item Collective motion of migrating and detached particles:
 
 In the case of multilayer deposits, particles located on top of the deposit are the ones exposed to the flow (and to the near-wall turbulent structures \cite{bristow2020secondary}). Thus, these particles are more susceptible to start moving, especially if they are protruding from the average deposit (leading to higher hydrodynamic forces and more chances to interact with near-wall structures) \cite{celik2014instantaneous, li2019fully}. Once a particle starts moving within a complex deposit, it can collide with its neighbors, thereby inducing a collective motion of particles within a multilayer deposit. As displayed in Fig.~\ref{fig:fig_jain_2021_collective}, this collective motion is related to the size and shape of particles \cite{jain2021impact}: in the case of Zingg ellipsoids, their collective motion results in the propagation of some sorts of sediment waves (or dunes) that slowly migrate downstream of the flow. Such collective motions have profound consequences on resuspension. First, the inter-particle collisions can lead to much higher and sustained resuspension rates (as new particles continuously start moving as they are impacted by other moving particles). Second, during saltation or reptation, particles detach from the surface and move in the near-wall region. Depending on how far away from the surface they travel during such hopping motion, their velocity upon impacting the deposit changes. In fact, saltating particles travel much further away and tend to impact the deposit with a higher velocity, thereby leading to a large number of particles being set in motion due to splashing effects. Meanwhile, reptating or creeping particles remain (very) close to the surface. As a result, when the amount of particles moving near the surface is large (as in Fig.~\ref{fig:fig_jain_2021_collective}), particles constantly collide with each other. Such collective motions are key in the distinction between fluvial and aeolian transport \cite{pahtz2020unification, pahtz2021unified}. 
 
 \begin{figure}[ht]
  \centering
  \includegraphics[width = 0.8\textwidth]{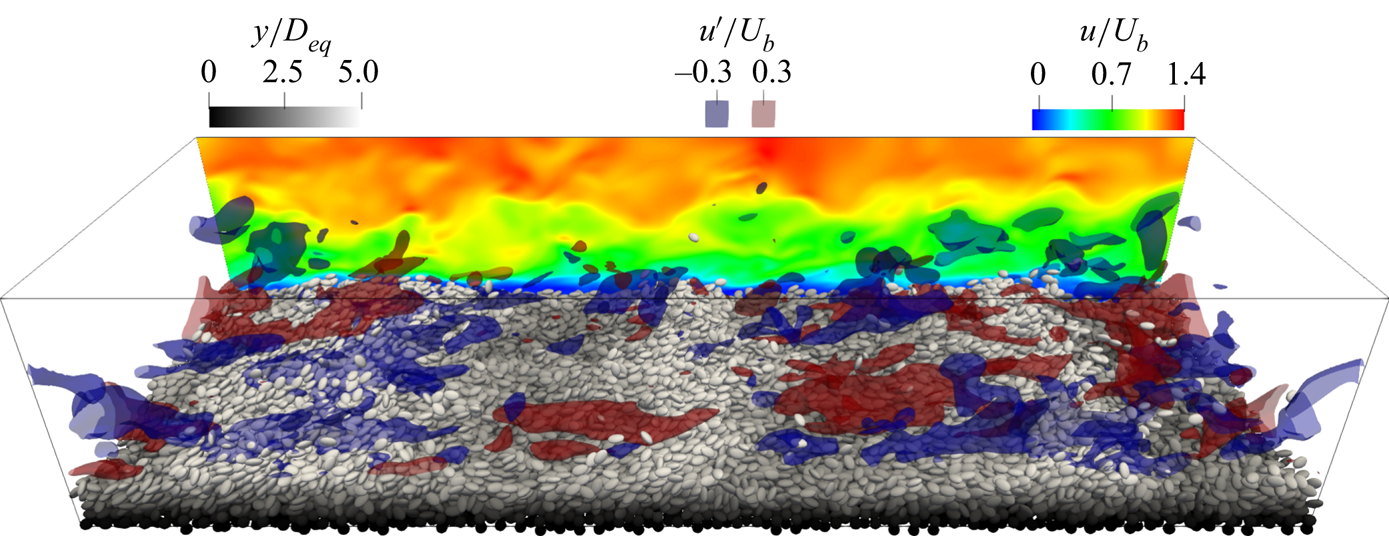}
  \caption{Instantaneous snapshot of a simulation with a complex multilayer deposit composed of Zingg ellipsoids. Ellipsoids are colored according to their wall-normal position. A contour plot of the streamwise velocity is shown on the back side of the domain, while 3D isosurfaces of the instantaneous streamwise velocity are plotted in red and blue. This shows the collective motion of ellipsoids, which form dunes that are slowly moving in the streamwise direction. Reprinted with permission from \cite{jain2021impact}. Copyright 2021, Cambridge University Press.}
  \label{fig:fig_jain_2021_collective}
 \end{figure}
 
 Models that accurately reproduce these collective effects are being developed and represent valuable steps towards unified models. This is where building new bridges between communities dealing with granular flows, which are specialized in dealing with such dense flows of particles \cite{forterre2008flows}, turns out to be quite constructive.

 \item Cluster resuspension
 
 Another challenge of multilayer resuspension is related to the fact that whole clusters can be resuspended during single resuspension events instead of only individual particles. In fact, when the deposits formed are not too compact, clusters can be identified as particles tightly bonded to each other (thereby forming a more compact deposit than the surrounding deposit, as displayed in Fig.~\ref{fig:sketch_multilayer}). Due to the high cohesion of such clusters, it becomes more probable that these clusters resuspend as whole blocs. This brings out several issues:
 
 \begin{itemize}
  \item Morphology of deposits:
  
  Cluster resuspension is directly related to local fluctuations in the deposit structure. In turn, the deposit morphology is a direct consequence of how particles accumulate on a surface. For instance, compact deposits are typically formed when dealing with large/heavy particles (such as sediments or gravels). This is due to the strong gravity forces, which favor the motion of particles until reaching mechanically stable positions (i.e., with isostatic constraints, such as on flat areas or on regions with three contacts). In contrast, when dealing with colloidal particles, cohesion forces dominate over gravity forces and the resulting morphology depends on the deposition scenario \cite{henry2012towards}. In fact, the two extreme cases are either loose structures (which occur when colloids attach immediately at the point of first contact) or very compact ones (obtained when colloids move until reaching stable positions). In reality, the deposit structure can take a range of values in-between these two extremes \cite{biegert2017collision}, possibly due to the role of collisions, surface roughness or non-sphericity of particles (which induces intermediate stable positions with isostatic constraints). Properly capturing the deposit structure and cohesion is thus all the more important \cite{comola2019cohesion, agudo2014neighbors}. In fact, particles/clusters trapped within the deposit bed can be much harder to remove than particles/clusters protruding from the bed, and this has led to the introduction of a `burial depth' in some resuspension models (see for instance \cite{martino2009onset, yager2018resistance}).
  
  Current measurements do not provide sufficient information to fully characterize deposit structures, due to the opacity of the deposits which prevents detailed analysis of their internal structure. Nevertheless, the authors believe that recent advances in experimental techniques now open the way for detailed analysis of deposit structures. In fact, X-ray computed tomography techniques allow to see right through opaque particles \cite{stannarius2019high, hodge2020xray}. Recent studies have started to use this technique to monitor the motion of a 3D pack composed of 1055 particles and analyze how the critical shear stress for incipient motion changes according to various parameters (like grain protrusion). These techniques could also provide information on restructuring events during time.
    
 \begin{figure}[ht]
  \centering
  \includegraphics[scale=0.2]{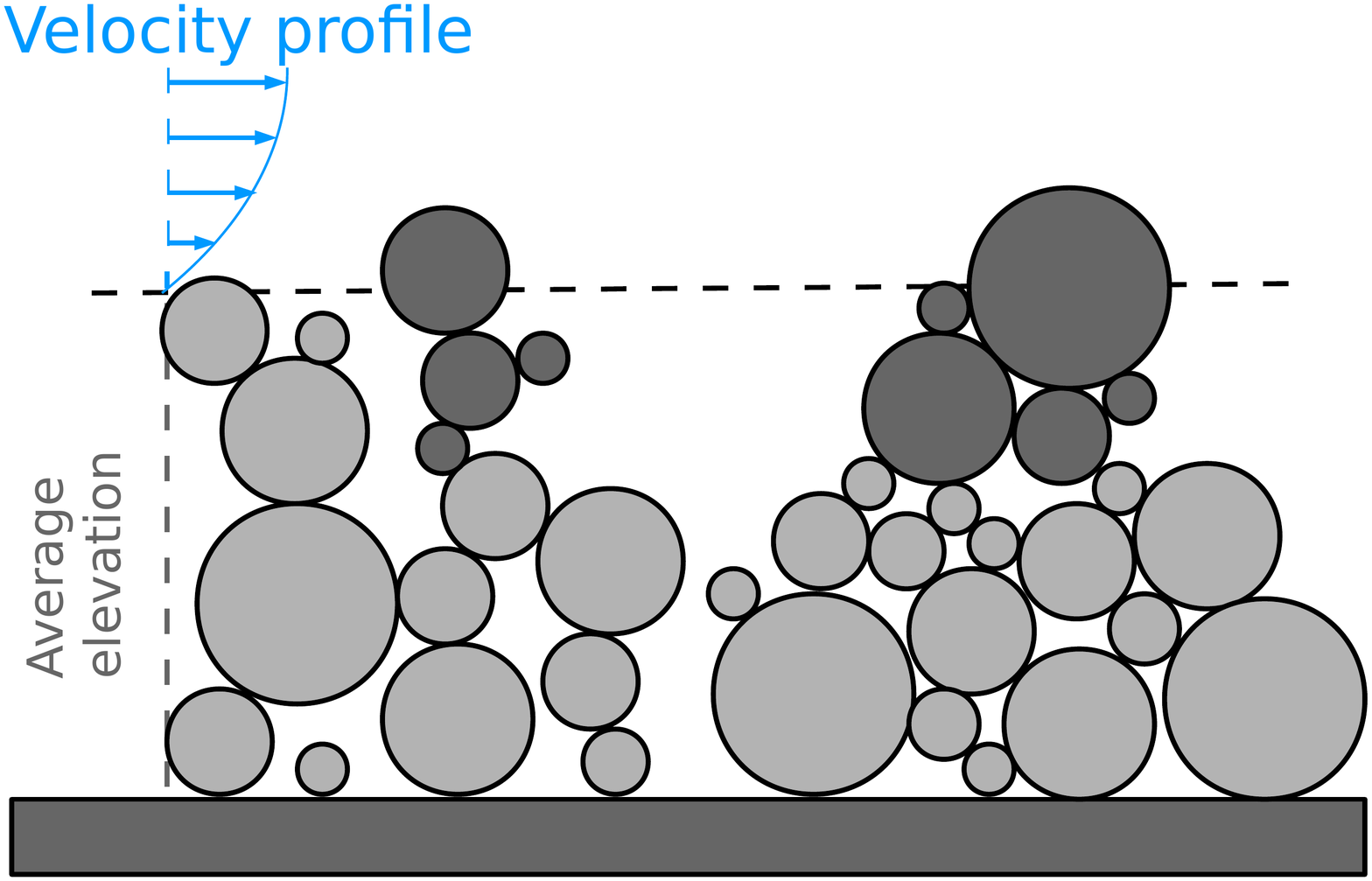}
  \caption{Sketch of a multilayer deposit exposed to a fluid flow: darker particles constitute clusters of particles that can be resuspended altogether.}
  \label{fig:sketch_multilayer}
 \end{figure}
  
  \item Consolidation/ageing effects
  
  Cohesion forces within a cluster/deposit are not always constant in time. In the case of solid particles, aging effects (such as sintering) can lead to cohesion forces increasing with time \cite{abd2007influence}. Sintering occurs at high temperature and leads to the creation of stronger bonds between contacting particles due to the precipitation occurring preferentially near the contact area between surfaces. Such modifications of particle bond energies with time are generally referred to as `consolidation processes' and have been shown to be responsible for lower resuspension rates on heated surfaces after long exposure time \cite{abd2007influence}. When they take place, these consolidation effects have thus a profound impact on long-term resuspension. For instance, in the case of alpine snow, higher temperature leads to snow melting and the resulting higher water content decreases the cohesion forces in snow: this process is responsible for wet loose-snow avalanches \cite{mcclung2006avalanche}.
  
 \end{itemize}

 Reproducing accurately the resuspension of such complex deposits requires to properly take into account the morphology of the deposit formed. This can be obtained naturally with individual particle tracking approaches (provided that the adhesion/cohesion forces are properly estimated). Alternatively, statistical models can be developed by identifying clusters of particles within complex deposits (as in \cite{friess2002modelling, iimura2009simulation}) and possibly by treating the motion of parcels. Parcels correspond to numerical particles which represent a large number of real particles. It allows to reduce the computational costs associated with individual particle tracking approaches, making such approaches tractable in realistic situations. However, in turn, such simplifications might require additional modeling. In fact, even if there exists a single cluster within a region, this does not mean that individual particles within this cluster cannot interact with each other. Hence, if such interactions have to be accounted for, new models have to be introduced to reproduce these effects (like filtering in large-eddy simulations of turbulent flows). 
 
 \item Mixed deposit (or binary mixtures):
 
  In real situations, deposits are often composed of particles with a range of sizes, shapes and possibly even composition. For instance, as displayed in Fig.~\ref{fig:fig_lamb_2016_gravel_sand}, beach sediments are usually formed by grains with a range of sizes (typically within a few millimeters). In river transport, the range of sizes can be even more pronounced, with the presence of both millimeter-size sand grains and centimeter-size gravels. In addition, river beds can be composed of a mixture of sand/gravels as well as plastic debris (as sketched in Fig.~\ref{fig:fig_frei_2019_plastic_bed}).  
 
 \begin{figure}[ht]
  \centering
  \captionsetup[subfigure]{justification=centering}
  \begin{subfigure}{0.45 \linewidth}
   \centering
   \includegraphics[width=1.0\textwidth, trim = 5cm 0cm 25.0cm 19cm, clip]{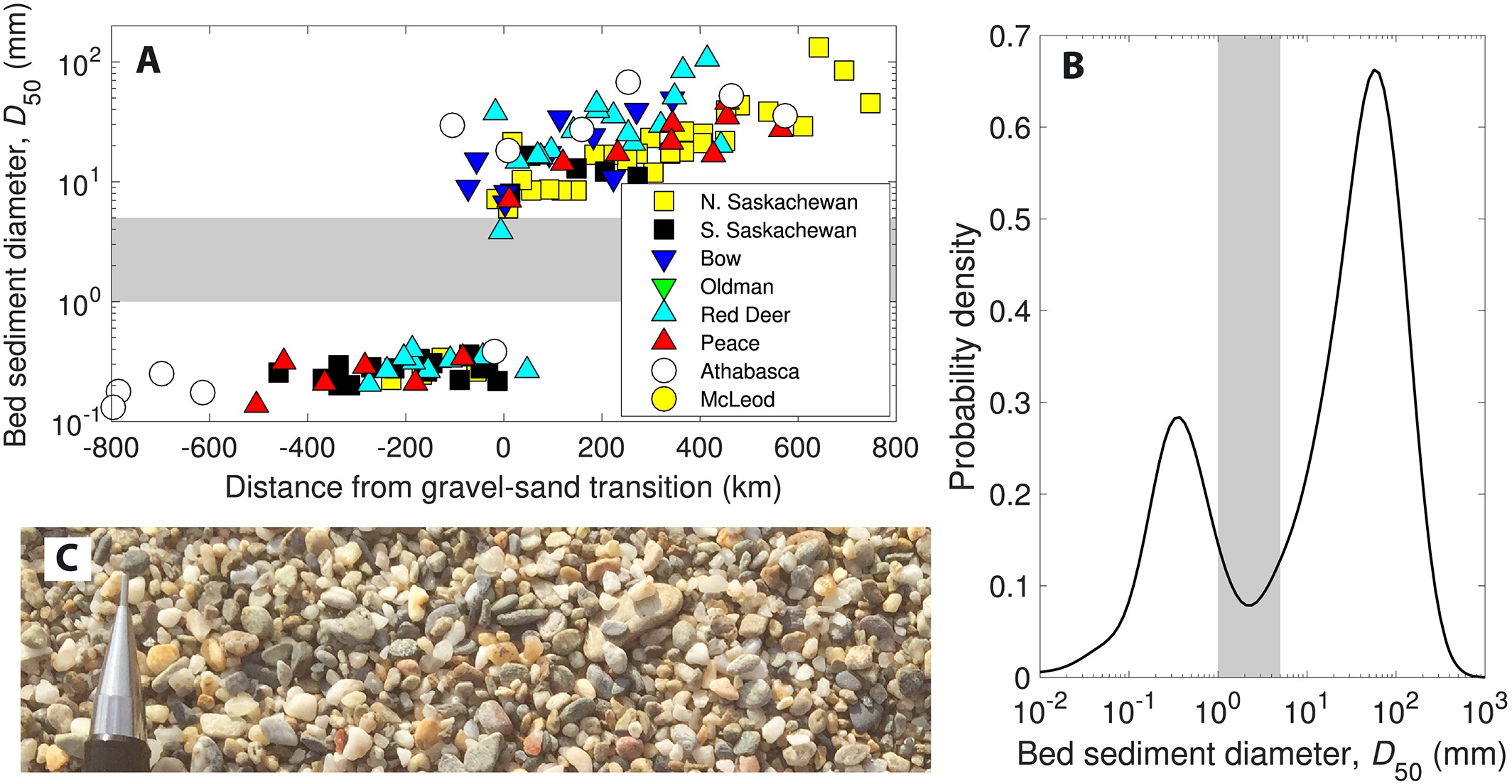}
   \caption{Image of beach sediments from Githio (Greece) showing grains with diameters ranging from 1 to \SI{5}{mm}. Reprinted with permission from \cite{lamb2016grain}. Copyright 2016, John Wiley \& Sons.}
   \label{fig:fig_lamb_2016_gravel_sand}
  \end{subfigure}
  \hspace{5pt} 
  \begin{subfigure}{0.45 \linewidth}
   \centering
   \includegraphics[width=1.0\textwidth, trim = 0cm 0cm 0cm 0cm, clip]{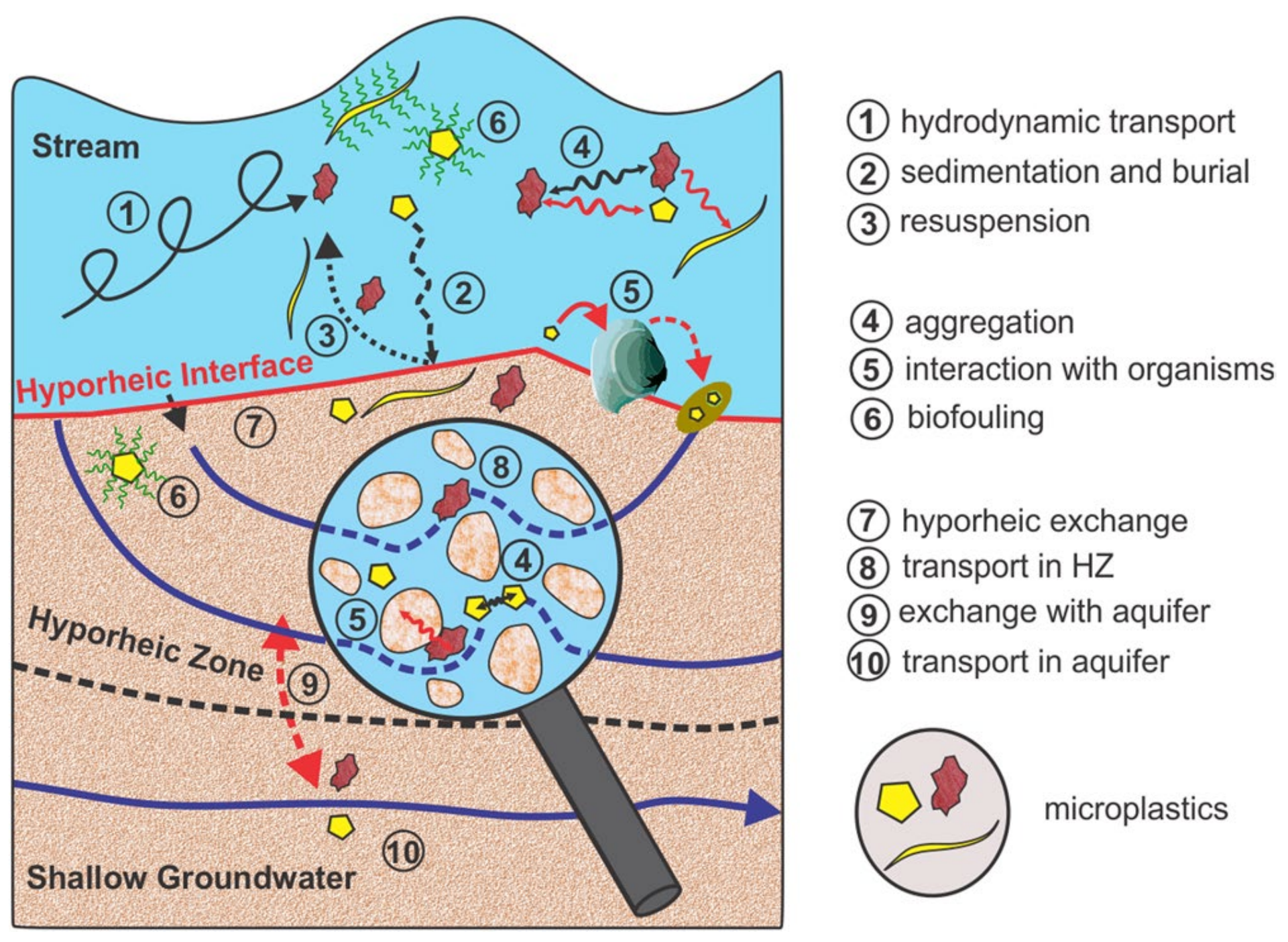}
   \caption{Sketch showing microplastic debris in river sediment beds together with the potential processes controlling its transport. Reprinted with permission from \cite{frei2019occurence}. Copyright 2019, Springer Nature.}
   \label{fig:fig_frei_2019_plastic_bed}
  \end{subfigure}
  \caption{Images of particles within mixed beds, i.e., formed by the accumulation of several types of particles in a complex multilayer system. These particles can vary according to their size, shape, density, composition, etc.}
  \label{fig:illustr_mixed_bed}
 \end{figure}
 
 The presence of particles with different sizes, shapes and composition can severely affect resuspension. In such cases, one of the issues is to properly define what has to be measured. One could define a resuspension rate for each type of particle present within the deposit. This allows to distinguish if some of the particles within the mixture resuspend more easily than others. Yet, due to the collective motion of nearby particles, the resuspension rates might depend strongly on local particle velocities. If such collective phenomena are significant, one has to consider measuring individual resuspension events and correlating these events to the local fluid and particle velocities. As a result, resuspension rates averaged over all particles of the same type might be irrelevant unless conditioned on local measures of collective motion (e.g., mean particle velocity). This opens a whole new range of interesting discussions as to the information that should be contained in the measured quantities. Alternatively, one can also rely on a resuspension rate averaged for the whole mixture. In that case, the drawback is that no information is available on how exactly the different grains behave. Another difficulty is related to the motion of particles once they are detached from the surface, which depends on the particle size since small particles tend to undergo saltation while large particles are more prone to reptation (as in megaripples \cite{tholen2022megaripple}).
 
 In terms of modeling approaches, capturing the resuspension of complex mixtures is relatively straightforward when resorting to particle-tracking approaches (which naturally follow each individual particle with its given size and properties). New models would be required for statistical approaches (such as those relying on the tracking of parcels instead of individual particles). Nevertheless, precise simulations might prove useful to investigate the role of armoring, where some heavy and large particles prevent the resuspension of lighter and smaller particles located beneath them (thereby shielding the particles easier to resuspend). In addition, such complex deposits with multiple types of particles can display very different splashing effects, since each size of particles would have a different probability to be resuspended due to the impaction of a saltating particle \cite{chen2019experimental}.
 
\end{itemize}

 \subsection{The road ahead: a personal viewpoint on future models}
 \label{sec:next_model:opinion}
  
Devising generic models that not only perform as the state-of-the-art ones (outlined in Section~\ref{sec:models:approach}) but also accounting for the new phenomena discussed throughout the present section is likely to be pursued for some time. Following the guidelines set forth in Section~\ref{sec:art_model}, it is essential to keep in mind that such extended models should be realistic while remaining tractable in practical situations. To wrap up this section, we would like to share personal views on which modeling approaches appear as promising candidates for this task.

As described in Section~\ref{sec:next_model:complete}, more realistic models are expected to account for local variations both in particle (e.g., shape, size, density) and surface properties (e.g., contaminated or inhomogeneous surfaces). This is where particle tracking approaches could shine, since they allow to treat in a rather straightforward manner any distribution of particle properties (size, shape, etc.). However, the drawback of these approaches is that they contain a high level of information. Hence, to remain tractable in practical situations, work is still needed to come up with accurate but reduced descriptions of the key phenomena (such as using dumbbells/trumbbells for elongated flexible fibers). In contrast, customized models would require more extensive adaptations to be applicable to complex particles. In particular, empirical models would require to have a set of data for the resuspension of each type of particles. 

In addition, as discussed in Section~\ref{sec:next_model:unify}, unified models can be developed provided that all the mechanisms are properly included (including inter-particle collisions and the collective motion that results from continuous and sustained collisions). In the authors' opinion, such unified models are now within reach, thanks to numerous developments for both monolayer and multilayer resuspension that could be gathered in a generic model spanning the range of deposits encountered in real applications. One issue is, however, to develop models that remain flexible enough so that additional mechanisms/forces can be introduced/changed depending on the application which is aimed for. Such a flexibility is needed to apply the same model to predict resuspension induced by various mechanisms or even a combination of them, as is the case for walking-induced resuspension (which is triggered by the airflow, by vibrations due to the foot tapping on the nearby surface and by electrostatic forces \cite{lai2017experimental}). Here also, we believe that particle-tracking approaches are more adapted to treat such complex issues while remaining flexible enough to easily add/remove/change some of the mechanisms/phenomena/forces at play. In our view, the key challenge is related to the amount of information that should be resolved by such models. In fact, tractable models cannot realistically track the motion of each individual particle while solving exactly the hydrodynamic forces and adhesion forces for each particle (as in \cite{vowinckel2016entrainment}). Consequently, simplifications are in order and careful considerations about the consistency between the various levels of description entering each sub-model are needed. This is where dynamic PDF approaches appear as particularly interesting candidates. 
Indeed, we are of the opinion that dynamic PDF models allow to treat with a high-enough accuracy some of the key phenomena (like transport, polydisperse and/or multimodal particle properties) while still remaining tractable in realistic situations thanks to adequate models that capture the phenomena that are not explicitly solved. 

 \section{Conclusions and perspectives}
  \label{sec:concl}

Particle resuspension is a process in which a set of discrete elements, or particles, adhering to a wall surface are set in motion and eventually carried away from that surface by a fluid flow. As such, a cursory look at this problem might give the impression that the physics is already well understood and that what remains to be done is to work out simple rules for more engineering concerns. This is not so and this review has been written with two main objectives in mind.

The first main objective was to reveal the rich complexity of this phenomena and shed new light on the key physical processes involved. To fulfill that purpose, the approach chosen in this work has consisted in bringing out the key mechanisms through which particles deposited on a surface are resuspended and, then, in analyzing the physical forces involved in terms of fundamental interactions (particle-fluid, particle-particle and particle-surface). Central to this endeavor is the issue of coming up with universal descriptions. This leads to revisiting not just the formulation but the very essence of a model of particle resuspension. What makes up a model and what is the nature of the information to be resolved are questions worth pondering on. Once a consistent modeling framework was set up, it appeared, however, that the present picture needs to be extended. A specific section was therefore dedicated to proposing new areas of developments, in which two directions where significant progress is expected were suggested. The first direction is at the level of each particle and consists in accounting for more complex individual objects in order to go beyond the current assumption of small spherical particles. By complex, it is meant here particles with more complex shapes but also potentially deformable or flexible. Once a mechanical description of these particles is chosen in terms of a selected number of variables attached to each particle and whose evolution in time characterizes the dynamics of that particle, a satisfactory resuspension model should be able to accommodate such particle properties. The second direction of progress is at the level of description of the collective behavior, in particular to reproduce key aspects of multilayer resuspension. These two directions operate at two different levels but it is interesting to note that we need to capture at the same time more complex dynamics of individual particles as well as more complex collective phenomena. In other words, the quest for a truly unified picture continues and it is hoped that the present review is helpful to point out potential new roads. 

This corresponds to the second main objective of this work which is basically a call for new contributions since addressing these challenging tasks would greatly benefit from physicists with a keen eye for both particle dynamics and collective behavior. While trying to encourage colleagues to work at the nanoscale (in fact, decades before nanotechnology actually picked up), Richard P. Feynman once formulated his invitation by ``there is plenty of room at the bottom''. It is believed that the same is true even at more macroscopic levels and that resuspension is an example that ``there is also plenty of room at the top''.

%
%

\section*{Declaration of competing interest}
The authors declare that they have no known competing financial interests or personal relationships that could have appeared to influence the work reported in this review.



\section*{Acknowledgement}
CH acknowledges Gr\'{e}gory L\'{e}crivain (HZDR, Germany) for useful and constructive discussions, especially on collision-induced resuspension from dense monolayer deposits. 

SB. Research presented in this article was supported by the Laboratory Directed Research and Development program of Los Alamos National Laboratory under project number 20210204DR.

\bibliography{biblio}

\end{document}